\documentclass[a4paper, twoside,openright]{report}

\usepackage[a4paper,top=3cm,bottom=3cm,left=3cm,right=3cm]{geometry} 
\usepackage[fontsize=12pt]{scrextend}
\usepackage[english]{babel}
\usepackage[utf8]{inputenc} 
\usepackage[fixlanguage]{babelbib}
\usepackage{babel,csquotes,xpatch} 
\usepackage[T1]{fontenc}
\usepackage{lipsum}
\usepackage{rotating}
\usepackage{fancyhdr}
\usepackage{extramarks}
\usepackage{helvet}
\usepackage{midpage}
\usepackage{textcomp}
\usepackage{gensymb}
\usepackage{float}
\usepackage{caption}
\usepackage{subfig}
\usepackage{multicol}
\DeclareUnicodeCharacter{2212}{-}
\usepackage{amssymb}
\usepackage{amsmath}
\usepackage{amsthm}         
\usepackage{amsfonts}
\usepackage{graphicx}
\usepackage[dvipsnames]{xcolor}  
\usepackage{listings}          
\usepackage{hyperref}     
\usepackage[normalem]{ulem}
\usepackage{ragged2e}
\usepackage{appendix}
\usepackage{stackengine}
\usepackage{arydshln}
\usepackage{breqn}
\usepackage[most]{tcolorbox}
\usepackage{xcolor}
\usepackage{bm}
\makeatletter
\newcommand*{\rom}[1]{\expandafter\@slowromancap\romannumeral #1@}
\makeatother

\usepackage{changepage}
\makeatletter
\newenvironment{chapabstract}{%
    \begin{center}%
        \bfseries Chapter Abstract
    \end{center}}%
   {\par}
\makeatother

\usepackage{xfrac}
\usepackage{tensor}
\usepackage{braket}
\usepackage{revsymb}

\newcommand{\duevirg}[1]{``#1''}

\usepackage{mathrsfs}
\usepackage{comment}
\newenvironment{preface}{\thispagestyle{empty}\topskip0pt\vspace*{\fill}\small}{\vspace*{\fill}}

\definecolor{BluPantone}{RGB}{0, 60, 113}

\hypersetup{
    colorlinks,
    linkcolor=BluPantone,
    citecolor=BluPantone
}

\newtheorem{theorem}{Theorem}[section]

\newenvironment{remark}[1][Remark]{\begin{trivlist}
\item[\hskip \labelsep {\bfseries #1}]}{\end{trivlist}}

\begin{document}

\begin{titlepage}
\begin{figure}[!htb]
    \centering
    \includegraphics[keepaspectratio=true,scale=0.4]{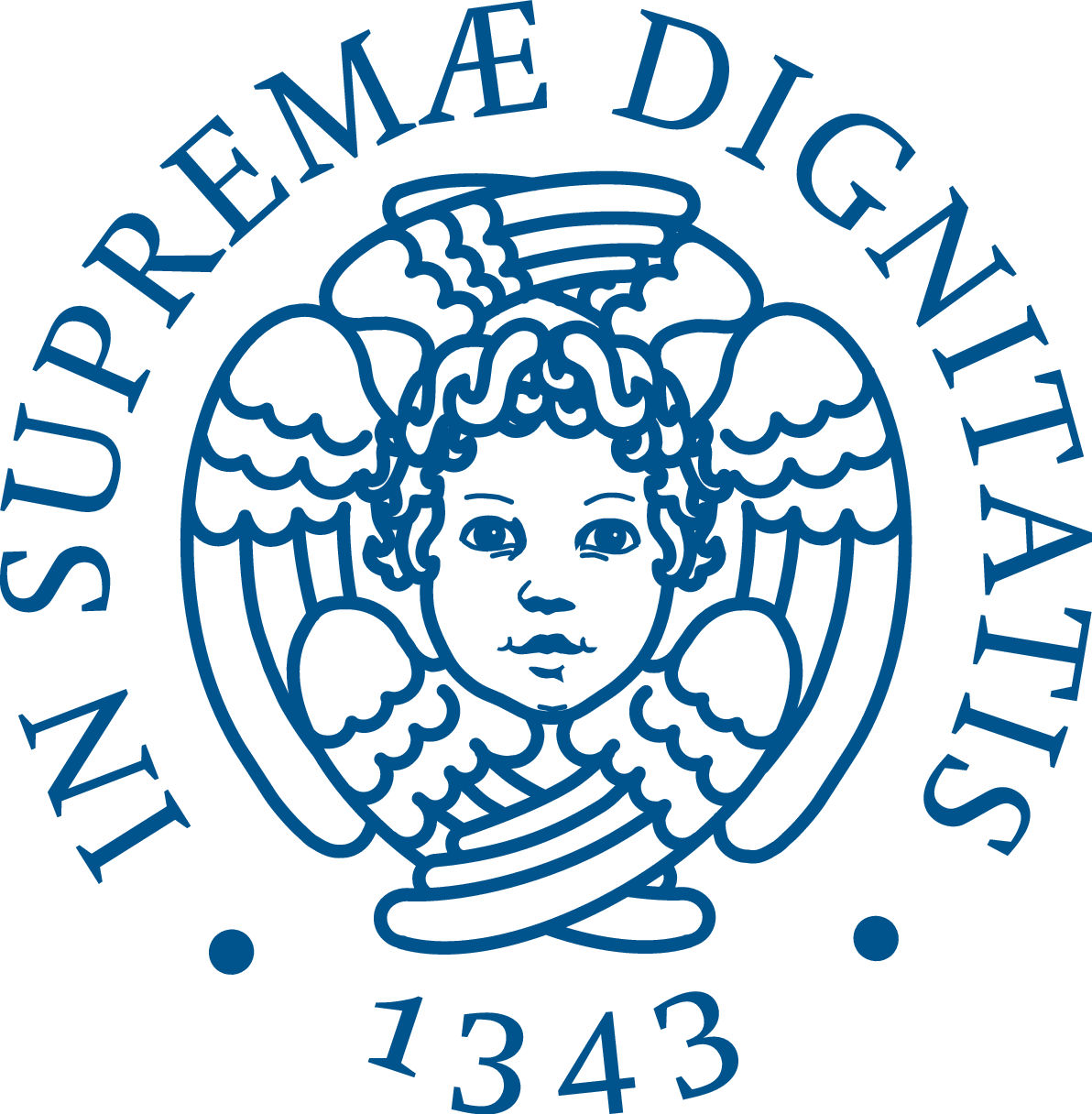}
\end{figure}

\begin{center}
   \huge{U}\LARGE{NIVERSITY OF} \huge{P}\LARGE{ISA}
    \vspace{5mm}
    \\\Large{D}\large{EPARTMENT OF} \Large{P}\large{HYSICS} \Large{E}\large{.} \Large{F}\large{ERMI}
    \vspace{5mm}
    \\ \Large{M}\large{ASTER'S} \Large{T}\large{HESIS}
\end{center}

\vspace{13mm}
\begin{center}
    {\LARGE{\bf Gravitational Waves and Black Hole \\ \vspace{5mm} perturbations in Acoustic Analogues}}
    
    
\end{center}
\vspace{30mm}

\begin{minipage}[t]{0.50\textwidth}
	{ \textbf{\large{Supervisors:}}{\normalsize\vspace{3mm}
	\\ \large{Prof. Massimo Mannarelli}  \normalsize\vspace{0.3mm} \\ \large{Prof. Stefano Liberati}\normalsize\vspace{0.3mm} \\ \large{Prof.ssa Maria Luisa Chiofalo}  \normalsize\vspace{0.3mm}\\ \large{Prof. Dario Grasso} } }
\end{minipage}
\hfill
\begin{minipage}[t]{0.50\textwidth}\raggedleft
	{\textbf{\large{Candidate:}}{\normalsize\vspace{3mm} \\ \large{Chiara Coviello}}}
\end{minipage}

\vspace{30mm}
\hrulefill
\\\centering{\large{ACADEMIC YEAR 2022/2023}}

\end{titlepage}

\pagestyle{plain}
\newpage
~\newpage
\begin{preface}
\begin{flushright}
\textit{
\textbf{\large{A Sara, Claudia e Pasquale}}}
\end{flushright}\end{preface}

\tableofcontents


\chapter*{Introduction}
\markboth{Introduction}{Introduction}
\addcontentsline{toc}{chapter}{Introduction}

Analogue gravity is a research field dedicated to exploring models (typically but not always based on condensed matter physics) that replicate various aspects of General Relativity and Quantum Field Theory in curved spacetime, to obtain lessons of potential relevance on the roads towards a theory of Quantum Gravity. Historically analogies have played a very important role in physics; to quote Johannes Kepler: \duevirg{\textit{And I cherish more than anything else the Analogies, my most trustworthy masters. They know all the secrets of Nature, and they ought to be least neglected in Geometry.}}. In 1981 Bill Unruh established a powerful analogy between hydrodynamic flow with a supersonic region and a black hole, showing how different physics can be described mathematically in the same way and initiating the research field of Analogue Gravity. One possibility to exploit this hydrodynamics/gravity analogy is to create analogue black holes within Bose-Einstein condensates. 
Indeed, it is possible to show that the fluctuations around the macroscopic classical wavefunction, representing the Bose gas in the mean field approximation, can be described as a massless scalar field on an emergent acoustic metric tensor, which is determined by the characteristics of the condensate. These fluctuations, called phonons, are the low-energy collective excitations of the system. Specifically, they propagate as a massless scalar field on an acoustic black hole metric if there is a transonic flow, and concepts like the event horizon become applicable. The analogy is pushed even further: there is an analogous of the Hawking radiation for the condensate. Indeed, at the acoustic horizon, quantum fluctuations result in a thermal radiation of phonons: the acoustic equivalent of Hawking radiation. Remarkably, this emission near the acoustic horizon has been simulated numerically and verified experimentally with atomic Bose-Einstein condensates.\\
This thesis builds upon the above field of research in Analogue Gravity. In particular, the goal is to design a system where an acoustic horizon is excited by a gravitational wave-like perturbation. To achieve this, we have firstly studied the regime in which a Bose-Einstein condensate should be arranged so that the emergent acoustic metric corresponds to a gravitational wave on a flat Minkowski background. This first result of the thesis is achieved by exploiting the invariance under coordinate transformations of General Relativity. Indeed, after deriving the general form of a metric perturbation of the background acoustic metric, we sought an appropriate gauge in which we could express a gravitational wave such that it becomes possible to compare the two metric perturbations: the acoustic and gravitational ones. We have also checked that the condensate quantities found through this comparison satisfy hydrodynamic equations and thus the obtained system is physical. Noticeably, the obtained perturbation of the acoustic metric, that is what we call the analogue gravitational wave, satisfies Einstein's equations in vacuum. Therefore we have derived, for the first time, an expression for a gravitational wave-like perturbation which propagates upon an acoustic background. This is something that has never be done before: none of the previously studied acoustic models were able to reproduce the dynamics of General Relativity, but they represented only static solutions. The second result of the thesis comes straightforwardly from the first one: we have expanded the analogue gravitational wave solution to realize an impinging gravitational wave-like perturbation to an acoustic horizon. We have opted for a cylindrical geometry rather than a simpler two-dimensional plane horizon. This choice facilitates a more general calculation and is experimentally feasible in the laboratory, as we have confirmed the validity of hydrodynamic equations. Certainly, this configuration is unrelated to observed astrophysical black holes. Such a system could be used by further studies to investigate the response of the acoustic horizon to a gravitational wave-like perturbation. \\
Our analogue model indeed paves the way for the theoretical and experimental determination of the reflectivity, the shear viscosity and the entropy of an acoustic horizon determined by an external perturbation akin to a gravitational wave. Hence, it may allow to test the conjectured Kovtun-Son-Starinets lower bound for the shear viscosity coefficient to entropy density ratio.\\
The thesis is structured in two parts. In the first one (Chapters \ref{chap:GRandGW}-\ref{chap:analoguegravity}) we set the stage by discussing the various theoretical frameworks and the corresponding methods necessary for the development of the thesis: General Relativity, gravitational waves, black holes, Quantum Field Theory in curved spacetime and Analogue Gravity. Deriving the main results of the thesis required the development of a comprehensive and rigorous understanding of various concepts and techniques within these frameworks, along with the ability to implement them in practical calculations. The second part (Chapters \ref{chap:simulationGWMinkowski}-\ref{chap:GWinBH}) contains the elaboration and presentation of the original results of the thesis. More in detail, the thesis is organized as follows:
\begin{description}
    \item[Chapter 1:] \textbf{General Relativity and Gravitational Waves.} We offer a concise overview of the theory of General Relativity, including one of its predictions: gravitational waves. Additionally, we explore the energy conditions for the energy-momentum tensor.
    \item[Chapter 2:] \textbf{Black Holes in General Relativity.} We present a brief review of the Schwarzschild black hole in General Relativity and delve into various spacetime horizon definitions. In the process, an overview of the geometry of hypersurfaces, the formalism of geodesic congruences and trapped surfaces is necessary. The chapter concludes with black hole thermodynamics.
    \item [Chapter 3:] \textbf{Quantum Field Theory in Curved Spacetime.} We offer a brief review of Quantum Field Theory in curved spacetime. We derive the Unruh effect and present the Hawking radiation. We also present the spacetime thermodynamics, both in the equilibrium and out of equilibrium, and discuss the shear viscosity over entropy density ratio.
    \item [Chapter 4:] \textbf{Analogue Gravity.} We give a brief review of Analogue Gravity, starting with an historical perspective. We derive the acoustic metric for non relativistic fluids and non relativistic Bose-Einstein condensates. The acoustic Schwarzschild black hole and a vortex geometry fluid flow are also presented. The chapter concludes with an overview of Analogue Gravity experiments.
    \item [Chapter 5:] \textbf{Gravitational Wave acoustic metric.} We present the first result of this thesis: the reproduction of a gravitational wave perturbation on a flat background acoustic metric in a Bose-Einstein condensate, i.e. an analogue gravitational wave. Remarkably this perturbation satisfies Einstein's equations. To obtain this result we have used some notions learned from Chapter 1: the gauge invariance of General Relativity, Einstein's equations and gravitational waves. We have also used the concept of acoustic metric learned in Chapter 4. All the necessary steps to achieve the propagation of the analogue gravitational wave are reported. 
    \item [Chapter 6:]\textbf{Acoustic Black Hole with a Gravitational Wave-like perturbation.} We extend the result of Chapter 5 and present the second original result of this thesis: the perturbation of an acoustic black hole through a gravitational wave-like perturbation. After illustrating the designed system, we study how the acoustic horizon is perturbed and compute its generators. To achieve these results the notions of horizons learned in Chapter 2 are needed, as well as their translation into the Analogue Gravity language, seen in Chapter 4. In particular, we employ the vortex geometry introduced in Chapter 4 that provides a non-trivial acoustic horizon. We show that the geometry of hypersurfaces and the congruence of geodesics notions presented in Chapter 2 are powerful tools for studying the perturbed acoustic horizon. 
    \item[]\textbf{Conclusions and future perspectives.} We summarize the original results of this thesis, outlining the methods employed to achieve them. Additionally, we emphasize the future perspectives that this thesis opens up and their significance.
\end{description}

\part{Theoretical frameworks and methods}

\chapter{General Relativity and Gravitational Waves}
\label{chap:GRandGW}

\begin{chapabstract}
    \begin{adjustwidth}{1cm}{1cm}
        A brief review of the theory of General Relativity is presented, focusing also on one of its consequences: gravitational waves \cite{1973grav.book.Misner}-\cite{maggiore2018gravitational}. 
    \end{adjustwidth}
\end{chapabstract}

\section{Introduction}
For more than two hundred years, the gravitational force between masses had been described by the Newton's law of universal gravitation. It was formulated in Newton's 1687 work \textit{Philosophi\ae Naturalis Principia Mathematica} and states that every mass $m_1$ attracts every other mass $m_2$ through a force $\mathbf{F}$, directed along the line connecting their centres of mass $\mathbf{e}_{(r)}$, given by:
\begin{equation*}
    \label{eq:Newton}
    \mathbf{F}=G\frac{m_1 m_2}{r^2}\mathbf{e}_{(r)}
\end{equation*}
where $r$ is the distance between the centers of mass of the two objects and $G$ is the universal gravitational constant. Equivalently, using the language of the gravitational potential $\phi$:
\begin{equation*}
    \nabla^2 \phi =4\pi G\rho_m,
\end{equation*}
where $\rho_m$ is the mass density. Newton's theory of gravity brilliantly predicted the trajectories of celestial bodies orbiting in space, as well as freely falling objects on Earth. However, within a century of its formulation, flaws have emerged: careful astronomical observation revealed inexplicable discrepancies between theory and observations, like the fact that it gave the wrong prediction for the precession of the perihelion of Mercury's orbit. A new theory of space, time and gravitation was thus required, and it was only in 1915, when Einstein proposed the theory of General Relativity, that observations ---such as the perihelion precession of Mercury's orbit,
the deflection of light by the Sun,
the gravitational redshift of light--- and experiments have been found to be in excellent agreement with theoretical predictions. General Relativity reduces to Newtonian gravity when considering slow-moving particles in weak, unchanging gravitational fields and it has been historically based on two fundamental principles: the equivalence principle and Mach's principle. The former is expressed in the Newtonian theory of gravitation and states that the gravitational force on a body is proportional to its inertial mass. All bodies are influenced by gravity and since motion is independent on the nature of the bodies, the paths of freely falling objects define a preferred set of curves in spacetime: there is always a locally inertial frame that can be chosen at any location in spacetime in a gravitational field to make the gravitational field disappear. This suggests the possibility that the structure of spacetime itself may possess gravitational field properties. The second principle is a much more imprecise idea and it can be formulated as the fact that local physical laws are determined by the large-scale structure of the universe, so that concepts like the local definition of non-accelerating and non-rotating should have no meaning in a universe devoid of matter. Einstein's development of the general theory of relativity was guided by Mach's idea: the local structure of spacetime is influenced by the global distribution of matter. General relativity states that the intrinsic observer-independent properties of spacetime are described by a spacetime metric, which does not need to be globally flat as in special relativity: the deviation of the spacetime metric from flatness is measured by the curvature of spacetime. This curvature explains the physical phenomena typically attributed to a gravitational field ---in other words gravity is the manifestation of the curvature of spacetime--- and is related to the stress-energy-momentum tensor of the matter content of spacetime via Einstein's equations partially in accordance to Mach's ideas. Moving on from the historical perspective, it is more precise to state that the foundations of the theory of General Relativity lie in the equivalence principle along with the principle of general covariance. The latter states that a physical law is true if it is valid in the absence of gravity and if it is generally covariant, i.e. it is preserved in form under an arbitrary coordinate transformation. According to this principle, if a tensor equation is valid in the absence of gravity, it remains valid in the presence of an arbitrary gravitational field. The equations governing the gravitational field, thus Einstein's equations, are tensor equations. Gravitational waves and black holes are two different solutions of the Einstein field equations. Remarkably, they are no longer speculative predictions: the first gravitational wave has been detected in 2015 and at the moment about one-hundred events from the coalescence of compact binaries have been detected by the LIGO-Virgo interferometers \cite{2015GW}. Notably, those gravitational observations and also the electromagnetic observations of the shadow around a compact object (in 2019 the first image of the light ring around a compact object was released following observations by the Event Horizon Telescope \cite{2019bhImage}) are compatible with black holes.

\section{Notations and definitions}
\label{sec:notation-subsection}
In this second section, we summarize notations and definitions used throughout the thesis.
\begin{itemize}

    \item
    Greek indices, such as $\alpha,\beta,\mu,\nu,...$ take the values $0,...,3$; while Latin letters $i,j,...$ denotes spatial indices $1,2,3$. Repeated upper and lower indices are summed over. We indicate the symmetric part of a tensor by using round brackets on indices, while the antisymmetric part is indicated with square brackets.
    
    \item 
    The flat space metric is
    $$
    \eta_{\mu\nu}=(-,+,+,+).
    $$
    
    \item
    We define
    $$
    \begin{aligned}
        & x^{\mu}=\left(x^0,\mathbf{x}\right), \quad x^0=ct,\\
        & \partial_\mu=\frac{\partial}{\partial     x^{\mu}}=\left(\frac{1}{c}\partial_t,\partial_i\right),
    \end{aligned}
    $$
    and 
    $$
    d^4 x=dx^0 d^3 x=c dt d^3 x.
    $$

    \item
    We denote the curved spacetime metric by $g_{\mu\nu}(x)$ and its determinant by $g$. The spacetime interval is
    $$
    ds^2=g_{\mu\nu}dx^{\mu}dx^{\nu}.
    $$
    
    \item
    The covariant derivatives of vectors fields are defined as
    $$
    \begin{aligned}
    & \mathcal{D}_{\nu} A^{\mu}= \partial_{\nu} A^{\mu}+\Gamma^{\mu}{}_{\nu\rho} A^{\rho} \\
    & \mathcal{D}_{\nu} B_{\mu}= \partial_{\nu} B_{\mu}-B_{\rho}\Gamma^{\rho}{}_{\mu\nu},
    \end{aligned}
    $$
    where the Christoffel symbol is the Levi-Civita connection
    $$
    \Gamma^\rho{}_{\mu \nu}=\frac{1}{2} g^{\rho \sigma}\left(\partial_\mu g_{\sigma \nu}+\partial_\nu g_{\sigma \mu}-\partial_\sigma g_{\mu \nu}\right),
    $$
    which gives a covariant derivative that is equal to the partial derivative in the local inertial frame. The covariant divergence is
    $$
    \mathcal{D}_{\mu}A^{\mu}=\partial_{\mu}A^{\mu} +\frac{1}{2} \frac{\partial_{\mu}g}{g}A^{\mu}=\frac{1}{\sqrt{|g|}}\partial_{\mu}\left( \sqrt{|g|} A^{\mu}\right).
    $$
    
    \item
    The parallel transport of the vector $A^{\mu}$ along the curve $x^{\mu}(\lambda): \mathbb{R}\to \mathcal{M}$ (where $\mathcal{M}$ is a manifold and $\lambda$ is the parameter along the curve that expresses each point of the curve itself) is
    $$
    \frac{d x^{\mu}}{d\lambda} \mathcal{D}_{\mu}A^{\nu}=f(\lambda) A^{\nu}
    $$
    with $f$ an arbitrary function of $\lambda$.
    
    \item
    The geodesics equation, thus the free particle's trajectory in a given metric, is a curve $x^\mu(\lambda)$ ($\lambda$ parameterized the curve) whose tangent vector is parallel propagated along itself: 
    $$
    u^{\nu}\mathcal{D}_{\nu} u^{\mu}=f(\lambda) u^{\mu},
    $$
    where $u^{\mu}\equiv d x^{\mu}/d\lambda$. A natural choice of parameterization is the so called affine parameterization: 
    $$
    u^{\nu}\mathcal{D}_{\nu} u^{\mu}=0.
    $$
    It is possible to show that given a curve which satisfies $u^{\nu}\mathcal{D}_{\nu} u^{\mu}=f(\lambda) u^{\mu}$, it is always possible to reparameterize it so that it satisfies $u^{\nu}\mathcal{D}_{\nu} u^{\mu}=0$. Negative, positive and null norms define respectively timelike, spacelike and null geodesics. For a timelike geodesics the affine parameterization corresponds to $\lambda \propto \tau + \text{constant}$, with $\tau$ the proper time on the curve. We can rewrite $u^\nu\mathcal{D}_\nu u^\mu=0$ also as
    $$
    \frac{d u^{\mu}}{ds}+ \Gamma^{\mu}{}_{\nu\rho}u^{\nu}u^{\rho}=0
    $$
    with $s$ the affine parameter, and interpret $-m\Gamma^{\mu}{}_{\nu\rho}u^{\nu}u^{\rho}$ as a gravitational four-force. The action for a particle of mass $m$ moving on a timelike curve $\mathcal{C}$ is 
    $$
    \mathcal{S}=-m c \int_{\mathcal{C}} d\tau,
    $$ 
    where $d\tau=\sqrt{-ds^2}=\sqrt{-u^\mu u^\nu g_{\mu\nu}}d\lambda$ with $\lambda$ an arbitrary parameter on $\mathcal{C}$. The particle worldline will be such that $\delta \mathcal{S}/\delta x(\lambda)=0$: this is a geodesics. 

    \item
    The Riemann tensor is defined as
    $$
    R^\mu_{\phantom{\mu}\nu \rho \sigma}=\partial_\rho \Gamma^\mu{}_{\nu \sigma}-\partial_\sigma \Gamma^\mu{}_{\nu \rho}+\Gamma^\mu{}_{\alpha \rho} \Gamma^\alpha{}_{\nu \sigma}-\Gamma^\mu{}_{\alpha \sigma} \Gamma^\alpha{}_{\nu \rho}.
    $$
    The Ricci tensor is $R_{\mu \nu}=R^\alpha_{\phantom{\alpha}\mu \alpha \nu}$, and the Ricci scalar is $R=g^{\mu \nu} R_{\mu \nu}$. Let $n$ be the spacetime dimension, the Weyl tensor (that is different from zero only for $n>3$) is
    $$
    \begin{gathered}
    C_{\mu \nu \rho \sigma}=R_{\mu \nu \rho \sigma}+\frac{1}{n-2}\left(R_{\mu \sigma} g_{\nu \rho}-R_{\mu \rho} g_{\nu \sigma}+R_{\nu \rho} g_{\mu \sigma}-R_{\nu \sigma} g_{\mu \rho}\right) \\
    +\frac{1}{(n-1)(n-2)} R\left(g_{\mu \rho} g_{\nu \sigma}-g_{\mu \sigma} g_{\nu \rho}\right) .
    \end{gathered}
    $$

    \item 
    Let $\gamma_s(t)$ denote a smooth one-parameter ($s$) family of geodesics, where $t$ is the affine parameter along the geodesics themselves, then the vector field $t^\alpha=(\partial/\partial t)^\alpha$ is tangent to it
    $$
    t^\alpha \mathcal{D}_\alpha t^\beta=0.
    $$
    The vector field $\xi^\alpha=(\partial/\partial s)^\alpha$, that is called deviation vector, represents the displacement to an infinitesimally nearby geodesics. Since $\xi^\alpha$ and $t^\alpha$ are coordinate vector fields, they commute:
    $$
    t^\beta \mathcal{D}_\beta \xi^\alpha=\xi^\beta \mathcal{D}_\beta t^\alpha,
    $$
    and so it is possible to demonstrate that $\xi^\alpha t_\alpha$ is constant along each geodesic. The quantity $v^\alpha=t^\beta \mathcal{D}_\beta \xi^\alpha$ can be interpreted as the relative velocity of an infinitesimally nearby geodesic: it provides, along a geodesic, the rate of change of the displacement to an infinitesimally close one. Likewise, the quantity $a^\alpha=t^\gamma \mathcal{D}_\gamma v^\alpha $ could be interpret as the relative acceleration of an infinitesimally nearby geodesic in the family. It is possible to show that the following equation, which is the geodesic deviation equation, holds:
    $$
    a^\alpha=-R^\alpha_{\phantom{\alpha}\delta\gamma\beta} \xi^\beta t^\gamma t^\delta.
    $$
    Hence, $a^\alpha=0$ for all families of geodesics if and only if the Riemann tensor vanishes, while some geodesics will accelerate toward or away from each other if and only if the Riemann tensor is different from zero. The geodesic deviation equation thus relates the tendency of geodesics to accelerate toward or away from each other to the curvature of the manifold. 

    \item
    Under an infinitesimal diffeomorphism, that is a transformation which is invertible, differentiable and with a differentiable inverse,
    $$
    x^{\prime\mu}(x^\nu)=x^\mu+ \epsilon \zeta^\mu
    $$
    with $\epsilon\ll1$, the metric changes as:
    $$
    g^{\prime}_{\mu\nu}=g_{\mu\nu}-\epsilon(\mathcal{D}_\mu \zeta_\nu+\mathcal{D}_\nu \zeta_\mu).
    $$
    If $g^\prime_{\mu\nu}=g_{\mu\nu}$, so if the Killing equation
    $$
    \mathcal{D}_\mu \zeta_\nu+\mathcal{D}_\nu \zeta_\mu=0
    $$
    is valid, then the diffeomorphism generated by $\zeta^{\mu}$, which is called Killing vector field, is a symmetry of the metric. It is possible to show that if in some coordinates the metric coefficients are independent of one of them, let us say $x^1$, then $\zeta^\mu=\delta^\mu{}_1$ is a Killing vector.
    \item
    The action for a particle of mass $m$ moving on a timelike curve is invariant under a diffeomorphism generated by a Killing vector. Thus $\zeta^\mu$ is associated with a symmetry of the particle action and so with a conserved charge $\mathcal{Q}$. This charge is
    $$
    \mathcal{Q}=\zeta^\mu p_\mu=m\zeta^\mu \frac{dx^\nu}{d\tau}g_{\mu\nu},
    $$
    with $p_\mu$ the particle's four-momentum, and $\tau$ the proper time.
\end{itemize}

\section{Einstein field equations}
The Einstein's equations, that are non linear second-order differential equations, read
\begin{equation}
    \label{eq:einstein1}
    G_{\mu\nu}=\frac{8 \pi G}{c^4} T_{\mu \nu},
\end{equation}
where $G_{\mu\nu}=R_{\mu \nu}-\frac{1}{2} g_{\mu \nu} R$ is the Einstein's tensor and $T_{\mu\nu}$ the energy-momentum tensor. The Einstein's equations can be derived by functional variation of the action $\mathcal{S}=\mathcal{S}_{\text{grav}}+\mathcal{S}_{\text{mat}}$ where $\mathcal{S}_{\text{grav}}$ is the Einstein-Hilbert action (the gravitational action) and $\mathcal{S}_{\text{mat}}$ is the covariant integral of the lagrangian $\mathcal{L}_{\text{mat}}$ (the matter action):
\begin{equation}
    \label{eq:Einstein-Hilbert_action}
    \mathcal{S}_{\text{grav}}= \frac{c^3}{16 \pi G}\int d^4x \sqrt{|g|}R
\end{equation}
\begin{equation}
    \mathcal{S}_{\text{mat}}=\int d^4x \sqrt{|g|}\mathcal{L}_{\text{mat}}.
\end{equation}
The energy-momentum tensor $T^{\mu \nu}$ is defined from the variation of the matter action $\mathcal{S}_{\text{mat}}$ under a change of the metric $g_{\mu \nu} \rightarrow g_{\mu \nu}+\delta g_{\mu \nu}$, according to
\begin{equation}
    \delta \mathcal{S}_{\text{mat}}=\frac{1}{2 c} \int d^4 x \sqrt{|g|} T^{\mu \nu} \delta g_{\mu \nu} .
\end{equation}
For a detailed analysis of the Einstein-Hilbert action's boundary terms see Appendix \ref{app:Gibbons-Hawking-York}. Taking the trace of Equation (\ref{eq:einstein1}) with $g^{\mu\nu}$, we can write Einstein field equations as
\begin{equation}
    \label{eq:einstein2}
    R_{\mu \nu}=\frac{8 \pi G}{c^4} \left( T_{\mu \nu} -\frac{1}{2} g_{\mu \nu} T^{\rho}{}_{\rho}\right).
\end{equation}
It is thus clear that if $T_{\mu \nu}=0$ then $R_{\mu\nu}=0$ and so in vacuum the Ricci tensor vanishes, but in general the curvature is different from zero and it is contained in the Weyl's tensor (only when the dimension $n$ of the spacetime is $n\ge 4$, otherwise when $n\le 3$ in vacuum the spacetime is locally flat). Einstein's equations are ten and with them we need to determine the ten components of the metric, $g_{\mu\nu}$: it seems that these equations determine all the spacetime geometry. However, this is not what happens because the following four equations
\begin{equation}
    G_{\mu 0}=\frac{8 \pi G}{c^4} T_{\mu 0}
\end{equation}
do not contain terms in $\partial_0^2 g_{\mu\nu}$, thus they are only constraints on $g_{\mu\nu}(\underline{x}^0, \mathbf{x})$ and $\partial_0 g_{\mu \nu}(\underline{x}^0, \mathbf{x})$ initial conditions (at the initial time $\underline{x}^0$), not dynamical equations. The six equations
\begin{equation}
    G_{i j}=\frac{8 \pi G}{c^4} T_{i j}
\end{equation}
are instead dynamical. Therefore, specifying $g_{\mu\nu}$ and $\partial_0 g_{\mu \nu}$ at an initial time $\underline{x}^0$ does not give an unique solution: the four functions given by all the possible coordinate transformation $x^{\prime\mu}(x)$ that leave the initial section unchanged, can give a family of solutions that satisfy both Einstein's equations and the same initial conditions. In order to have a single solution, we have to fix the gauge, similarly to when we choose a gauge in electromagnetism. The gauge symmetry of General Relativity is thus the invariance under the group of all possible coordinate transformations
\begin{equation}
    \label{eq:diffeo_inv_GR}
    x^{\mu}\to x^{\prime\mu}(x),
\end{equation}
where $x^{\prime\mu}(x)$ is an arbitrary diffeomorphism. Under this transformation the metric transforms as:
\begin{equation}
    g_{\mu\nu}(x)\to g'_{\mu\nu}(x')=\frac{\partial x^{\rho}}{\partial x^{\prime\mu}}\frac{\partial x^{\sigma}}{\partial x^{\prime\nu}}g_{\rho\sigma}(x).
\end{equation}

\section{Energy conditions}
\label{sec:energycondition-sec}
Now we focus on the matter part of Einstein's equations. In the actual universe it would be impossibly complicated to describe the exact energy-momentum tensor, because of the large number of different matter fields contributions. Thus it might seem that one has little hope in solving Einstein's equations since their right-hand side is not known. However, there are certain inequalities that are physically reasonable to assume for the energy-momentum tensor: these will be discussed in this section \cite{1973lsss.book.Hawking}. In order to present them, it is useful to decompose the $T^{\alpha\beta}$ of a possibly anisotropic bu not dissipative system as:
\begin{equation}
    \label{eq:Tmunu_energy}
    T^{\alpha \beta}=\rho c^2\hat{e}_0^\alpha \hat{e}_0^\beta+p_1 \hat{e}_1^\alpha \hat{e}_1^\beta+p_2 \hat{e}_2^\alpha \hat{e}_2^\beta+p_3 \hat{e}_3^\alpha \hat{e}_3^\beta,
\end{equation}
where the vectors $\hat{e}_\mu^\alpha$ form an orthonormal basis and satisfy 
\begin{equation}
    g_{\alpha \beta} \hat{e}_\mu^\alpha \hat{e}_\nu^\beta=\eta_{\mu \nu}, \quad g^{\alpha \beta} =\eta^{\mu \nu}\hat{e}_\mu^\alpha \hat{e}_\nu^\beta.
\end{equation}
Thus, we notice that the energy density $\rho c^2$ and principal pressures $p_i$ are eigenvalues of the $T^{\alpha\beta}$, while $\hat{e}_\mu^\alpha$ are the normalized eigenvectors. If we consider a perfect fluid, then $p_1=p_2=p_3$ and we call them $p$, and so:
\begin{equation}
    \label{eq:Tmunu_perfect}
    T^{\alpha \beta}=p g^{\alpha \beta}+(\rho c^2+p) \hat{e}_0^\alpha \hat{e}_0^\beta.
\end{equation}
In this case $c\hat{e}_0^\alpha$ is the four-velocity of the perfect fluid. \\
In general, we define $v^\alpha$ as the normalized and future-directed four-velocity of an arbitrary observer in spacetime and decompose it as ($\gamma$ is chosen so that $v^2=-c^2$):
\begin{equation}
    \label{eq:observer_vel}
    v^\alpha=\gamma\left(\hat{e}_0^\alpha+a \hat{e}_1^\alpha+b\hat{e}_2^\alpha+d\hat{e}_3^\alpha\right), \quad \gamma=c\left(1-a^2-b^2-d^2\right)^{-1 / 2}
\end{equation}
where the $a,\,b$ and $d$ are functions of the coordinates and determine the space dependence of the velocity. They are such that $a^2+b^2+d^2<1$ because $v^\alpha$ is a timelike vector. We also define an arbitrary, future-directed null vector $k^\alpha$:
\begin{equation}
    \label{eq:null_vec}
    k^\alpha=\hat{e}_0^\alpha+a'\hat{e}_1^\alpha+b'\hat{e}_2^\alpha+d'\hat{e}_3^\alpha,
\end{equation}
where $a',\,b'$ and $d'$ are arbitrary functions of the coordinates such that $a^{\prime 2}+b^{\prime 2}+d^{\prime 2}=1$ so that $k^2=0$. \\
Now, we present the energy conditions that typically hold for classical matter, while they can be violated by quantized matter fields \cite{1973lsss.book.Hawking}.

\subsubsection{Weak energy condition}
The weak energy condition states that the energy density of any matter distribution, as measured by any observer, must be non-negative. Therefore, since an observer with four-velocity $v^\alpha$ measures the energy density to be $T_{\alpha \beta} v^\alpha v^\beta$, we must have
\begin{equation}
    \label{eq:WEC}
    T_{\alpha \beta} v^\alpha v^\beta \geq 0.
\end{equation}
This assumption is equivalent to say that the energy density as measured by any observer is non-negative. If we substitute the definitions given before, thus Equations (\ref{eq:Tmunu_energy}) and (\ref{eq:observer_vel}), we obtain:
\begin{equation}
    \rho c^2+a^2 p_1+b^2 p_2+d^2 p_3 \ge 0.
\end{equation}
So, if $a=b=d=0$ then $\rho \ge 0$, while if $b=d=0$ then $\rho c^2+p_1 \ge 0$ and similarly for $p_2$ and $p_3$. The weak energy condition thus implies
\begin{equation}
    \rho \geq 0, \quad \rho c^2+p_i>0 \;\; \forall i.
\end{equation}

\subsubsection{Null energy condition}
The null energy condition makes the same statement as the weak form, except that $v^\alpha$ is replaced by an arbitrary future-directed null vector $k^\alpha$. Thus,
\begin{equation}
    \label{eq:NEC}
    T_{\alpha \beta} k^\alpha k^\beta \geq 0.
\end{equation}
This condition plays an important role in cosmology, see for instance \cite{nullEnergy_cosmology}. If we substitute Equations (\ref{eq:Tmunu_energy}) and (\ref{eq:null_vec}), then we get
\begin{equation}
    \rho c^2+a^{\prime 2} p_1+b^{\prime 2}p_2+ d^{\prime 2}p_3 \ge 0.
\end{equation}
Choosing $b'=d'=0$ enforces $a'=1$ and thus $\rho c^2+p_1\ge0$, and similarly for $p_2$ and $p_3$. The null energy condition therefore implies
\begin{equation}
    \rho c^2+p_i\ge 0 \;\; \forall i.
\end{equation}
Notice that the weak energy condition implies the null one.

\subsubsection{Strong energy condition}
The strong energy condition states
\begin{equation}
    \left(T_{\alpha \beta}-\frac{1}{2} T^\rho{}_\rho g_{\alpha \beta}\right) v^\alpha v^\beta \geq 0
\end{equation}
or $T_{\alpha \beta} v^\alpha v^\beta \geq- T^\rho{}_\rho/2$. Because of Einstein's equations, $T_{\alpha \beta}-g_{\alpha \beta} T^\rho{}_\rho /2=R_{\alpha \beta} c^4/ (8 \pi G)$, the strong energy condition is a statement about the Ricci tensor. This condition is related to Dark Energy \cite{noteLiberati}. Substituting Equations (\ref{eq:Tmunu_energy}) and (\ref{eq:observer_vel}) we obtain
\begin{equation}
    \gamma^2\left(\rho c^2+a^2 p_1+b^2 p_2+d^2 p_3\right) \geq \frac{1}{2}\left(\rho c^4-p_1 c^2-p_2 c^2-p_3 c^2\right).
\end{equation}
Choosing $a=b=d=0$ then $\rho c^2+p_1+p_2+p_3\ge0$, while if $b=d=0$ we get $\rho c^2+p_1+p_2+p_3 \geq a^2\left(p_2+p_3-\rho c^2-p_1\right)$. Since this must be valid for any $a^2<1$, the condition becomes $\rho c^2+p_1 \geq 0$ (and analogous relations for $p_2$ and $p_3$). The strong energy condition thus implies
\begin{equation}
    \rho c^2+p_1+p_2+p_3\geq 0, \quad \rho c^2+p_i \geq 0 \;\; \forall i.
\end{equation}
We can note that the strong energy condition does not imply the weak one.

\subsubsection{Dominant energy condition}
The idea that matter should flow along timelike or null world lines is embodied by the dominant energy condition, that states:
\begin{equation}
    -T^\alpha{}_\beta v^\beta \text { is a future-directed, timelike or null vector field}
\end{equation}
where $-T^\alpha{}_\beta v^\beta$ is the matter's momentum density as measured by an observer with four-velocity $v^\alpha$. Hence, the physical interpretation of this condition is that to any timelike observer the local energy density appears non-negative, and there are no superluminal fluxes, being the energy flux timelike or null. This holds for all known forms of matter. Substituting Equations (\ref{eq:Tmunu_energy}) and (\ref{eq:observer_vel}) we obtain
\begin{equation}
    \rho^2 c^4-a^2 p_1^2-b^2p_2^2-d^2p_3^2 \geq 0 .
\end{equation}
If we choose $a=b=d=0$ then we get $\rho \geq 0$, while choosing $b=d=0$ gives $\rho^2 c^4 \geq a^2 p_1^2$. Since this is valid for any $a^2<1$, we have $\rho c^2 \geq\left|p_1\right|$ (and similar relations for $p_2$ and $p_3$). So, the dominant energy condition implies
\begin{equation}
    \rho \geq 0, \quad \rho c^2 \geq |p_i|\;\;\forall i.
\end{equation}

\subsubsection{Averaged energy conditions}
Averaged versions of the energy conditions can be formulated: we require the same condition but mediated along a curve. For example, the averaged null energy condition states that
\begin{equation}
    \int_\gamma T_{\alpha\beta}k^\alpha k^\beta \mathrm{d}\lambda \geq 0,
\end{equation}
where $\gamma$ is a null geodesic.

\section{Gravitational Waves}
\label{sec:GWs-section}
In this section we focus on a particular solution of Einstein field equations in linearized theory: gravitational waves. Being in linearized theory means that we consider the metric as expanded around the flat one
\begin{equation}
\label{eq:linexpGW_1}
    g_{\mu\nu}=\eta_{\mu\nu}+\epsilon h_{\mu\nu},
\end{equation}
with $\epsilon\ll 1$, and expand the equations to linear order in $\epsilon$. In linearized theory indices are raised and lowered with the flat metric. A symmetry of this theory consists in a coordinate transformation of the form
\begin{equation}
    \label{eq:gaugetrasf_lin}
    x^\mu \to x^{\prime\mu}=x^\mu+\epsilon\zeta^\mu(x),
\end{equation}
which leads to a transformation of the metric, that at the lowest order is
\begin{equation}
    \label{eq:gaugetrasf_hmunu}
    \eta_{\mu\nu}+\epsilon h_{\mu\nu}(x)\to \eta_{\mu\nu}+\epsilon h^{\prime}_{\mu\nu}(x^\prime)=\eta_{\mu\nu}+\epsilon \left( h_{\mu\nu}-\partial_\mu\zeta_\nu-\partial_\nu \zeta_\mu \right).
\end{equation}
Thus, slowly varying infinitesimal diffeomorphisms are a symmetry of linearized theory, while the full General Relativity has the complete invariance under coordinate transformations, not just in the infinitesimal version. Moreover, it is possible to show that linearized theory is invariant under finite Poincaré transformations (the group formed by translations and Lorentz transformations), while the full General Relativity does not have this symmetry, since the flat spacetime metric does not play any special role. In linear theory the Riemann tensor, up to the first order in $\epsilon$, is:
\begin{equation}
\label{eq:linearizedRiemann}
    R_{\mu\nu\rho\sigma}=\epsilon\frac{1}{2}\left( \partial_\nu \partial_\rho h_{\mu\sigma}+\partial_\mu\partial_\sigma h_{\nu\rho}-\partial_\mu\partial_\rho h_{\nu\sigma}-\partial_\nu\partial_\sigma h_{\mu\rho}\right).
\end{equation}
It is invariant under the gauge transformation in Equation (\ref{eq:gaugetrasf_lin}), while in the non-linearized theory the Riemann tensor is covariant under arbitrary diffeomorphisms. It is possible to show that defining 
\begin{equation}
    h=\eta^{\mu\nu}h_{\mu\nu}, \quad \Bar{h}_{\mu\nu}=h_{\mu\nu}-\frac{1}{2}\eta_{\mu\nu}h
\end{equation}
the linearized Einstein's equations read
\begin{equation}
    \label{eq:Einstein_lin}
    \mathcal{O}\Bar{h}_{\mu\nu}=-\frac{16 \pi G}{c^4}T_{\mu\nu},
\end{equation}
where the operator $\mathcal{O}$ is
\begin{equation}
    \mathcal{O}\Bar{h}_{\mu\nu}\equiv \square \Bar{h}_{\mu\nu} + \eta_{\mu\nu} \partial^{\rho}\partial^{\sigma} \Bar{h}_{\rho\sigma}-\partial^{\rho}\partial_{\nu} \Bar{h}_{\mu\rho}-\partial^{\rho}\partial_{\mu} \Bar{h}_{\nu\rho}.
\end{equation}
To find solutions to this equation, we first choose to work in the Lorentz gauge ---using the symmetry in Equation (\ref{eq:gaugetrasf_lin}) it possible to show that we can always choose this gauge---:
\begin{equation}
    \partial^\nu \Bar{h}_{\mu\nu}=0
\end{equation}
and since it gives four conditions, it reduces from ten to six the independent components of $h_{\mu\nu}$. In this gauge Equation (\ref{eq:Einstein_lin}) becomes
\begin{equation}
    \label{eq:linEinstein_lorentz}
    \square \Bar{h}_{\mu\nu}=-\frac{16 \pi G}{c^4}T_{\mu\nu}.
\end{equation}
This equation together with the gauge condition imply
\begin{equation}
    \label{eq:Tmunu_conserved_lin}
    \partial^\nu T_{\mu\nu}=0.
\end{equation}
Thus, in the linearized theory the energy-momentum tensor is conserved and equation (\ref{eq:Tmunu_conserved_lin}) leads to a conserved four-vector: $P^\mu=\int dV T^{0\mu}$; while in the full theory the conservation reads $\mathcal{D}^\nu T_{\mu\nu}=0$ and we do not have a conserved quantity.

\subsection{Gravitational waves in vacuum}
If we want to study the propagation of gravitational waves, we need to focus on Equation (\ref{eq:linEinstein_lorentz}) outside the source:
\begin{equation}
    \label{eq:Einstein_lin_nosource}
    \square \Bar{h}_{\mu\nu}=0,
\end{equation}
which has plane wave solutions and implies that gravitational waves travel at the speed of light $c$ ($\square=-\partial_t^2/c^2+\nabla^2$). In this case, even if we have already used the symmetry of the theory to go in the Lorentz gauge, we can do another coordinate transformation such that $\square \zeta_\mu=0$, since it does not spoil the $\partial^\nu \Bar{h}_{\mu\nu}=0$ condition. Hence, we can choose the functions $\zeta_\mu$ to impose four conditions on $h_{\mu\nu}$: this further reduces the degrees of freedom of $h_{\mu\nu}$ to two. We choose $\zeta^0$ such that $h=0$, so $\Bar{h}_{\mu\nu}=h_{\mu\nu}$; $\zeta^i(x)$ are chosen so that $h^{0i}(x)=0$, thus the Lorentz condition is $\partial^0 h_{00}=0$ and, as far as the gravitational wave is concerned, this means $h_{00}=0$ ---since gravitational waves are not the static part of the gravitational interaction---. This defines the transverse-traceless (TT) gauge:
\begin{equation}
    h^{0\mu}=0, \quad h^i{}_i=0,\quad \partial^j h_{ij}=0.
\end{equation}
The solution of Equation (\ref{eq:Einstein_lin_nosource}) in the TT gauge for a wave moving along the $z$ axis is:
\begin{equation}
    \label{eq:gw_tt}
    h_{\mu\nu}^{TT}(t,z)=
    \begin{pmatrix}
     0 &0&0&0\\
     0& h^{TT}_{+}(t,z) & h^{TT}_{\times}(t,z) &0 \\
     0& h^{TT}_{\times}(t,z)& -h^{TT}_{+}(t,z) &0 \\
     0 &0&0&0\\
    \end{pmatrix}=
    \begin{pmatrix}
     0 &0&0&0\\
     0& h_{+} & h_{\times} &0 \\
     0& h_{\times}& -h_{+} &0 \\
     0 &0&0&0\\
    \end{pmatrix}
    \cos \left(\omega(t-z/c)\right)
\end{equation}
where $h_{+}$ and $h_{\times}$ are called the amplitudes of the ``plus'' and ``cross'' polarizations of the wave. Under a rotation of $\pi/4$ with respect to $z$, these polarizations transform one into the other. Lorentz transformations that leave invariant the propagation direction, that in general we call $\hat{\mathbf{n}}$, are the rotation around the $\hat{\mathbf{n}}$ axis and the boosts in the $\hat{\mathbf{n}}$ direction. We can notice that in (2+1)-dimensions gravitational waves do not exist: a symmetric $2\times 2$ traceless matrix in the perpendicular region to the propagation direction ---that here is 1-dimension--- is vanishing. This is due to the fact that Einstein's equations in vacuum are $R_{\mu\nu}=0$, and under the three spatial dimensions the Ricci tensor uniquely determines the Riemann one: imposing $R_{\mu\nu}=0$ means that the spacetime is flat (while this is not true from the (3+1)-dimensions on: there is information stored in Riemann but not in Ricci). We now ask what it physically means to be in the TT gauge. To do that, we consider a test mass at rest at $\tau=0$ and see how it is influenced by the propagating gravitational wave in the TT gauge. Since $dx^i/d\tau=0$ at $\tau=0$ for the test particle, its geodesic equation is
\begin{equation}
   \left. \frac{d^2 x^i}{d \tau^2}\right|_{\tau=0}=-\left[ \Gamma^i_{00}\left( \frac{dx^0}{d\tau}\right)^2\right]_{\tau=0}
\end{equation}
where 
\begin{equation}
    \Gamma_{00}^i=\frac{1}{2}(2\partial_0 h_{0i}-\partial_i h_{00}).
\end{equation}
In the TT gauge both $h_{0i}$ and $h_{00}$ are identically equal to 0, therefore if at time $\tau=0$  $dx^i/d\tau=0$, then also $d^2 x^i/d \tau^2=0$ and so $dx^i/d\tau=0$ remains valid at all times. Thus, in the TT gauge even after the gravitational wave has arrived, particles that were at rest before it comes are still at rest: the position of the free test masses that were initially at rest remains unchanged as a result of this frame's coordinates stretching in response to the gravitational wave's arrival.

\subsection{The energy-momentum tensor of gravitational waves}
Gravitational waves are themselves a source of spacetime curvature, and their effect on the background curvature is formally equivalent to that of matter with an energy-momentum tensor $t^{\mu\nu}$, that at large distances from the source is \cite{Maggiore:2007ulw}
\begin{equation}
    \label{eq:t_munu_GW_esplicito}
    t_{\mu\nu}=\epsilon^2\frac{c^4}{32\pi G}\langle \partial_\mu h_{\alpha\beta}\partial_\nu h^{\alpha\beta}\rangle,
\end{equation}
where $\langle..\rangle$ denotes a spatial average over many reduced wavelengths $\lambdabar=\lambda/2\pi$. From Equation (\ref{eq:t_munu_GW_esplicito}) it is possible to show that \cite{Maggiore:2007ulw}
\begin{equation}
    t^{00}=\epsilon^2\frac{c^2}{16\pi G}\langle \dot h_+^2+\dot h_\times^2\rangle,
\end{equation}
where $h_+$ and $h_\times$ are the plus and cross polarization of the wave defined in the gauge $\text{TT}$ (see Equation (\ref{eq:gw_tt})). \\
To understand from where this result comes from, we need to introduce how we can separate gravitational waves from the background. Indeed, to study whether gravitational waves generate a curvature, we must allow the curved background spacetime $\Bar{g}_{\mu\nu}(x)$ to be dynamical and write
\begin{equation}
    g_{\mu\nu}(x)=\Bar{g}_{\mu\nu}(x)+\epsilon h_{\mu\nu}(x).
\end{equation}
However, this leads to the problem of how we decide what is the background and what the fluctuation. When there is a distinct scale gap, gravitational waves and the spacetime background naturally split apart. For example, $\epsilon h_{\mu\nu}$ represents small ripples on a smooth background if 
\begin{equation}
    \lambdabar \ll L_B,
\end{equation}
where $L_B$ is the typical scale of spatial variation of $\Bar{g}_{\mu\nu}$ and $\lambdabar$ is the reduced wavelength of the small amplitude perturbations. Alternatively, the distinction occurs in frequency space if
\begin{equation}
    f\gg f_B,
\end{equation}
where $f$ is the frequency around which $h_{\mu\nu}$ is peaked and $f_B$ is the maximum value of $\Bar{g}_{\mu\nu}$'s frequencies. In this case a static or slowly varying background is being perturbed at high-frequency by $\epsilon h_{\mu\nu}$. Now that we know how to separate the perturbation from the background, we can expand Einstein's equations around the background metric $\Bar{g}_{\mu\nu}$. \\
Now, we show that when dealing with a perturbation of the background metric, for a consistent expansion we have to confine our analysis to linear order in $\epsilon$. \\
Remembering that we have two small parameters ($\epsilon$ and one of $\lambdabar/L_B$ and $f_B/f$), suppose we expand the Ricci tensor to $\mathcal{O}(\epsilon^2)$ ($R_{\mu\nu}\simeq \Bar{R}_{\mu\nu}+R_{\mu\nu}^{(1)}+R_{\mu\nu}^{(2)}$, so $\Bar{R}_{\mu\nu}$ is constructed only with $\Bar{g}_{\mu\nu}$, $R_{\mu\nu}^{(1)}$ is linear in $\epsilon $ and $R_{\mu\nu}^{(2)}$ is quadratic in $\epsilon$), splitting Einstein's equations as written in Equation (\ref{eq:einstein2}) into the low and high frequency parts, we get \cite{Maggiore:2007ulw}
\begin{equation}
    \label{eq:Einstein_scaleseparation}
    \begin{aligned}
        &\Bar{R}_{\mu\nu}=-[R_{\mu\nu}^{(2)}]^{\text{low}}+\frac{8\pi G}{c^4}\left(T_{\mu\nu}-\frac{1}{2}g_{\mu\nu}T\right)^{\text{low}}\\
        &R_{\mu\nu}^{(1)}=-[R_{\mu\nu}^{(2)}]^{\text{high}}+\frac{8\pi G}{c^4}\left(T_{\mu\nu}-\frac{1}{2}g_{\mu\nu}T\right)^{\text{high}},
    \end{aligned}
\end{equation}
where ``low'' (``high'') indicates the projection on long (short) wavelengths or low (high) frequencies. $\Bar{R}_{\mu\nu}$ contains only low-frequency modes, since it is constructed from $\Bar{g}_{\mu\nu}$, while $R_{\mu\nu}^{(1)}$ by definition is linear in $\epsilon h_{\mu\nu}$ and therefore contains only high-frequency modes. The point now is that $R^{(2)}_{\mu\nu}$ is quadratic in $\epsilon h_{\mu\nu}$ and it actually contains both high and low frequencies. For this reason in Equation (\ref{eq:Einstein_scaleseparation}) we have equated terms of different orders in the $\epsilon$-expansion. Indeed, we have a second small parameter, $\lambdabar/L_B$ or $f_B/f$, which can compensate for the smallness in $\epsilon$: the non-linearity of Einstein's equations mixes up these two expansions. The Einstein's equations themselves establish the relative strengths of these parameters. For example, if we set $T_{\mu\nu}=0$, we have that the first equation in (\ref{eq:Einstein_scaleseparation}) reads
\begin{equation}
    \Bar{R}_{\mu\nu}\sim \epsilon^2(\partial h_{\mu\nu})^2
\end{equation}
because one finds that $[R_{\mu\nu}^{(2)}]^{\text{low}}$ is of order $\epsilon^2(\partial h_{\mu\nu})^2$ \cite{Maggiore:2007ulw}.
At the same time, taking $\Bar{g}_{\mu\nu}=\mathcal{O}(1)$ in $\epsilon$, since the scale variation of $\Bar{g}_{\mu\nu}$ is $L_B$ and that of $ h_{\mu\nu}$ is $\lambdabar$, we have:
\begin{equation}
    \label{eq:order_gmunu_h}
    \partial \Bar{g}_{\mu\nu} \sim \frac{1}{L_B}, \quad \partial h_{\mu\nu} \sim \frac{1}{\lambdabar}.
\end{equation}
Moreover, since $\Bar{R}_{\mu\nu}$ is constructed from the second order derivatives of the background metric, from Equation (\ref{eq:order_gmunu_h}) we have
\begin{equation}
    \Bar{R}_{\mu\nu}\sim \frac{1}{L_B^2}.
\end{equation}
Putting all these three equations together we have:
\begin{equation}
    \label{eq:condition_smallness_GW_1}
    \epsilon\sim\frac{\lambdabar}{L_B}.
\end{equation}
Now we consider $T_{\mu\nu}$ non vanishing, with the contribution of gravitational waves to the background negligible with respect to the contribution of matter sources. In this case 
\begin{equation}
    \label{eq:condition_smallness_GW_2}
    \frac{1}{L_B^2}\sim \frac{\epsilon^2}{\lambdabar^2}+\text{matter contribution}\gg \frac{\epsilon^2}{\lambdabar^2},
\end{equation}
since the curvature is dominated by matter. This explains why the linearized approximation seen at the beginning of this section cannot be extended beyond linear order. Indeed, if the background metric is $\eta_{\mu\nu}$, it means that we are forcing $1/L_B$ to be strictly equal to zero. Because of that, any infinitesimally small but finite value of $\epsilon$ inevitably contradicts the condition in Equation (\ref{eq:condition_smallness_GW_2}) and the expansion in powers of $\epsilon$ has no domain of validity. In addition from Equations (\ref{eq:condition_smallness_GW_1}) and (\ref{eq:condition_smallness_GW_2}) we understand that the notion of gravitational wave has a clear definition only when dealing with small amplitudes, where $\epsilon \ll 1$. If $\epsilon$ approaches the order of one, then also $\lambdabar/L_B$ is at least of order one. However, since the separation between $\lambdabar$ and $L_B$ is fundamental to the definition of gravitational waves, when $\epsilon$ reaches the order of one, the distinction between gravitational waves and the background disappears. \\
At this point, with the aim of finding the expression for the energy-momentum tensor $t^{\mu\nu}$ of gravitational waves, we focus on the first equation in (\ref{eq:Einstein_scaleseparation}) (the second describes the propagation of $h_{\mu\nu}$ on the background spacetime). When there is a clear separation of scales, projections along the long wavelength modes can be done by averaging (we denote it with $\langle...\rangle$) over a spatial volume with side $\Bar{\ell}$ such that $\lambdabar \ll \Bar{\ell}\ll L_B$ (if $h_{\mu\nu}$ is a high-frequency perturbation of a quasi-static background, we average over the time $\Bar{t}$ with $1/f\ll \Bar{t}\ll 1/f_B$ to project along small frequencies). In this way, we have integrated out the fluctuations that take place on a length-scale smaller than $\Bar{\ell}$ from the fundamental equations of the theory. Now we can define an effective energy-momentum tensor of matter $\Bar{T}^{\mu\nu}$ such that
\begin{equation}
    \langle T_{\mu\nu}-\frac{1}{2}g_{\mu\nu}T\rangle =\Bar{T}_{\mu\nu}-\frac{1}{2}\Bar{g}_{\mu\nu}\Bar{T},
\end{equation}
where $\Bar{T}=\Bar{g}_{\mu\nu}\Bar{T}^{\mu\nu}$ is the trace. Thus $\Bar{T}^{\mu\nu}$ is a purely long wavelength or low frequency quantity. We also express the energy momentum tensor of the gravitational waves, see Equation (\ref{eq:t_munu_GW_esplicito}), as
\begin{equation}
    \label{eq:t_munu_implicito}
    t_{\mu\nu}=-\frac{c^4}{8\pi G}\langle R^{(2)}_{\mu\nu}-\frac{1}{2}\Bar{g}_{\mu\nu}R^{(2)}\rangle,
\end{equation}
with $R^{(2)}=\Bar{g}^{\mu\nu}R^{(2)}_{\mu\nu}$, and its trace as
\begin{equation}
    t=\Bar{g}^{\mu\nu}t_{\mu\nu}=\frac{c^4}{8\pi G}\langle R^{(2)}\rangle.
\end{equation}
In this way the first equation in (\ref{eq:Einstein_scaleseparation}) can be written as \cite{Maggiore:2007ulw}:
\begin{equation}
    \Bar{R}_{\mu\nu}-\frac{1}{2}\Bar{g}_{\mu\nu}\Bar{R}=\frac{8\pi G}{c^4}\left(\Bar{T}_{\mu\nu}+t_{\mu\nu}\right).
\end{equation}
Therefore, the dynamics of $\Bar{g}_{\mu\nu}$ is determined by $\Bar{T}_{\mu\nu}$ and $t_{\mu\nu}$, which only depend on the gravitational field itself. Hence, the energy-momentum tensor of gravitational waves acts on the background metric in the same way of the matter energy-momentum tensor.

\subsection{Gravitational waves production}
Now, we look for solutions of Equation (\ref{eq:linEinstein_lorentz}), thus we focus on the source region. The most interesting astrophysical sources of gravitational waves are neutron stars and black holes in compact binary systems. Since we are in linearized theory, this means that we see how gravitational waves are produced considering that the bodies that acts as source move in flat spacetime ---thus the gravitational field generated by the source is supposed to be sufficiently weak and we describe the source dynamics using Newtonian gravity---.
\begin{figure}[ht]
    \centering
    \includegraphics[width=0.8\textwidth]{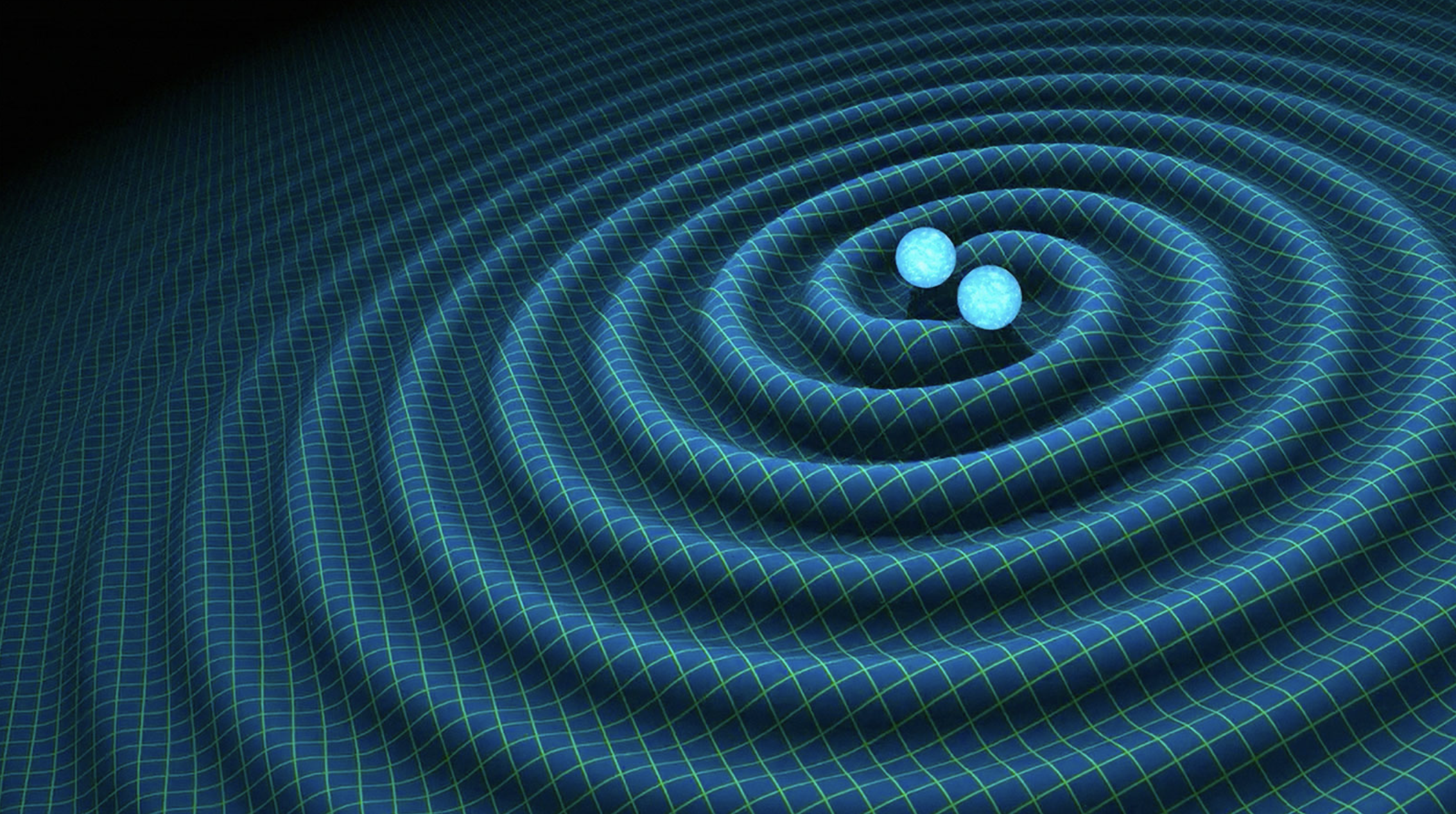}
    \caption{\textit{An artist's impression of gravitational waves generated by binary neutron stars.} Credits: R. Hurt/Caltech-JPL.}
    \label{fig:GW_binaryNS}
\end{figure}
It is possible to show that in the non-relativistic approximation ($v\ll c$ with $v$ the velocity of objects that are part of the source; this condition can also be written as $\lambda \gg \ell$ where $\lambda$ is the gravitational wave wavelength and $\ell$ is the dimension of the source) and in the radiation zone approximation ($|\mathbf{r}-\mathbf{r^\prime}|\gg \lambda, \ell$; where $|\mathbf{r}-\mathbf{r^\prime}|$ is the distance between the source and the observer) we get:
\begin{equation}
    \Bar{h}_{ij}=-\frac{2}{3}\frac{G}{c^4}\frac{1}{r}\partial_t^2 \mathcal{Q}_{ij}
\end{equation}
where $\mathcal{Q}_{ij}$ is the quadrupole momentum of the energy density of the matter distribution of the source:
\begin{equation}
    \mathcal{Q}_{ij}=\int d^3 x\left(3x_i x_j-r^2 \delta_{ij}\right)\frac{T_{00}}{c^2}.
\end{equation}
Hence, there is neither monopole nor dipole radiation for gravitational waves: the leading term of the multipole expansion is the mass quadrupole. Moreover, we understand that $\Bar{h}_{ij}$ depends on the second derivative in time of $\mathcal{Q}_{ij}$. The absence of monopole and dipole radiation can be seen as the expression of the fact that the graviton ---the conjectured particle that mediates gravity--- is a massless particle with helicity $\pm 2$ (that is the projection of the total angular momentum on the direction of motion). Therefore, it is impossible to put it in a state with total angular momentum $j=0$ or $j=1$ \cite{Maggiore:2007ulw}. For a brief review of graviton as a spin-2 field see Appendix \ref{app:graviton_spin2}.
\vspace{7mm}\\
In summary, this chapter introduced General Relativity and one of its predictions: gravitational waves. In the next chapter, we shift our focus to another intriguing solution of Einstein's equations: black holes.

\chapter{Black Holes in General Relativity}
\begin{chapabstract}
    \begin{adjustwidth}{1cm}{1cm}
        We introduce black holes physics and their thermodynamics, from the point of view of General Relativity \cite{1973grav.book.Misner}-\cite{2003gieg.book.Hartle},\cite{2004rtmb.book.Poisson}-\cite{Pathintegral.bhThermodynamics:HartleHawking},\cite{Penrose_singularities}-\cite{BH.lessons.Jacobson},\cite{Damour_membrane},\cite{1986bhmp.book.Thorne}.
    \end{adjustwidth}
\end{chapabstract}

\section{Schwarzschild metric}
\label{sec:SchwarzschildMetric-sec}
A comprehensive understanding of a non-rotating black hole is essential for our objective of perturbing an analogue black hole. The most straightforward application of a theory of gravity involves a spherically symmetric gravitational field. This scenario is particularly relevant when describing, for instance, the gravitational fields produced by objects like the Earth or the Sun. Beyond its practical significance, solving this problem in General Relativity leads us to fascinating solutions that describe novel phenomena: black holes. Thus now, we investigate the case of vacuum solutions of Einstein's equations exhibiting perfect spherical symmetry. Within the framework of General Relativity, the unique spherically symmetric vacuum solution ($T_{\mu\nu}=0$) is the Schwarzschild metric. It has been derived by Schwarzschild in 1916 and it describes the gravitational field exterior to a static, spherically symmetric, non-rotating body. Noticeably it represents the first exact solution of Einstein's equations ever found. The Schwarzschild metric is:
\begin{equation}
    \label{eq:Schwarzschild}
    ds^2=-\left(1-\frac{r_s}{r}\right)(dx^{0})^2+\left(1-\frac{r_s}{r}\right)^{-1} dr^2+ r^2 d\Omega^2.
\end{equation}
$r_s$ is the Schwarzschild radius
\begin{equation}
    r_s=\frac{2G M}{c^2},
\end{equation}
and $M$ is the mass of the gravitating object: in General Relativity it is just a parameter; we associate to it the meaning of the mass of the object by doing the Newtonian limit. For $M\to 0$ we recover the flat Minkowski metric $\eta_{\mu\nu}$. Moreover, the metric becomes progressively Minkowskian as $r\to\infty$: this property is known as asymptotic flatness. We notice that the fact that the Schwarzschild metric is static does not imply that the object that generates the metric have to be static too. Indeed, when deriving the Schwarzschild metric, there is no requirement for the source to be static itself \cite{carroll.book.}. It could be a collapsing star, as long as the collapse maintains the spherical symmetry. There are two important theorems concerning the Schwarzschild metric:
\begin{theorem}[Birkhoff's Theorem]
    Any spherically symmetric solution of vacuum Einstein field equations must be given by the Schwarzschild metric.
\end{theorem}
\begin{theorem}[Israel's Theorem]
    The only static (stationary and non-rotating) and asymptotically-flat vacuum spacetime that possesses a regular horizon is the Schwarzschild solution.
\end{theorem}
The Schwarzschild metric has two singularities: one in $r=0$ ($g_{00}=\infty$) and one in $r=r_s$ ($g_{rr}=\infty$). The question of what type of coordinate-independent signal to look for as an indication that something is out of control with the geometry remains an open topic. We do not delve into this issue in detail, but instead, we focus on a simple criterion for identifying when a problem arises: when the curvature becomes infinitely large. To be more precise, we can derive various scalar quantities from the curvature, and since scalars are coordinate-independent, we can meaningfully state when they become infinite. The simplest such scalar quantity is the Ricci scalar $R=g^{\mu\nu}R_{\mu\nu}$, but we can also construct higher-order scalars such as $R^{\mu\nu}R_{\mu\nu}$, $R^{\mu\nu\rho\sigma}R_{\mu\nu\rho\sigma}$, and so on. If any of these scalars tends to infinity as we approach a certain point that is not infinitely distant, we consider that point as a singularity in the curvature. For the Schwarzschild metric, a direct calculation reveals that:
\begin{equation}
    R^{\mu\nu\rho\sigma}R_{\mu\nu\rho\sigma}=\frac{48 G^2 M^2}{r^6}.
\end{equation}
This is sufficient to convince us that $r=0$ represents a genuine singularity. For what concerns $r=r_s$, none of the curvature invariants blows up: $r=r_s$ is just a coordinates singularity and it is disposable with a coordinate change (see Sections \ref{se:EddingtonFinkelstein-sec}-\ref{sec:CarterPenrose-sec}). We now remark that the Schwarzschild solution is valid only in vacuum, thus it holds outside a spherical body such as a star. Nevertheless, when considering the Sun, we are addressing a celestial body with a radius extending to
\begin{equation}
    R_\odot =\frac{10^6 G M_\odot}{c^2}.
\end{equation}
Therefore, $r=r_s=2GM_\odot/c^2$ is far inside the solar interior: the Schwarzschild metric does not apply there because it is not the solution of the Einstein's equation inside the star. Indeed, solutions for the interiors of stars involve the process of connecting the external Schwarzschild metric to an interior metric that exhibits complete smoothness at the origin. However, there exist objects for which the entire Schwarzschild metric is required: black holes.

\section{Spacetime horizons}
At this point, we introduce the concepts of event horizon (Section \ref{sec:eventHorizon-sec}) and apparent horizon (Section \ref{sec:apparentHorizon-sec}). This introduction is crucial as we will later need to translate them into the language of Analogue Gravity. Additionally, we present the notion of a local horizon (Section \ref{sec:LocalHorizon-sec}), which is essential for studying spacetime thermodynamics. Furthermore, we delve into the concept of a Killing horizon, accompanied by the related idea of surface gravity (Section \ref{sec:killing_horizon-section}), which holds significance in the context of Hawking radiation. To understand these spacetime horizon concepts, we present trapped surfaces (Section \ref{sec:trappedSurfaces}) and conduct a review of hypersurface geometry (Section \ref{sec-geometry_hyper}) and the formalism of geodesic congruences (Section \ref{sec:congruenceGeodesics-sec}). This formalism is also necessary for deriving the equations that describe the kinematics of a horizon, known as the Raychaudhuri equations. Lastly, we provide a brief overview of the membrane paradigm (Section \ref{sec:membrane-section}), offering a different perspective on the event horizon. We start by establishing the following concepts (see \cite{1973grav.book.Misner} for more details):
\begin{description}
    \item[Time orientable spacetime] A spacetime $(\mathcal{M},g_{\mu\nu})$ is said to be time orientable if at each point we can locally distinguish a future and a past lightcone. In this case, there must exists a continuous and everywhere non-zero globally defined timelike vector field $v^\mu$.
    \item[Space orientable spacetime] A spacetime is space orientable if there exists a continuous three-form $\epsilon_{[\mu\nu\lambda]}\not=0$ everywhere and if $v^\mu\epsilon_{[\mu\nu\lambda]}=0$ for any $v^\mu$ timelike.
    \item[Orientable spacetime] A spacetime is orientable if and only if there is a continuous globally defined 4-form everywhere non-zero: it is the volume form. Stating that spacetime is both time and space orientable is equivalent to say that it is orientable.
    \item[Chronological future $\textbf{\textit{I}}^{\textbf{+}}\textbf{(p)}$] The chronological future of a point $p$, $I^+(p)$, is the set of points $q$ of the manifold $\mathcal{M}$ such that there exists a future directed timelike curve $\gamma$ with $\gamma(0)=p$ and $\gamma(1)=q$. We can also define the chronological future of a subset $\mathcal{U}\subset\mathcal{M}$ of the manifold: $I^+(\mathcal{U})=\bigcup\limits_{p\in \mathcal{U}} I^+(p)$.
    \item[Chronological past $\textbf{\textit{I}}^{\textbf{-}}\textbf{(p)}$] The chronological past of a point $p$, $I^-(p)$, is the set of points $q$ of the manifold $\mathcal{M}$ such that there exists a past directed timelike curve $\gamma$ with $\gamma(0)=p$ and $\gamma(1)=q$. We can also define the chronological past of a subset $\mathcal{U}$ of the manifold: $I^-(\mathcal{U})=\bigcup\limits_{p\in \mathcal{U}} I^-(p)$.
    \item[Causal future $\textbf{\textit{J}}^{\textbf{+}}\textbf{(p)}$] The causal future of a point $p$, $J^+(p)$, is the set of points $q$ of the manifold $\mathcal{M}$ such that there exists a future directed causal curve $\gamma$ with $\gamma(0)=p$ and $\gamma(1)=q$. We can also define the causal future of a subset $\mathcal{U}$ of the manifold: $J^+(\mathcal{U})=\bigcup\limits_{p\in \mathcal{U}} J^+(p)$.
    \item[Causal past $\textbf{\textit{J}}^{\textbf{-}}\textbf{(p)}$] The causal past of a point $p$, $J^-(p)$, is the set of points $q$ of the manifold $\mathcal{M}$ such that there exists a past directed causal curve $\gamma$ with $\gamma(0)=p$ and $\gamma(1)=q$. We can also define the causal past of a subset $\mathcal{U}$ of the manifold: $J^-(\mathcal{U})=\bigcup\limits_{p\in \mathcal{U}} J^-(p)$.
    \item[Future timelike infinity $\boldsymbol{\iota}^{\textbf{+}}$] All timelike trajectories that are outside the black hole end up in the future timelike infinity: it is the point where we end up at fixed $r>r_s$ for $t\to\infty$.
    \item[Past timelike infinity $\boldsymbol{\iota}^{\textbf{-}}$] All timelike trajectories at fixed $r$ that are outside the black hole ($r>r_s$) end up in the past timelike infinity for $t\to-\infty$.
    \item[Spatial infinity $\boldsymbol{\iota}^{\textbf{0}}$] All trajectories at $t$ fixed (tachyons' trajectories) end up in the spatial infinity for $r\to\infty$.
    \item[Future null infinity $\mathscr{I}^{\textbf{+}}$] It indicates the region of spacetime toward which light signals travel; geometrically, it is reached through null geodesics towards the future.
    \item[Past null infinity $\mathscr{I}^{\textbf{-}}$] It indicates the region of spacetime from which light signals travel; geometrically, it is reached through null geodesics towards the past.
\end{description}

\subsection{Event horizon}
\label{sec:eventHorizon-sec}
In this subsection we introduce the concept of event horizon. Firstly, we define a black hole region $\mathcal{B}_{\text{bh}}$ as:
\begin{equation}
    \label{eq:BH_definition}
    \mathcal{B}_{\text{bh}}=\mathcal{M}-J^-(\mathscr{I}^+).
\end{equation}
This definition means that there exists a set of events within our spacetime that will never be causally connected with $\mathscr{I}^+$. Now, we define $\Bar{J}^-(\mathcal{U})$ (with $\mathcal{U}\subset\mathcal{M}$) to be the topological closure of $J^-(\mathcal{U})$, thus it is the intersection of all closed sets containing $J^-(\mathcal{U})$, therefore it includes all the limiting points. We also define the boundary of $\Bar{J}^-(\mathcal{U})$ as: $\dot J^-(\mathcal{U})=\Bar{J}^-(\mathcal{U})-J^-(\mathcal{U})$. We define the future event horizon as the boundary of the closure of the causal past of $\mathscr{I}^+$:
\begin{equation}
    \mathcal{H}^+=\dot J^-(\mathscr{I}^+).
\end{equation}
Notice that $\iota^0$ and $\mathscr{I}^-$ are contained in $J^-(\mathscr{I}^+)$ and thus they are not part of $\mathcal{H}^+$. In the Schwarzschild metric, the event horizon is located at $r=r_s$. From Equation \ref{eq:Schwarzschild}, we observe that $r=r_s$ is a singularity in the spacetime. However, it is only an apparent singularity, as explored in Section \ref{sec:apparent_sing_EH}, where we delve into understanding how to address and resolve this singularity. \\
Having defined the black hole region and its event horizon by using $\mathscr{I}^+$, we need to know a priori what is the infinite future evolution of our spacetime in order to properly define the event horizon. Because of that, when we want to describe physical processes and interactions with black holes we will use more practical notions of horizon. All of these concepts can also be extended to a white hole, which is defined as:
\begin{equation}
    \mathcal{B}_{\text{wh}}=\mathcal{M}-J^+(\mathscr{I}^-).
\end{equation}
Analogously we can define a past event horizon:
\begin{equation}
    \mathcal{H}^-=\dot J^+(\mathscr{I}^-).
\end{equation}
Event horizons are null hypersurfaces: they can be seen as a collection of null geodesics which are their generators.  It is important to notice that on the horizon, any two points cannot be separated by a timelike interval; it will always be a null interval. This is locally evident due to the fundamental nature of the horizon as a null hypersurface, while from a global perspective, this fact can be established through a proof by contradiction \cite{noteLiberati}. The evolution of the horizon is described by Raychaudhuri equations and in order to understand them we have to introduce the geometry of hypersurfaces (Section \ref{sec-geometry_hyper}) and the congruence of geodesics formalism (Section \ref{sec:congruenceGeodesics-sec}) where for simplicity we set $c=1$. To introduce the concept of apparent horizon we need to present trapped surfaces (Section \ref{sec:trappedSurfaces}).

\subsection{Geometry of hypersurfaces}
\label{sec-geometry_hyper}
Now, we review the formalism related to the geometry of hypersurfaces. An hypersurface $\Sigma$ is a $(n-1)$-submanifold of the $n$-manifold $\mathcal{M}$ and can be defined in two ways. The first one is the implicit way: the points $x^{\mu}$ of an hypersurface $\Sigma$ are the ones in which a function of the coordinates $\phi(x^{\mu})$ has a constant value. The second is the embedding way: we start with an $(n-1)$-manifold and we map it into a $n$-manifold by a coordinate transformation, so $x^{\alpha}=x^{\alpha}(y^a)$ with $y^a \in \Sigma$. In general, we can introduce the unit normal vector $n^{\alpha}$ to a hypersurface:
\begin{equation}
    n^{\alpha}n_{\alpha}= \varepsilon = \begin{cases}
        -1 \;\;\;\text{if $\Sigma$ is spacelike}\\
        +1 \;\;\; \text{if $\Sigma$ is timelike} 
    \end{cases}
\end{equation}
and when $n^{\alpha}$ is null, then $\Sigma$ is a (n-2)-submanifold. By construction, each vector $v^{\alpha}\in T_p\Sigma$, where $T_p\Sigma$ is the tangent space to $\Sigma$, is orthogonal to $n^{\alpha}$:
\begin{equation}
    \label{eq:TangSigma}
    n^\alpha v_\alpha=0.
\end{equation}
We require that $n^{\alpha}$ is directed along the direction in which $\phi$ is growing: $n^{\alpha}\partial_{\alpha}\phi >0$. It is possible to show that 
\begin{equation}
    n_{\alpha}=\frac{\varepsilon \partial_{\alpha}\phi}{|g^{\mu\nu}\partial_{\mu}\phi \partial_{\nu}\phi|^{1/2}}.
\end{equation}
Indeed $\partial_{\alpha} \phi$ automatically selects derivatives perpendicular to the hypersurface, since on the hypersurface $\phi$ is constant. From Equation (\ref{eq:TangSigma}) it is clear that, since for a null hypersurface $n^{\alpha}n_\alpha=0$, then we have $n^{\alpha}\in T\Sigma$. We can thus take a parameterized family of null curves $x^\mu(\lambda)$ in $\Sigma$ and write $n^\mu=dx^\mu/d\lambda$; it can be shown that these curves are null geodesics \cite{1997gr.qc.Townsend}. In general, the vector fields normal to the hypersurface $S=\text{constant}$, with $S(x)$ a smooth function of the spacetime coordinates $x^\mu$, are
\begin{equation}
    \label{eq:vectorField_normalHyp}
    l=\Tilde{f}(x)(g^{\mu\nu}\partial_\nu S)\frac{\partial}{\partial x^\mu},
\end{equation}
where $\Tilde{f}$ an arbitrary non-zero function. If 
\begin{equation}
    \label{eq:vectorField^2_normalHyp}
    l^2=g^{\mu\nu}\partial_\mu S \partial_\nu S \Tilde{f}^2
\end{equation}
is equal to zero for a particular hypersurface $\Sigma$ in the family, then $\Sigma$ is a null hypersurface. Now we use the embedding $x^{\alpha}=x^{\alpha}(y^a)$, with $y^a$ the coordinates belonging to $\Sigma$, to define a metric on the hypersurface. We consider a four-dimensional manifold $\mathcal{M}$ (coordinates in $\mathcal{M}$ are indicated with Greek indices; while the ones in $\Sigma$ with Latin indices) and define 
\begin{equation}
    \theta^{\alpha}_a= \frac{\partial x^{\alpha}}{\partial y^a}.
\end{equation}
These three vectors are tangent to the hypersurface by construction: $n_\alpha \theta^\alpha_a=0\,\forall a$. The tangent space to $\Sigma$ is
\begin{equation}
    T\Sigma=\left\{\frac{\partial x^{\alpha}}{\partial y^a} e_{\alpha}=\frac{\partial }{\partial y^a}\right\}_{a=1,2,3} \subset T\mathcal{M},
\end{equation}
while a basis for $T\mathcal{M}$ at a point $\in \Sigma$ is: $\{n^{\alpha}, \theta^{\alpha}_a\}$.
This makes possible to construct a metric with the $y^a$ coordinates:
\begin{equation}
    m_{ab}=g_{\alpha\beta}\theta^{\alpha}_a \theta^{\beta}_b.
\end{equation}
The invariant element evaluated at $\Sigma$ is then:
\begin{equation}
    ds^2|_{\Sigma}=g_{\alpha\beta} d x^\alpha dx^\beta|_\Sigma=g_{\alpha\beta}\theta^\alpha_a \theta^\beta_b dy^a dy^b=m_{ab}dy^a dy^b.
\end{equation}
$m_{ab}$ is the so called induced metric, or first fundamental form. We can also define a symmetric tensor on the four-dimensional spacetime, $m_{\alpha\beta}$, that is not a proper metric since it is degenerate in one direction ($m_{\alpha\beta}n^{\beta}=0$):
\begin{equation}
\label{eq:trasversemetric_geometry}
    m_{\alpha\beta}=g_{\alpha\beta}-\varepsilon n_{\alpha} n_\beta.
\end{equation}
Thus $m_{\alpha\beta}$ is a projector operator: it acts on vectors of the spacetime and projects them on the hypersurface. Moreover it provides the induced metric on $\Sigma$:
\begin{equation}
    m_{ab}=m_{\alpha\beta}\theta^\alpha_a \theta^\beta_b.
\end{equation}
The infinitesimal volume element on an hypersurface $\Sigma$ is 
\begin{equation}
d\Sigma=\sqrt{|m|} d^3 y \quad \text{with} \;\;  m=\text{det}(m_{ab})
\end{equation}
and the one oriented is
\begin{equation}
    \label{eq:infinitesimal_oriented}
    d \Sigma_\mu =\varepsilon n_\mu d \Sigma= \varepsilon n_\mu \sqrt{|m|} d^3y.
\end{equation}
We define the trace of the extrinsic curvature, or second fundamental form, $\textit{k}$, as:
\begin{equation}
    \textit{k}=m^{\alpha\beta} \mathcal{D}_\alpha n_\beta.
\end{equation}
If we consider an infinitesimal variation of $\textit{k}$, since $\delta n^\alpha=\delta n_\alpha=\delta g_{\alpha\beta}|_\Sigma=0$ (in the variational principle the boundary hypersurface does not change), then
\begin{equation}
    \label{eq:extrcurv_variation}
    \delta \textit{k}= \frac{1}{2}m^{\alpha\beta}n^\mu \mathcal{D}_\mu \delta g_{\alpha\beta}.
\end{equation}
This concludes our brief overview of the geometry of hypersurfaces, and now we are in a position to introduce the definition of the congruence of geodesics.


\subsection{Congruence of geodesics}
\label{sec:congruenceGeodesics-sec}
At this point, we introduce the formalism of the congruence of geodesics and the Raychaudhuri equations. These equations are important since they express the action of gravity on a congruence of geodesics, such as those describing the trajectory of a group of photons or a cluster of massive particles. \\
Let $\mathcal{M}$ be a manifold and $O \subset \mathcal{M}$ submanifold. A congruence in $O$ is a family of curves such that through each $p \in O$ there passes precisely one curve in this family. The tangents to a congruence generate a vector field in $O$, and conversely every continuous vector field yields a congruence of curves. If this vector field is smooth, then the congruence is said to be smooth.\\
If we consider a spacetime $(\mathcal{M}, g_{\mu\nu})$ foliated by the hypersurfaces $\Sigma_t$ (each hypersurface is identified by a specific value of $t$) and on $\Sigma_t$ we take some coordinates $y^a$, there exists only one curve of the congruence for every $y^a$. Since the congruence of curves covers all the spacetime, we can parameterize the spacetime as $x^\alpha\to(t,y^a)$: we indicate a point in the spacetime by saying that we are at a fixed $t$ (a specific $\Sigma_t$) and then we give $y^a$ (we are in a particular curve of the congruence). In these coordinates we loose invariance under reparameterizations. The curves of the congruence have a tangent vector $t^\alpha$ with 
\begin{equation}
    \label{eq:t-alpha_vector}
    t^\alpha \partial_\alpha t =t^\alpha \mathcal{D}_\alpha t =1 \quad \rightarrow \quad t^\alpha =\left.\frac{\partial x^\alpha}{\partial t }\right|_{y^a}.
\end{equation}
We define $n^\alpha$ as the vector perpendicular to $\Sigma_t$. In general $t^\alpha \not = n^\alpha$ (if $t^\alpha  = n^\alpha$ it means that we are in flat spacetime), while $\partial_\alpha t$ is orthogonal to $\Sigma_t$ and thus parallel to $n^\alpha$, but not equal because of the normalization. A congruence of geodesics is a particular case of this description. \\
In the following for simplicity we denote the covariant derivative as: $\mathcal{D}_\alpha u_\beta=u_{\beta;\alpha}$. \\
Firstly, we focus on a congruence of timelike geodesics, so $u^\alpha u_\alpha=-1$ with $u^\alpha$ the vector field of tangents. We want to determine how such a congruence evolves in time. To do that, we define 
\begin{equation}
    B_{\alpha\beta} = u_{\alpha;\beta}.
\end{equation}
It determines the evolution of the deviation vector $\xi^\alpha$:
\begin{equation}
    u^\beta\xi^\alpha{}_{;\beta}=B^\alpha{}_\beta \xi^\beta.
\end{equation}
where $\xi^\alpha$ is given in Section \ref{sec:notation-subsection}. Thus, $B^\alpha{}_\beta$ measure the failure of $\xi^\alpha$ to be parallel transported along the congruence. We may decompose $B_{\alpha\beta}$ into trace, symmetric trace-free and antisymmetric parts (considering the dimension of the spacetime $n=4$):
\begin{equation}
    B_{\alpha\beta}= \sigma_{\alpha\beta}+ \omega_{\alpha\beta}+\frac{1}{3}m_{\alpha\beta}\theta
\end{equation}
with
\begin{equation}
    m_{\alpha\beta}=g_{\alpha\beta}+u_\alpha u_\beta
\end{equation}
the transverse metric (see Equation (\ref{eq:trasversemetric_geometry})). \\
$\theta$ is the expansion scalar:
\begin{equation}
    \theta =B^{\alpha}{}_\alpha= u^\alpha{}_{;\alpha}
\end{equation}
it describes how the congruence isotropically dilates and shrinks. In particular, the congruence will be diverging if $\theta>0$ while it will be converging if $\theta<0$.\\
$\sigma_{\alpha\beta}$ is the shear tensor:
\begin{equation}
    \sigma_{\alpha\beta}=B_{(\alpha\beta)}-\frac{1}{3}\theta m_{\alpha\beta} 
\end{equation}
it gives information on the relative shearing of the geometry of the cross-sectional area orthogonal to the flow lines: how things get elongated or squashed.\\
$\omega_{\alpha\beta}$ is the rotation tensor:
\begin{equation}
    \omega_{\alpha\beta}=B_{[\alpha\beta]}
\end{equation}
it gives information on how geodesics rotate one on the other. It is possible to show that the following Equation holds \cite{2004rtmb.book.Poisson}:
\begin{equation}
    \label{eq:Ray_expansion}
    \frac{\mathrm{d} \theta}{\mathrm{d} \tau}=-\frac{1}{3}\theta^2-\sigma_{\alpha\beta}\sigma^{\alpha\beta}+\omega_{\alpha\beta}\omega^{\alpha\beta}-R_{\alpha\beta}u^\alpha u^\beta.
\end{equation}
This is the Raychaudhuri's equation for the expansion rate for a congruence of timelike geodesics. It is the key equation used in the proof of the singularity theorems. The Raychaudhuri equation is completely geometric in its nature and the presence of the term proportional to the Ricci tensor implies that it can be used in conjunction with the Einstein's equations to provide insights into how the geometry behaves based on the conditions imposed on the matter stress-energy tensor (see Section \ref{sec:energycondition-sec}). Therefore, it is natural to replace the Ricci tensor with the stress-energy tensor (refer to Equation (\ref{eq:einstein2})) and depending on the nature of the congruence introduce some reasonable conditions to predict the behavior of matter as described by the Raychaudhuri equations. \\
From Equation (\ref{eq:Ray_expansion}) follows the focusing theorem:
let us assume the strong energy condition holds (see Section \ref{sec:energycondition-sec}), and thus being on shell in Einstein field equations we have $R_{\alpha\beta}u^\alpha u^\beta \geq 0$. Additionally, let a congruence of timelike geodesics be hypersurface orthogonal, meaning it is everywhere orthogonal to a family of spacelike hypersurfaces that foliates the spacetime. This is expressed by $\omega_{\alpha\beta}=0$ \cite{2004rtmb.book.Poisson}. Under these conditions, the Raychaudhuri equation for the expansion rate implies 
\begin{equation}
    \frac{\mathrm{d}\theta}{\mathrm{d}\tau}=-\frac{1}{3}\theta^2 -\sigma^{\alpha\beta}\sigma_{\alpha\beta}-R_{\alpha\beta}u^\alpha u^\beta \leq 0.
\end{equation}
Therefore, the focusing theorem expresses that the expansion must decrease during the congruence's evolution. The physical interpretation of this theorem is that, when the strong energy condition holds, gravitation is an attractive force and as a result of this attraction the geodesics get focused. Moreover, under the conditions of the theorem, we get that if the congruence is initially converging ($\theta(0)<0$), then within a finite proper time $\tau\leq 3/|\theta(0)|$ we have $\theta(\tau)\to -\infty$ \cite{2004rtmb.book.Poisson}. So the congruence develops a point at which some of the geodesics come together: a caustic, that is a singularity of the congruence.\\
Finally, we turn to the case of null geodesics. We denote its tangent vector field as $k^\alpha$: the difference from the previous case it that $k^\alpha$ is null. In this case, in order to isolate the transverse part of the metric we need to introduce an auxiliary null vector field $N_\alpha$, such that
\begin{equation}
    N_\alpha k^\alpha=-1.
\end{equation}
The transverse metric, that is not unique since $N^\alpha$ is not uniquely determined, is then
\begin{equation}
    m_{\alpha\beta}=g_{\alpha\beta}+k_\alpha N_\beta+ N_\alpha k_\beta
\end{equation}
and it is purely transverse and effectively two-dimensional. Like what we have done for the congruence of timelike geodesics, we introduce
\begin{equation}
    B_{\alpha\beta}=k_{\alpha;\beta}.
\end{equation}
Since 
\begin{equation}
    \label{eq:Bab_failureparalleltransp}
    k^\beta\xi^\alpha{}_{;\beta}=B^\alpha{}_\beta\xi^\beta
\end{equation}
$B_{\alpha\beta}$ is a measure of the failure of $\xi^\alpha$ to be parallel transported along the congruence. However, $B_{\alpha\beta}$ is orthogonal only to $k^\alpha$ but not to $N^\alpha$, thus we should remove the non transverse component from Equation (\ref{eq:Bab_failureparalleltransp}). After some calculations it is possible to show that \cite{2004rtmb.book.Poisson}:
\begin{equation}
    \label{eq:deformation_tensor-null}
    \Tilde{B}_{\alpha\beta}=m^\mu{}_\alpha m^\nu{}_\beta B_{\mu\nu}
\end{equation}
is the purely transverse part of $B_{\mu\nu}$. Now, we decompose $\Tilde{B}_{\alpha\beta}$ into its irreducible parts, like what we have done for the timelike congruence:
\begin{equation}
    \label{eq:deformation_decomposition-null}
    \Tilde{B}_{\alpha\beta}=\frac{1}{2}\theta m_{\alpha\beta}+\sigma_{\alpha\beta}+\omega_{\alpha\beta}
\end{equation}
with similar definitions of $\theta$, $\sigma_{\alpha\beta}$ and $\omega_{\alpha\beta}$ as in the previous case of timelike geodesics. \\
In this case, the Raychaudhuri equation for the expansion scalar is \cite{2004rtmb.book.Poisson}:
\begin{equation}
    \label{eq:Raychaudhuri_null}
    \frac{\mathrm{d}\theta}{\mathrm{d}\lambda}=-\frac{1}{2}\theta^2-\sigma^{\alpha\beta}\sigma_{\alpha\beta}+\omega^{\alpha\beta}\omega_{\alpha\beta}-R_{\alpha\beta}k^\alpha k^\beta.
\end{equation}
Notice that it is invariant under a change of auxiliary null vector $N^\alpha$ ($\theta=k^\alpha{}_{;\alpha}$).\\
The focusing theorem in the case of null geodesics congruence states: let the null energy condition hold (see Section \ref{sec:energycondition-sec}), so---being on shell in Einstein field equations---$R_{\alpha\beta}k^\alpha k^\beta \geq 0$, and let a congruence of null geodesics be hypersurface orthogonal, thus $\omega_{\alpha\beta}=0$, then the Raychaudhuri equation for the expansion scalar implies
\begin{equation}
    \frac{\mathrm{d}\theta}{\mathrm{d}\lambda}=-\frac{1}{2}\theta^2-\sigma^{\alpha\beta}\sigma_{\alpha\beta}-R_{\alpha\beta}k^\alpha k^\beta \leq 0.
\end{equation}
Therefore, the geodesics are focused during the evolution of the congruence. Under the conditions of this theorem we get that if the congruence is initially converging ($\theta(0)<0$), then within a finite affine parameter $\lambda \leq 2/|\theta(0)|$ we have $\theta(\lambda)\to-\infty$. This generally signals the occurrence of a caustic. The focusing theorem is used in the proof of Penrose singularity theorem.

\subsection{Trapped surfaces}
\label{sec:trappedSurfaces}
\begin{figure}[ht]
    \centering
    \includegraphics[width=0.65\textwidth]{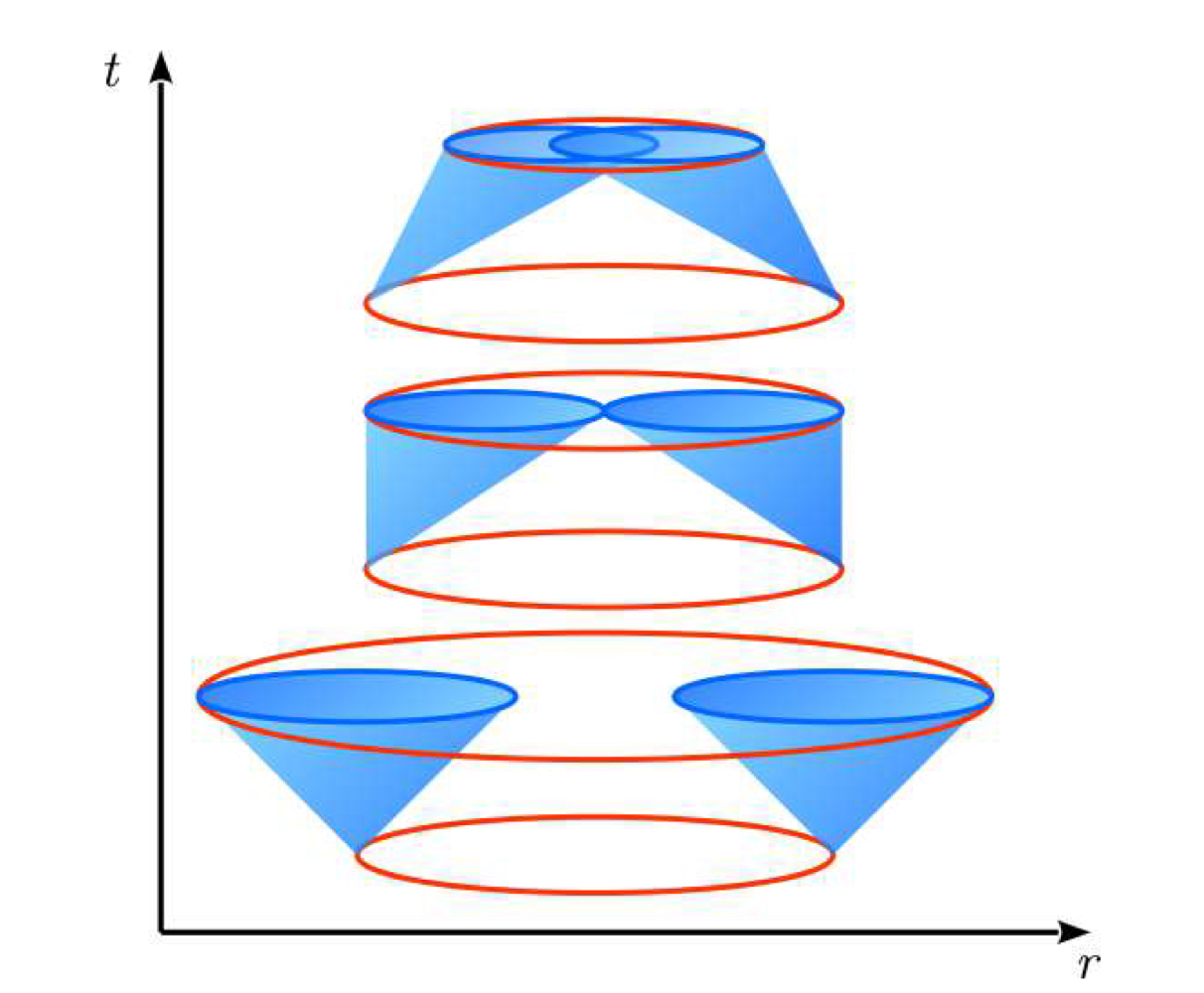}
    \caption{\textit{Trapped surfaces.} Illustration of the the evolution of a light cone in the presence of gravity, resulting in the emergence of a trapped surface. From \cite{noteLiberati}.}
    \label{fig:trapped_region}
\end{figure}

Now, we introduce the concept of trapped surfaces needed to the definition of the apparent horizon. In an Eddington–Finkelstein diagram (as shown in Figure \ref{fig:Eddington_Finkelstein}), the light-cones appear straight at a significant distance from the horizon, but they start to incline as they approach the horizon. This observation serves as the basis for defining the concept of a trapped region. In essence, timelike observers are forced to remain within their own light cones, and if during a gravitational collapse these cones become increasingly tilted over time the 2-surface $\mathcal{S}^2$ will become causally connected with progressively smaller areas, as illustrated in Figure \ref{fig:trapped_region}.
Therefore, in a trapped region both the expansion of the ingoing light rays $\theta_n$ and the expansion of the outgoing null congruence $\theta_l$ are negative. \\
A trapped surface is a closed spacelike 2-dimensional surface, whose ingoing and outgoing null normal congruences are both converging: the expansion of both ingoing and outgoing congruences orthogonal to the 2-surface is negative. The limit case between having a trapped surface and a non-trapped surface, characterized by $\theta_l=0$, is referred to as a marginally trapped surface. \\
Trapped surfaces, together with the focusing theorem, are used in Penrose singularity theorem \cite{Penrose_singularities}\cite{HawkingPenrose_singularity}. It states that if there is a trapped surface, if $R_{\alpha\beta}k^\alpha k^\beta\geq 0$ is valid for all null $k^\alpha$ and if the spacetime has the form $\mathcal{M}=\mathbb{R}\times \Sigma$, with $\Sigma$ a non-compact, connected, Cauchy surface, then a singularity is unavoidable. Penrose's theorem at the beginning was formulated as the fact that the existence of a trapped surface implies the existence of a singularity on the boundary of its future. This theorem explains that the singularity of the Schwarzschild metric at $r=0$ is not just an artifact of the spherical symmetry, but that also an asymmetric collapse will not prevent the formation of a singularity: once collapse reaches a certain point, evolution to a singularity is inevitable. The conclusion appears to be that typical time dependent solutions usually result in singularities (or begin in them). However, even though General Relativity does not resolve the singularities, we can take solace in the belief that singularities are hidden behind event horizons. This is the cosmic censorship conjecture: naked singularities (the ones that are not hidden behind an event horizon) cannot form in gravitational collapse from generic and initially non-singular states in an asymptotically flat spacetime obeying the dominant energy condition (see Section \ref{sec:energycondition-sec}). A consequence of cosmic censorship is the fact that classical black holes only grow bigger, they never shrink. This is the Hawking's area theorem: assuming the cosmic censorship and the weak energy condition, the area of a future event horizon in an asymptotically flat spacetime is non-decreasing. This theorem holds significant importance in the realm of black hole thermodynamics, a concept we will delve into in more detail in Section \ref{sec:BHThermodynamics-section}.

\subsection{Apparent horizon}
\label{sec:apparentHorizon-sec}
Now we introduce a new concept of horizon: an apparent horizon. The apparent horizon $\mathcal{A}$ is an outer marginally trapped surface (see Section \ref{sec:trappedSurfaces}). In the case of a stationary black hole, the apparent horizon and the event horizon coincides. However, during a collapse, the apparent horizon is contained within the event horizon. Thus the apparent horizon is seen as the instantaneous surface of a black hole, and it is a local notion, and as such in principle observable. Hence, the distinction between the apparent horizon and the event horizon can be summarized as follows: the apparent horizon marks the boundary of a region from which at a specific time nothing can escape, while the event horizon is the boundary of a region from which at any time nothing can escape.

\subsection{Local horizon}
\label{sec:LocalHorizon-sec}
Another concept that we introduce is the local notion of horizon, that will be useful in the analysis of the spacetime thermodynamics (see Sections \ref{sec:spacetime_thermod_eq} and \ref{sec:spacetime_therm_noneq}). In analogy with the global definition of a horizon as the boundary of the past of future null infinity (see Equation (\ref{eq:BH_definition})), one can typically consider a local horizon at point $p$ within a generic spacetime as one side of the boundary of the past of a spacelike 2-dimensional surface patch $\mathcal{P}$ including $p$. As a result, near $p$ the local horizon will be formed by the congruence of null geodesics that are perpendicular to $\mathcal{P}$. These geodesics are defined by the past-oriented tangent null vector $k^\alpha$. Moreover, with respect to the point $p$, we can introduce a local inertial frame through local Lorentz invariance of spacetime \cite{2003gieg.book.Hartle}. This is allowed as long as we confine our considerations to a region of dimensions much smaller than $\mathcal{R}(p)^{-1}$, where $\mathcal{R}(p)$ represents the smaller scale associated with the radius of curvature at point $p$. Within this region the metric will be approximately Minkowski: $g_{\mu\nu}=\eta_{\mu\nu}+\mathcal{O}(\epsilon^2)$. An explicit realization of the local inertial frame is done in Riemann normal coordinates \cite{Maggiore:2007ulw}.

\subsection{Killing horizon and surface gravity}
\label{sec:killing_horizon-section}
We now introduce another concept related to horizons: the Killing horizon. This is particularly significant as it gives rise to the concept of surface gravity, which plays a crucial role in defining the Hawking temperature (see Section \ref{sec:Hawking_rad}). If we consider a Schwarzschild spacetime, we can notice that the vector $t^\alpha=\partial x^\alpha/\partial t$, defined in Equation (\ref{eq:t-alpha_vector}), is a Killing vector of that spacetime. Since
\begin{equation}
    g_{\alpha\beta}t^\alpha t^\beta=1-\frac{r_s}{r}
\end{equation}
$t^\alpha$ is timelike outside the black hole, null on the horizon and spacelike inside. The surface $r=r_s$ can thus be called a Killing horizon: it is an hypersurface on which the norm of the Killing vector vanishes. Thus $t^\alpha$ is tangent to the horizon's null generators and, since it is orthogonal to itself on the horizon, it is also normal to the horizon. Hawking has proved that the event horizon of a stationary asymptotically flat spacetime is a Killing horizon \cite{1973lsss.book.Hawking}. In static black hole spacetimes, the event (see Section \ref{sec:eventHorizon-sec}), apparent (see Section \ref{sec:apparentHorizon-sec}) and Killing horizons all coincide. This implies that for stationary black holes it is possible to find a Killing vector orthogonal to the horizon. However, we are aware of the presence of another null vector on the surface and its trajectories serve as the generators of the null hypersurface. In other words, there must be a vector denoted as $l^\mu$ on $\mathcal{H}^+$ that is both null and affinely parameterized $l^\mu \mathcal{D}_\mu l^\nu |_{\mathcal{H}^+}=0$. Calling $t^\mu$ the Killing vector, since both $l^\mu$ and $t^\mu$ are null vectors, we can write $t^\mu=f(x) l^\mu$ and also $t^\mu \mathcal{D}_\mu t^\nu=\kappa t^\nu|_{\mathcal{H}^+}$. Thus, $\kappa$ is associated to the failure of the geodesic to be affinely parameterized.\\         
Now, focus on a Schwarzschild black hole. We choose $x^\mu =(v,r,\theta,\phi)$ ---thus we choose ingoing Eddington-Finkelstein coordinates--- and $t^\alpha=(1,0,0,0)$ then we get
\begin{equation}
    t^\beta \mathcal{D}_\beta t^\alpha=\left(\frac{1}{2}\partial_r f, \frac{1}{2}f \partial_r f, 0 , 0\right),
\end{equation}
where $f=1-r_s/r$. Therefore, at $r=r_s$, since $f=0$ we obtain:
\begin{equation}
    \left.t^\beta \mathcal{D}_\beta t^\alpha\right|_{r=r_s}=\left(\frac{1}{2}\partial_r f, 0,0,0\right)=\kappa \left.t^\alpha\right|_{r=r_s}.
\end{equation}
So $t^\alpha$ is parallel propagated along itself at the event horizon: it satisfy the geodesic equation. We have denoted by $\kappa$ the proportionality factor:
\begin{equation}
    \kappa=\left.\frac{1}{2}\frac{\partial f}{\partial r}\right|_{r=r_s}=\frac{c^2}{4GM}=\frac{GM}{c^2 r_s^2},
\end{equation}
where we have restored $c$ and $G$ to show the dimensions of $\kappa$. We name $\kappa$ surface gravity. This definition is justified by the fact that it can be demonstrated that $\kappa$ is the force required of an observer at infinity to hold a particle of unit mass stationary at the event horizon \cite{2004rtmb.book.Poisson}. \\
It is possible to generalize this definition to an arbitrary black hole spacetime, as long as an event horizon exists, and $\kappa$ is linked to a Killing vector that becomes parallel propagated along itself right at the Killing horizon $\Sigma$. From this definition it is possible to show that:
\begin{equation}
    \kappa^2=-\frac{1}{2}\mathcal{D}^\beta t^\alpha \mathcal{D}_\beta t_\alpha.
\end{equation}
where the right-hand side is evaluated at $\Sigma$. In a static asymptotically flat spacetime, the normalization of the time-translation Killing vector ($t_\mu t^\mu(r\to\infty)=1$) fixes the surface gravity. The same happens in a stationary spacetime. \\
At this point, in order to understand the meaning of the name ``surface gravity'', we consider a static observer, thus its four-velocity $u^\mu$ is such that
\begin{equation}
    t^\mu=V(x) u^\mu
\end{equation}
and because of the $u^\mu$ normalization, we have
\begin{equation}
    \label{eq:redshift_factor}
    V=\sqrt{-t_\mu t^\mu}.
\end{equation}
So $V$, that is called redshift factor, ranges from 0 at the Killing horizon to 1 at infinity. We can rewrite the four-acceleration $a^\mu=u^\nu \mathcal{D}_\nu u^\mu$ in terms of $V$:
\begin{equation}
    a_\mu=\mathcal{D}_\mu \ln V.
\end{equation}
So its magnitude is
\begin{equation}
    a=\sqrt{a_\mu a^\mu}=V^{-1}\sqrt{\mathcal{D}_\mu V \mathcal{D}^\mu V}.
\end{equation}
Thus at the Killing horizon $a$ goes to infinity, but an observer at infinity will see the acceleration as redshifted by a factor $V$: this turns out to be the surface gravity. Therefore, at the horizon $\Sigma$:
\begin{equation}
    \kappa=Va=\sqrt{\mathcal{D}_\mu V\mathcal{D}^\mu V}.
\end{equation}
We can understand now that in a static asymptotically flat spacetime, the surface gravity is the acceleration of a static observer near the horizon as measured by a static observer at infinity. The zeroth law of black hole mechanics (see Section \ref{sec:BHThermodynamics-section}) states that the surface gravity $\kappa$ of a stationary black hole is constant over the event horizon \cite{1973Blackholemachanics.Bardeen}. As a final observation, we note that if $\mathcal{H}^+$ is the event horizon of a Killing vector $t^\mu$, then it is also the event horizon for a vector $\Bar{t}^\mu=C t^\mu$. In such a situation, we would have $\Bar{\kappa}=C \kappa$ implying that the normalization of the surface gravity can be chosen arbitrarily. However, typically, this normalization is adjusted so that at infinity, the Killing vector precisely matches that of Minkowski spacetime, which has a norm of -1. This leaves us with an arbitrary choice regarding the sign of the Killing vector, but this can be resolved by imposing the condition that the Killing vector is directed towards the future.

\subsection{The membrane paradigm}
\label{sec:membrane-section}
We conclude this section on spacetime horizons by offering a different viewpoint of the event horizon: the one presented in the context of the membrane paradigm. The membrane paradigm, introduced by Thorne, Prince and Macdonald, is a different viewpoint for black hole physics and, as long as we focus on the region outside the horizon, it is mathematically equivalent to the standard general relativistic theory of black holes, while it loses its validity inside the horizon \cite{1986bhmp.book.Thorne}. It is the result of the combination between the 3+1 formulation of General Relativity and the Damour's membrane-horizon formalism \cite{Damour_membrane}. The 3+1 formulation consists in decomposing the four-dimensional spacetime into a three-dimensional space plus one-dimensional time: we choose a preferred family of 3-dimensional spacelike hypersurfaces in spacetime and treat them as though they were a three-dimensional space that evolves as time passes. Damour has rewritten the equations that govern the evolution of a black hole horizon so that it is possible to identify terms that look like electric conductivity, shear and bulk viscosity, temperature, entropy, surface pressure etc.; thus he has first noticed a connection between horizon dynamics and fluid dynamics. In the membrane paradigm the event horizon is seen as a two-dimensional membrane that resides in a three-dimensional space, made by a two-dimensional fluid that has finite entropy and temperature but which cannot conduct heat. Hence, interactions between the horizon and the external world are described with fluids equations, thus we have a thermodynamical description. We use a timelike stretched horizon (a timelike surface located at short distance outside the true event horizon) to approximate the null horizon: the shear and expansion of the horizon (seen for the congruence of null geodesics) are interpreted to be viscous terms in the law describing how the horizon area/entropy changes. However, it is not clear whether the relation between the dynamics of causal horizons and hydrodynamics is deeper than an analogy.

\section{The event horizon in Schwarzschild metric}
\label{sec:apparent_sing_EH}
At this point, to conclude our overview of the Schwarzschild metric, we delve into the event horizon's coordinate singularity and explore how to solve it. To achieve this, we present the Schwarzschild metric in Eddington-Finkelstein coordinates (Section \ref{se:EddingtonFinkelstein-sec}), Kruskal-Szekeres coordinates (Section \ref{sec:KruskalSezekeres-sec}), and Carter-Penrose coordinates (Section \ref{sec:CarterPenrose-sec}).

\subsection{Schwarzschild coordinates}
\label{sec:geodesicSCHWarz}
Now, we want to see what happens to a test particle (that means the particle does not change the geometry of spacetime and hence it has no influence on the metric) near the event horizon of a Schwarzschild metric. The symmetry group of Schwarzschild solution is $\mathbb{R}\times SO(3)$ (time translations and space rotations), thus there are four Killing vectors and four constant of the motion for a free particle: if $K^\mu$ is a Killing vector, then $K_\mu u^\mu=\text{constant}$ with $u^\mu=d x^\mu/d \lambda$ and $\lambda$ the affine parameter. The invariance under time translations leads to the conservation of energy, while the invariance under spatial rotations leads to the conservation of the three components of angular momentum. We can think of the angular momentum as a three-dimensional vector characterized by its magnitude (one component) and direction (two components). The conservation of angular momentum's direction implies that the particle's motion will occur within a specific plane. We can conveniently select this plane to be the equatorial plane of our coordinate system. If the particle is not situated in this plane, we can adjust the coordinates through rotation until it aligns with the equatorial plane. Consequently, the two Killing vectors responsible for conserving the angular momentum's direction imply that, in the case of a single particle, we can set $\theta=\pi/2$. The two remaining Killing vectors correspond to energy and the magnitude of angular momentum. Indeed, the time translation invariance give rise to the time Killing vector
\begin{equation}
    K^\mu=(1,0,0,0)\quad K_\mu=K^\nu g_{\mu\nu}=\left(-\left(1-\frac{r_s}{r}\right), 0,0 ,0\right),
\end{equation}
and so the associated conserved quantity $\eta$ is:
\begin{equation}
    \label{eq:Schwcons2}
    K_\mu u^\mu=-\left(1-\frac{r_s}{r}\right)\frac{dx^0}{d\lambda}=\eta.
\end{equation}
We choose $\lambda$ such that $\eta=1$. The other left Killing vector is
\begin{equation}
    L^\mu=(0 ,0, 0,1)\quad L_\mu=L^\nu g_{\mu\nu}=( 0 ,0 ,0 ,r^2 \sin^2\theta ),
\end{equation}
that, since we have chosen $\theta=\pi/2$, implies the conservation of the quantity $\ell$:
\begin{equation}
    \label{eq:Schwcons3}
    L_\mu u^\mu= r^2 \frac{d\phi}{d\lambda}=\ell.
\end{equation}
Moreover, we have an additional condition on the $u^\mu$ normalization (implied by the geodesic equation), which is one conserved quantity along the path:
\begin{equation}
    \label{eq:Schwcons1}
    -g_{\mu\nu}u^\mu u^\nu=\delta
\end{equation}
with $\delta$ constant and equal to $1,0,-1$ respectively for timelike (massive particles), null (massless particles), spacelike geodesics. We do not choose $\lambda$ as the proper time so that it is possible to treat simultaneously massive and massless particles. Putting Eqs. (\ref{eq:Schwcons2}), (\ref{eq:Schwcons3}) and (\ref{eq:Schwcons1}) together, we end up with:
\begin{equation}
    \left(1-\frac{r_s}{r}\right)^{-1}\left(\frac{dr}{d\lambda}\right)^2+U(r)=1-\delta
\end{equation}
with
\begin{equation}
    U(r)=\frac{\ell^2}{r^2}-\frac{r_s}{r-r_s}.
\end{equation}
At long distances $U(r)$ goes like the Newton potential, while near $r_s$ it is dominated by the term $r_s/(r-r_s)$, independently on how much $\ell$ is big. Near the horizon all the trajectories are the same independently on $\delta$ because the $r_s/(r-r_s)$ term dominates. Hence, we study massless particles since studying the case with $\delta=0$ is not a limitation near the $r_s$. Radial null geodesics ($ds^2=0$) are such that:
\begin{equation}
    c^2 dt^2=\left(1-\frac{r_s}{r}\right)^{-2}dr^2=dr^{\star 2}
\end{equation}
where $r^\star$ is the Regge-Wheeler radial coordinate:
\begin{equation}
    r^\star=r+r_s\ln \left| \frac{r-r_s}{r_s}\right|.
\end{equation}
Since $r\in(r_s,\infty)$ then $r^\star\in(-\infty,\infty)$. Trajectories written as $r(x^0)$ are then
\begin{equation}
    r(x^0)= r_s +e^{(\pm x^0+a)/r_s}
\end{equation}
with $a$ a constant and the plus sign for outgoing geodesics, while the minus for the ingoing ones. We notice that no particle reaches the horizon in a finite $x^0$. This is due to the fact that as $r\to r_s^+$ light cones tighten: at $r_s$ light cones degenerate (the horizon is a null surface), and thus going inside the horizon means to have non physical geodesics ($ds^2>0$). If instead of looking at $dr/dx^0$ we focus on $dr/d\lambda$, we get that it does not diverge near $r_s$: if $\delta=1$ and $\lambda=\tau$ then the particle reaches the horizon in a finite proper time. Thus the particle passes through the horizon but in the coordinate reference frame $(x^0,r,\theta,\phi)$ we do not see it. \\
At this point, we aim to introduce different coordinates for studying the Schwarzschild metric, allowing us to resolve the coordinate singularity at $r=r_s$.

\subsection{Eddington-Finkelstein coordinates}
\label{se:EddingtonFinkelstein-sec}
\begin{figure}[H]
    \centering
    \includegraphics[width=0.8\textwidth]{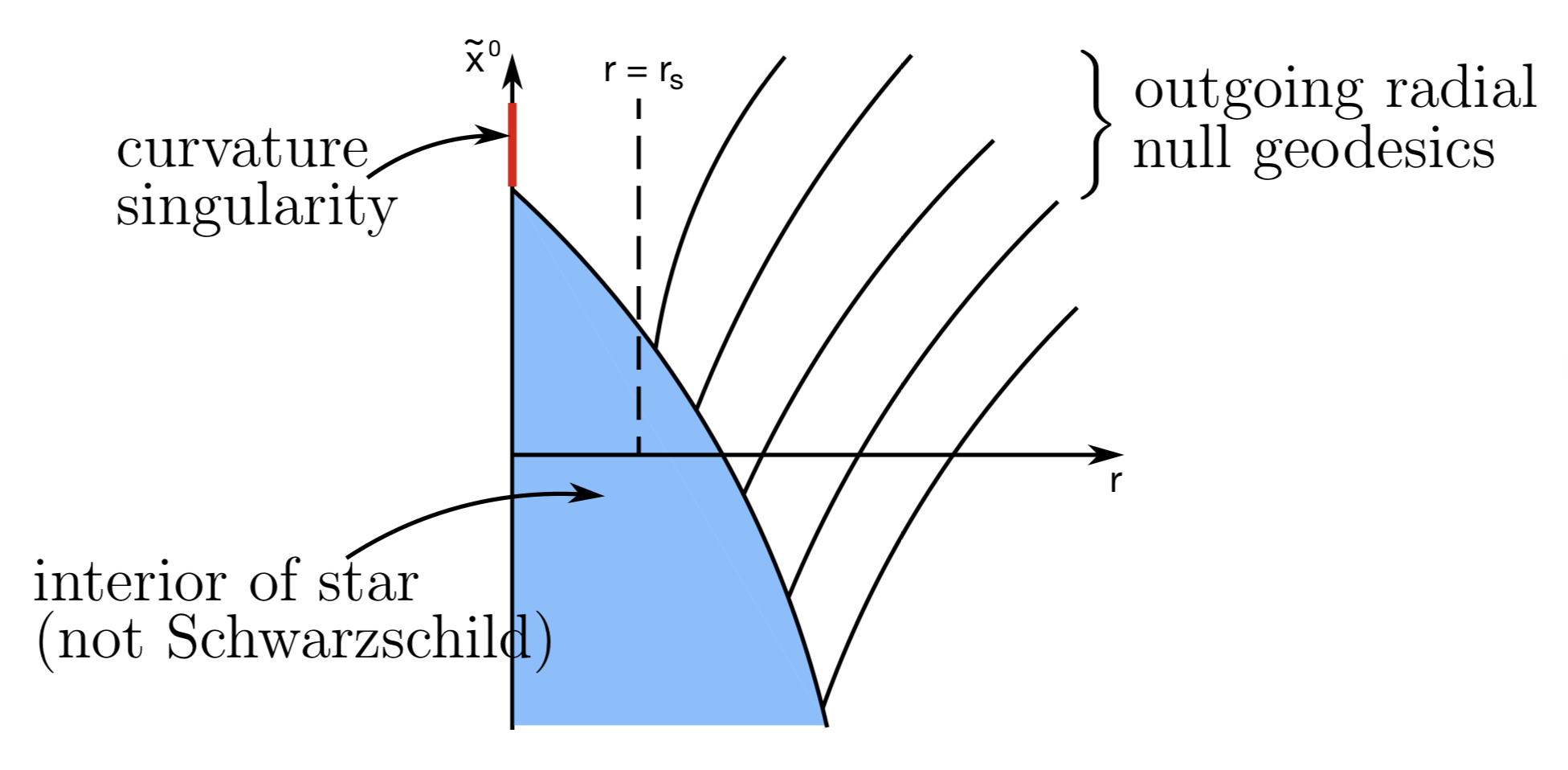}
    \caption{\textit{Collapsing star in Eddington-Finkelstein diagram.} The diagram shows a collapsing star generating a Schwarzschild black hole in Eddington-Finkelstein coordinates. The dashed curve is the event horizon at $r=r_s$; the red line represents the curvature singularity that arises from the collapse. The blue region is the star interior and its black contour line represents the surface of the star. Adapted from \cite{2017lectures.Reall}.}
    \label{fig:collapseEF}
\end{figure}

Now, we aim to express the Schwarzschild metric in a different coordinate reference frame to avoid the singularity at $r_s$. We start this discussion by examining the Eddington-Finkelstein coordinates. For doing it we notice that on radial null geodesics 
\begin{equation}
    d(ct \pm r^\star)=0.
\end{equation}
We define the ingoing radial null geodesics $v$, and the outgoing one $u$, as:
\begin{equation}
    \begin{aligned}
        &v=ct+r^\star \quad v\in(-\infty,\infty)\\
        &u=ct-r^\star \quad u\in(-\infty,\infty)
    \end{aligned}
\end{equation}
The Schwarzschild metric rewritten in the ingoing Eddington-Finkelstein coordinates $(v,r,\theta,\phi)$ is:
\begin{equation}
\begin{aligned}
    d s^2  &=\left(1-\frac{r_s}{r}\right)\left(-(d x^0)^2+d r^{* 2}\right)+r^2 d \Omega^2=\\
    &=-\left(1-\frac{r_s}{r}\right) d v^2+2 d r d v+r^2 d \Omega^2.
\end{aligned}
\end{equation}
Since the relationship between $v$ and $r$ is only established for $r>r_s$, this metric is first defined for $r>r_s$, but it can now be analytically continued to all $r>0$. The metric in these new coordinates is not singular at $r=r_s$: the singularity at the event horizon in Schwarzschild coordinates was actually a coordinate singularity. Therefore if we consider a star collapsing with spherical symmetry and forming a black hole, nothing at $r=r_s$ will stop the star from collapsing over the event horizon. To visualize better the solution in the ingoing Eddington-Finkelstein coordinates, we can do another change of coordinates:
\begin{equation}
    v\to \Tilde{x}^0=v-r
\end{equation}
and thus
\begin{equation}
    ds^2=\left(1-\frac{r_s}{r}\right)(d\Tilde{x}^0)^2-2\frac{r_s}{r}d\Tilde{x}^0 dr-\left(1+\frac{r_s}{r}\right) dr^2+r^2 d\Omega^2.
\end{equation}
\begin{figure}[ht]
    \centering
    \includegraphics[width=0.7\textwidth]{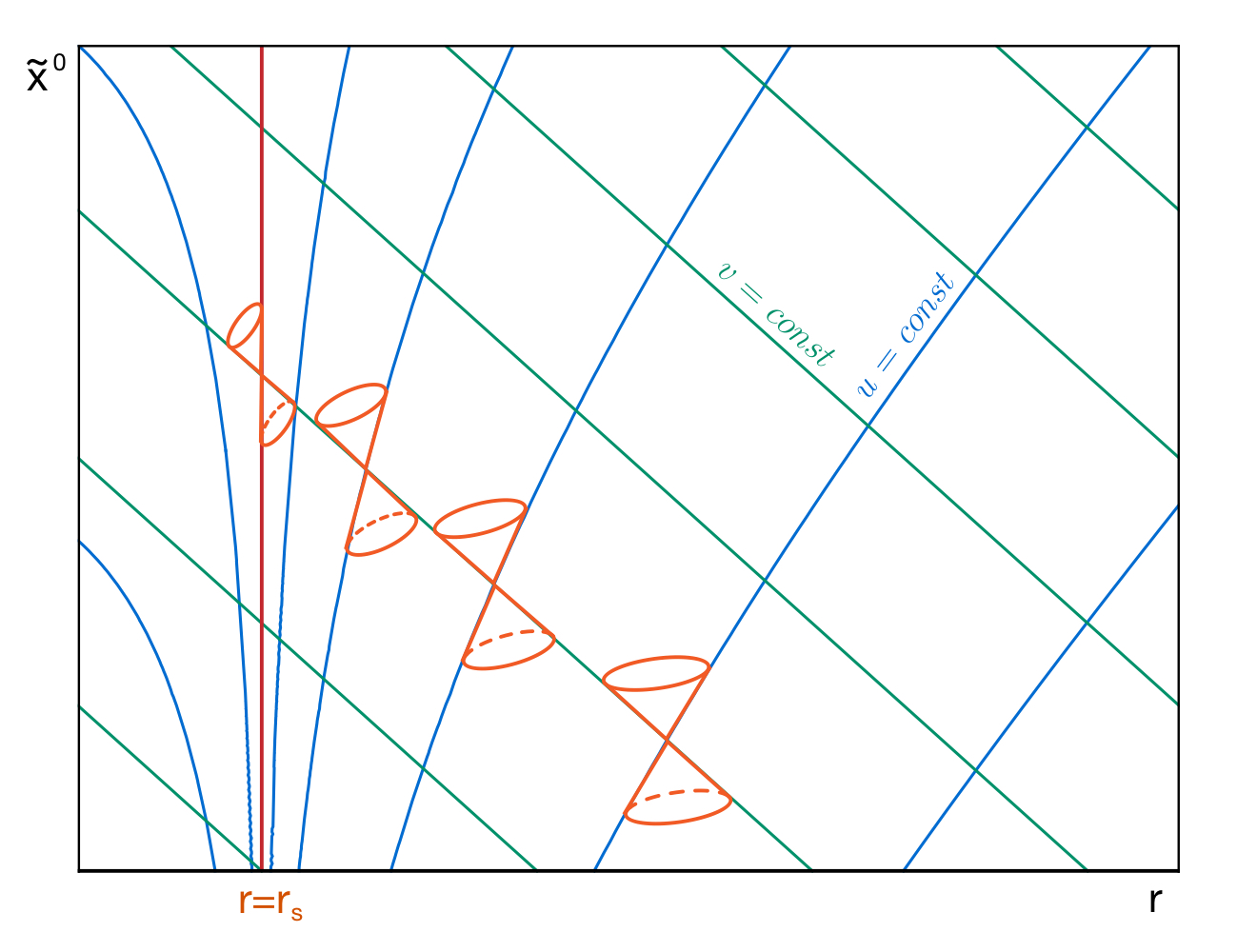}
    \caption{\textit{Eddington-Finkelstein diagram of the Schwarzschild metric.} This diagram shows the trajectories of ingoing (in green) and outgoing (in blue) light rays in a Schwarzschild metric expressed in the ingoing Eddington-Finkelstein coordinates. The red line indicates the event horizon, located at $r=r_s$. The diagram illustrates the distortion of light cones (in orange) near the black hole's horizon (with two angular dimensions suppressed). Adapted from \cite{noteLiberati}.}
    \label{fig:Eddington_Finkelstein}
\end{figure}
In these coordinates we see that light cones distort as $r\to r_s^+$ (see Figure \ref{fig:Eddington_Finkelstein}), so that the event horizon is tangent to light cones (thus it is a null curve) and that no future directed timelike or null worldline can reach $r>r_s$ from $r\le r_s$, while future directed timelike or null worldline can reach $r<r_s$ from $r\ge r_s$. The region $r<r_s$ is a black hole from which nothing can escape. In particular, no signal from the star's surface can escape to infinity once the surface has passed through the event horizon: the star has collapsed into a black hole. It is possible to redo the same reasoning we have followed here with $u$ instead of $v$, and in this case we end up with a region $r<r_s$ from which everything can escape but in which nothing can enter: it is a white hole, the time reversal of a black hole. Indeed the Schwarzschild metric written in the outgoing Eddington-Finkelstein coordinates is:
\begin{equation}
    ds^2=-\left( 1-\frac{r_s}{r}\right) du^2-2dr du +r^2 d\Omega^2.
\end{equation}
In this coordinates we can visualize the process of a spherically symmetric gravitational collapse (see Figure \ref{fig:collapseEF}): once a massive spherical star exhausts its nuclear fuel it will shrink under its own gravity. Because of the spherical symmetry of the system, the geometry outside the star is given by the Schwarzschild metric (see Section \ref{sec:SchwarzschildMetric-sec}).
If the star is not too massive, it may eventually stabilize as a white dwarf or neutron star; while if it is sufficiently massive, nothing can prevent its contraction until it reaches its Schwarzschild radius. As the star's surface approaches $r=r_s$, it will collapse and form a curvature singularity in finite proper time (it is possible to show that the total proper time along a timelike curve in $r\leq r_s$ cannot exceed $\pi M$ \cite{2017lectures.Reall}). Now that we have observed the resolution of the apparent singularity at $r=r_s$ in Eddington-Finkelstein coordinates, we proceed to describe the Schwarzschild metric in Kruskal-Szekeres coordinates, where we encounter the maximal extension of a Schwarzschild spacetime.

\subsection{Kruskal-Szekeres coordinates}
\label{sec:KruskalSezekeres-sec}
At this stage, we explore the Schwarzschild spacetime in another crucial set of coordinates: the Kruskal-Szekeres coordinates. In this coordinate system, the singularity at the event horizon is still avoided, and it is possible to achieve the maximal extension of the Schwarzschild spacetime. These coordinates are introduced from the ingoing and outgoing Eddington-Finkelstein coordinates defining for $r>r_s$:
\begin{equation}
    \begin{aligned}
        &V=e^{v/2r_s}\\
        &U=-e^{-u/2r_s}.   
    \end{aligned}
\end{equation}
The metric in terms of $(U,V,\theta,\phi)$ is:
\begin{equation}
    \begin{aligned}
    ds^2&=\frac{-4r_s^3}{r(U,V)}e^{-r(U,V)/r_s}\mathrm{d}U\mathrm{d}V+r^2 d\Omega^2=\\
    &=-\frac{32 M^3}{r(U,V)}e^{-r(U,V)/2M}\mathrm{d}U \mathrm{d}V+r^2(U,V)\mathrm{d}\Omega^2
    \end{aligned}
\end{equation}
where in the last equation we have used $c=G=1$. It is initially defined for $U<0$ and $V>0$, but we can extend it by analytic continuation to $U>0$ and $V<0$. 
\begin{figure}[ht]
    \centering
    \includegraphics[width=0.55\textwidth]{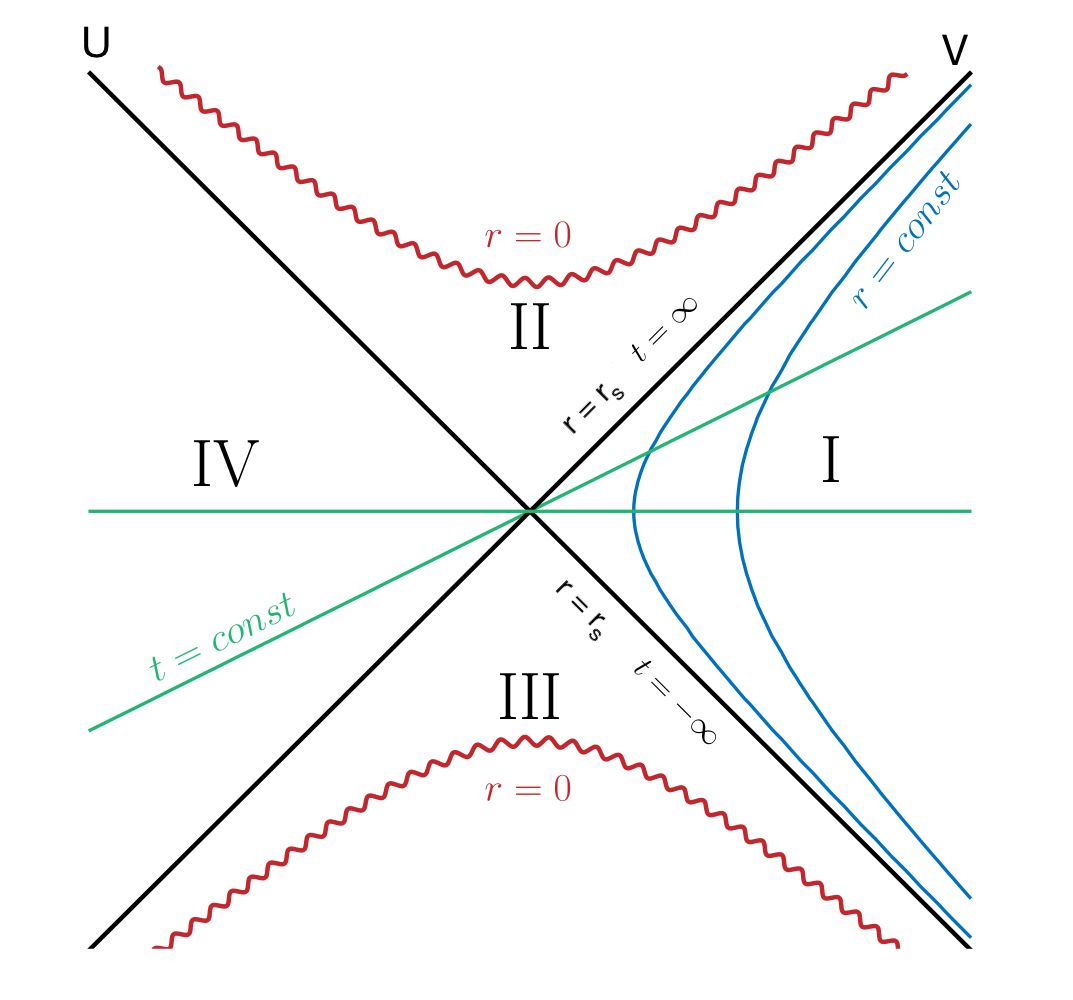}
    \caption{\textit{Kruskal-Szekeres diagram of the maximal extension of a Schwarzschild spacetime.} This diagram shows trajectories at fixed radius $r>r_s$ (blue lines), and the ones at fixed time $t$ (green lines) in a Schwarzschild metric expressed in the Kruskal-Szekeres coordinates. Region \rom{2} represents the black hole, while region \rom{3} the white hole. Region \rom{4} is an identical copy of region \rom{1}: they are asymptotically flat regions. Physical regions are only the \rom{1} and \rom{2} ones. The two red wavy lines indicates the singularity at $r=0$. Notice that two angular dimensions are suppressed. Adapted from \cite{noteLiberati}.}
    \label{fig:Kruskal_Szekeres}
\end{figure}
As we can see in Figure (\ref{fig:Kruskal_Szekeres}), curves at $r$ constant are now indicated by the ones at constant $UV$, since 
\begin{equation}
    UV=-\left(\frac{r}{r_s}-1\right) e^{r/r_s}
\end{equation}
while the ones at constant $x^0$ are now those at constant $\frac{V}{U}$, since
\begin{equation}
    \frac{V}{U}=-e^{x^0/r_s}.
\end{equation}
Therefore, $r=r_s$ corresponds to either $U=0$ or $V=0$, while the singularity at $r=0$ corresponds to $UV=1$. Nonetheless, diagram in Figure \ref{fig:Kruskal_Szekeres} does not yet represent the asymptotic geometric behavior accurately. To account for that, we will have to once more implement a coordinate transformation using the $\arctan$ function: we go in Carter-Penrose coordinates.

\subsection{Carter-Penrose coordinates}
\label{sec:CarterPenrose-sec}
\begin{figure}[ht]
    \centering
    \includegraphics[width=0.75\textwidth]{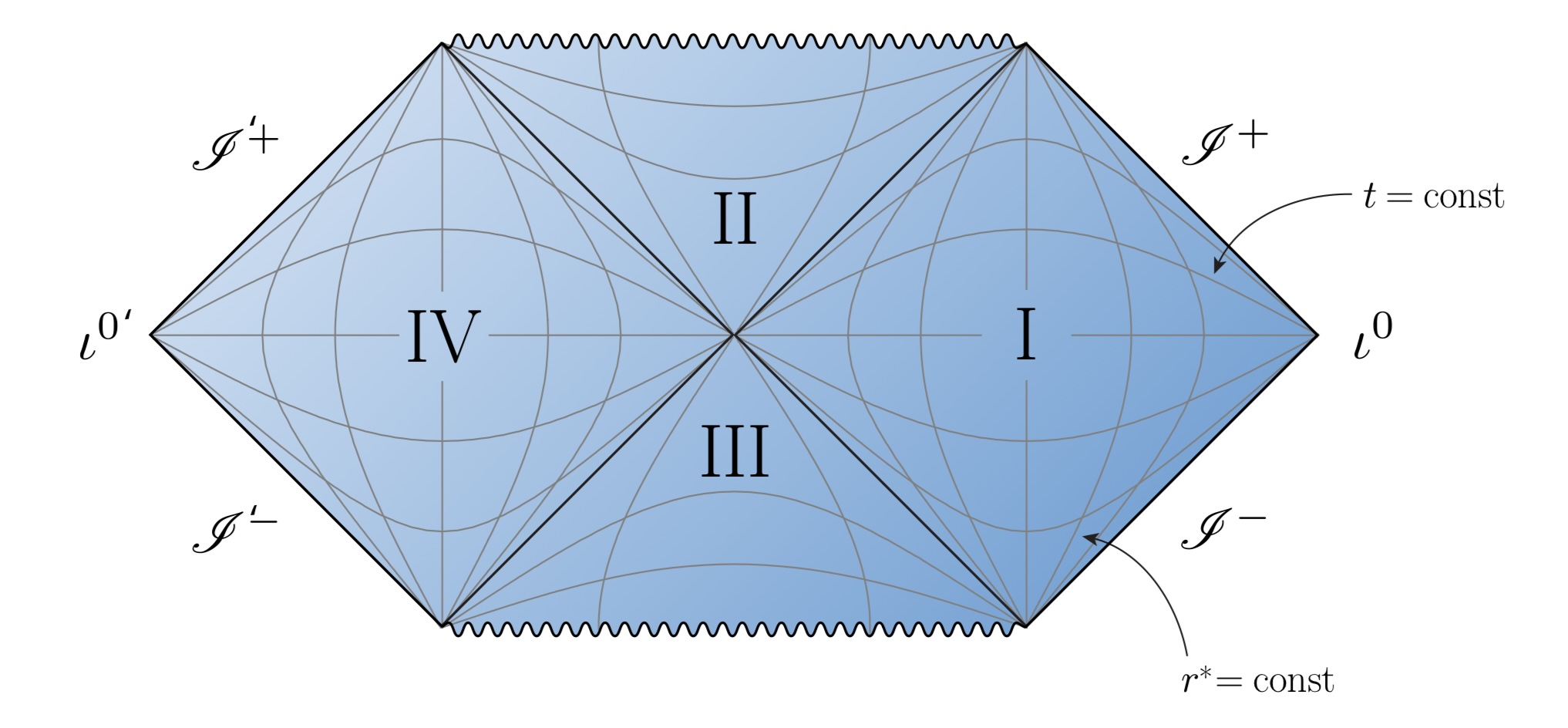}
    \caption{\textit{Carter-Penrose diagram of the maximal extension of a Schwarzschild spacetime.} Notice that two angular dimensions are suppressed in the diagram: each point is actually a 2-sphere. From \cite{noteLiberati}.}
    \label{fig:Carter-Penrose}
\end{figure}

Carter-Penrose coordinates are useful to describe the asymptotic geometric structure: they bring the infinite into a finite region. So, let us proceed by introducing a new set of coordinates (with $c=G=1$):
\begin{equation}
    \begin{aligned}
        &\mathcal{U}=\arctan \left(\frac{U}{\sqrt{M}}\right)\\
        &\mathcal{V}=\arctan \left(\frac{V}{\sqrt{2M}}\right)
    \end{aligned}
\end{equation}
which are in a finite range: $-\pi/2<U,V<\pi/2$. So, the entire infinite structure of spacetime is mapped into a finite region. By implementing one last coordinate transformation
\begin{equation}
    \begin{aligned}
        &T=\frac{\mathcal{V}+\mathcal{U}}{2}\\
        &X=\frac{\mathcal{V}-\mathcal{U}}{2}
    \end{aligned}
\end{equation}
we obtain a metric that, after an appropriate conformal rescaling, results in the conformal diagram displayed in Figure \ref{fig:Carter-Penrose}. It represents the maximal extension of an eternal Schwarzschild black hole, thus which has always existed and will always exist.
\begin{figure}[ht]
    \centering
    \includegraphics[width=0.46\textwidth]{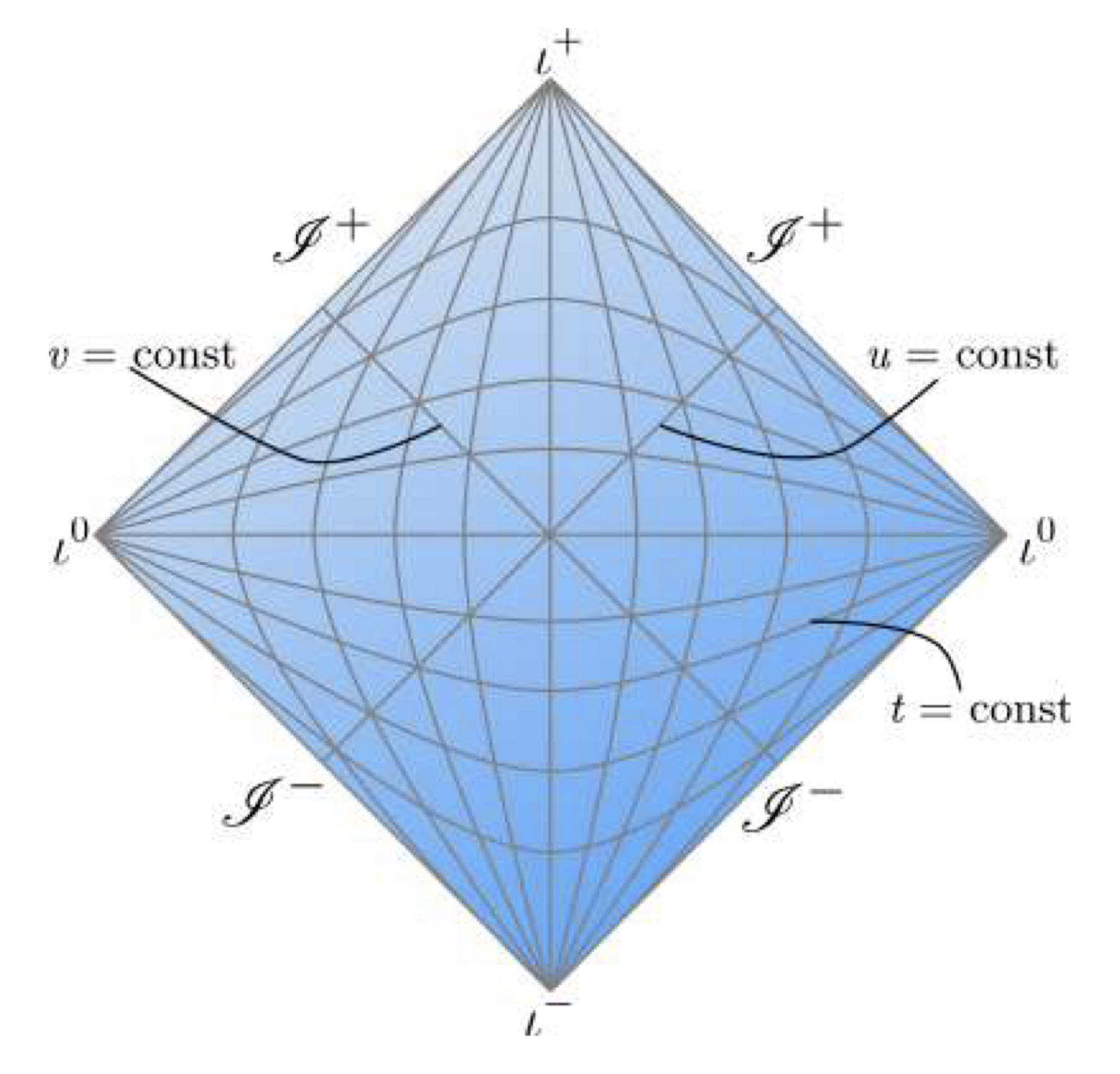}
    \caption{\textit{Carter-Penrose diagram of Minkowski spacetime.} $u=\text{const}$ and $v=\text{const}$ refer to outgoing and ingoing photons respectively. From \cite{noteLiberati}.}
    \label{fig:CP_Minkowski}
\end{figure}
\begin{description}
    \item[Region \rom{1} and \rom{4}] Region \rom{1} and region \rom{4} correspond to asymptotically flat regions, which represent the exterior of a black hole. In these regions, $\mathscr{I}^{\pm}\cup \iota^0$ is identical to what we find in Minkowski spacetime (the Carter-Penrose diagram of a Minkowski spacetime is shown in Figure \ref{fig:CP_Minkowski}). We define an asymptotically flat spacetime as a spacetime with an asymptotic structure resembling that of Minkowski spacetime. This implies that the Schwarzschild metric possesses the same Killing vectors as Minkowski spacetime at infinity. This is significant as it allows us to define a mass and an angular momentum on spacelike hypersurfaces. The two identical copies of the spacetime geometry near $r=r_s$ have been attached to each other at the circle with a radius of $r=r_s$. One face corresponds to region \rom{1}, while the other corresponds to region \rom{4}: see Figure \ref{fig:wormhole}. 
    \begin{figure}[ht]
        \centering
        \includegraphics[width=0.37\textwidth]{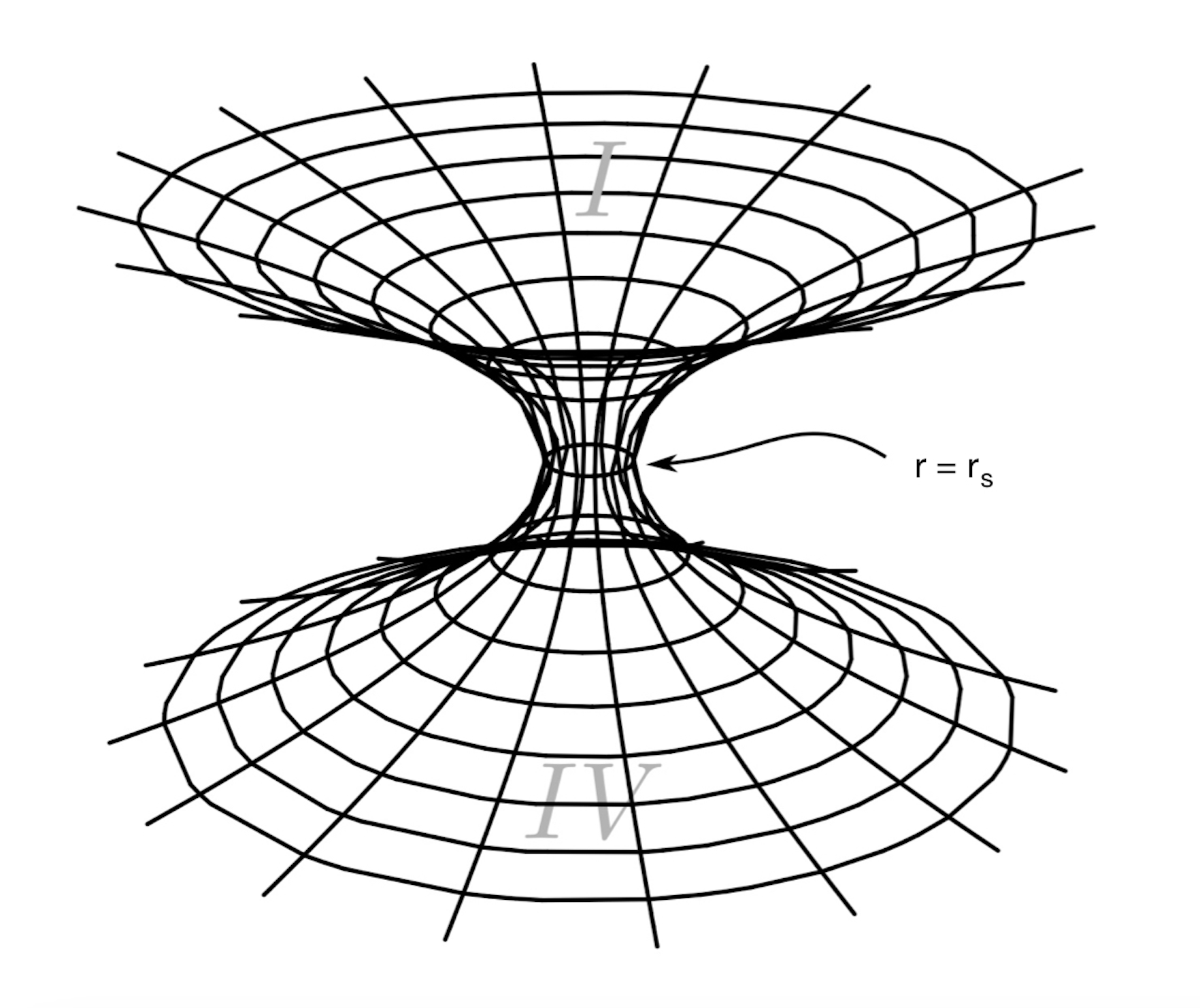}
        \caption{\textit{Representation of the Einstein-Rosen bridge.} This depiction illustrates the Einstein-Rosen bridge, an example of a wormhole, with one spatial dimension suppressed: each circle represents a 2-sphere. Adapted from \cite{Dowker.notes}.}
        \label{fig:wormhole}
    \end{figure}
    This structure is recognized as the Einstein-Rosen bridge, which is an example of a wormhole. It is important to note that no observer can ever cross the wormhole, as is evident both in the Kruskal diagram and in the Carter-Penrose one. Indeed, to cross the wormhole from region \rom{1} to \rom{4}, or vice versa, the observer's trajectory would need to be spacelike at some point.
    
    \item[Region \rom{2}] Region \rom{2} is the conventional region associated with a black hole, with the wavy line in Figure \ref{fig:Carter-Penrose} representing the black hole singularity located at $r=0$ in Schwarzschild spacetime. It is important to note that the only way to exit region \rom{2} is either by reaching the singularity at $r=0$ or by traveling faster than the speed of light to reach region \rom{1} or \rom{4}. Additionally, the singularity is spacelike.
    \item[Region \rom{3}] Region \rom{3} is known as a white hole, which is essentially the temporal inverse of a black hole. As already stated, in a white hole, nothing can enter, and everything must leave, in stark contrast to a black hole where nothing can exit, and everything must remain inside. However, time reversal is disrupted by initial conditions, such as when a black hole forms due to the collapse of a star. When a collapse occurs, the Carter-Penrose diagram of Schwarzschild spacetime takes on the appearance depicted in Figure \ref{fig:star_collapse}. 
    \begin{figure}[H]
    \centering
    \includegraphics[width=0.35\textwidth]{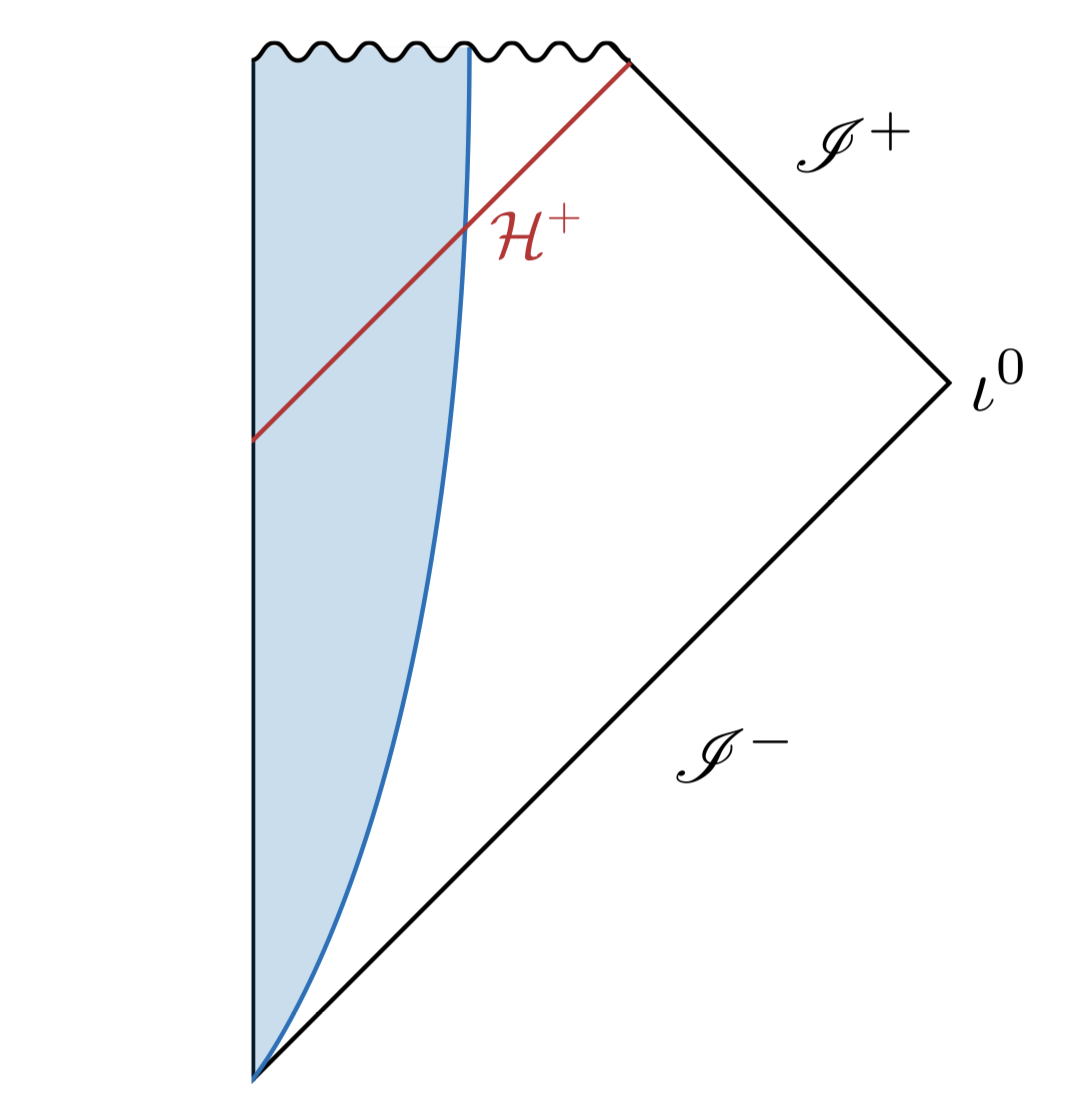}
    \caption{\textit{Collapsing star in Carter-Penrose diagram.} The diagram illustrates a collapsing star generating a Schwarzschild black hole in Carter-Penrose coordinates. The surface of the star is the blue line, while the red line indicates the horizon $\mathcal{H}^+$. From \cite{noteLiberati}.}
    \label{fig:star_collapse}
    \end{figure}
    In this scenario, we observe that the white hole is no longer in existence. Nevertheless, we can also observe that the causal structure of black holes formed from stellar collapse at late times would be indistinguishable from that of eternal black holes \cite{1973grav.book.Misner}.
\end{description}
Therefore, in Carter-Penrose coordinates, we not only avoid the apparent singularity at the event horizon but also describe the asymptotic geometric structure of the maximal extension of the Schwarzschild black hole.

\section{Black holes thermodynamics}
\label{sec:BHThermodynamics-section}
As a final paragraph on black holes in General Relativity, we focus on black holes thermodynamics. This is crucial for the present thesis as it paves the way for the existence of Hawking radiation. Its acoustic equivalent holds significance in the context of Analogue Gravity, aligning with the goals of this thesis. Bekenstein proposed the formulation of a thermodynamic theory of black holes comparing the black hole physics and the laws of thermodynamics: he proposed that black holes have an actual entropy \cite{BHentropy.Bekenstein}. He found inspiration in the analogous behaviors of black hole area and entropy, as both quantities exhibit an irreversible tendency to increase. Indeed in 1971 Hawking predicted that the total area of event horizons should never shrink: this is the area law \cite{HawkingareaTHEORETICAL}. Remarkably, in 2021 the Hawking's area theorem was confirmed for the first time through observations of gravitational waves \cite{HawkingareaEXPERIMENT}. Bardeen, Carter and Hawking have explored the idea further, publishing the four laws of black hole mechanics, which govern the behavior of black holes, and that are derived from the comparison to the four laws of thermodynamics \cite{1973Blackholemachanics.Bardeen}. At the beginning the possibility of seeing black holes as thermodynamical entities was just seen as a purely formal and coincidental analogy, but when Hawking discovers that quantum processes allow a black hole to emit a thermal flux of particles, then it became clear that the laws of black hole mechanics are nothing but a description of its thermodynamics \cite{BHexplosions.Hawking}. Indeed, Hawking identified an imbalance in the creation/annihilation process, leading to a spontaneous emission from vacuum: an observer in the asymptotic future would measure a thermal emission at a temperature that is inversely proportional to the black hole mass, and black holes would eventually evaporate. He demonstrated that, even though the local emission rate was minimal, over the course of the Universe's existence, the emissions could accumulate to become substantial in determining the black hole's mass, eventually leading to its complete evaporation \cite{ParticleCreationBH.Hawking}. However, classically black holes cannot radiate. In the following, in order to simplify the notation we consider $c=\hslash=G=k_B=1$, with $k_B$ the Boltzmann constant. We now present the four laws of black holes thermodynamics.

\begin{description}
    \item[Zeroth law]
    The surface gravity $\kappa$ of a stationary black hole is uniform over the entire event horizon. Thus $\kappa$ is constant along the horizon's null generators and does not vary from generator to generator.
\end{description}

\begin{description}
    \item[First law]
    In a quasi-static process the changes in mass $\delta M$, spin $\delta J$, and surface area $\delta A$ between the initial and final stationary black holes are related by:
    \begin{equation}
        \label{eq:first_law}
        \delta M=\frac{\kappa}{8\pi}\delta A+ \Omega_H \delta J.
    \end{equation}
    
\end{description}

\begin{description}
    \item[Second law](\textit{Area Theorem})
    If the null energy condition is satisfied, then the surface area $A$ of a black hole can never decrease
    \begin{equation}
        \delta A \geq 0.
    \end{equation}
    Thus, if two black holes merge, the area of the final black hole will exceed the total area of the two original black holes. 
\end{description}

\begin{description}
    \item[Third law] If the stress-energy tensor is bounded and satisfies the weak energy condition, then the surface gravity of a black hole $\kappa$ cannot be reduced to zero within a finite advanced time.
\end{description}
These four laws have a remarkable similarity to the thermodynamics ones. At rest a black hole has a energy $E=M$. Now, we examine a thermodynamic system possessing identical energy $E$ and angular momentum $J$ as the black hole. This system is governed by the first law of thermodynamics (with $\mu$ the chemical potential that enforces conservation of angular momentum):
\begin{equation}
    dE=TdS+\mu dJ.
\end{equation}
This is identical to the first law of black hole mechanics (Equation (\ref{eq:first_law})) when we establish the following equivalences:
\begin{equation}
    \label{eq:identifications}
    T=\lambda \kappa, \quad S=\frac{A}{8\pi\lambda},\quad \mu=\Omega_H
\end{equation}
for some constant $\lambda$. Furthermore, by making these identifications, we find that the zeroth law of thermodynamics (which states that the temperature remains constant in a body in thermodynamic equilibrium) becomes the zeroth law of black hole mechanics, and the second law of thermodynamics (which states that entropy increases with time) becomes the second law of black hole mechanics. This resemblance suggests that black holes might be thermodynamic objects. Another reason for considering this is that, without attributing entropy to black holes, one could potentially violate the second law of thermodynamics simply by introducing matter into a black hole. This would effectively lead to a decrease in the total entropy of the universe as observed by an external observer. This is what led Bekenstein in 1972 to propose that black holes possess an entropy proportional to their surface area, as mentioned earlier \cite{BHentropy.Bekenstein}. This proposition presents a serious problem: if Equation (\ref{eq:identifications}) is accurate, then a black hole possesses a temperature and, as a result, must radiate energy similar to any other hot object in a vacuum. However, according to its definition, a black hole cannot emit anything. The resolution to this problem needed a further step: Quantum Field Theory in curved spacetime. As already mentioned, these various concepts were reconciled into a coherent framework through Stephen Hawking's discovery in 1974 \cite{BHexplosions.Hawking}. He demonstrated that when matter is treated in a quantum mechanical manner, a black hole indeed emits radiation, characterized by a blackbody spectrum at the temperature known as the Hawking temperature (analytically found for the first time through the semi-classical calculation made by Hawking that shows that a distant observer will detect a thermal spectrum of particles coming from the black hole \cite{BHexplosions.Hawking}). Therefore, black holes do behave as thermodynamic systems: as we will see in next chapter, thanks to quantum Hawking radiation it became clear that the analogy is in fact an identity. The fact that black holes are thermodynamic systems ---thus their temperature and entropy are not less real than their masses--- reveals a deep connection between quantum mechanics, thermodynamics and gravitation, but at now this connection is still not understood. We do not know which is the microscopic, statistical origin of black hole entropy. A quantum theory of gravity might be necessary to provide the underlying explanation for the entropy of black holes.
\vspace{7mm}\\
In summary, in this chapter we analyzed a Schwarzschild black hole and explored spacetime horizons. Our study concluded with black hole thermodynamics, revealing a challenge: the inconsistency between black holes as depicted in General Relativity, where they cannot emit anything, and the prediction that they possess a temperature, consequently radiating energy. This discrepancy leads us to introduce the framework of Quantum Field Theory in curved spacetime in the next chapter, where this problem finds resolution.

\chapter{Quantum Field Theory in Curved Spacetime}
\label{chap:qftcs}
\begin{chapabstract}
    \begin{adjustwidth}{1cm}{1cm}
        We briefly introduce Quantum Field Theory in curved spacetime \cite{QFTinCurvedST.DeWitt}\cite{book.qfcs.Birrell.Davies} \cite{gr.qc.JacobsonQFT}\cite{2017lectures.Reall}\cite{noteLiberati}\cite{2015lectures.Hartman}, the Unruh effect \cite{Unruh.BHevaporation}, Hawking radiation \cite{BHexplosions.Hawking} and related open issues \cite{Blackholes.Harlow}\cite{noteLiberati}\cite{2015lectures.Hartman}. This holds significance for the purposes of the thesis, as we will need to introduce the acoustic equivalent of Hawking radiation within the context of Analogue Gravity. We also investigate the thermodynamics of spacetime \cite{Jacob_EinsteinEOS}-\cite{ChircoLiberati_noneqThermo} and the viscosity to entropy density ratio \cite{KSS_viscosity_stronglyInteracting}-\cite{overviewETA/S}. 
    \end{adjustwidth}
\end{chapabstract}

\section{Quantum field theory in curved spacetime}
The study of Quantum Field Theory in curved spacetime consists in seeing how a fixed metric ---thus we do not consider the dynamics of the metric in response to energy-momentum described by Einstein field equations--- influences quantum-mechanical matter fields. The epochal event in this study was the discovery that black holes emit thermal radiation at the Hawking temperature \cite{BHexplosions.Hawking}. \\
One relevant difference between Quantum Field Theory in flat spacetime and Quantum Field Theory in curved spacetime is that in a curved spacetime observers in different kinematical states will typically disagree on how many particles are around, while in a flat spacetime every inertial observer will agree on what is the vacuum state and on how many particles are observed. In order to see this explicitly, we study a real scalar field $\phi(t,\mathbf{x})$ in a curved spacetime. During all of this chapter we consider $\hslash=c=G=k_B=1$. In the minimal coupling, the action is
\begin{equation}
    \mathcal{S}=\int d^n x\sqrt{|g|}\left(-\frac{1}{2}g^{\mu\nu}\mathcal{D}_\mu \phi \mathcal{D}_\nu \phi -\frac{1}{2}m^2 \phi^2\right).
\end{equation}
where $\mathcal{D}_\mu$ is the covariant derivative defined in Section \ref{sec:notation-subsection}. The equation of motion of the scalar field is
\begin{equation}
    \label{eq:eom_scalarfield}
    \square\phi-m^2\phi=0,
\end{equation}
where $\square=g^{\mu\nu}\mathcal{D}_\mu \mathcal{D}_\nu$. The inner product on two solutions $\phi_1,\phi_2$ of Equation (\ref{eq:eom_scalarfield}) is
\begin{equation}
    \label{eq:scalarproductCST}
    \langle\phi_1,\phi_2\rangle=-i\int_\Sigma d^{n-1}x  \sqrt{\gamma} n^\mu \left(\phi_1 \mathcal{D}_\mu \phi_2^\ast-\phi_2^\ast \mathcal{D}_\mu \phi_1\right),
\end{equation}
where $\gamma$ is the determinant of $\gamma_{ij}$, that is the induced metric on the spacelike hypersurface $\Sigma$, with a unit normal vector $n^\mu$. The inner product is independent of the choice of $\Sigma$, and $n^\mu \mathcal{D}_\mu$ is the derivative with respect to the time parameter of the foliation of the spacetime that we are considering. It is possible to show that $\langle\phi_1,\phi_2\rangle$ is independent on time if $\phi_1,\phi_2$ are two solutions of the equation of motion and vanish at the lateral edge of the foliated spacetime. The difference from the flat spacetime case emerges here: it is possible to find a complete set of basis modes for solutions to Equation (\ref{eq:eom_scalarfield}), but there will generally be many of them, without a reason for favouring one over any others, and so the notion of vacuum will depend sensitively on the chosen set of basis modes. Indeed, we will generically not be able to classify modes as positive- or negative- frequency since there will not be in general any timelike Killing vector. We can always find a complete set of solutions to Equation (\ref{eq:eom_scalarfield}), $f_i(x^\mu)$ (where the index $i$ can be discrete or continuous), orthonormal with respect to the scalar product in Equation (\ref{eq:scalarproductCST}), and with the corresponding conjugate modes with negative norm. We may expand our field as:
\begin{equation} 
    \phi=\sum_i \left(\hat{a}_i f_i+\hat{a}_i^\dagger f_i^\ast\right)
\end{equation}
where $\hat{a}_i$ and $\hat{a}_i^\dagger$ are the annihilation and creation operators. Therefore, the vacuum state $\ket{0_f}$ is such that
\begin{equation}
    \hat{a}_i \ket{0_f}=0\quad \text{for all } i
\end{equation}
and the number operator $\hat{n}_{fi}$ for each mode is defined as
\begin{equation}
    \hat{n}_{fi}=\hat{a}_i^\dagger \hat{a}_i.
\end{equation}
Due to the basis modes non-uniqueness, we can consider another set of modes $g_i(x^\mu)$ and exactly like what we have done with the $f_i$ modes, we can decompose the field operator as
\begin{equation}
    \phi=\sum_i \left(\hat{b}_i g_i+\hat{b}_i^\dagger g_i^\ast\right)
\end{equation}
and define the vacuum state $\ket{0_g}$ and the number operator as:
\begin{equation}
     \hat{b}_i \ket{0_g}=0\quad \text{for all } i, \quad \hat{n}_{gi}=\hat{b}_i^\dagger \hat{b}_i.
\end{equation}
Now we expand each set of modes in terms of the other, through the so called Bogoliubov transformations:
\begin{equation}
    \begin{aligned}
        & g_i=\sum_f \left( \alpha_{ij} f_j+\beta_{ij}f_j^\ast\right)\\
        & f_i=\sum_f \left( \alpha_{ji}^\ast g_j-\beta_{ji}g_j^\ast\right)
    \end{aligned}
\end{equation}
where $\alpha_{ij}$ and $\beta_{ij}$ are the Bogoliubov coefficients , which can be used also to transform between operators:
\begin{equation}
    \begin{aligned}
        &\hat{a}_i=\sum_j \left(\alpha_{ji}\hat{b}_j+\beta_{ji}^\ast \hat{b}_j^\dagger\right)\\
        & \hat{b}_i=\sum_j \left(\alpha_{ij}^\ast\hat{a}_j-\beta_{ij}^\ast \hat{a}_j^\dagger\right).
    \end{aligned}
\end{equation}
Since $\beta_{ij}$ describes the admixture of creation operators from one basis into the annihilation operators in the other basis, then we expect that if any of the $\beta_{ij}$ are non vanishing then the vacuum states will not coincide. Indeed, if we consider that the system is in $\ket{0_f}$, and we compute the number of $g$-particles in the $f$-vacuum, we get:
\begin{equation}
    \braket{0_f|\hat{n}_{gi}|0_f}=\sum_j |\beta_{ij}|^2.
\end{equation}
Therefore, if any of the $\beta_{ij}$ are different from zero, then the vacuum states will not coincide.

\section{Unruh effect}
\label{sec:UnruhEffect-section}
One striking consequence of curved spacetime is the Unruh effect, historically derived as an attempt to understand the physics underlying the Hawking effect \cite{Unruh.BHevaporation}. It states that in the Minkowski vacuum state a uniformly accelerating observer will observe a thermal spectrum of particles. In other words: the number operator found with the expansion of the field in modes appropriate to the accelerated observer, evaluated in the traditional Minkowski vacuum, gives a thermal spectrum of particles. In order to explicitly see this result, we consider a massless scalar field $\phi$ in two dimensions, so the equation of motion is $\square\phi=0$. Thus, as a first step, we study a two-dimensional Minkowski spacetime (the metric in inertial coordinates is $ds^2=-dt^2+dx^2$) as seen by an observer moving at a uniform acceleration in the $x$-direction of magnitude $\alpha=\sqrt{a_\mu a^\mu}=\text{constant}$ (with $a^\mu=\mathrm{d}^2 x^\mu/\mathrm{d}\tau^2$). Its trajectory $x^\mu(\tau)$, with $\tau$ the proper time, is given by
\begin{equation}
    \label{eq:trajectory_Rindler}
        t(\tau)=\frac{1}{\alpha}\sinh(\alpha\tau), \quad x(\tau)=\frac{1}{\alpha}\cosh(\alpha\tau)
\end{equation}
and obeys
\begin{equation}
    x^2(\tau)=t^2(\tau)+\alpha^2.
\end{equation}
So the accelerated observer travels along an hyperboloid from past null infinity to future null infinity. We now choose new coordinates $(\eta,\xi)$ such that in the wedge $x>|t|$, that we call region \rom{1}, we have:
\begin{equation}
        t=\frac{1}{a}e^{a\xi}\sinh(a\eta), \quad
        x=\frac{1}{a}e^{a\xi}\cosh(a\eta)
\end{equation}
where $a$ is an arbitrary parameter, and $-\infty<\eta,\xi
<+\infty$ cover the wedge $x>|t|$. Thus for fixed $\xi$, trajectories in terms of $\eta$ are the same of the ones in Equation (\ref{eq:trajectory_Rindler}) written in terms of $\tau$: $\eta$ plays the same role of $\tau$. Indeed, in these coordinates, the constant-acceleration path is given by
\begin{equation}
\label{eq:eta_xi_Rindler}
        \eta(\tau) =\frac{\alpha}{a}\tau, \quad 
        \xi(\tau) =\frac{1}{a}\ln\left(\frac{a}{\alpha}\right)
\end{equation}
and therefore we can see that $\tau$ is proportional to $\eta$ and that $\xi$ is constant. Moreover, we can notice that in these coordinates the metric is:
\begin{equation}
    \label{eq:Rindler_metric}
    ds^2=e^{2a\xi}\left(-d\eta^2+d\xi^2\right).
\end{equation}
With this metric, the region \rom{1} is called Rindler space. 
\begin{figure}[ht]
    \centering
    \includegraphics[width=0.75\textwidth]{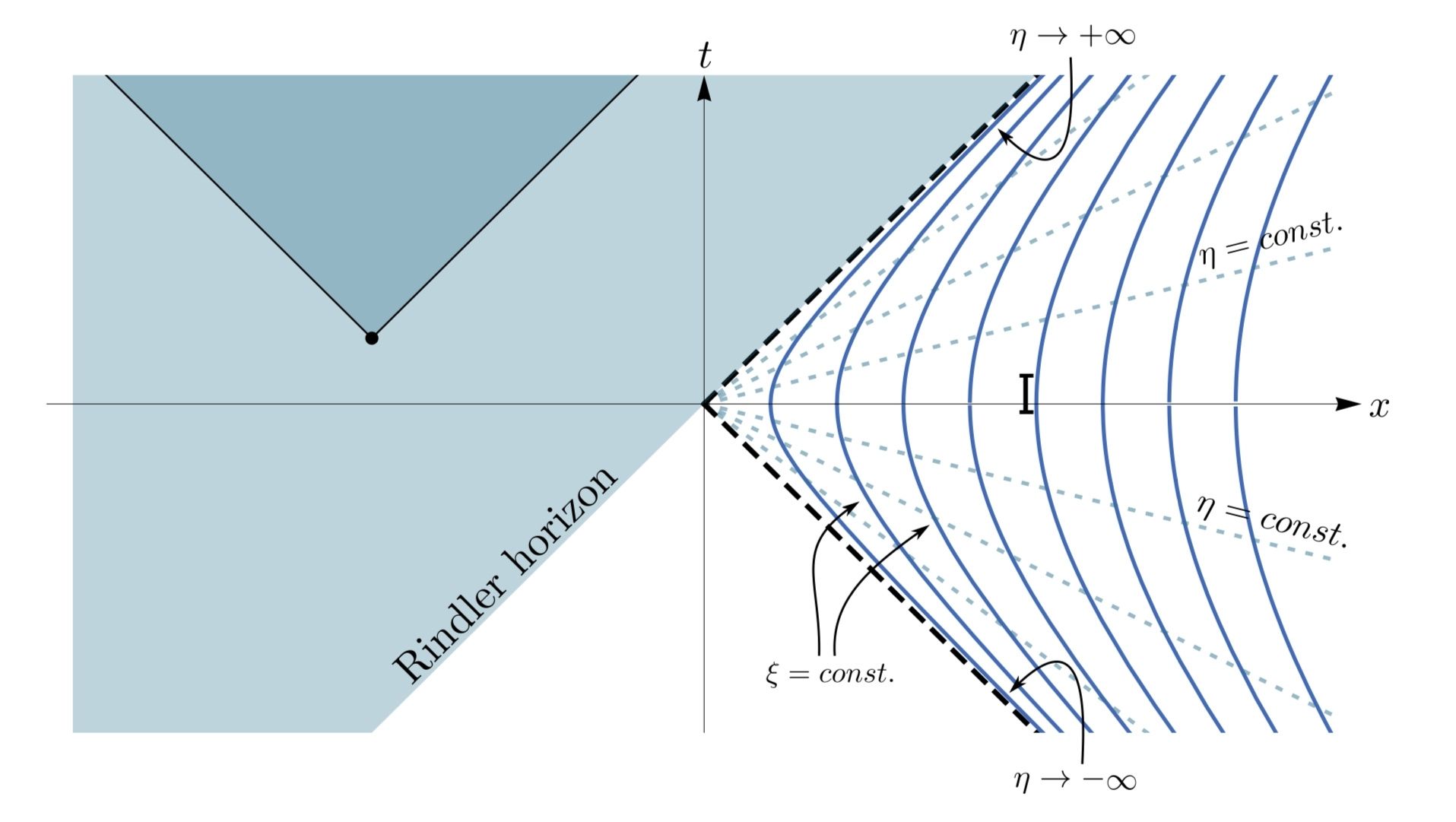}
    \caption{\textit{Rindler space.} Eternally accelerating observers in Minkowski spacetime: their worldlines are labelled by $\xi$ and shown in blue. Events in the shaded region are hidden to them. Region \rom{1} is the Rindler space. Adapted from \cite{Dowker.notes}.}
    \label{fig:Rindler_space}
\end{figure}
The causal structure of this space resembles the one of the region $r>r_s$ of the Schwarzschild solution: the null lines $x=\pm t$ resemble event horizons in the Kruskal diagram (see Figure \ref{fig:Kruskal_Szekeres}), with constant-acceleration paths in Rindler space being related to static observers in Schwarzschild. We can notice that the metric in Equation (\ref{eq:Rindler_metric}) is independent of $\eta$ and thus $\partial_\eta$ is a Killing vector (see Section \ref{sec:notation-subsection}). The Klein-Gordon equation in Rindler coordinates is then
\begin{equation}
    \square \phi=e^{-2a\xi}(-\partial^2_\eta+\partial^2_\xi)\phi=0
\end{equation}
which has normalized plane waves as solutions
\begin{equation}
    f_k=\frac{1}{\sqrt{4\pi \omega_k}}e^{-i\omega_k \eta+ik\xi},
\end{equation}
with $\omega_k=|k|$. Moreover, $f_k$ has positive frequency: $\partial_\eta=-i\omega_k f_k$. Thus $\{f_k,f^\ast_k\}$ is a complete basis that we can use to quantize the field in the Rindler wedge. Thus solutions to the Klein-Gordon equation in the Rindler wedge of a flat two-dimensional spacetime can be expanded both in terms of a Minkowski set of modes and in terms of the Rindler ones, giving  different interpretations as a Fock space of the same Hilbert space. The Rindler number operator in region \rom{1} is
\begin{equation}
    \hat{n}_R(k)=\hat{b}_k^\dagger \hat{b}_k
\end{equation}
and, called $\ket{0_M}$ the Minkowski vacuum, it is possible to show that:
\begin{equation}
    \braket{0_M|\hat{n}_R(k)|0_M}=\frac{1}{e^{2\pi\omega_k/\alpha}-1}\delta(0).
\end{equation}
Therefore we have obtained the Unruh effect: an observer that moves with uniform acceleration $\alpha$ through the Minkowski vacuum observes a Planck spectrum of particles with temperature
\begin{equation}
    \label{eq:Unruh_temperature}
    T=\frac{\alpha}{2\pi}.
\end{equation}
From Equation (\ref{eq:eta_xi_Rindler}) it is clear that an observer moving at $\xi=0$ has a constant acceleration $\alpha=a$, while any other observer moving on a path with $\xi=\text{constant}$ feels an acceleration $\alpha=ae^{-a\xi}$. Thus the temperature, as $\xi\to+\infty$, redshifts all the way to zero. Indeed, a Rindler observer at infinity will be nearly inertial: he would be like an ordinary Minkowski observer and so share the same vacuum state notion. In conclusion, the Unruh effect teaches us that vacuum and particles are observer-dependent notions.

\section{Hawking radiation}
\label{sec:Hawking_rad}
At this point, we are able to introduce the concept of Hawking radiation, which plays a crucial role in the entire line of research in Analogue Gravity. As we will discuss in detail in Section \ref{sec:intro_AG}, historically, it has been the leading motivation for the development of this research field. The Hawking radiation is a thermal radiation in presence of a black hole event horizon \cite{BHexplosions.Hawking}\cite{ParticleCreationBH.Hawking}\cite{Pathintegral.bhThermodynamics:HartleHawking}. Because of the similarity between the causal structure of a maximally extended Schwarzschild spacetime and the Rindler space's one, it is possible to derive the Hawking radiation as a consequence of the Unruh effect ---even though historically the Hawking effect has been discovered before the Unruh one--- for a Schwarzschild black hole \cite{BH.lessons.Jacobson}\cite{carroll.book.}. To do that, we consider a static observer in a Schwarzschild spacetime located at $r_1>r_s$ that moves along orbits of the timelike Killing vector $t^\alpha$. Now, we use what we have seen in Section \ref{sec:killing_horizon-section}, in particular the definition of the redshift factor (Equation 
(\ref{eq:redshift_factor})) as $V=\sqrt{-t_\alpha t^\alpha}$ and its value for static observers in Schwarzschild \cite{carroll.book.}:
\begin{equation}
    V=\sqrt{1-\frac{2M}{r}}.
\end{equation}
From this result, it is possible to show that the magnitude of the static observer acceleration is:
\begin{equation}
    a=\frac{M}{r^2\sqrt{1-2M/r}}.
\end{equation}
Therefore, observers very close to the event horizon located at $r_1$ have a very large acceleration $a_1=a(r_1)$ and so the associated timescale $a_1^{-1}$ is very small compared to $r_s$. Since $r_s$ sets the radius of curvature of the spacetime near the horizon, we have that on timescales $a_1^{-1}\ll r_s$ spacetime looks essentially flat. Now, we assume that close to the horizon the quantum state of a scalar field resembles the Minkowski vacuum as seen by a freely-falling observer. Thus the static observer at $r_1$ near the horizon, with $a_1\gg r_s^{-1}$, looks like an observer in a flat spacetime that moves with constant acceleration: he will detect an Unruh radiation at a temperature $T_1=a_1/2\pi$. If we now consider a static observer at infinity ($r_2\gg r_s$), we have that he should detect the propagated thermal radiation observed near the horizon, thus it is redshifted to a temperature
\begin{equation}
    T_2=\frac{V_1}{V_2}T_1=\frac{V_1}{V_2}\frac{a_1}{2\pi}.
\end{equation}
with $V_i=V(r_i)$. Because of the fact that $V_2 \xrightarrow{r_2 \to \infty}1$, the observed temperature is:
\begin{equation}
    T_2=\lim\limits_{r_1 \to r_s}\frac{V_1 a_1}{2\pi}=\frac{\kappa}{2\pi}
\end{equation}
where $\kappa$ is the surface gravity. Therefore, we have derived the Hawking effect: observers far from the black hole see a thermal radiation flux emitted from the black hole at a temperature proportional to its surface gravity. The difference with uniformly accelerating observers in flat spacetime is that in Schwarzschild spacetime at infinity the static Killing vector has a finite norm. Therefore, using Quantum Field Theory in curved spacetime we have found the definition of the black hole temperature, which was missing in classical theory but required for the consistency of the black hole thermodynamics. Hence, black holes have a well defined temperature (analytically found for the first time through the semi-classical calculation made by Hawking that shows that a distant observer will detect a thermal spectrum of particles coming from the black hole \cite{BHexplosions.Hawking}), that is called Hawking temperature
\begin{equation}
    \label{eq:Hawking_temp}
    T_{\text{BH}}=\frac{\kappa}{2\pi}.
\end{equation}
For a Schwarzschild black hole 
\begin{equation}
    T_{\text{Schwarzschild}}=\frac{1}{8\pi  M}= 1.2 \times 10^{26} K \left(\frac{1g}{M}\right)\simeq 6\times 10^{-8} K \left( \frac{M_\odot}{M}\right)
\end{equation}
where $M_\odot\sim 10^{30} kg$ is the mass of the Sun. So the radiation from an astrophysical black hole would be hopelessly observable since it is at a much lower temperature than the one of the Cosmic Microwave Background ($\simeq 2.7 K$). We note also that black holes are thermodynamically unstable, they have a negative heat capacity: the more they evaporate, the more their mass decreases, the more their temperature increases, the more they emit. Moreover, the lifetime of a black hole goes like \cite{gr.qc.JacobsonQFT}:
\begin{equation}
    t_{\text{evap}} \propto M^3.
\end{equation}
Therefore for $M=M_\odot$ we have $t_{\text{evap}} \simeq 10^{67} \,\text{years}$; while for instance for primordial black holes, that are hypothetical black holes that formed soon after the Big Bang not from a star collapse, of mass $M=10^{11} kg$ we have $t_{\text{evap}} \simeq 3 \cdot 10^{9}\, \text{years}$: we could see their explosion since the age of the Universe is $\simeq 13.7 \cdot 10^9 \, \text{years}$ \cite{arxivPrimordialBH}. With this proportionality between black hole's temperature and the surface gravity, it is possible to find from the black hole mechanics laws that the entropy of black holes is
\begin{equation}
    \label{eq:entropy_BH}
    S_{\text{BH}}=\frac{A}{4} \simeq \frac{A}{10^{-69} m^2}
\end{equation}
where $A$ is the horizon surface and $m$ indicates meters. This is known as the Bekenstein-Hawking entropy for black holes. The microscopic interpretation for this entropy is subject of ongoing investigation, in particular one would like to understand which degrees of freedom are responsible for it.

\subsection{Open issues with Hawking radiation}
\label{sec:openissueHawking_entropy}
The discovery of Hawking radiation led to new issues: the trans-Planckian problem and the information loss problem.

\subsubsection{Trans-Planckian problem}
The trans-Planckian problem refers to an issue in Hawking's initial calculation for the derivation of Hawking radiation: it involves quantum particles with wavelengths that, close to the black hole's event horizon, are shorter than the Planck length $l_P\sim 10^{-33}\,cm$. This phenomenon arises because of the unique behavior near the horizon. Indeed, when we trace a particle emitted from a black hole with a finite frequency back to the horizon, it must have an infinite frequency, implying a trans-Planckian wavelength (see for example \cite{noteLiberati}). Since frequencies are not covariant objects, one might think that boosting to a suitable Lorentz frame can remove the issue. Nevertheless, we are specifically referring to UV s-waves originating from the black hole, and there is no way to eliminate these by boosting to a different frame.

\subsubsection{The information loss problem}
\label{sec:informationLoss-sec}
Work of Israel, Carter, Hawking, Wheeler and others implies the no-hair theorem: a stationary black hole can be completely characterised by only three independent externally observable classical parameters: its mass $M$, its electric charge $q$ and its angular momentum $J$. Consequently, the no-hair theorem presents a paradox. When a thermodynamic system vanishes beyond a black hole's event horizon, its entropy is lost to an external observer. Hence, at an initial glance, it appears that this physical process violates the second law of thermodynamics, which asserts that entropy cannot decrease. As the black hole swallows the system, the area of the event horizon usually increases, thus could this increase in area be seen as a form of compensation for the loss of entropy associated with the matter that disappeared into the black hole? The answer is yes and we already know it, since we have seen that black holes possess an entropy which is proportional to their area (Equation (\ref{eq:entropy_BH})). Moreover, Bekenstein proposed a generalized version of the second law of thermodynamics \cite{BHentropy.Bekenstein}\cite{Generalizedsecondlaw.Bekenstein}: the sum of black hole entropy, $S_{\text{BH}}$, and the ordinary entropy of matter and radiation fields in the black hole exterior region, $S_{\text{matter}}$, cannot decrease. This hypothesis thus posits that processes involving black holes are characterized by the following relationship:
\begin{equation}
    \Delta (S_{\text{BH}}+S_{\text{matter}})\ge 0.
\end{equation}
The generalized second law of thermodynamics establishes a distinctive connection between thermodynamics, gravity, and quantum theory, providing us with a rare glimpse into the elusive realm of Quantum Gravity. Nevertheless, there is presently no comprehensive proof, one rooted in the fundamental microscopic principles of quantum gravity, for the validity of this principle. The information loss problem is related to the nature of entropy in this context. Indeed because of the Hawking radiation's prediction that a black hole will gradually evaporate within a finite time, a black hole cannot be eternal and thus at a certain point in time the singularity vanishes. This suggests that information is either truly lost within black holes, or that some mechanism within Hawking radiation has preserved this information. As black holes emit Hawking radiation, the particles that escape are entangled with those that have entered the black hole. This entanglement gives rise to an associated entropy known as the von Neumann entropy $S_{\text{vN}}$ (or entropy of entanglement), that captures the concept of information. For a state characterized by the density matrix $\rho$ the von Neumann entropy is:
\begin{equation}
    S_{\text{vN}}=-\text{tr}(\rho \ln\rho).
\end{equation}
Decomposing the density matrix as $\rho=\sum\limits_i \lambda_i \ket{i}\bra{i}$, where $\lambda_i$ is the set of eigenvalues of $\rho$, we get
\begin{equation}
    S_{\text{vN}}=-\sum\limits_i \lambda_i \ln \lambda_i.
\end{equation}
We can notice that $S_{\text{vN}}$ is invariant under a unitary time evolution. Moreover a pure, separable state would yield $S_{\text{vN}}=0$; while a maximally entangled state would result in $S_{\text{vN}}=\ln N$ with $N$ the dimension of the Hilbert space. Once a black hole has completely evaporated, all that remains is the radiation, which is entangled and consequently in a mixed state. On the other hand, the black hole could have formed from a pure state, since it is usually formed by the gravitational collapse of matter. However, in quantum mechanics evolving from a pure state to a mixed state is impossible under unitary evolution, hence the paradox. Indeed, it seems that black holes can exactly do this. There are many attempts to solve this problem (for a list of them see for example \cite{noteLiberati}), and in order to do that, it is important to understand the nature of $S_{\text{BH}}$. One possibility is to interpret it as a Boltzmann entropy, so $S_{\text{BH}}$ arises from the counting of the microstates responsible for the black hole configuration. In this proposal $S_{\text{BH}}=\log \mathcal{W}$, where $\mathcal{W}$ is the number of microscopic configurations that gives the same $M$, $J$ and $q$ values. However, it is still not clear what are these degrees of freedom and how this proposal can address the loss of information problem. Another possibility is to relate $S_{\text{BH}}$ to the entropy of entanglement. With this approach we can find $S_{\text{BH}}=A/\Lambda^2$, where $\Lambda$ is a UV cutoff needed to make the result finite. Nevertheless this result is only valid for a single field living around the black hole, what happens for the other Standard Model fields is still an open issue. One possibility is that they introduce extra divergent terms, and demanding a finite result will establish a renormalization condition for the Newton constant.

\section{Equilibrium spacetime thermodynamics}
\label{sec:spacetime_thermod_eq}
\begin{figure}[ht]
    \centering
    \includegraphics[width=0.6\textwidth]{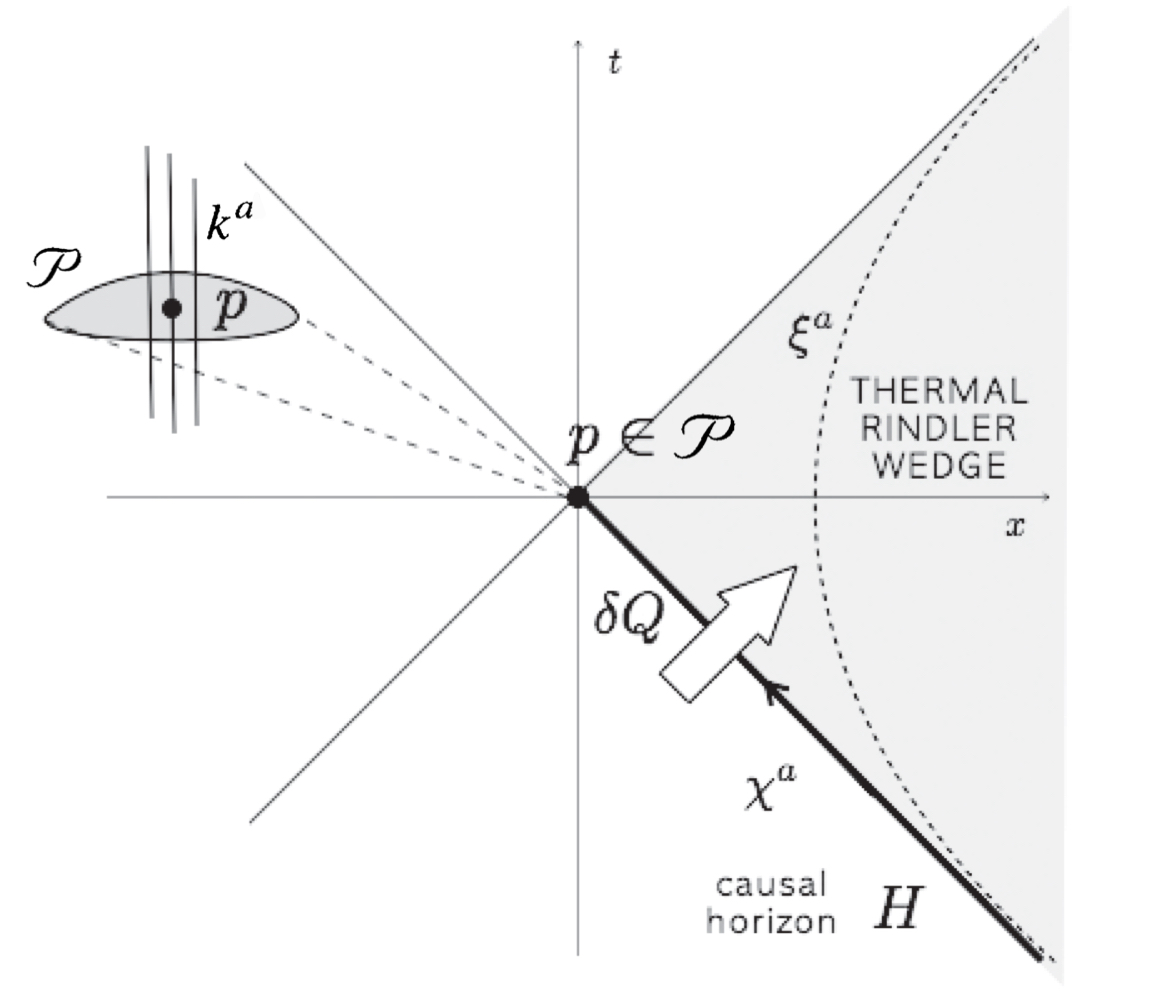}
    \caption{\textit{Thermal Rindler wedge.} Spacetime diagram illustrating the heat flow $\delta \mathcal{Q}$ through the local Rindler horizon $H$ of a 2-surface element $\mathcal{P}$. On the upper left there is a zoom of the region close to $p$, where the grey region $\mathcal{P}$ is a 2-surface element, while $k^a$ is the past pointing tangent null vector. In the diagram, each point represents a two-dimensional spacelike surface. The hyperbola $\xi^a$ is a uniformly accelerated trajectory, and $\chi^a$ represents the approximate boost Killing vector on $H$. Adapted from \cite{ChircoLiberati_noneqThermo}.}
    \label{fig:HeatFlux}
\end{figure}

Now that we have established that black holes possess both temperature and entropy, we can explore whether these properties are intrinsic to horizons in general and how they relate to the causal structure of spacetime. To examine the broader applicability of this framework, we study the thermodynamics of spacetime. Jacobson has derived Einstein field equations from the proportionality of entropy and the horizon area applied to local Rindler causal horizons, together with the fundamental relation $\delta \mathcal{Q}=TdS$ \cite{Jacob_EinsteinEOS}. A local horizon at $p$ has been defined in Section \ref{sec:LocalHorizon-sec}: it is constituted by the congruence of null geodesics, characterized by the past pointing tangent null vector $k^\alpha$, that are orthogonal to a spacelike 2-surface $\mathcal{P}$ including $p$. On the local inertial frame around $p$, we can construct a local Rindler frame by the usual coordinate transformations (see Section \ref{sec:UnruhEffect-section}), see Figure \ref{fig:HeatFlux}. 
We define heat in spacetime dynamics as energy that flows across a causal horizon (see Figure \ref{fig:HeatFlux}), and we consider the degrees of freedom beyond that causal horizon as our system. It is possible to show that the entropy of that system is proportional to the area of that horizon (it is the entanglement entropy). Moreover, we consider the Unruh temperature $T=\kappa/2\pi$ associated to a uniformly accelerated observer just inside the horizon ---the same observer should be used to measure the heat flow--- as the temperature of the system. Since the acceleration diverges as the accelerated worldline approaches the horizon, both the energy flux and the Unruh temperature diverge, but their ratio tends to a finite limit: we analyze the thermodynamics in this limit. Because of the fact that we want to apply the equilibrium thermodynamic relation $\delta \mathcal{Q}=TdS$, we need to further specify the system. Indeed, as we have defined it so far, in general it could be out of equilibrium because of the expansion, contraction or shear of the horizon. For simplicity, we assume that the past directed null normal congruence to $\mathcal{P}$ to have a vanishing expansion and shear at $p$. Let $\chi^\mu$ be the future pointing to the inside past of $\mathcal{P}$ local Killing vector generating boosts orthogonal to $\mathcal{P}$ (see Figure \ref{fig:HeatFlux}). Then the heat flux to the past of $\mathcal{P}$ is
\begin{equation}
    \delta \mathcal{Q}=\int T_{\mu\nu}\chi^\mu \mathrm{d}\Sigma^\nu.
\end{equation}
Considering $k^\mu$ as the tangent vector to the horizon, $\lambda$ as the affine parameter and being $\mathrm{d}\mathcal{A}$ the area element on a cross section of the horizon, we can rewrite:
\begin{equation}
    \label{eq:deltaQ}
    \delta \mathcal{Q}=-\kappa \int \lambda T_{\mu\nu}k^\mu k^\nu \mathrm{d}\lambda \mathrm{d}\mathcal{A}.
\end{equation}
Assuming $\mathrm{d}S=\eta \delta \mathcal{A}$ with $\eta$ an undetermined dimensional constant and since the area variation is $\delta \mathcal{A}=\int \theta \mathrm{d}\lambda \mathrm{d}\mathcal{A}$ with $\theta$ the expansion of the horizon generators, then $\delta \mathcal{Q}=T\mathrm{d}S$ means that the presence of the energy flux is associated with a focusing of the horizon generators. Integrating the Raychaudhuri equation (Equation (\ref{eq:Raychaudhuri_null})) where we neglect the $\theta^2$, $\sigma_{\alpha\beta}\sigma^{\alpha\beta}$ and $\omega_{\alpha\beta}\omega^{\alpha\beta}$ terms ---we do that because we have chosen the local Rindler horizon to be instantaneously stationary at $\mathcal{P}$--- for sufficiently small $\lambda$ we obtain:
\begin{equation}
    \theta=-\lambda R_{\alpha\beta}k^\alpha k^\beta,
\end{equation}
thus
\begin{equation}
\label{eq:deltaA}
    \delta \mathcal{A}=-\int \lambda R_{\alpha\beta}k^\alpha k^\beta \mathrm{d}\lambda \mathrm{d}\mathcal{A}.
\end{equation}
From Equations (\ref{eq:deltaQ}) and (\ref{eq:deltaA}), we see that $\delta \mathcal{Q}=T\eta \delta \mathcal{A}$ can only be valid if for all null $k^\alpha$:
\begin{equation}
    \label{eq:preEinstein-eq}
    T_{\alpha\beta}k^\alpha k^\beta=\frac{\eta}{2\pi}R_{\alpha\beta}k^\alpha k^\beta.
\end{equation}
Thus for some function $f$:
\begin{equation}
    \frac{2\pi}{\eta}T_{\alpha\beta}=R_{\alpha\beta}+fg_{\alpha\beta},
\end{equation}
and from local conservation of energy and momentum and contracted Bianchi identity, we get $f=-R/2+\Lambda$, with $\Lambda$ some constant. In conclusion Einstein's equation holds:
\begin{equation}
    R_{\mu\nu}-\frac{1}{2}Rg_{\mu\nu}+\Lambda g_{\mu\nu}=\frac{2\pi}{\eta}
\end{equation}
with $\eta$ defining Newton's constant: $G=(4\pi\eta)^{-1}$ and so $\eta^{-1/2}$ is identified as twice the Planck length. Thus, the Einstein's equation can be derived as a local constitutive equation for an equilibrium spacetime thermodynamics. This result has greatly strengthen the idea that black holes thermodynamics should be in fact extendable to some more general spacetime thermodynamics. A further step would be that of studying non-equilibrium spacetime thermodynamics.

\section{Non-equilibrium spacetime thermodynamics}
\label{sec:spacetime_therm_noneq}
After Jacobson's thermodynamical derivation of the Einstein's equations, it became clear that a generalization to a non-equilibrium thermodynamical setting was needed \cite{ElingGuedensJacob_noneqThermo}\cite{Eling_HydroST}. Chirco and Liberati have shown that the horizon kinematics define the intrinsic spacetime properties involved in the irreversible processes \cite{ChircoLiberati_noneqThermo}. In the non-equilibrium thermodynamics, the rate of the change in entropy is due both to the rate of entropy exchange with the surroundings $d_e S$ and to the one due to internal processes $d_i S$:
\begin{equation}
    dS=d_e S+d_i S.
\end{equation}
The generalized Clausius relation, $d S=\delta \mathcal{Q}/T+\delta N$ where $\delta N$ is the so called uncompensated heat, gives:
\begin{equation}
    \begin{aligned}
        & d_eS=\frac{\delta \mathcal{Q}}{T} \quad \text{at the reversible level}\\
        & d_iS=\delta N \quad \text{at the irreversible level.}
    \end{aligned}
\end{equation}
Since the null geodesic congruence kinematics around $p$ is expected to be related to the internal spacetime variables (all the thermal information of the Rindler wedge vacuum is recorded on the horizon boundary), we can use the analogy between that congruence and a classical fluid (see Section \ref{sec:membrane-section}) in order to study the non-equilibrium process. Thus, we approximate the local Rindler horizon with a stretched horizon (a timelike hypersurface), that is formally equivalent to 2+1 fluid living on the spacelike two-dimensional cross section of the stretched horizon and moving with velocity $v^\mu$ (unit timelike vector tangent to the hypersurface, that tends to the Killing vector $\xi^\mu$ in the limit in which the Rindler horizon is recovered). From a kinematical point of view, $\nabla_\mu v^\nu$ is the equivalent of the deformation tensor of the congruence $\Tilde{B}_{\alpha\beta}$ defined in Equation (\ref{eq:deformation_tensor-null}). Similarly to the decomposition of $\Tilde{B}_{\alpha\beta}$ into its trace and symmetric trace free-parts (that we have derived in Equation (\ref{eq:deformation_decomposition-null}), but here we consider a null rotation tensor) we have:
\begin{equation}
    \Tilde{B}_{\alpha\beta}=\frac{1}{2}\theta m_{\alpha\beta}+\sigma_{\alpha\beta}.
\end{equation}
We can do a decomposition of $\nabla_\nu v_\mu$:
\begin{equation}
    V_{\mu\nu}=\frac{1}{3}\left(\nabla^\rho v_\rho\right)\delta_{\mu\nu}+\Tilde{V}_{\mu\nu}
\end{equation}
where $V_{\mu\nu}$ is its symmetric component, with $\Tilde{V}_{\mu\nu}$ the symmetric traceless tensor $\Tilde{V}_{\mu\nu}=\left(\nabla_\mu v_\nu+\nabla_\nu v_\mu\right)/2$. The presence of velocity gradients is responsible for the production of internal entropy, since it is related to dissipative processes:
\begin{equation}
    d_iS=-\frac{1}{T}P_{\mu\nu}V^{\mu\nu}
\end{equation}
with $P_{\mu\nu}=p\delta_{\mu\nu}+\Tilde{P}_{\mu\nu}$ the viscous pressure, respectively decomposed in its bulk and traceless components. Under a local equilibrium hypothesis, the following constitutive equations are valid:
\begin{equation}
    \begin{aligned}
        &p=-\zeta\nabla^\rho v_\rho\\
        &\Tilde{P}_{\mu\nu}=-2\eta \Tilde{V}_{\mu\nu}    \end{aligned}
\end{equation}
with $\zeta$ and $\eta$ respectively the bulk and shear viscosity of the fluid, which are coefficients that describes the out-of-equilibrium hydrodynamics of a fluid. The shear viscosity tends to restore the uniform flow reducing the fluid velocity gradient in the direction orthogonal to the fluid flow; 
while the bulk viscosity describes how a fluid with a  velocity gradient in the direction longitudinal to the fluid flow restores the homogeneous flow. These coefficients in the framework of linear response theory are related to the time dependent of equilibrium fluctuations of the pressure tensor through the Green-Kubo formulas \cite{HOHENBERG}\cite{Forster.book}. The internal entropy production reads:
\begin{equation}
    d_i S=\frac{\zeta}{T}\left( \nabla^\rho v_\rho\right)^2+\frac{2\eta}{T}\lVert \Tilde{V} \rVert^2,
\end{equation}
with $\lVert \Tilde{V} \rVert^2=\Tilde{V}_{\mu\nu}\Tilde{V}^{\mu\nu}$. Associating the bulk term $\nabla^\rho v_\rho$ to the expansion $\theta$ and $\Tilde{V}_{\mu\nu}$ to the shear $\sigma_{\alpha\beta}$ of the congruence, we expect to have:
\begin{equation}
    \label{eq:d_iS1}
    d_i S=\frac{1}{T}\int  \left(\zeta \theta^2 +2\eta \lVert \sigma \rVert^2 \right)d\mathcal{A} dt
\end{equation}
where $t$ is the time label along the horizon null hypersurface ($\chi^\mu \nabla_\mu t=1$ where $\chi^\mu$ is the approximate Killing vector). Thus, the kinematical quantities of the horizon congruence, turned on by the local spacetime curvature, identify the irreversible viscous transfer of momentum into the system. Now we want to derive in another way the expression for the internal entropy: assuming $dS=\alpha \delta \mathcal{A}$ we get, up to $\mathcal{O}(\lambda^2)$ (using Raychaudhuri equation):
\begin{equation}
    dS=\alpha \int d\mathcal{A}d\lambda \left[\theta-\lambda \left(\frac{1}{2}\theta^2+ \lVert \sigma \rVert^2 +R_{\mu\nu}k^\mu k^\nu \right)  \right]_p
\end{equation}
where $\lambda$ is the null congruence affine parameter and $\lVert \sigma \rVert^2=\sigma_{\mu\nu}\sigma^{\mu\nu}$. From this equation at the first order in $\lambda$ we get (at the zeroth order we get: $\theta=0$ since the heat flux at $p$ is zero):
\begin{equation}
  T_{\mu\nu}k^\mu k^\nu =\frac{\alpha}{2\pi} \left(R_{\mu\nu}k^\mu k^\nu+\lVert \sigma \rVert^2\right)
\end{equation}
which differs from Equation (\ref{eq:preEinstein-eq}) because of $\lVert \sigma \rVert^2$. We can associate this extra term with $d_i S$:
\begin{equation}
    \label{eq:d_iS2}
    d_i S=-\alpha\int \lambda \lVert \sigma \rVert^2 d\mathcal{A}d\lambda=\frac{\alpha}{\kappa}\int d\mathcal{A}dt \lVert \sigma \rVert^2 \geq 0.
\end{equation}
Comparing this equation with Equation (\ref{eq:d_iS1}), we can interpret the expression in Equation (\ref{eq:d_iS2}) as the standard entropy production term for a fluid with shear viscosity $\eta$ defined as:
\begin{equation}
    \frac{2\eta}{T}=\frac{\alpha}{\kappa}
\end{equation}
thus
\begin{equation}
    \eta=\frac{\alpha}{4\pi}.
\end{equation}
This result is consistent (see Section \ref{sec:KSS-sec}) with the universal relation for the shear viscosity to entropy ratio found in the AdS/CFT context \cite{eta/s_Maldacena}\cite{KSS_holography-hydrodynamics}.

\section{Shear viscosity to entropy ratio}
\label{sec:KSS-sec}
The shear viscosity $\eta$ to entropy density $s$ ratio is conjectured to satisfy the Kovtun-Son-Starinets (KSS) bound: in fact, $1/4\pi$ seems the lower bound for $\eta/s$ of any fluid in nature \cite{overviewETA/S}\cite{KSS_holography-hydrodynamics}. Whether this is an actual lower bound is still unclear. This limit has been firstly derived in the context of AdS/CFT correspondence, that is the gauge/gravity duality which is also a strong/weak duality (it allows us to describe strongly interacting gauge theories in terms of weakly coupled gravitational systems) \cite{eta/s_Maldacena}. Furthermore, the development of non-equilibrium spacetime thermodynamics implies that Rindler spacetime saturates the KSS lower bound. Additionally, Policastro, Son and Starinets have demonstrated that the same universal ratio applies to a $N=4$ supersymmetric Yang-Mills theory dual to a gravitational description \cite{2001Son_eta_supersymm}. In general, the shear viscosity coefficient is given by the Green-Kubo relation valid in the case of small disturbances from equilibrium characterized by long wavelengths and small frequencies with respect to the mean free path and collision time respectively (i.e., the hydrodynamic limit) \cite{Forster.book}\cite{HOHENBERG}\cite{KHALATNIKOV}. The essential aspect of deriving these relationships involves incorporating a microscopic description of a system into the resolution of the macroscopic governing equation. Let us consider the following quantities
\begin{equation}
    \begin{aligned}
        n&(\mathbf{r},t),\\
        \mathbf{g}(\mathbf{r},t)&=m n \mathbf{v}(\mathbf{r},t),\\
        \varepsilon&(\mathbf{r},t),
    \end{aligned}
\end{equation}
with $n$ the number density, $\mathbf{g}$ the momentum density, $\mathbf{v}$ the velocity, and $\varepsilon$ the energy density. They are associated with local conservation laws, which assert that the change in time in the amount of the specific density at a point or within a volume can only be determined by the quantity entering or leaving the volume:
\begin{equation}
    \begin{aligned}
    &\partial_t n(\mathbf{r},t)+\nabla \cdot \frac{\mathbf{g}(\mathbf{r},t)}{m}=0,\\
    &\partial_t g_i(\mathbf{r},t)+\nabla_j \pi_{ij}(\mathbf{r},t)=0,\\
    &\partial_t \varepsilon(\mathbf{r},t)+\nabla \cdot \mathbf{j}_\varepsilon (\mathbf{r},t)=0,
    \end{aligned}
\end{equation}
with $\pi_{ij}$ the momentum tensor current and $\mathbf{j}_\varepsilon$ the energy current flow. These equations express the conservation laws for the system's densities of mass, momentum and energy, respectively. Indeed, their time change is locally compensated by the divergence of a corresponding current, that acts like a force. The system spatial symmetries dictate the tensor structure of the acting forces. Then, adding explicit considerations of entropy production leads to the so-called constitutive relations, given by:
\begin{equation}
    \begin{aligned}
    &\mathbf{g}(\mathbf{r},t)=m n  \mathbf{v}(\mathbf{r},t),\\
    \pi_{ij}(\mathbf{r},t)=p(\mathbf{r},t)\delta_{ij}-\zeta \nabla \cdot \mathbf{v}(&\mathbf{r},t)\delta_{ij}-\eta\left[\nabla_i v_j(\mathbf{r},t)+\nabla_j v_i (\mathbf{r},t)-\frac{2}{3}\nabla \cdot \mathbf{v}\delta_{ij}\right],\\
    \mathbf{j}_\varepsilon=(&\varepsilon+p) \mathbf{v}(\mathbf{r},t)-\kappa \nabla T(\mathbf{r},t).
    \end{aligned}
\end{equation}
The first terms on the right-hand side of each equation, called reactive terms, come about after requiring that no entropy-density $s$ is produced over time, i.e. $\mathrm{d}s(\mathbf{r},t)/\mathrm{d}t=0$, neither directly ($\partial s/\partial t=0$) nor via its spatial gradients ($\mathrm{d}s/\mathrm{d}t\to \nabla s \cdot \mathbf{v}$). These are the total fluid momentum $mn\mathbf{v}$ in terms of the velocity field $\mathbf{v}(\mathbf{r},t)$, the work per unit volume $p\delta_{ij}$, via the average pressure $p$ at equilibrium, and the energy flow $(\varepsilon+p)\mathbf{v}(\mathbf{r},t)$ in terms of $p$ and of the average energy $\varepsilon$ at equilibrium. The other terms are called dissipative, since they result from requiring that entropy be produced, i.e. $\mathrm{d}s/\mathrm{d}t>0$, according to the second law of thermodynamics, which fixes the only possible forces which can couple to the corresponding currents so to ensure odd parity under time reversal. Specifically $\nabla_i v_j$ for the mass flow, and $\nabla T$ for energy flow driven by temperature $T$ gradients. The dissipative terms in the constitutive equations for the energy stress tensor $\pi_{ij}$ and the energy-current vector $\mathbf{j}_\varepsilon$ encompass the transport coefficients: $\zeta$ and $\eta$ are the bulk and shear viscosities, already seen before, while $\kappa$ the thermal conductivity. The missing dissipative term in the equation for $\mathbf{g}$ simply expresses Galilean invariance. We notice that the bulk part contains the divergence of the velocity, while in the shear one only transverse currents will count. Indeed, as previously stated, shear viscosity functions to reestablish a uniform flow by decreasing the gradient of fluid velocity perpendicular to the flow direction. On the other hand, bulk viscosity characterizes how a fluid, with a velocity gradient along the longitudinal direction of the fluid flow, returns to a state of homogeneous flow. Now, we use the constitutive relations and thermodynamic relations into the conservation laws, so that we obtain hydrodynamic equations. Their validity is limited to long-wavelengths ($k\to 0$) and low frequency ($\omega\to 0$) conditions. By solving them and comparing those solutions for the densities to what we get from linear response theory in the same $k\to 0$ $\omega\to 0$ limit, we can derive Green-Kubo relations, which provide us with microscopic expressions for the dissipative transport coefficients \cite{KADANOFFMartin}. In linear response theory the response function for the operator $A_i$ is:
\begin{equation}
    \chi''_{ij}(\mathbf{r} \mathbf{r'}, t-t')=\frac{1}{2}\langle [A_i(\mathbf{r}, t), A_j(\mathbf{r'}, t')]\rangle_{eq},
\end{equation}
where the expectation value of the commutator is calculated in the equilibrium ensemble. Decomposing $\mathbf{g}$ in its longitudinal $\mathbf{g}_L$ and transverse $\mathbf{g}_T$ parts, so such that $\nabla \times \mathbf{g}_L=0$ and $\nabla\cdot \mathbf{g}_T=0$, and introducing the heat flux $q(\mathbf
r,t)$, the Green-Kubo relations are \cite{Forster.book}:
\begin{equation}
    \begin{aligned}
        &\eta=\lim_{\omega\to 0}\lim_{k\to 0}\frac{\omega}{k^2}\chi''_{g_T g_T}(\mathbf{k}, \omega),\\
        \frac{4}{3}\eta+\zeta=\lim_{\omega\to 0}\lim_{k\to 0} &\frac{\omega}{k^2}\chi''_{g_L g_L}(\mathbf{k}, \omega)=\lim_{\omega\to 0}\lim_{k\to 0} \frac{m^2\omega^3}{k^4}\chi''_{nn}(\mathbf{k}, \omega),\\
        &k=\lim_{\omega\to 0}\lim_{k\to 0} \frac{\omega}{k^2}\chi''_{qq}(\mathbf{k}, \omega).
    \end{aligned}
\end{equation}
Notice the $k\to 0$ first, $\omega\to 0$ afterwards order of the limits. Reverse order of the limits applied to $\chi_{ij}(\mathbf{k},\omega)$ provide us with microscopic expressions for the thermodynamic derivatives, i.e. the static susceptibilities, that are connected to the reactive terms. 
\vspace{7mm}\\
In summary, in this chapter, we introduced Quantum Field Theory in curved spacetime and explored spacetime thermodynamics, which led us to investigate the ratio of shear viscosity to entropy density. This ratio holds significance as a prospective avenue for this thesis. In the next chapter, we will delve into Analogue Gravity, a research field where models replicating some properties of gravitational systems are studied. In this context, analogues of effects predicted by Quantum Field Theory in curved spacetime and reviewed in this chapter---the Unruh effect and Hawking radiation---have been observed experimentally.

\chapter{Analogue Gravity}
\label{chap:analoguegravity}

\begin{chapabstract}
    \begin{adjustwidth}{1cm}{1cm}
        A brief review of Analogue Gravity is presented, focusing on analogue models in classical non-relativistic fluids and in Bose-Einstein condensates \cite{Barcelò_analoguegravityReview}-\cite{Garay_sonicBH}. We also discuss the results of two remarkable experiments of this line of research: the Steinhauer's one \cite{Balbinot_DensityCorr}-\cite{2021Kolobov}, and the one of Hu \textit{et al.} \cite{2019HuChin}. 
    \end{adjustwidth}
\end{chapabstract}

\section{Introduction}
\label{sec:intro_AG}
\begin{figure}[ht]
    \centering
    \includegraphics[width=0.6\textwidth]{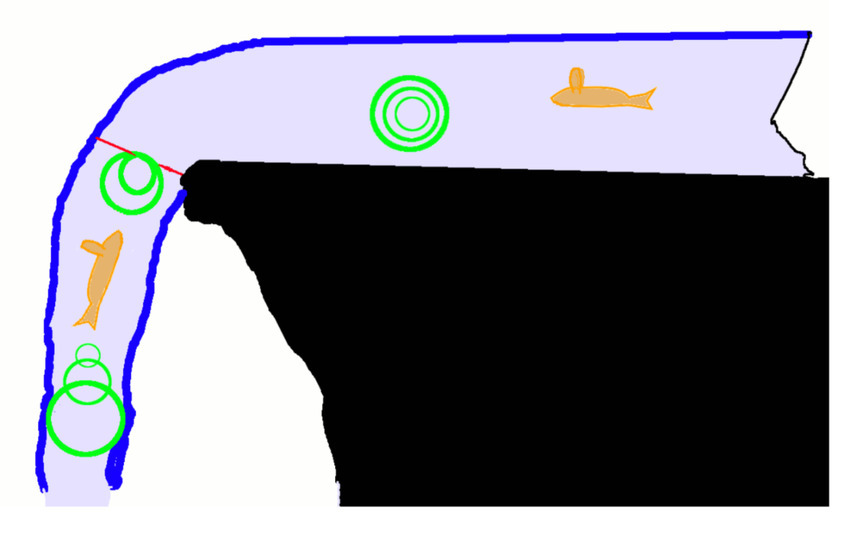}
    \caption{\textit{Supersonic region and analogue black holes.} Unruh's illustration presented in \cite{Unruh_HawkingRadMeasure}. The sound emitted by the falling fish cannot reach the other fish in the river: the moving fluid tips the sound cones (in green). In particular, supersonic flow tips them past a line, that is what becomes the acoustic horizon. Notice the analogy with what happens for light at the event horizon of a black hole.}
    \label{fig:fish}
\end{figure}

Analogue Gravity is a research program that investigates models that mimic different aspects of General Relativity and Quantum Field Theory in curved spacetime. These models are typically, but not always, based on condensed matter physics and some of them can and are realized in the laboratory. Historically, Analogue Gravity models were initially researched to drive forward the theory of black holes or improve their visualization. \\
One example of analogy which helps to understand how black holes trap light was made by Unruh in 1972 at a colloquium \cite{AG_historical}. He compared the mechanism that traps light inside a black hole's event horizon to that of a fish falling into a waterfall, when the water is moving faster than sound (see Figure \ref{fig:fish}). In this case, the fish's noises would not be heard over the waterfall similarly to what happens to light signals inside a black hole.
When Hawking stated that all black holes should radiate in 1974, the situation changed: they were described also by thermodynamics and quantum field theory, not only by classical gravitational theory \cite{BHexplosions.Hawking}. Then, it was understood that the hydrodynamic analogy should be improved to emulate such phenomenon. Note that studying analogies is more than a mere comparison to understand astrophysical scenario: the Hawking radiation cannot be observed at the moment (as seen in Chapter \ref{chap:analoguegravity} the Hawking temperature is very small compared to the CMB one), thus similar quantum phenomena have been duplicated in analogous systems, providing indirect proof for their equivalent occurrences in astrophysics. Clearly, an analogy does not imply identity, and we are not claiming that the analogue models we examine are entirely identical to General Relativity. Our assertion is solely that there exist analogue models, that effectively encompass and faithfully portray some geometrical and dynamical key characteristics of General Relativity. \\
The pursuit of proof for the Hawking effect using analogue experiments became feasible due to the recognition of black holes as thermodynamic entities: while gravitational theories could solely be tested through observations, the realms of thermodynamics and quantum field theory permitted experimentation. Around 1980, Unruh further investigated his fish analogy in an effort to seek experimental proof of Hawking radiation within a laboratory setting \cite{Unruh_experimentalBHev}. He considered the motion of sound waves in a convergent, irrotational fluid flow and found that the equation of motion for this massless scalar field was the same as for the motion of a scalar wave within a specific geometry, whose metric resembles the Schwarzschild's one. This configuration is commonly referred to as a sonic black hole. \\
This acoustic analogue of black holes provided physicists with the prospect of realizing an experimental proof for an astrophysical phenomenon. It did not definitively resolve the inquiries arising from the concept of black hole radiation, such as where are the particles emitted from, what is the source of the radiation, how does this process obey causality and energy conservation, how does Hawking radiation relate to the concept of black hole information, and so on. However, it did offer a valuable and tangible path for theoretical predictions. \\
Following years of theoretical advancements aimed at measuring the Hawking effect in laboratory settings, technology finally advanced to a point where the first experiments emerged in the late 2010s. They can be separated into fluid-based experiments and the optical ones, but both of them observe the Hawking effect as the scattering of paired excitations on either side of the horizon \cite{2008Rousseaux_exp}\cite{2011Weinfurtner_exp}. In 2019 Hawking correlations across the horizon in an atomic Bose-Einstein condensate have been observed \cite{2019MunozdeNova}. As of now, there remains an ongoing discussion regarding whether spontaneous emission from the quantum vacuum has been detected in atomic experiments. A potential method to distinguish between thermal amplification and quantum noise would involve measuring the entangled characteristics of observed Hawking correlations. While the confirmation of spontaneous emission is still pending, the existence of these correlations does indicate paired interactions at the black hole horizon due to the Hawking effect, thus verifying this aspect of Quantum Field Theory in curved spacetime within a laboratory context. \\
Present-day investigations in this field encompass three primary aspects that extend beyond the initial aim of Analogue Gravity ---the observation of the Hawking effect---: exploring novel phenomena within field theory, advancing our comprehension of condensed matter systems, and forecasting new field-theoretic phenomena. The experiments also allow to envision the Hawking effect and related phenomena in different geometries. For example, in \cite{Visser_acousticBH} Visser generalized Unruh's analogy, discussing non-spherical geometries, like $(2+1)$-dimensions vortex flow. Furthermore, as previously mentioned, drawing parallels to the astrophysical/cosmological context is presently aiding in enhancing our understanding of condensed matter systems. For instance, the observation of the vortex's relaxation in Quasi-Normal Modes (QNMs) in 2020 paved the way for the application of techniques from relativity to condensed matter systems \cite{2020vortex_qnm}. In astrophysics, the spectrum of QNMs is independent of the properties of the object that caused the black hole disturbance but it relies solely on the black hole's mass and angular momentum \cite{maggiore2018gravitational}\cite{2015GW}. In fact, according to the principles of general relativity, along with the electric charge, these are the only three parameters required to completely describe a black hole, see the discussion in Section \ref{sec:informationLoss-sec}. Consequently, the observation of QNMs permits the comprehensive characterization of the perturbed black hole through a process known as black hole spectroscopy \cite{BHspectroscopy}. This same principle applies to vortex flows, which can also be fully characterized through the spectroscopic examination of their QNMs, which can be referred to as "analogue black hole spectroscopy" \cite{ABHspectroscopy}\cite{vortex_spectrosocpy}. This example shows how analogy with astrophysical phenomena, independently on their empirical validations, can help to advance the theory in condensed matter systems.

\section{Acoustic metric for non relativistic fluids}
Let us consider a fluid that is barotropic, i.e. the number density $n$ is a function only of the pressure $p$, inviscid, i.e. with zero viscosity flow, and with irrotational, i.e. vorticity free, and possibly time-dependent flow. We define $\theta$ as the velocity potential $\mathbf{v}=\nabla \theta/m$ with $\mathbf{v}$ the velocity of the fluid and $m$ the mass of particles composing it. Now, we consider $n$, $p$ and $\theta$ as made by a background quantity and a fluctuation on it:
\begin{equation}
    \label{eq:fluctFLuids}
    \begin{aligned}
        &n=n_0+\epsilon n_1 +\mathcal{O}(\epsilon^2)\\
        &p=p_0+\epsilon p_1+\mathcal{O}(\epsilon^2)\\
        &\theta=\theta_0+\epsilon \theta_1 +\mathcal{O}(\epsilon^2).\\
    \end{aligned}
\end{equation}
It is possible to show, and we will do it later, that the equation of motion for $\theta_1$, that describes an acoustic disturbance, is identical to the d’Alembertian equation of motion for a minimally-coupled massless scalar field propagating in a (3+1)-dimensional Lorentzian geometry
\begin{equation}
    \label{eq:FieldEq_barotropicFluid}
    \square \theta_1=\frac{1}{\sqrt{|g|}}\partial_\mu \left(\sqrt{|g|}g^{\mu\nu}\partial_\nu \theta_1\right)=0.
\end{equation}
Therefore, the propagation of sound is governed by an acoustic effective metric $g_{\mu\nu}(t,\mathbf{x})$, whose signature is $(-,+,+,+)$. The metric is completely specified by three scalars: the two background quantities $n_0$ and $\theta_0$ defined in Equation (\ref{eq:fluctFLuids}) and local speed of sound $c_s$. We remark now that the fluctuations defined in Equation (\ref{eq:fluctFLuids}) do not change the metric, but they have to be interpreted as fluctuations on the top of the metric. Actually, the three degrees of freedom per point in spacetime of the metric are reduced to two because of the equation of continuity. In $(t,x,y,z)$ coordinates the emergent metric $g_{\mu\nu}(t,\mathbf{x})$ is:
\begin{equation}
    \label{eq:analogueMetric_barotropicFluid}
    g_{\mu\nu}(t,\mathbf{x})=\frac{n}{mc_s}
    \begin{pmatrix}
        -(c_s^2-v^2)&-v_x&-v_y&-v_z\\
        -v_x&1&0&0\\
        -v_y&0&1&0\\
        -v_z&0&0&1
    \end{pmatrix}
    .
\end{equation}
In general, when the fluid is non-homogeneous and flowing, the acoustic Riemann tensor associated with this Lorentzian metric will be non-zero.

\begin{remark}
    Sound waves do not ``see'' the physical spacetime Minkowski metric $\eta_{\mu\nu}$, which couples to the fluid particles: we consider Galilean relativity for the underlying fluid mechanics. However, they couple to the effective acoustic metric $g_{\mu\nu}$. The acoustic geometry naturally acquires the crucial characteristic of ``stable causality'' from the underlying Newtonian time parameter: there exists a global time function $t$ whose gradient is everywhere timelike and future-directed. The acoustic metric is related to the distribution of matter in a simple algebraic fashion (not by Einstein equations). Furthermore, the conformal factor in Equation (\ref{eq:analogueMetric_barotropicFluid}) may not be uniform, but it does not alter the causal structure of the emergent curved spacetime.
\end{remark}

Now, we want to derive Equations (\ref{eq:FieldEq_barotropicFluid}) and (\ref{eq:analogueMetric_barotropicFluid}). The fundamental equations of fluid dynamics are:
\begin{equation}
    \begin{aligned}
        m\partial_t n+\nabla \cdot (mn \mathbf{v})=0 \quad &\text{continuity equation}\\
        mn \frac{\mathrm{d}\mathbf{v}}{\mathrm{d}t}=mn [\partial_t \mathbf{v}+(\mathbf{v}\cdot \nabla)\mathbf{v}]=\mathbf{f} \quad &\text{Euler's equation}
    \end{aligned}
\end{equation}
Under the assumption of zero viscosity, the force density is:
\begin{equation}
    \mathbf{f}=-\nabla p.
\end{equation}
Now, as said we consider the fluid to be barotropic, so we can define the specific enthalpy (an intensive property) as only a function of $p$:
\begin{equation}
    h(p)=\int_0^p \frac{\mathrm{d}p'}{n(p')}, \quad \nabla h=\frac{1}{n} \nabla p.
\end{equation}
Thus, using that $\mathbf{a}\times (\nabla \times \mathbf{b})=\nabla (\mathbf{a}\cdot \mathbf{b})-(\mathbf{a}\cdot \nabla)\mathbf{b}$ for $(\mathbf{v}\cdot\nabla)\mathbf{v}$, and considering the flow to be locally irrotational, Euler's equation becomes:
\begin{equation}
    \partial_t \theta + h+ \frac{1}{2m}(\nabla \theta)^2=0.
\end{equation}
Now we linearise the continuity and the Euler's equations around some assumed background. In this way we define (linear) sound and more generally acoustical disturbances as the linearised fluctuations in the dynamical quantities like in Equation (\ref{eq:fluctFLuids}). We can thus decompose the exact motion characterized by exact variables $(n,p,\theta)$ into two components: an average bulk motion $(n_0,p_0,\theta_0)$ and small amplitude acoustic disturbances denoted by $(\epsilon n_1,\epsilon p_1,\epsilon\theta_1)$. Typically, disturbances with sufficiently low frequencies and long wavelengths are grouped together with the overall bulk motion, whereas disturbances with higher frequencies and shorter wavelengths are categorized as acoustic disturbances. The particular physical assumption that forms the basis for the linearization program, limiting the applicability of our analysis, involves the necessity for the high-frequency, short-wavelength disturbances to have a small amplitude. The linearized continuity equation gives rise to a set of two equations:
\begin{equation}
\label{eq:con1}
    \begin{aligned}
        &m\partial_t n_0+\nabla \cdot(mn_0\mathbf{v}_0)=0\\
        &m\partial_t n_1+m\nabla \cdot(n_1 \mathbf{v}_0+n_0 \mathbf{v}_1)=0.
    \end{aligned}
\end{equation}
The linearized Euler's equation instead leads to:
\begin{equation}
\label{eq:eul1}
    \begin{aligned}
        &\partial_t \theta_0+h_0 +\frac{1}{2m}(\nabla \theta_0)^2=0\\
        &\partial_t \theta_1+\frac{p_1}{n_0}+\mathbf{v}_0\cdot \nabla \theta_1=0.
    \end{aligned}
\end{equation}
Rearranging (\ref{eq:eul1}) and substituting it in (\ref{eq:con1}), we obtain \cite{Barcelò_analoguegravityReview}:
\begin{equation}
    \label{eq:waveeq_velocitypot_nometrica}
    -\partial_t \left(\frac{\partial n}{\partial p}n_0\left(\partial_t \theta_1+\frac{1}{m}\nabla\theta_0\cdot \nabla \theta_1\right)\right)+\nabla \cdot \left(\frac{n_0}{m}\nabla \theta_1-\frac{\partial n}{\partial p} \frac{n_0}{m}\nabla\theta_0\left(\partial_t \theta_1+\frac{1}{m}\nabla\theta_0\cdot \nabla \theta_1\right)\right)=0.
\end{equation}
The wave equation presented here describes the propagation of the linearized scalar potential $\theta_1$. Moreover, once it is established, using Euler's equation we can subsequently determine $n_1$: this wave equation provides a comprehensive description of the propagation of acoustic disturbances. The background fields are required to satisfy the equations of fluid motion for a barotropic, inviscid, and irrotational flow. However, beyond these restrictions, they can exhibit unrestricted temporal and spatial variations. Now we notice that, being the local speed of sound $c_s$ defined as 
\begin{equation}
    c_s^{-2}=\frac{\partial m n}{\partial p},
\end{equation}
if we consider the matrix
\begin{equation}
    f^{\mu\nu}(t, \mathbf{x})=\frac{n_0}{m c_s^2}\begin{pmatrix}\begin{array}{c;{2pt/2pt}cc}
        -1 &-v_0^j\\
        \hdashline[3pt/4pt]
        -v_0^i&(c_s^2\delta^{ij}-v_o^i v_0^j)
    \end{array}
    \end{pmatrix}
\end{equation}
and introduce (3+1)-dimensional spacetime coordinates, $x^\mu=(t, x, y ,z)$, then Equation (\ref{eq:waveeq_velocitypot_nometrica}) is written as:
\begin{equation}
    \partial_\mu(f^{\mu\nu}\partial_\nu \theta_1)=0.
\end{equation}
By comparing this equation to the equation of motion for a scalar field $\phi$ propagating in a (3+1)-dimensional Lorentzian geometry
\begin{equation}
    \square \phi=\frac{1}{\sqrt{|g|}}\partial_\mu \left(\sqrt{|g|}g^{\mu\nu}\partial_\nu \phi\right)=0,
\end{equation}
we can rewrite the wave Equation (\ref{eq:waveeq_velocitypot_nometrica}) as $\square\theta_1=0$, provided that we identify
\begin{equation}
    \sqrt{|g|} g^{\mu\nu}=f^{\mu\nu}.
\end{equation}
This implies
\begin{equation}
    g^{\mu\nu}(t, \mathbf{x})=\frac{m}{n_0 c_s}\begin{pmatrix}
        \begin{array}{c;{2pt/2pt}cc}
        -1 &-v_0^j\\
        \hdashline[3pt/4pt]
        -v_0^i&(c_s^2\delta^{ij}-v_o^i v_0^j)
    \end{array}
    \end{pmatrix}
    .
\end{equation}
Inverting this matrix we can determine the acoustic metric itself
\begin{equation}
    g_{\mu\nu}(t, \mathbf{x})=\frac{n_0}{ mc_s}\begin{pmatrix}
        \begin{array}{c;{2pt/2pt}cc}
        -(c_s^2-v_0^2) &-(v_0)_j\\
        \hdashline[3pt/4pt]
        -(v_0)_i&\delta_{ij}
    \end{array}
    \end{pmatrix}
    .
\end{equation}
The acoustic interval is then:
\begin{equation}
    ds^2=g_{\mu\nu}\mathrm{d}x^\mu \mathrm{d}x^\nu=\frac{n_0}{mc_s}\left[-c_s^2 \mathrm{d}t^2 + \left(\mathrm{d} x^i-v_0^i \mathrm{d}t\right)\delta_{ij}\left(\mathrm{d} x^j-v_0^j \mathrm{d}t\right)\right].
\end{equation} 

Recognizing how to establish a horizon in the context of an acoustic metric is straightforward. The fluid flow remains subcritical when its speed is slower than the speed of sound, $v_0^2<c_s^2$, and it becomes supercritical when the flow exceeds this speed, $v_0^2>c_s^2$. At the interface between these regions, an acoustic horizon forms. This horizon prevents perturbations from crossing in both directions: the speed at which these acoustic perturbations propagate is the speed of sound, making it impossible for them to traverse the horizon from the supercritical region to the subcritical one. Near the acoustic horizon, the behavior of perturbations will resemble that of a massless scalar field near a trapping horizon in the spacetime of a gravitational black hole.

\section{Acoustic metric for non relativistic Bose-Einstein condensates}
One of the primary objectives in researching analogue models of gravity is to simulate semi-classical gravity phenomena like the Hawking radiation effect. Systems exhibiting a high level of quantum coherence, extremely low temperatures, and a slow speed of sound offer the most promising test environment. \\
There exist two categories of quantum systems consisting of identical particles: bosonic and fermionic. As per the spin-statistic theorem, bosons possess integer spins, leading to a completely symmetric wavefunction in a many-body bosonic system. In contrast, fermions have half-integer spins, resulting in a completely antisymmetric wavefunction in a many-body fermionic system \cite{Forster.book}-\cite{PitaevskiiBEC}. The connection between spin and symmetry ultimately arises from how states transform under Lorentz transformations \cite{Schwinger_quantizedFields}. The second quantization formalism is used to describe these many-body systems, and the symmetry or antisymmetry of the wavefunctions is incorporated into the construction of the Fock space. This is achieved by defining the appropriate commutation relations between the ladder operators: bosonic operators have commutative relations, whereas fermionic operators have anticommutative relations. These commutation relations lead to a direct implication: in the case of fermions, any 1-particle state can only be either unoccupied or occupied by one fermion, which is known as the Pauli exclusion principle; whereas for bosons, there is no limit to the number of bosons that can occupy the same 1-particle state and this leads to the concept of condensation. Indeed, focusing on bosons, we have that when we cool the system below a critical temperature and a large (macroscopic) number of particles occupy the ground state, then the system enters a phase referred to as Bose-Einstein condensation. This framework characterize not only fundamental particles but also, with reasonable accuracy, cold atoms. \\
In 2016, Steinhauer conducted the first experimental detection of the Hawking effect within an acoustic black hole using a Bose-Einstein condensate \cite{2016Steinhauer}. Moreover, in 2019, Hu \textit{et al.} have observed the Unruh effect in a disk shaped Bose-Einstein condensate \cite{2019HuChin}. A brief overview of these two experiments is given in Section \ref{sec:experiments-sec}.\\
For a Bose gas in the dilute gas approximation, so when the average spacing between the particles in a gas is greater than the scattering length $a$, we can approximate the true interaction potential between particles by a Fermi pseudo-potential: at sufficiently low temperatures, where the de Broglie wavelength is much longer than the range of boson-boson interaction, the scattering process can be well approximated by the s-wave scattering and therefore the potential is spherically symmetric. The replacement of the true two-body interaction potential by a pseudo-potential is a crucial step in connecting mean-field treatments with many body theory. In this approximation and using the Hartree-Fock approximation, the total wavefunction of the system of $N$ bosons is taken as a product of single-particle functions $\hat{\Psi}$. We can then describe a Bose gas by a quantum field $\hat{\Psi}$ satisfying the Gross-Pitaevskii equation, where the interaction potential is represented by the Fermi pseudo-potential \cite{RevBEC}:
\begin{equation}
\label{eq:GrossQunatumFiledEntire}
    i\hslash \frac{\partial}{\partial t}\hat{\Psi}=\biggl(-\frac{\hslash^2}{2m}\nabla^2 + V_{\text{ext}}(\mathbf{x})+\gamma(a) \hat{\Psi}^{\dag}\hat{\Psi}\biggr) \hat{\Psi}
\end{equation}
where $V_{\text{ext}}$ is the external potential and $\gamma(a)=\frac{4\pi a\hslash^2}{m}$ parameterises the strength of the interactions between the different bosons in the gas.\\
The quantum field can be divided into a macroscopic (classical) condensate $\psi(x)=\langle\hat{\Psi}(x)\rangle$, and a fluctuation $\hat{\varphi}(x)$: 
\begin{equation}
    \label{eq:psi_decomp}
    \hat{\Psi}(x)=\psi(x)+\hat{\varphi}(x),
\end{equation}
where the expectation value is defined as the trace of the density matrix times the operator itself. Thus this approach enables us to include the excitations of the condensate. This splitting indicates a division of scales, with the fluctuation $\hat{\varphi}(x)$ being linked to rapid and small scale changes and the background field $\psi(x)$ to long-distance and long-time scales. Neglecting back-reactions of the quantum fluctuations on the background, the equation for the classical wave function coming from Equation (\ref{eq:GrossQunatumFiledEntire}) becomes closed and the mean-field is described by the solution of the Gross-Pitaevskii equation \cite{PitaevskiiBEC}
\begin{equation}
\label{eq:gross1}
    i\hslash \frac{\partial}{\partial t}\psi=\biggl(-\frac{\hslash^2}{2m}\nabla^2 + V_{\text{ext}}(\mathbf{x})+\gamma(a) n_c \biggr) \psi
\end{equation}
with $n_c\equiv |\psi(t,\mathbf{x})|^2$ is the condensate density. Equation (\ref{eq:gross1}) is the most commonly adopted expression to approximate the dynamics of the condensate wavefunction. The linearized quantum fluctuations, on the other hand, follow the Bogoliubov-de Gennes equation
\begin{equation}
    i\hslash \frac{\partial}{\partial t}\hat{\varphi}(t,\mathbf{x})=\biggl(-\frac{\hslash^2}{2m}\nabla^2 + V_{\text{ext}}(\mathbf{x})+\gamma(a)2 n_c \biggr)\hat{\varphi}(t,\mathbf{x})+\gamma(a) \psi^2(t,\mathbf{x}) \hat{\varphi}^{\dag}(t,\mathbf{x})
\end{equation}
which is coupled to the mean-field equation. We remark that this set of coupled equations are written as a consequence of having neglected back-reactions within the self consistent mean-field approximation. The Bogoliubov-de Gennes equation is a linear differential operator equation for $\hat{\varphi}$ and $\hat{\varphi}^\dagger$. In principle, it can be diagonalized by pairing it with its corresponding conjugate equation, allowing us to determine the propagation modes and their associated dispersion relations. These new propagating modes represent quasi-particles, which propagate as non-interacting particles in the new effective theory established by the Bogoliubov-de Gennes equation over the background provided by the solution of the Gross-Pitaevskii equation. For an homogeneous condensate, it is feasible to examine in detail the Bogoliubov-de Gennes equation, and the investigation of quasi-particles is well established. In this scenario, the condensate wavefunction is presumed to undergo a linear phase variation over time, while its number density remains a constant and uniform real value. From the Fourier transformed in space and time of the Bogoliubov-de Gennes equation and its conjugate, we can derive the dispersion relation of the quasi-particles \cite{Bogolyubov:eq}\cite{landau_pitaevskii}:
\begin{equation}
    \label{eq:dispersion_relation}
    \begin{gathered}
    \omega(k)=\sqrt{\frac{\gamma(a) n_c}{m}k^2}\sqrt{1+\frac{k^2}{4\gamma(a)n_c m}}
    =c_s|k| \sqrt{1+\frac{1}{4}\xi^2k^2}=c_s|k|(1+\mathcal{O}(k^2))
     \end{gathered}
\end{equation}
with 
\begin{equation}
\label{eq:speedSound_BEC}
    c_s=\sqrt{\frac{\gamma(a) n_c}{m}}
\end{equation}
\begin{equation}
\label{eq:healing_length_BEC}
     \xi=\frac{1}{\sqrt{\gamma(a)n_c m}},
\end{equation}
the speed of sound and healing length, respectively. We observe that the dispersion relation (Equation (\ref{eq:dispersion_relation})) appears nearly linear when momenta are small, resembling the dispersion relation of photons in a vacuum. However, when momenta become higher, this linearity breaks down because the higher-order terms involving momentum $k$ can no longer be neglected. This scenario with the uniform condensate is quite straightforward yet illustrative. The dispersion relation we have identified, which is linear for small momenta, forms the foundation of the concept of Analogue Gravity in Bose-Einstein condensates. Within this linear regime, the speed at which Bogoliubov quasi-particles propagate through the condensate is analogous to the speed of light for the propagation of photons in empty space. This observation is the initial indication that Bose-Einstein condensates can potentially replicate the characteristics of massless fields in spacetime. \\
In Equation (\ref{eq:dispersion_relation}) the parameter $c_s$ corresponds to the velocity of sound within the condensate, and it establishes the connection between lengths and time intervals in the linear dispersion relation for low momenta. On the other hand, $\xi$ represents the healing length, representing the smallest scale below which the linear dispersion relation approximation becomes inadequate. \\
At this point, going back to the equation for the condensate wavefunction and the fluctuation one, we can notice that when the mean-field is described by the Gross-Pitaevskii equation, the linearized fluctuation can be solved separately after that the solution of the Gross-Pitaevskii equation is found. Instead, when we include the back-reaction of the quantum fluctuation in the dynamics of the condensate wavefunction, the equations for the mean-field and the quantum fluctuation cannot be separated. Therefore, when we consider back reactions the modified Gross-Pitaevskii equation for the condensate wavefunction and the modified Bogoliubov de Gennes equation for fluctuations are \cite{Barcelò_analoguegravityReview}
\begin{equation}
    i \hslash\frac{\partial}{\partial t}\psi (t,\mathbf{x})=\biggl(-\frac{\hslash^2}{2m}\nabla^2+V_{\text{ext}}(\mathbf{x})+\gamma n_c\biggr)\psi(t,\mathbf{x})+\gamma (2 \Tilde{n} \psi(t,\mathbf{x})+\Tilde{m} \psi^*(t,\mathbf{x}))
\end{equation}
\begin{equation}
    i \hslash\frac{\partial}{\partial t}\hat{\varphi}(t,\mathbf{x})=\biggl(-\frac{\hslash^2}{2m}\nabla^2+V_{\text{ext}}(\mathbf{x})+\gamma 2n_T\biggr)\hat{\varphi}(t,\mathbf{x})+\gamma m_T \hat{\varphi}^\dagger(t,\mathbf{x})
\end{equation}
respectively, with
\begin{equation}
    \Tilde{n}=\langle \hat{\varphi}^\dagger \hat{\varphi}\rangle, \quad \Tilde{m}=\langle \hat{\varphi}\hat{\varphi}\rangle, \quad m_c=\psi^2, \quad n_T=n_c+\Tilde{n}, \quad m_T=m_c+\Tilde{m}.
\end{equation}
$\Tilde{n}$ and $\Tilde{m}$ are called anomalous terms: $\Tilde{n}$ is essentially the normal condensate density, and $\Tilde{m}$ the off-diagonal term.\\
Now, we want tot see that it is possible to describe a phonon, i.e. a phase fluctuation of the condensate, as a massless scalar field propagating on an emergent acoustic metric. We consider a Bose-Einstein condensate and neglect the back-reactions of the quantum fluctuation in the dynamics of the condensate wavefunction. We adopt the Madelung representation for the wave function of the condensate
\begin{equation}
    \psi(t,\mathbf{x})=\sqrt{n_c(t,\mathbf{x})}e^{-i\theta(t,\mathbf{x})/\hslash},
\end{equation}
and the quantum acoustic representation for quantum perturbations 
\begin{equation}
    \hat{\varphi}(t,\mathbf{x})=e^{-i\theta/\hslash}\left(\frac{1}{2\sqrt{n_c}}\hat{n}_1-i\frac{\sqrt{n_c}}{\hslash}\hat{\theta}_1\right).
\end{equation}
Then, it is possible to demonstrate that, introducing (3+1)-dimensional spacetime coordinates $x^{\mu}\equiv (t; x^i)$, the wave equation for $\hat{\theta}_1$, i.e. the phonon, is written as the one of a quantum scalar field over a curved background \cite{Barcelò_analoguegravityReview}:
\begin{equation}
    \square \hat{\theta}_1=\frac{1}{\sqrt{|g|}}\partial_{\mu}(\sqrt{|g|}g^{\mu\nu}\partial_{\nu}\hat{\theta}_1)=0
\end{equation}
with
\begin{equation}
\label{eq:analogue_metric_BEC}
    g_{\mu\nu}(t,\mathbf{x})= \frac{n_c}{mc_s }
\begin{pmatrix}
\begin{array}{c;{2pt/2pt}cc}
-\left[c_s^2-v^2\right] & -v_{j} \\ \hdashline[3pt/4pt]
    -v_{i} & \delta_{ij}   
\end{array}
\end{pmatrix}
\end{equation}
$c_s$ is given in Equation (\ref{eq:speedSound_BEC}) representing the speed of sound of the phonons in the medium, while $\mathbf{v}$ is given by 
\begin{equation}
    \label{eq:velocityBEC}
    \mathbf{v}=\frac{\nabla \theta}{m}
\end{equation}
and it is the irrotational velocity field of the condensate.\\ 
Now, we focus on the case of a spherically symmetric analogue metric. When $n_c$ and $\mathbf{v}$ are only functions of radius, it is possible to show that the quasi-particles moves in a spherically symmetric acoustic metric whose line element is \cite{Visser_acousticBH}:
\begin{equation}
    \mathrm{d}s^2=\frac{n_c}{m c_s}\left(-c_s^2\mathrm{d}t^2+\left(\mathrm{d}r-v_r \mathrm{d}t\right)^2+r^2\left(\mathrm{d}\theta^2+sin^2\theta \mathrm{d}\phi^2\right)\right).
\end{equation}
An horizon thus forms at $r_H$ where the radial velocity equals the speed of sound:
\begin{equation}
    \left.v_r\right|_{r_H}=\left.c_s\right|_{r_H}.
\end{equation}
Having understood how the acoustic metric emerges for phonons in a Bose-Einstein condensate, we now want to investigate more deeply the analogy with the black hole metric.

\section{Acoustic Schwarzschild black hole}
Having discussed the mean-field treatment of the analogue system, we now want to analyze better how the acoustic metric can reproduce a black hole. In particular we see how well the acoustic metric can approximate the Schwarzschild geometry. In order to do that, we need to introduce one of the unconventional representations of the Schwarzschild geometry, known as the Painlevé-Gullstrand line element, and in doing that we set $c=1$. This representation essentially involves an atypical selection of coordinates within the Schwarzschild spacetime and it has been rediscovered several times (see for example \cite{painleve_diKW}). The Painlevé-Gullstrand coordinates are related to the Schwarzschild ones (indicated by the subscript $s$) by 
\begin{equation}
\label{eq:pinleve1}
    t=t_s \pm \left[ 4 M \tanh^{-1} \left( \sqrt{\frac{2GM}{r}}\right)-2\sqrt{2GMr}\right]
\end{equation}
or, equivalently,
\begin{equation}
\label{eq:painleve2}
    \mathrm{d}t=\mathrm{d}t_s \pm \frac{\sqrt{2GM/r}}{1-2GM/r}\mathrm{d}r.
\end{equation}
Let us consider the ingoing ($+$) and outgoing ($-$) coordinates, as those that cover the future horizon and the black hole singularity and the past horizon and the white-hole singularity, respectively. In modern notation, the Schwarzschild geometry cast in Painlevé-Gullstrand coordinates reads: 
\begin{equation}
    \mathrm{d}s^2=-\mathrm{d}t^2+\left( \mathrm{d}r\pm \sqrt{\frac{2GM}{r}}\mathrm{d}t\right)^2+r^2(\mathrm{d}\theta^2+\sin^2\theta \mathrm{d}\phi^2)
\end{equation}
or, equivalently,
\begin{equation}
    \mathrm{d}s^2=-\left(1-\frac{2GM}{r}\right) \mathrm{d}t^2 \pm \sqrt{\frac{2GM}{r}}\mathrm{d} r \mathrm{d} t + \mathrm{d}r^2+r^2(\mathrm{d}\theta^2+\sin^2\theta \mathrm{d}\phi^2).
\end{equation}
As pointed out by Kraus and Wilczek, the Painlevé-Gullstrand line element exhibits several instructive characteristics \cite{painleve_diKW}. Notably, the spatial slices at constant time are entirely flat, meaning that there is no curvature in space. Instead, all the spacetime curvature of the Schwarzschild geometry is concentrated in the time-time and time-space components of the metric. When working with the Painlevé-Gullstrand line element, it may appear straightforward to force the acoustic metric in this form by setting $\rho$ and $c_s$ as constants and defining $v=\sqrt{2GM/r}$. However, while this approach indeed aligns the acoustic metric with the Painlevé-Gullstrand form, a significant issue arises: the continuity equation $\nabla \cdot (\rho \mathbf{v})=0$, which was used in deriving the acoustic equations, is no longer satisfied. The most feasible approach consists in considering the speed of sound $c_s$ as a constant normalized to unity ($c_s=1$) and setting $v = \sqrt{2GM/r}$. Then, using the continuity equation we deduce that $\rho |\mathbf{v}| \propto 1/r^2$, leading to $\rho \propto r^{-3/2}$. Since the speed of sound remains constant, integrating the relationship $c_s^2=\mathrm{d}p/\mathrm{d}\rho$ we get that the equation of state must be $p=p_\infty +c^2\rho$, so the background pressure satisfies $p-p_\infty \propto c_s^2 r^{-3/2}$. As a result, the acoustic metric now takes the following form:
\begin{equation}
    \mathrm{d}s^2 \propto r^{-3/2}\left[-\left(1-\frac{2GM}{r}\right) \mathrm{d}t^2 \pm \sqrt{\frac{2GM}{r}}\mathrm{d} r \mathrm{d} t + \mathrm{d}r^2+r^2(\mathrm{d}\theta^2+\sin^2\theta \mathrm{d}\phi^2)\right].
\end{equation}
Therefore the acoustic metric is conformal to the Schwarzschild one cast in the Painlevé-Gullstrand form, but not identical to it. For many practical purposes, this is still satisfactory. Indeed, we have an event horizon, we can establish surface gravity, and we can study Hawking radiation. As both surface gravity and Hawking temperature are conformal invariants, this suffices for examining basic features of the Hawking radiation process \cite{T_conformalInvariance}. The sole manner in which the conformal factor may influence Hawking radiation is through back-scattering off the acoustic metric \cite{Visser_acousticBH}. However if we concentrate on the region near the event horizon, we can take the conformal factor to be a constant.

\section{Real-analogue black holes dictionary}
Now that we have seen to what extent we can reproduce the Schwarzschild spacetime geometry, we examine how the definitions of horizons, surface gravity, and Hawking temperature, as seen in the previous chapters, translate into the language of Analogue Gravity.
\begin{itemize}
    \item \textbf{Trapped surface}: Take any closed two-dimensional surface. If, across the entire surface, the fluid velocity consistently points inward, and if the normal component of the fluid velocity is higher than the local speed of sound, then regardless of the direction in which a sound wave travels, it will be carried inward by the fluid flow and become confined within the surface. Such a surface is called ``outer-trapped''. Inner-trapped surfaces instead, can be characterized by requiring that the fluid flow consistently points outward with a normal component exceeding the speed of sound. These definitions are made possible primarily due to the natural definition of ``at rest'' provided by the background Minkowski metric, which allows us to adopt this uncomplicated and direct criterion. The acoustic trapped region is determined as the region that includes outer trapped surfaces.
    
    \item \textbf{Apparent horizon}: The acoustic apparent horizon is the two-dimensional surface where the normal component of the fluid velocity is equivalent to the local speed of sound.
    
    \item \textbf{Event horizon}: The event horizon is a null surface, that is generated by null geodesics. The future event horizon is the boundary of the region from which null geodesics (phonons) cannot escape. A past event horizon is defined as the boundary of the region that cannot be reached by incoming phonons. In stationary geometries, the apparent and event horizons are the same. However, in time-dependent geometries, this distinction is often significant.

    \item \textbf{Surface gravity}: In the context of acoustics, there exists a specific way to describe the null geodesics responsible for generating the event horizon, which is considered the most intuitive and natural. This parameterization is based on the Newtonian time coordinate of the underlying physical metric. This enables us to clearly establish a concept of surface gravity, even in the case of non-stationary (time-dependent) acoustic event horizons. This is achieved by quantifying how much the natural time parameter deviates from being an affine parameter for the null generators of the horizon. It is possible to show that the surface gravity $g_H$ in a static acoustic spacetime is \cite{Barcelò_analoguegravityReview}:
    \begin{equation}
        g_H=\left.\frac{1}{2}\frac{\partial (c_s^2-v^2)}{\partial n}\right|_H=c_{sH} \left.\frac{\partial |c_s-v|}{\partial n}\right|_H,
    \end{equation}
    where $\partial/\partial n$ is a normal derivative with respect to the horizon, which is denoted by $H$ and $c_{sH}$ is the speed of sound evaluated at the horizon. This expression of $g_H$ can also be written as
    \begin{equation}
        g_H=\left|1-\frac{1}{2}\frac{\partial^2 p}{\partial \rho^2}\frac{\partial \rho}{\partial p}\rho \right| \lVert \mathbf{a} \rVert,
    \end{equation}
    where $\mathbf{a}$ is the fluid's acceleration. This indicates that the surface gravity is directly connected to the acceleration of the fluid when it crosses the horizon, up to a dimensionless factor that relies on the equation of state. We note that for a Bose-Einstein condensate (on which we will focus on later), $p=\frac{1}{2}k\rho^2+C$ implying $c_s^2=k\rho$, we have $g_H=\frac{1}{2}\lVert \mathbf{a} \rVert$. In a stationary (non-static) acoustic spacetime, splitting up the fluid flow into normal $\mathbf{v}_\perp$ and tangential components $\mathbf{v}_\parallel$ in the vicinity of the horizon, it is possible to show that \cite{Barcelò_analoguegravityReview}
    \begin{equation}
        g_H=\left.\frac{1}{2}\frac{\partial (c_s^2-v_\perp^2)}{\partial n}\right|_H=c_{sH} \left.\frac{\partial |c_s-v_\perp|}{\partial n}\right|_H.
    \end{equation}

    \item \textbf{Hawking temperature} The emergence of an acoustic horizon entails the amplification of quantum vacuum fluctuations: there should be a thermal flux of phonons escaping towards infinity in the region where $v<c_s$. The temperature associated with this radiation can be defined by analogy with the Hawking temperature in gravity (see Equation (\ref{eq:Hawking_temp})). It is related to both the surface gravity $g_H$ and the speed of sound at the horizon $c_{sH}$ by \cite{Visser_acousticBH}:
    \begin{equation}
        k_B T_H=\frac{\hslash g_H}{2\pi c_{sH}}=\left.\frac{\hslash}{2\pi}\frac{\partial |c_s-v|}{\partial n}\right|_H.
    \end{equation}
    with $k_B$ the Boltzmann constant. In \cite{Unruh_experimentalBHev} Unruh found the original form for the Hawking temperature, but there it is derived considering $c_s$ constant.
\end{itemize}

\section{Vortex geometry}
\label{sec:ABH_cylindrical}
Due to the transversal nature of gravitational waves, their reproduction in the analogue system ---that we aim at--- now requires us to shift our focus to a `draining bathtub' fluid flow, which has a different geometry compared to the spherically symmetric acoustic black hole \cite{Visser_acousticBH}. We model it as a (2+1)-dimensional flow configuration with a sink at the origin. The continuity equation dictates that, for the fluid's radial velocity component, the following equation must hold:
\begin{equation}
    \rho\, v_r \propto \frac{1}{r}.
\end{equation}
A possible solution to this constraint is that the background density $\rho$ remains uniform, without dependence on position, throughout the entire flow. This, in turn, implies that the background pressure $p$ and the speed of sound $c_s$ also remain constant across the fluid flow.
Moreover, the condition of the flow being vorticity free implies, via Stokes' theorem:
\begin{equation}
    v_\theta \propto \frac{1}{r}.
\end{equation}
Regarding the background velocity potential (see Equation (\ref{eq:velocityBEC})) it must be written as a linear combination of the angle $\theta$ and $\ln r$: it is not a monodromic function due to its discontinuity when $\theta$ transitions across $2\pi$. Instead, the velocity potential should be understood as being defined patch-wise on overlapping regions surrounding the vortex core at $r=0$. The fluid flow is \cite{Visser_acousticBH}:
\begin{equation}
    \mathbf{v}=\frac{(A \hat{r}+B\hat{\theta})}{r}.
\end{equation}
with $A$ and $B$ having the dimensions of length$^2$/time. Without considering an overall prefactor space independent, the acoustic metric for a draining bathtub is explicitly as follows:
\begin{equation}
    \mathrm{d}s^2=-c_s^2\mathrm{d}t^2+\left( \mathrm{d}r-\frac{A}{r}\mathrm{d}t \right)^2+\left( r \mathrm{d}\theta-\frac{B}{r}\mathrm{d} t\right)^2.
\end{equation}
The acoustic event horizon forms when the fluid velocity's radial component exceeds the speed of sound, that is at:
\begin{equation}
    \label{eq:acHOR_posit_vortex}
    r_H=\frac{|A|}{c_s}.
\end{equation}
The sign of $A$ makes a difference. When $A<0$, we are referring to a future acoustic horizon (acoustic black hole). Conversely, when $A>0$, we are referring to a past event horizon (acoustic white hole). While this construction has been explained within the framework of (2+1)-dimensions, it is important to note that we have the flexibility to introduce an additional dimension by transitioning to (3+1)-dimensions. In doing so, we can interpret the result as a combination of a conventional vortex filament and a line source (or line sink):
\begin{equation}
    \label{eq:cylindricalBH_metric}
    \mathrm{d} s^2=-c_s^2 \mathrm{d}t^2+\left( \mathrm{d}r-\frac{A}{r}\mathrm{d} t \right)^2+\left(r \mathrm{d}\theta -\frac{B}{r}\mathrm{d}t\right)^2+\mathrm{d}z^2.
\end{equation}

\section{Analogue Gravity Experiments}
\label{sec:experiments-sec}
In this section we briefly present the general setup for Analogue Gravity experiments with Bose-Einstein condensates. Moreover, we discuss some details of the experiments that have observed the analogous of the Hawking radiation and of the Unruh temperature.

\subsection{Bose-Einstein condensates in the laboratory}
\label{sec:BECexperimental-sec}
In 1995 the first Bose-Einstein condensates were realized in the Wieman and Cornell group at JILA with $^{87} \text{Rb}$ atoms, and by the Ketterle group at MIT with $^{23}\text{Na}$ \cite{BEC1}\cite{BEC2}. Later on, also Fermi superfluids have been realized: with fermionic isotopes of $^{40}\text{K}$ atoms at JILA in the group of Jin and of $^6 \text{Li}$ atoms in the group of Ketterle at MIT \cite{FERMI1}\cite{Fermi2}. Bose-Einstein condensates have become significant due to their unique capability to independently manipulate various characteristics with the precision of atomic physics. It is possible to control modifications in different parameters, such as condensate temperature, system dimensionality, and two-body interaction strength (see below). First of all, referring to the condensate temperature, we have that with atomic gases it is possible to reach temperatures of tens of $nK$, through a cooling process. This process achieves high efficiency by combining laser and evaporative cooling techniques with appropriate trapping methods. Laser cooling can lower temperatures to the $\mu K$ range by slowing down atomic motion through photon absorption. Subsequently, temperatures as low as tens of $nK$ are achieved through magnetic or optical trapping of atoms, which increases their density, enhancing elastic collision rates. The most effective evaporative cooling process follows this, where atoms with higher energies in the trap are selectively removed using magnetic or optical fields, depending on the trapping method. As a result, the effective elastic collisions redistribute energy, leading to thermalization and a decrease in temperature down to $nK$. Achieving these extremely low temperatures is essential in detecting and investigating the acoustic analogue of the Hawking effect. Consequently, it is not unexpected that atomic gases have served as the pioneering analogue systems for measuring the Hawking temperature. Secondly, the laser cooling and trapping serve also to control the dimensionality of the sample. It is worth noting that significant advancements have been made in the past decade in creating well-suited confining shapes for experiments. This progress involves employing counter-propagating standing laser light waves in perpendicular directions to create for example planar 2D confinements \cite{BEC2d}. Additionally, sheets of laser light have been utilized to achieve box-like trapping configurations \cite{BECinaBox}\cite{HadzibabicBOX}\cite{ZwierleinBOX}. Finally, for what concerns the interaction strength, one relevant mechanism to tune it is the Fano-Feshbach resonance \cite{Feshbah}. Essentially, the Fano-Feshbach resonance allows precise control of the interactions between ultracold atoms using an external magnetic field. In analogue models, the interaction length serves as the primary tool for simulating various spacetimes, as indicated in Equations (\ref{eq:analogue_metric_BEC}) and (\ref{eq:speedSound_BEC}). Consequently, manipulating the scattering length, theoretically we have a wide array of possibilities for creating acoustic metrics.

\subsection{Observation of the Hawking effect}
\label{sec:Steinhauer-sec}
Since the concept of analogue Hawking radiation was introduced, there has been extensive theoretical exploration of numerous potential analogue black holes. It was predicted that the observation of Hawking radiation could be achieved through the analysis of density correlations between Hawking particles (the emitted ones) and their partners (the trapped ones) \cite{Carusotto_2008}\cite{Balbinot_DensityCorr}. Moreover, it has also been explained that the density correlations could be used to observe the entanglement \cite{2015Steinhauer}. Based on these theoretical studies, Steinhauer has observed in 2016 the Hawking radiation in a Bose-Einstein condensate \cite{2016Steinhauer}. He has used a Bose-Einstein condensate of $8.000\;\,^{87}\text{Rb}$ atoms confined in a focused laser beam. Because of the fact that the radial trap frequency is greater than the maximum interaction energy, then the behavior of the system is effectively 1-dimensional. To mimic a black hole event horizon, an additional short-wavelength laser beam is used. It creates a very sharp potential step, i.e. very narrow with respect to the healing length: the flow velocity increases as the step is approached and this leads to the creation of a supersonic region. There is a changing also in the speed of sound, that depends on the radial trapping frequency and the density profile which are space-dependent functions. 
\begin{figure}[H]
    \centering
    \includegraphics[width=0.72\textwidth]{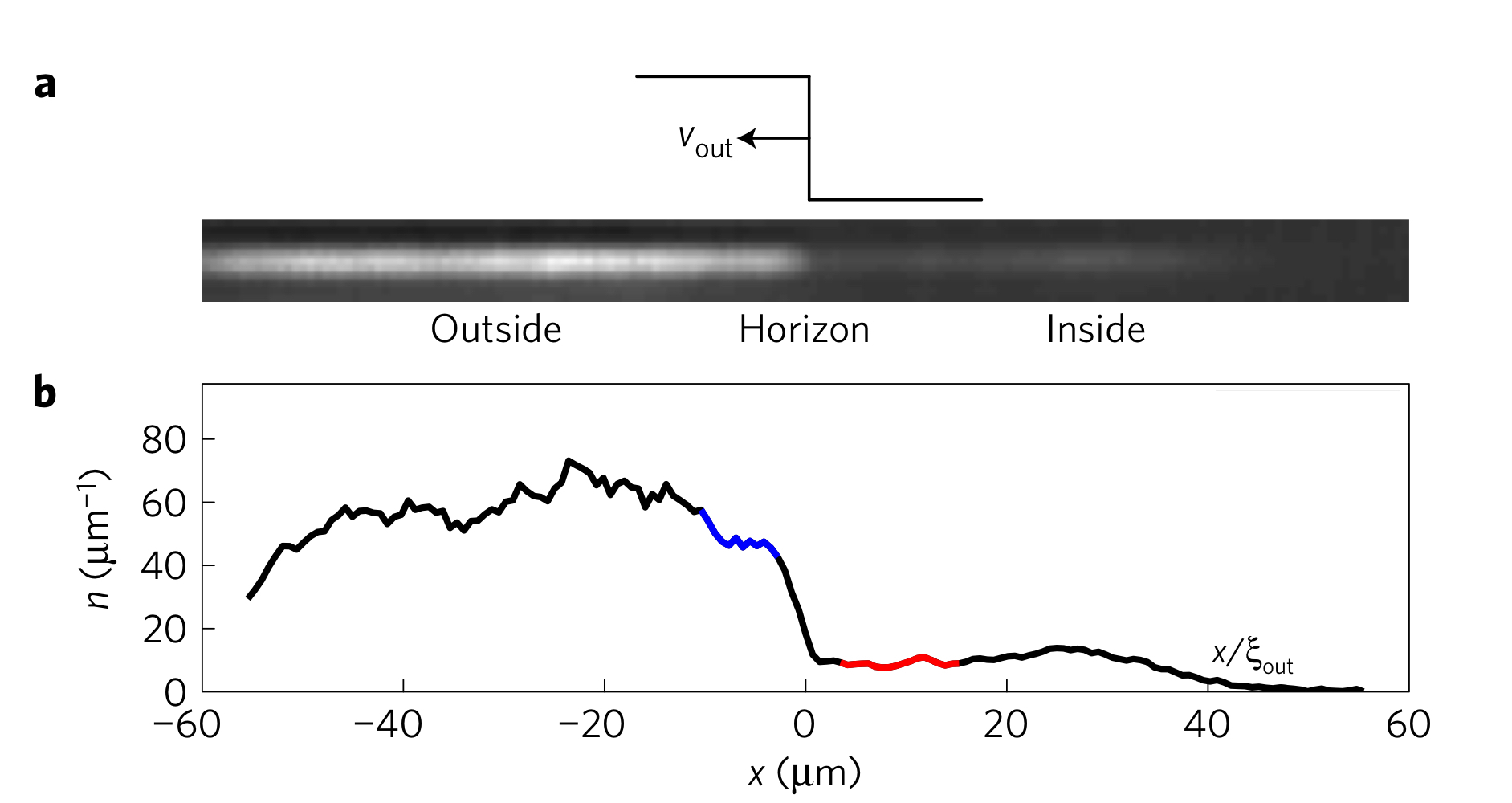}
    \caption{\textit{Structure of the black hole analogue realized with a Bose-Einstein condensate} \cite{2016Steinhauer}. \textbf{a}: Cigar-shaped, effectively 1-dimensional Bose-Einstein condensate with an imprinted step-like potential which traps phonons in the inside region to the right of the horizon. \textbf{b}: Density profile of \textbf{a}. The blue and red lines indicate respectively the region outside and inside the analogue black hole, where the Hawking/partner correlations are observed. Picture from \cite{2016Steinhauer}.}
    \label{fig:Steinhauer_system}
\end{figure}
Therefore, we have the creation of an acoustic black hole: outside the acoustic horizon the speed of sound is $c_{s_{\text{out}}}=0.57\, \text{mm}\,s^{-1}$ and the velocity flow $v_{\text{out}}=0.24\, \text{mm}\,s^{-1}$, while inside $c_{s_{\text{in}}}=0.25 \,\text{mm}\,s^{-1}$ and $v_{\text{in}}=1.02 \,\text{mm}\,s^{-1}$. This is displayed in Figure \ref{fig:Steinhauer_system}, where the horizon is considered at the origin of the coordinate system. In the left side of Figure \ref{fig:Steinhauer_system}, the flow is subsonic, while on the right, thus inside the analogue black hole, the flow is supersonic. Here, phonons are confined in a manner analogous to how photons are unable to reach the event horizon within a black hole. With this experimental set up, the Hawking and partner particles have been observed through the two-body density-density correlation function between points on opposite sides of the horizon given by:
\begin{equation}
    G^{(2)}(x,x')=\sqrt{\frac{\xi_{\text{out}}\xi_{\text{in}}}{n_{\text{out}}n_{\text{in}}}}\langle \delta n(x) \delta n(x')\rangle.
\end{equation}
Here, $n(x)$ is the number density of the condensate: $n_{\text{out}}$ and $n_{\text{in}}$ are the densities outside and inside the acoustic black hole, respectively; $\xi_{\text{out}}$ and $\xi_{\text{in}}$ are the healing lengths (see Equation (\ref{eq:healing_length_BEC})) outside and inside the analogue black hole, respectively; the positions $x$ and $x'$ are expressed in units of $\xi=\sqrt{\xi_{\text{out}}\xi_{\text{in}}}=1.8\,\mu\text{m}$. By Fourier transforming the correlations between points outside and inside the acoustic black hole it is possible to obtain the $k$-space correlation spectrum $\langle \hat{b}_{k_{\text{H}}} \hat{b}_{k_{\text{P}}}\rangle$:
\begin{equation}
    S_0\langle \hat{b}_{k_{\text{H}}} \hat{b}_{k_{\text{P}}}\rangle =\sqrt{\frac{\xi_{\text{out}}\xi_{\text{in}}}{L_{\text{out}}L_{\text{in}}}}\int \mathrm{d}x \mathrm{d}x' e^{ik_{\text{H}}x}e^{ik_{\text{P}}x'}G^{(2)}(x,x'),
\end{equation}
where $\hat{b}_{k_{\text{{H}}}}$ is the annihilation operator for a Hawking particle with wavenumber $k_{\text{{H}}}$ in units of $\xi^{-1}$ localized outside the analogue black hole, while $\hat{b}_{k_{\text{{P}}}}$ is the annihilation operator for a partner particle localized inside the acoustic black hole with wavenumber $k_{\text{{P}}}$ in units of $\xi^{-1}$. $L_{\text{out}}$ and $L_{\text{in}}$ are the length of the region outside and inside the analogue black hole, respectively, while $S_0$ is the zero-temperature static structure factor. The integral is performed over the region in the correlation function bounded by $-L_{\text{out}}/\xi<x<0$ and $0<x'<L_{\text{in}}/\xi$. 
\begin{figure}[ht]
    \centering
    \includegraphics[width=1\textwidth]{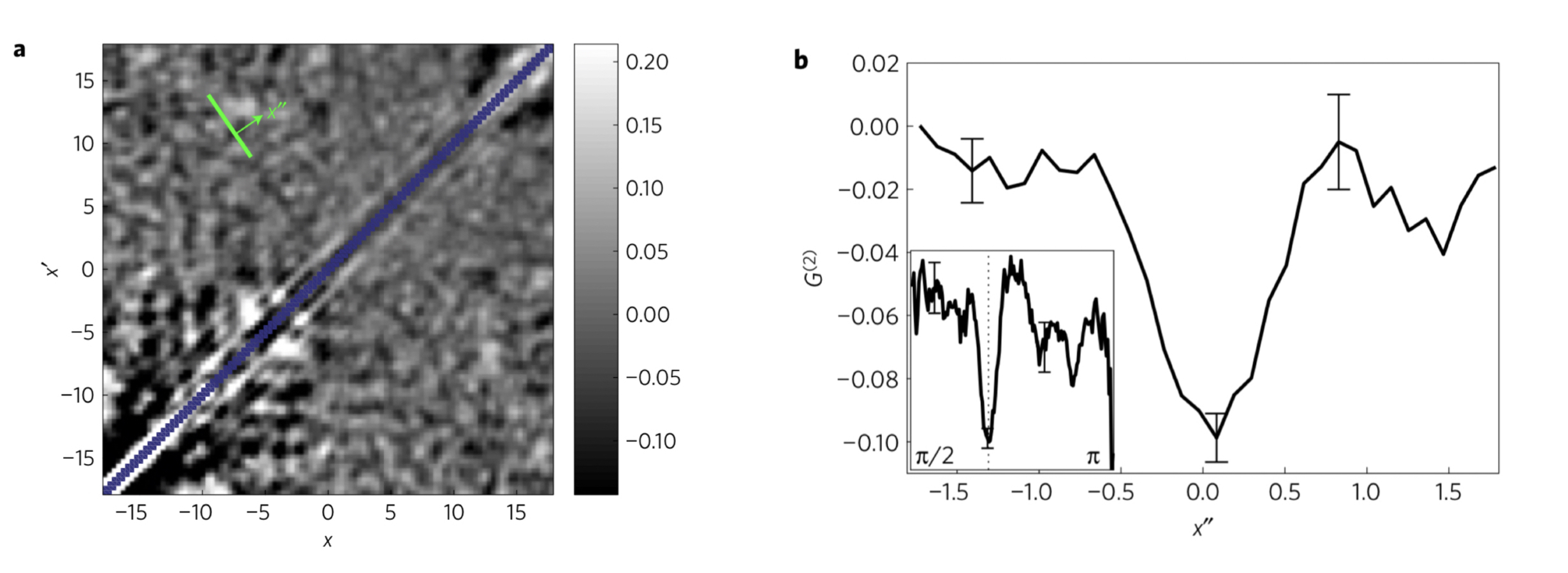}
    \caption{\textit{Observation of the Hawking effect in an experiment with a Bose-Einstein condensate} \cite{2016Steinhauer}. \textbf{a}: The density-density correlation function for every pair of points $(x,x')$, with $-L_{\text{out}}/\xi<x<0$ and $0<x'<L_{\text{in}}/\xi$ and the horizon located at the origin. The dark band extending from the centre of the figure represents the correlations between the Hawking and partner particles. The green rectangle is the area used for the two-dimensional Fourier transform. \textbf{b}: The profile of the correlation band along the $x''$ direction in \textbf{a}, averaged over the length of the band. The inset shows the angular profile which gives the angle of the Hawking-partner correlation band, as indicated by the dotted line. From \cite{2016Steinhauer}.}
    \label{fig:Steinhauer_obs}
\end{figure}
In the experiment the Hawking prediction is tested by measuring $\langle \hat{b}_{k_{\text{H}}} \hat{b}_{k_{\text{P}}}\rangle$, where Hawking radiation is posited to be the dominant source of such correlations, and comparing it to $(|\beta|^2+1)|\beta|^2$ where $|\beta^2|$ is the Planck distribution at the predicted Hawking temperature \cite{2019MunozdeNova}. The measured correlation function between pairs of points $(x,x')$ along the analogue black hole is shown in Figure \ref{fig:Steinhauer_obs}. The upper-left and lower-right quadrants of Figure \ref{fig:Steinhauer_obs} \textbf{a} show the correlations between points on opposite sides of the horizon. The dark side of these points is clearly visible emanating from the horizon: this is the observation of the Hawking and partner particles. By Fourier transforming the $G^{(2)}(x,x')$ in Figure \ref{fig:Steinhauer_obs} \textbf{b} in the region outlined in green in Figure \ref{fig:Steinhauer_obs} \textbf{a}, the plot of $S_0^2|\langle \hat{b}_{k_{\text{H}}} \hat{b}_{k_{\text{P}}}\rangle|^2$ is obtained: see Figure \ref{fig:bhbpcorr}. In the plot, the thick solid line, representing $S_0^2|\langle \hat{b}_{k_{\text{H}}} \hat{b}_{k_{\text{P}}}\rangle|^2$, for large $k$ values is in agreement with the thin light-grey curve, that is the theoretical prediction $S_0^2(|\beta|^2+1)|\beta|^2$ computed at $k_B T_H=0.36 m c^2_{s_{\text{out}}}$. Therefore the Hawking radiation is thermal at these high energies. However, for smaller values of $k$, the observed spectrum significantly deviates from the theoretical curve. This indicates that the long-wavelength Hawking pairs exhibit less correlation than expected or are generated in lower quantities, approximately 3.6 times less than expected. Another possible explanation is that the low-frequency waves did not have sufficient time to form, or other issues may have arisen during the experiment. Alternatively, it could be that certain effects occurring at low momenta, such as interactions with the medium, which are not integrated into the theoretical model, may contribute to the reduction in correlation. 
\begin{figure}[ht]
    \centering
    \includegraphics[width=0.6\textwidth]{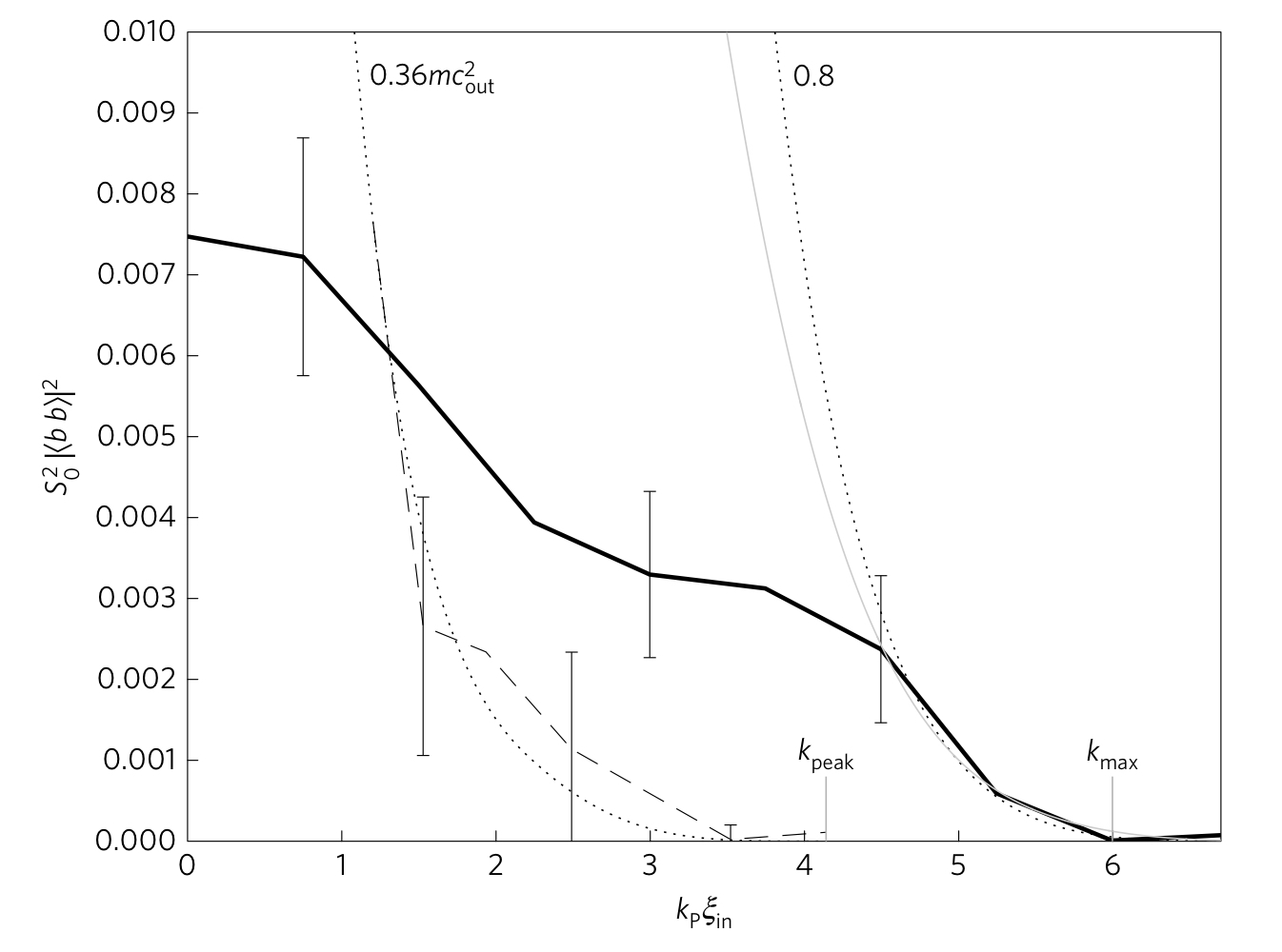}
    \caption{\textit{Observation of the Hawking effect in an experiment with a Bose-Einstein condensate} \cite{2016Steinhauer}. The thick solid line indicates $S_0^2|\langle \hat{b}_{k_{\text{H}}} \hat{b}_{k_{\text{P}}}\rangle|^2$ between the Hawking and partner particles. It is found from the Fourier transform of $G^{(2)}(x,x')$ in Figure \ref{fig:Steinhauer_obs} \textbf{b}. The thin solid grey curve indicates the maximally entangled $S_0^2(|\beta|^2+1)|\beta|^2$ curve calculated for $k_B T_H=0.36 m c^2_{s_{\text{out}}}$. The dashed curve is the measured phonon population squared. The solid curve exceeding the dashed curve corresponds to entanglement \cite{2015Steinhauer}. The dotted lines show the theoretical curves for phonons population observed at $k_B T_H=0.36 m c^2_{s_{\text{out}}}$ and $k_B T_H=0.8 m c^2_{s_{\text{out}}}$. From \cite{2016Steinhauer}.}
    \label{fig:bhbpcorr}
\end{figure}

In any case, Figure \ref{fig:bhbpcorr} shows that there is entanglement (solid curve exceeding the dashed one \cite{ParticleCreationBH.Hawking}) and that the radiation is thermal at sufficiently high momentum, thus at small distances in the two point correlation function. This is an indication of the fact that the observed radiation is the Hawking one. Steinhauer's work marked a significant step in the study of Analogue Gravity and the validation of Hawking radiation, offering insights into the behavior of quantum fields in the vicinity of event horizon analogues. This experiment has been improved in 2019 by de Nova, Golubkov, Kolobov and Steinhauer \cite{2019MunozdeNova}. We report here their result for the correlation spectrum of the Hawking radiation:
\begin{figure}[ht]
    \centering
    \includegraphics[width=0.6\textwidth]{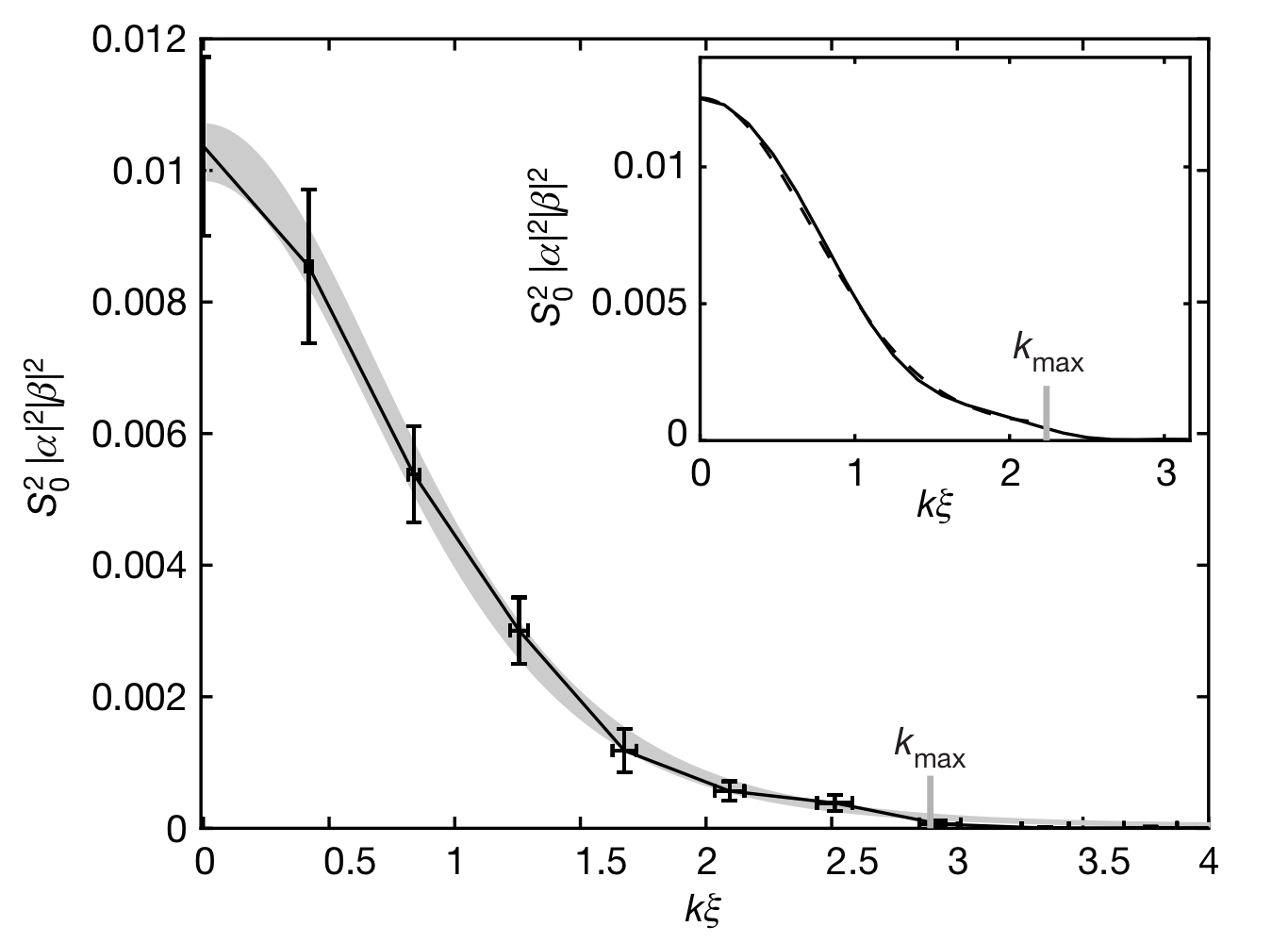}
    \caption{\textit{Observation of the Hawking effect in an experiment with a Bose-Einstein condensate} \cite{2019MunozdeNova}. In this plot the correlation spectrum between the Hawking and partner particle is presented. Black curve is the measured values of $S_0^2|\langle \hat{b}_{k_{\text{H}}} \hat{b}_{k_{\text{P}}}\rangle|^2$, while the grey curve represents the theoretical prediction at $T_H=0.351(4) nK=0.125(1)mc_{out}^2$: $S_0^2|\alpha|^2|\beta|^2$ with $\alpha=|\beta|^2+1$. The uncertainties in the theoretical curve reflects the uncertainties in the speed of sound and the flow velocities. The inset shows the correlation spectrum from a numerical simulation, where the dashed curve is the predicted one. From \cite{2019MunozdeNova}.}
    \label{fig:deNova}
\end{figure}
In Figure \ref{fig:deNova} very good agreement between the measured and predicted spectrum of Hawking radiation can be seen. This is quantified by the temperature assigned to the measured spectrum: $T_H^{meas}=0.124(6)mc_{out}^2$, which differs from the prediction $T_H=0.125(1)mc_{out}^2$ by $-1(5)\%$. Therefore, this work gives quantitative confirmation of the temperature and thermal nature of Hawking radiation. 

\subsection{Quantum simulation of Unruh radiation}
\label{sec:Hu-sec}
\begin{figure}[ht]
    \centering
    \includegraphics[width=1\textwidth]{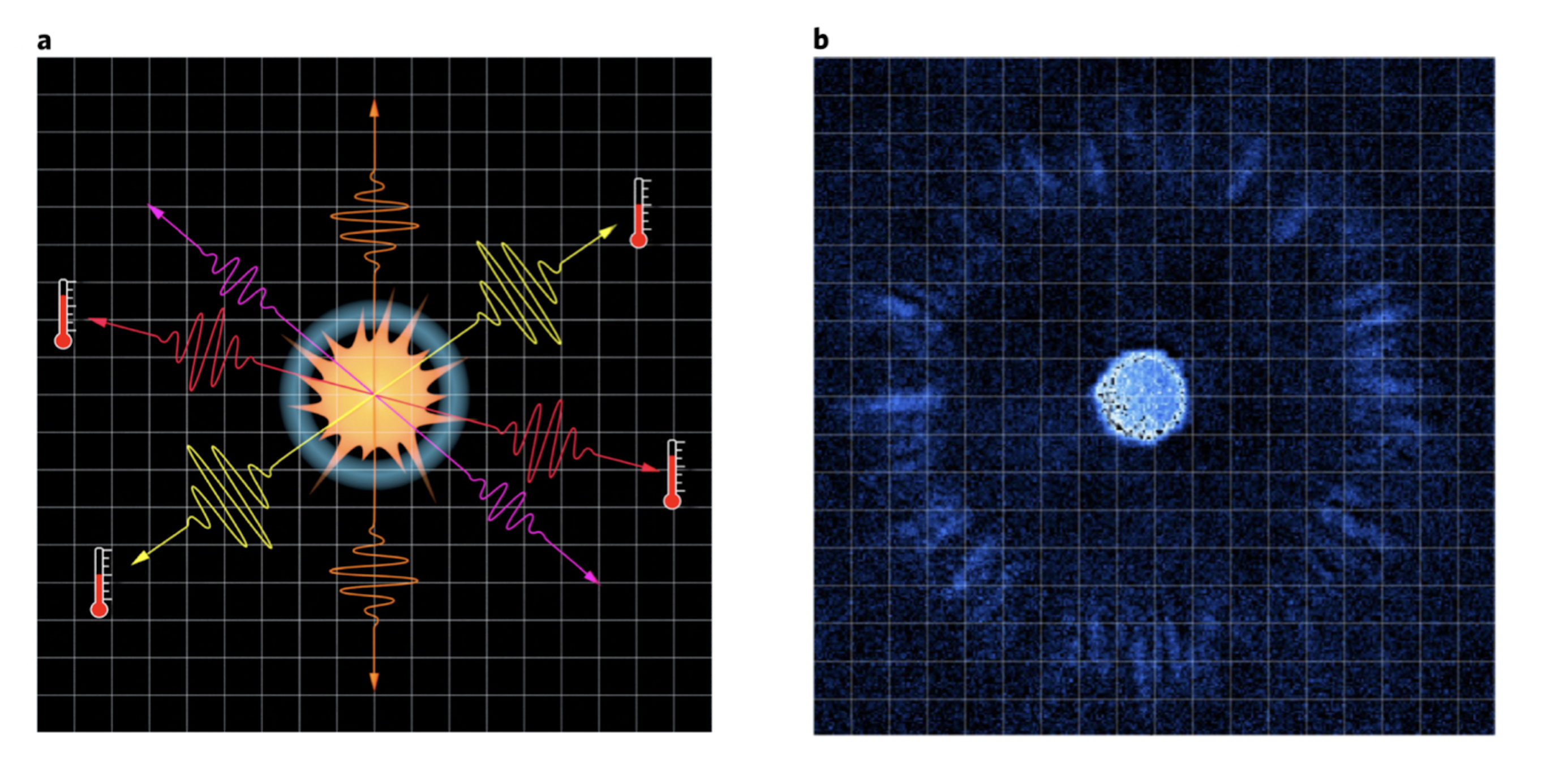}
    \caption{\textit{Observation of the Unruh effect in an experiment with a Bose-Einstein condensate} \cite{2019HuChin}. \textbf{a}: Disk-shaped Bose-Einstein condensate subjected to a modulation of the scattering lengths. A vacuum state in the inertial frame appears identical to a thermal state with the Unruh temperature to an observer in an accelerating frame. \textbf{b}: The observed Unruh radiation. It is made by a pair-creation process in a driven condensate, whose evolution is equivalent to a coordinate transformation to an accelerating frame. From \cite{2019HuChin}.}
    \label{fig:UnruhRad_obs}
\end{figure}

\begin{figure}[ht]
    \centering
    \includegraphics[width=0.65\textwidth]{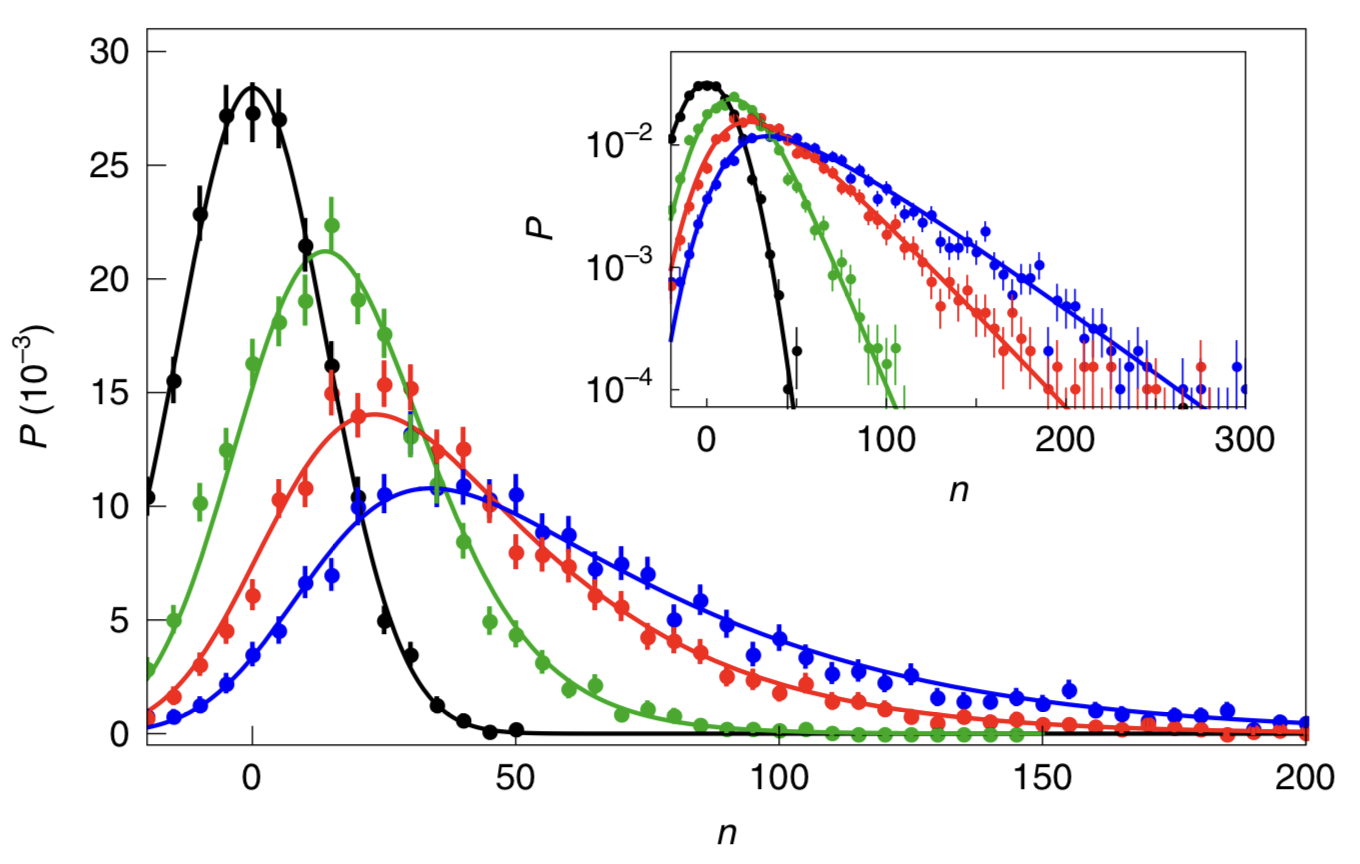}
    \caption{\textit{Probability distribution of atoms at different times} \cite{2019HuChin}. The measured probability distribution $P(n)$ of $n$ atoms within a 2° slice of the emission pattern after evolution times $\tau = 0, 3.36, 4.8$, and $6.24$ $ms$ (black, green, red and  blue circles). The solid lines are fits based on a thermal model; while the inset shows the data in the log scale. From \cite{2019HuChin}.}
    \label{fig:probabilityUnruh_obs}
\end{figure}

\begin{figure}[ht]
    \centering
    \includegraphics[width=0.65\textwidth]{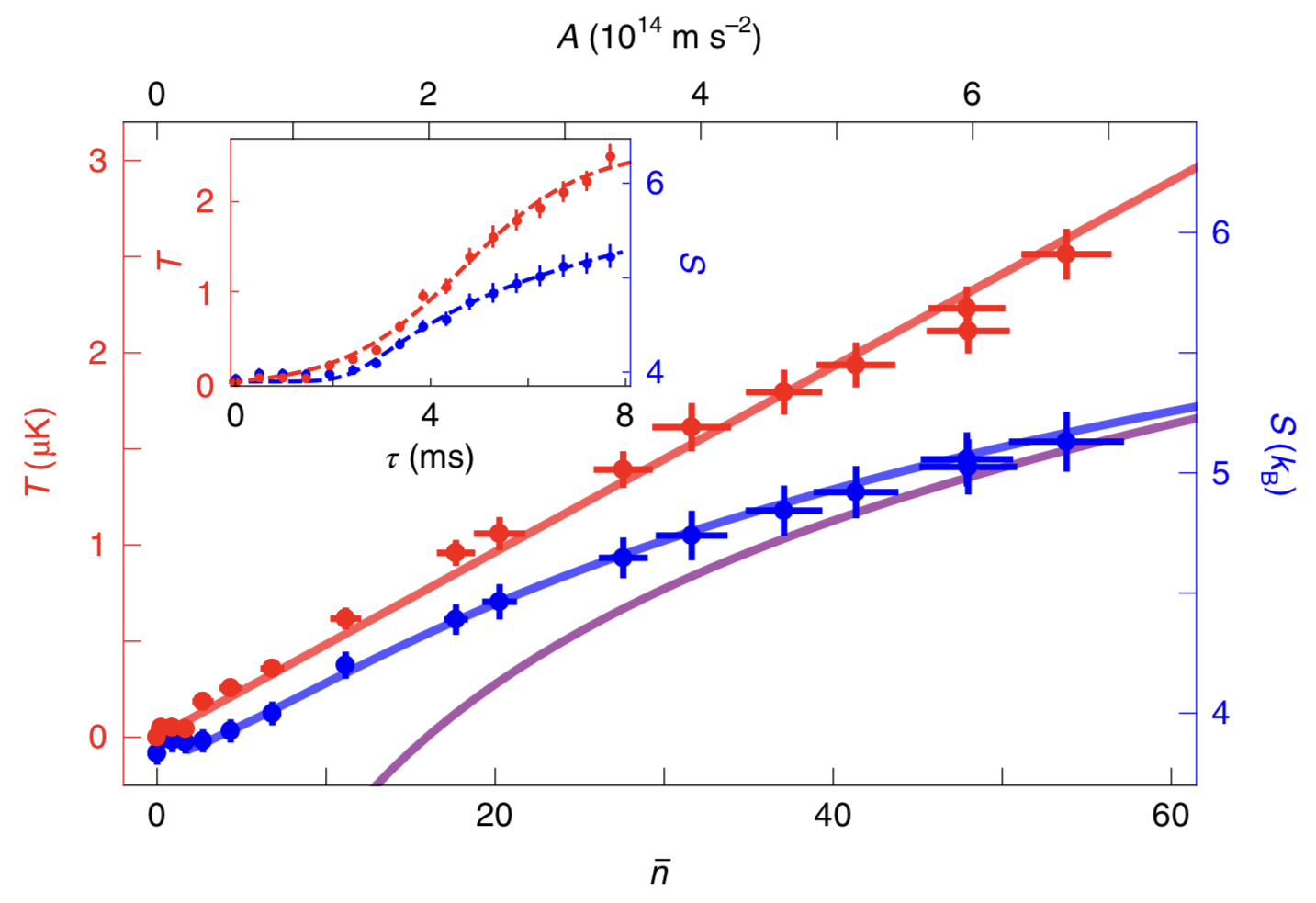}
    \caption{\textit{Thermal behavior of the matter-wave emission} \cite{2019HuChin}. Temperature $T$ (red circles) and entropy $S$ (blue circles) as functions of the mean population per mode $\Bar{n}$. The acceleration $A$ is shown  on the top. The red line is a fit of $T=\kappa A/c$. The blue line represents the prediction for $S$ that takes into account the detection noise, whereas the purple line represents the prediction that does not consider the noise. The inset shows the evolution of $T$ and $S$. From \cite{2019HuChin}.}
    \label{fig:UnruhTemp_fit}
\end{figure}

In 2019, Hu \textit{et al.} have experimentally observed the thermal fluctuations of a matter field that agree with the Unruh effect (see Figure \ref{fig:UnruhRad_obs}) \cite{2019HuChin}: according to Equation (\ref{eq:Unruh_temperature}), an accelerating observer will detect a thermal radiation from a Minkowski vacuum characterized by the Unruh temperature
\begin{equation}
    T_U=\frac{\hslash A}{2\pi c k_B},
\end{equation}
where $A$ is the acceleration and $c$ the speed of light. 
The experiment uses a Bose-Einstein condensate of $\sim 60000$ $^{133}\text{Cs}$ atoms confined in a disc-shaped trap: atoms homogeneously fill the trap of radius $R=13 \mu m$ and vertical thickness $z=1 \mu m$. The time evolution operator $\hat{U}(\tau)=\exp(-i\hat{\mathcal{H}}\tau/\hslash)$ of the system, with $\tau$ the evolution time, is used to simulate the transformation to an accelerating frame $\hat{R}_A$, with acceleration $A$. Given a quantum state $\psi_0$, the equivalence of the time evolution and the Rindler transformation is used:
\begin{equation}
    \hat{U}(\tau)\psi_0=\hat{R}_A \psi_0.
\end{equation}
Using this approach, it is possible to simulate physics in a highly accelerating frame based on a table-top experiment without the necessity of physically accelerating the sample. Indeed, experimentally observing the Unruh effect is extremely challenging, as it requires an enormous acceleration of $A=2.5\times 10^{14}\,m\,s^{-2}$ to generate Unruh radiation at a temperature of only $T_U=1\,\mu K$. Hence, the significance of the current approach is evident in its ability to align the evolution of the Bose-Einstein condensate with the frame transformation. In order to generate the frame boost for any quantum state $\psi_0$, the required Hamiltonian is found to be (in terms of the coupling constant $g_k$, the annihilation and creation operators $a_k$ and $a_k^\dagger$ with wavenumber $k$)
\begin{equation}
    \label{eq:Hu_Hamiltonian}
    \mathcal{H}=i\hslash \sum_k g_k (a^\dagger_k a^\dagger_{-k}-a_ka_{-k})
\end{equation}
and it describes the pair creation of excitations with opposite momenta. The acceleration of the related frame transformation is give by
\begin{equation}
    A=\frac{\pi E_k c}{2\hslash \ln \coth(g_k\tau)},
\end{equation}
with $E_k$ the energy of the excitation. Using this expression for $A$ in the equation for $T_U$, we get:
\begin{equation}
    k_B T_U=\frac{E_k}{4\ln \coth(g_k \tau)}.
\end{equation}
In the experiment, only momentum modes with the same amplitude $|\mathbf{k}|=k_f$ are addressed. The pair-creation Hamiltonian in Equation (\ref{eq:Hu_Hamiltonian}) is prepared by modulating the interactions of the Bose-Einstein condensate with frequency $\omega$ through a Feshbach resonance \cite{Feshbah}. Then the Hamiltonian reduces to:
\begin{equation}
    \mathcal{H}=i\hslash g \sum_{|\mathbf{k}|=k_f}(a^\dagger_k a^\dagger_{-k}-a_ka_{-k})
\end{equation}
with $k_f=\sqrt{\frac{m\omega}{\hslash}}$ and $m$ the atomic mass. Therefore, the Unruh radiation is simulated in the quantum system by parametrically modulating the interactions of the atomic condensate through a Feshbach resonance, that converts the oscillating magnetic field into an oscillating scattering length. The interparticle $s$-wave scattering length is oscillated by modulating the magnetic field close to a Feshbach resonance. As seen, the evolution of the atomic condensate is equivalent to a coordinate transformation to an accelerating frame without actually accelerating the system. A two dimensional, jet-like emission of atoms with momentum $|\mathbf{k}|=k_f$ is observed (see Figure \ref{fig:UnruhRad_obs} \textbf{b}). The emission pattern is divided evenly into 180 angular slices. The atom number $n$ within each slice is extracted to build the probability distribution $P(n)$ at different times $\tau$ (see Figure \ref{fig:probabilityUnruh_obs}). 
The probability $P(n)$ shows a thermal-like distribution: $P(n)\propto \exp(-n E/k_B T)$, with $E=\hslash\omega/2$ the kinetic energy of an emitted atom. The effective temperature $T$ has been extracted and it is possible to see that it depends linearly on the simulated acceleration $T=\kappa A/c$. By fitting the data at a fixed $A$, see Figure \ref{fig:UnruhTemp_fit}, they get $\kappa=1.17 (7) pK\, s$, which is consistent with the Unruh prediction $\kappa=\hslash/2\pi k_B\approx 1.22 pK\, s$. 
Hence, observations within the system suggest a thermal distribution, but unlike incoherent black-body radiation, Unruh radiation is expected to exhibit both spatial and temporal coherence, indicating its quantum origin. In \cite{2019HuChin} they have also observed long-range phase coherence and temporal reversal of matter-wave radiation. In summary, this work shows a new quantum simulation method for exploring quantum phenomena within a non-inertial reference frame. This approach has the potential to be broadened to include other frame transformations through the spatial or temporal control of interaction modulation.

\vspace{7mm}
This overview of Analogue Gravity marks the conclusion of the first part of the thesis. Now, we have all the necessary tools to describe the results obtained in this study.

\part{Results}

\begin{preface}
\textit{\large{At this point, we have all the necessary tools to describe the results of this thesis. \\
The first one, reported in Chapter \ref{chap:simulationGWMinkowski}, is a gravitational wave-like propagation on the top of an acoustic metric. Notably, in the past only static solutions of general relativity were designed in analogue models. Here, we present a system where analogue gravitational waves propagate satisfying Einstein's equations.\\
The second result, in Chapter \ref{chap:GWinBH}, consists in designing a system in which an acoustic horizon is excited by a gravitational wave-like perturbation. To this purpose, we choose a cylindrical acoustic black hole, where we introduce an appropriate extension in the geometry of the previously found gravitational wave-like perturbation.}}
\end{preface}

\chapter{Gravitational Wave acoustic metric}
\label{chap:simulationGWMinkowski}

\begin{chapabstract}
    \begin{adjustwidth}{1cm}{1cm}
        We present the first original result of this thesis work: the reproduction of a gravitational wave perturbation on the top of a flat background acoustic metric in a Bose-Einstein condensate.
    \end{adjustwidth}
\end{chapabstract}

\section{Method}
The aim is to study the propagation of a gravitational wave-like perturbation on the top of an acoustic flat metric. We choose to work with Bose-Einstein condensates due to the rapid progress being made within this quantum technology platform, as stated in Section \ref{sec:experiments-sec} of Chapter \ref{chap:analoguegravity}. We want to cast the acoustic metric that emerges for phonons in a Minkowski background with an imprinted gravitational wave perturbation. Notice that we will consider two types of different fluctuations: the background fluctuations, that will be related to the metric perturbations, and the phonons fluctuations, that are the modes which propagate on the emergent metric. In order to achieve this goal we follow the steps:
\begin{enumerate}
    \item \textbf{Perturbed acoustic metric}: We perturb the acoustic metric in Equation (\ref{eq:analogue_metric_BEC}) and write it as a background $\eta_{\mu\nu}^{(an)}$ plus a perturbation $h_{\mu\nu}^{(an)}$: $g_{\mu\nu}^{(an)}=\eta_{\mu\nu}^{(an)}+\epsilon h_{\mu\nu}^{(an)}$. See Section \ref{sec:perturbedacousticmetric-sec}.
    \item \textbf{Gravitational waves in different gauges}: We exploit the gauge symmetry of General Relativity, i.e. the invariance under coordinate transformations, to write the metric of a gravitational wave that propagates in vacuum $h_{\mu\nu}^{(grav)}$ (so the background gravitational metric $\eta^{(grav)}_{\mu\nu}$ is the Minkowski one) in an appropriate gauge so that it can be matched to $h^{(an)}_{\mu\nu}$. See Section \ref{sec:GW_newgauges-sec}.
    \item \textbf{Comparison between a gravitational wave and the perturbed acoustic metric}: By comparing $\eta^{(grav)}_{\mu\nu}$ and $h^{(grav)}_{\mu\nu}$ with $\eta^{(an)}_{\mu\nu}$ and $h^{(an)}_{\mu\nu}$, respectively, we identify the condensate characteristics that place the system in a regime where the acoustic metric resembles that of a gravitational wave in vacuum. See Section \ref{sec:comparisonGW-acoustic-sec}.
    \item \textbf{Physical systems:} We check whether with that choice of the condensate quantities the system is physical, so if the continuity equation, the Euler equation and the irrotational condition are satisfied. See section \ref{sec:physicalsystems-sec}.
\end{enumerate}
This approach is inspired by the work in \cite{Hartley_GWsimulation}, where a propagating homogeneous gravitational wave is simulated in a Bose-Einstein condensate. The main difference is that they do not consider the spatial dependence of the gravitational wave, while we do. This introduces lots of technical complications, since having the space dependence does not make it possible to work in Fermi normal coordinates for the step 2, i.e. the inertial limit of the proper detector frame, as done in \cite{Hartley_GWsimulation}. Because of that, we need to introduce different suitable gauges, which are not the usual coordinate reference frames. In the following, we present our results for all of the four steps indicated above, omitting the apex $^{(grav)}$ as it will be evident which metric we are referring to.

\section{Perturbed acoustic metric}
\label{sec:perturbedacousticmetric-sec}
We perturb $g_{\mu\nu}(t,\mathbf{x})^{(an)}$ written in Equation (\ref{eq:analogue_metric_BEC}) in order to write it as $g_{\mu\nu}^{(an)}=\eta_{\mu\nu}^{(an)}+\epsilon h_{\mu\nu}^{(an)}$. For simplicity, from now on in this section, we omit the apex $^{(an)}$ since we are only working with the acoustic metric. We consider small perturbations to the background quantities ($\epsilon \ll 1$):
\begin{equation}
    \begin{aligned}
        &n_c \to n_c+\epsilon\delta n_c\\
        &v_{i} \to v_{i}+\epsilon\delta v_{i}\\
        &c_s \to c_s +\epsilon \delta c_s
    \end{aligned}
\end{equation}
where all the quantities depend on space and time, and $\delta n_c$, $\delta v_i$, and $\delta c_s$ are the fluctuations of the condensate density, $i$-component of the condensate velocity, and sound speed on top of their corresponding averages $n_c$, $v_i$ and $c_s$, respectively. We remark that these perturbations are long-wavelength fluctuations of the background. Thus they are not associated to phonons excitations, as in Equation (\ref{eq:fluctFLuids}), but should be understood as the analogue of the metric fluctuations discussed in General Relativity, see Equation (\ref{eq:linexpGW_1}). Notice that $\delta c_s$ can be due both to a perturbation of the condensate density (determined by some additional external interaction) as well as to a perturbation of the scattering length, that as we have already seen can be achieved through a Feshbach resonance \cite{Feshbah}: see Equation (\ref{eq:speedSound_BEC}). From Equation (\ref{eq:speedSound_BEC}) we have
\begin{equation}
    \delta c_s^2= \frac{4 \pi \hslash^2}{m}\frac{n_c}{m}\delta a +\frac{4\pi a\hslash^2}{m}\frac{1}{m}\delta n_c.
\end{equation}
Thus, since $\delta c_s^2=2 c_s\delta c_s$, multiplying for $\frac{1}{2 c_s^2}$ we get
\begin{equation}
    \label{eq:deltaCS_funzione}
    \frac{\delta c_s}{c_s}=\frac{1}{2}\frac{\delta a }{a}+\frac{1}{2}\frac{\delta n_c}{n_c}.
\end{equation}
The acoustic metric thus becomes (with $v$ the modulus of $\mathbf{v}$):
\begin{equation}
    \label{eq:analog_pert}
    g_{\mu\nu}(t,\mathbf{x})\equiv \frac{n_c+\epsilon\delta n_c}{m (c_s+\epsilon\delta c_s)}
\begin{pmatrix}
    -\{(c_s+\epsilon\delta c_s)^2-(v+\epsilon\delta v)^2\} & -(v_{j}+\epsilon\delta v_{j}) \\
     -(v_{i}+\epsilon\delta v_{i}) & \delta_{ij}   \\
\end{pmatrix}
,
\end{equation}
with $i,j=1,2,3$, forming a $4\times 4$ symmetric matrix. Our aim is to write this $g_{\mu\nu}(t,\mathbf{x})$ as a background metric $\eta_{\mu\nu}(t,\mathbf{x})$ plus a perturbation $h_{\mu\nu}(t,\mathbf{x})$. We thus expand the expression obtained in Equation (\ref{eq:analog_pert}) at first order in $\epsilon$. We notice that the prefactor is:
\begin{equation}
     \frac{n_c\left(1+\epsilon\frac{\delta n_c}{n_c}\right)}{m c_s\left(1+\epsilon\frac{\delta c_s}{c_s}\right)}\simeq\frac{n_c}{m c_s}\biggl(1+\epsilon\frac{\delta n_c}{n_c}\biggr)\biggl(1-\epsilon\frac{\delta c_s}{c_s}\biggr)\simeq \frac{n_c}{m c_s}+\epsilon\frac{n_c}{m c_s}\biggl(\frac{\delta n_c}{n_c}-\frac{\delta c_s}{c_s}\biggr).
\end{equation}
Now we focus on the term that appears in $g_{00}$, not considering the prefactor:
\begin{equation}
    -\left\{\left[c_s\left(1+\epsilon\frac{\delta c_s}{c_s}\right)\right]^2-\left[v\left(1+\epsilon\frac{\delta v}{v}\right)\right]^2\right\}\simeq -(c_s^2-v^2)+\epsilon\left(2 v^2 \frac{\delta v}{v}-2 c_s^2 \frac{\delta c_s}{c_s}\right).
\end{equation}
Using these results we find:
\begin{equation}
    g_{00}=\frac{n_c}{m c_s}[-(c_s^2-v^2)]+\epsilon\frac{n_c}{m c_s} \biggl[-(c_s^2-v^2)\frac{\delta n_c}{n_c}-(c_s^2+v^2)\frac{\delta c_s}{c_s} +2v^2\frac{\delta v}{v}\biggr]
\end{equation}
\begin{equation}
    g_{0i}=\frac{n_c}{m c_s}(-v_{i})+\epsilon\frac{n_c}{m c_s}\biggl[-v_{i}\frac{\delta n_c}{n_c}+v_{i}\frac{\delta c_s}{c_s}-v_{i}\frac{\delta v_{i}}{v_{i}}\biggr]
\end{equation}
\begin{equation}
    g_{ij}=\frac{n_c}{m c_s}\delta_{ij}+\epsilon\frac{n_c}{m c_s}\left(\frac{\delta n_c}{n_c}-\frac{\delta c_s}{c_s}\right)\delta_{ij}.
\end{equation}
Therefore, we can write $g_{\mu\nu}=\eta_{\mu\nu}+\epsilon h_{\mu\nu}$ by identifying the background metric $\eta_{\mu\nu}$ as
\begin{equation}
    \label{eq:eta_an}
    \eta_{\mu\nu}(t,\mathbf{x})= \frac{n_c}{m c_s}
\begin{pmatrix}
    -(c_s^2-v^2) & -v_{x} & -v_{y} & -v_{z}\\
    -v_{x} &1&0&0\\
    -v_{y} &0&1&0\\
    -v_{z} &0&0&1\\
\end{pmatrix}
,
\end{equation}
and the perturbation metric $h_{\mu\nu}$ as
\begin{equation}
    \label{eq:h_an}
    h_{\mu\nu}(t,\mathbf{x})= \frac{n_c}{m c_s}
\begin{pmatrix}
    h_{00} & h_{01} & h_{02} & h_{03}\\
    h_{01} &h_{11}&0&0\\
    h_{02} &0&h_{22}&0\\
    h_{03} &0&0&h_{33}\\
\end{pmatrix}
,
\end{equation}
with
\begin{equation}
    h_{00}=-(c_s^2-v^2)\frac{\delta n_c}{n_c}+2v^2\frac{\delta v}{v}-(c_s^2+v^2)\frac{\delta c_s}{c_s}
\end{equation}
\begin{equation}
    h_{0i}=-v_{i}\frac{\delta n_c}{n_c}+v_{i}\frac{\delta c_s}{c_s}-v_{i}\frac{\delta v_{i}}{v_{i}}
\end{equation}
\begin{equation}
    h_{ij}=\left(\frac{\delta n_c}{n_c}-\frac{\delta c_s}{c_s}\right)\delta_{ij}.
\end{equation}
At this point, we explicit the dependence of $\frac{\delta c_s}{c_s}$ on $\frac{\delta a}{a}$ and $\frac{\delta n_c}{n_c}$ in the $h_{\mu\nu}$ terms, as computed in Equation (\ref{eq:deltaCS_funzione}):
\begin{equation}
    h_{00}=-\frac{1}{2}\left(3 c_s^2-v^2\right)\frac{\delta n_c}{n_c}-\frac{(c_s^2+v^2)}{2}\frac{\delta a }{a}+2v^2\frac{\delta v}{v}
\end{equation}
\begin{equation}
    h_{0i}=-v\frac{\delta v}{v}-\frac{1}{2}v\frac{\delta n_c}{n_c}+\frac{1}{2}v\frac{\delta a }{a}
\end{equation}
\begin{equation}
    h_{ij}=\left(\frac{1}{2}\frac{\delta n_c}{n_c}-\frac{1}{2}\frac{\delta a}{a}\right)\delta_{ij}.
\end{equation}
Now, in order to be able in next steps to compare this acoustic perturbed metric with that of a gravitational wave in vacuum, we want to express it in the coordinates $x^{\mu}=(c_s t,x,y,z)$. Notice that all what we have written so far was expressed in $x^{\mu}=(t,x,y,z)$. In these new coordinates $\eta_{\mu\nu}$, see Equation (\ref{eq:eta_an}), becomes:
\begin{equation}
    \label{eq:analog_eta_dx0=csdt}
    \eta_{\mu\nu}= \frac{n_c}{m c_s}
    \begin{pmatrix}
        -\left(1-\frac{v^2}{c_s^2}\right) &  -\frac{v_{j}}{c_s}\\
        -\frac{v_{i}}{c_s} & \delta_{ij}\\
    \end{pmatrix}
    .
\end{equation}
To obtain this expression we have used that $dt\rightarrow dx^0= c_s dt$ and so since $dt=dx^0/c_s$, we thus have that $\eta_{00}dt^2\to (\eta_{00}/c_s^2)(dx^0)^2$, $\eta_{0i}dtdx^i\to (\eta_{0i}/c_s)dx^0 dx^i$, while $\eta_{ij}$ are unchanged. For what concerns $h_{\mu\nu}$, in these new coordinates we find:
\begin{equation}
    \label{eq:analog_h_dx0=csdt}
    h_{\mu\nu}= \frac{n_c}{m c_s}
    \begin{pmatrix}
        2\frac{v^2}{c_s^2}\frac{\delta v}{v}-\frac{1}{2}\frac{(c_s^2+v^2)}{c_s^2}\frac{\delta a}{a}-\frac{1}{2}\frac{3c_s^2-v^2}{c_s^2}\frac{\delta n_c}{n_c} & -\frac{1}{2}\frac{v_{j}}{c_s}\frac{\delta n_c}{n_c}+\frac{1}{2}\frac{v_{j}}{c_s}\frac{\delta a }{a}-\frac{v_{j}}{c_s}\frac{\delta v_{j}}{v_{j}} \\
        -\frac{1}{2}\frac{v_{i}}{c_s}\frac{\delta n_c}{n_c}+\frac{1}{2}\frac{v_{i}}{c_s}\frac{\delta a }{a}-\frac{v_{i}}{c_s}\frac{\delta v_{i}}{v_{i}} & \left(\frac{1}{2}\frac{\delta n_c}{n_c}-\frac{1}{2}\frac{\delta a}{a}\right)\delta_{ij} \\  
    \end{pmatrix}
    .
\end{equation}
We want to compare Equation (\ref{eq:analog_eta_dx0=csdt}) with the Minkowski metric and Equation (\ref{eq:analog_h_dx0=csdt}) with a gravitational wave metric. Apart from a prefactor, $\eta_{\mu\nu}$ written in Equation (\ref{eq:analog_eta_dx0=csdt}) is the flat spacetime metric if $v_i=0\, \forall i$. With this choice, the perturbation of the acoustic metric becomes:
\begin{equation}
    \label{eq:acousticPerturbMetric}
    h_{\mu\nu}=\frac{n_c}{m c_s}
    \begin{pmatrix}
        -\frac{3}{2}\frac{\delta n_c}{n_c}-\frac{1}{2}\frac{\delta a}{a}& -\frac{\delta v_j}{c_s}\\
        -\frac{\delta v_i}{c_s}& \left(\frac{1}{2}\frac{\delta n_c}{n_c}-\frac{1}{2}\frac{\delta a}{a}\right)\delta_{ij}\\
    \end{pmatrix}
    .
\end{equation}
From now on we reintroduce the notation with $^{(an)}$ to refer to the acoustic metric. At this point we want to find a gauge in which the gravitational wave metric is written in the same form of the metric in Equation (\ref{eq:acousticPerturbMetric}), so that we can compare them.

\section{Gravitational waves in different gauges}
\label{sec:GW_newgauges-sec}
In this section, we proceed to find an expression for $h_{\mu\nu}$, i.e. the metric of a real gravitational wave, that is comparable to $h_{\mu\nu}^{(an)}$. To this aim, we use the symmetry of the linearized theory of General Relativity written in Equation (\ref{eq:gaugetrasf_lin}), i.e.
\begin{equation}
    \label{eq:gaugetrasf_riportatoin5}
    x^\mu \to x^{\prime\mu}=x^\mu +\epsilon \zeta^\mu,
\end{equation}
that gives rise to the transformation of $h_{\mu\nu}$ written in Equation (\ref{eq:gaugetrasf_hmunu}), that we report here for convenience:
\begin{equation}
\label{eq:gaugetrasfHMUNU_riportato5}
    \epsilon h_{\mu\nu}(x^\rho)\to \epsilon h'_{\mu\nu}(x^{\prime\rho})=\epsilon[ h_{\mu\nu}-(\partial_\mu \zeta_\nu+\partial_\nu \zeta_\mu)].
\end{equation}
Indeed, the usual coordinate frame that is used to write a gravitational wave, thus the $\text{TT}$ gauge, is not useful for our purpose, since a gravitational wave in the $\text{TT}$ gauge (see Equation \ref{eq:gw_tt}) is not comparable with $h_{\mu\nu}^{(an)}$. We remember here that a gravitational wave moving along the $z$ axis in the $\text{TT}$ gauge can be expressed as:
\begin{equation}
    \label{eq:gwTT_ridettoCap6}
    h_{\mu\nu}^{TT}(t,z)=
    \begin{pmatrix}
     0 &0&0&0\\
     0& h^{TT}_{+}(t,z) & h^{TT}_{\times}(t,z) &0 \\
     0& h^{TT}_{\times}(t,z)& -h^{TT}_{+}(t,z) &0 \\
     0 &0&0&0\\
    \end{pmatrix}=
    \begin{pmatrix}
     0 &0&0&0\\
     0& h_{+} & h_{\times} &0 \\
     0& h_{\times}& -h_{+} &0 \\
     0 &0&0&0\\
    \end{pmatrix}
    \cos \left(\omega(t-z/c)\right).
\end{equation}
In order to obtain a convenient expression, we begin with considering the Fermi normal coordinates with the assumption that time-varying Newtonian gravitational forces are of negligible magnitude, so that only gravitational waves contribute to the Riemann tensor. The metric in these coordinates is thus:
\begin{equation}
    \mathrm{d}s^2=-c^2 \mathrm{d}t^2(1+ R_{0i0j}x^i x^j)-2 c\,\mathrm{d}t\,\mathrm{d}x^i\left(\frac{2}{3}R_{0jik} x^j x^k \right) +\mathrm{d}x^i \mathrm{d} x^j \left(\delta_{ij}-\frac{1}{3}R_{ikjl}x^k x^l\right).
\end{equation}
Now, we use that the linearized Riemann tensor in Equation (\ref{eq:linearizedRiemann}) is invariant under gauge transformation in Equation (\ref{eq:gaugetrasf_riportatoin5}) \cite{Maggiore:2007ulw}. Due to this, we can compute it in the reference frame of our preference, assuming that we are performing reference frame transformations according to the Equation (\ref{eq:gaugetrasf_riportatoin5}). We choose to calculate it in the $\text{TT}$ gauge for a gravitational wave moving along the $z$ direction. In the calculation performed in \cite{Hartley_GWsimulation}, the space dependence of the gravitational wave was neglected, ending up with
\begin{equation}
    \label{eq:FermiNormalCoord_nospace}
    \mathrm{d}s^2=-c^2 \mathrm{d} t^2 \left[1-\epsilon \frac{1}{2c^2} ((x^2-y^2)\partial_t^2 h_+^{\text{TT}}+2xy \partial_t^2 h_\times^{\text{TT}})\right]+\delta_{ij}\mathrm{d}x^i \mathrm{d} x^j,
\end{equation}
that is the Minkowski metric plus a perturbation that enters only in the 00 component. However, in our study we do not want to neglect the space dependence of the gravitational wave, and with this choice we get:
\begin{equation}
    \begin{aligned}
        \mathrm{d}s^2&=-c^2\mathrm{d}t^2\left[1-\epsilon \frac{1}{2c^2} ((x^2-y^2)\partial_t^2 h_+^{\text{TT}}+2xy \partial_t^2 h_\times^{\text{TT}})\right]-\epsilon\mathrm{d}t \mathrm{d}x\frac{4}{3}\left[yz\partial_z \partial_t h_{\times}^{\text{TT}}+\right.\\
        &+\left.xz\partial_z \partial_t h_+^{\text{TT}} \right] -\epsilon \mathrm{d}t\mathrm{d}y\frac{4}{3}\left[xz\partial_z \partial_t h_{\times}^{\text{TT}}-yz\partial_z \partial_t h_+^{\text{TT}} \right]+\epsilon\mathrm{d}t \mathrm{d}z\frac{4}{3}\left[ (x^2-y^2)\partial_z \partial_t h_+^{\text{TT}}+\right.\\
        &+\left. 2xy\partial_z\partial_t h_\times^{\text{TT}}\right] +\mathrm{d}x^2\left[1+\epsilon\frac{1}{6}z^2\partial_z^2 h_+^{\text{TT}}\right]+\mathrm{d}y^2\left[1-\epsilon \frac{1}{6}z^2\partial_z^2 h_+^{\text{TT}}\right]+\mathrm{d}z^2 \biggl[1+\\
        &\left.+\epsilon \frac{1}{6}((x^2-y^2)\partial_z^2 h_+^{\text{TT}}+2xy \partial_z^2 h_\times^{\text{TT}})\right]+\epsilon\mathrm{d}x\mathrm{d}y\frac{2}{3}z^2\partial_z^2 h_\times^{\text{TT}}-\epsilon\mathrm{d}x\mathrm{d}z\frac{2}{3}\left[xz\partial_z^2 h_+^{\text{TT}}+\right.\\
        &+\left.yz\partial_z^2 h_\times^{\text{TT}}\right]-\epsilon \mathrm{d}y\mathrm{d}z\frac{2}{3}\left[xz\partial_z^2 h_\times^{\text{TT}}-yz\partial_z^2 h_+^{\text{TT}}\right].
    \end{aligned}
\end{equation}
We notice that this expression reduces to the one in Equation (\ref{eq:FermiNormalCoord_nospace}) when we neglect the space dependence of $h^{\text{TT}}_{\mu\nu}$, so when we neglect all the partial derivatives with respect to $z$. Moreover, we notice that the metric in this coordinate frame cannot be compared to $h_{\mu\nu}^{(an)}$ in Equation (\ref{eq:acousticPerturbMetric}). Due to this, we must transition to a different reference frame, distinct from both the $\text{TT}$ gauge and the proper detector frame, where through Equation (\ref{eq:gaugetrasfHMUNU_riportato5}) we get an $h'_{\mu\nu}(x^{\prime\rho})$ comparable to $h_{\mu\nu}^{(an)}(x^\rho)$. For simplicity, in the following we look for suitable coordinate transformations starting from the $\text{TT}$ gauge, in which we know the explicit form of a gravitational wave. Thus, we look for a coordinate frame transformation from the $\text{TT}$ gauge of the form
\begin{equation}
    \label{eq:CoordTrasf_daTT}
    x_{\text{TT}}^\mu \to x^{\prime\mu} =x^\mu_{\text{TT}}+\epsilon \zeta^\mu,
\end{equation}
i.e. we look for appropriate $\zeta^\mu$, from which we get a transformation of the gravitational wave metric as follows (the Minkowski one is unchanged)
\begin{equation}
    \label{eq:htrasf_daTT}
    h^{\text{TT}}_{\mu\nu}(x^\rho_{\text{TT}})\to h'_{\mu\nu}(x^{\prime \rho})=h^{\text{TT}}_{\mu\nu}-(\partial_\mu\zeta_\nu+\partial_\nu \zeta_\mu).
\end{equation}
Since our goal is to find a $h'_{\mu\nu}(x^{\prime \rho})$ that is written in the ``same way'' as $h_{\mu\nu}^{(an)}$, all their elements should be the same. We notice from Equation (\ref{eq:acousticPerturbMetric}) that $h_{\mu\nu}^{(an)}$ is diagonal in the spatial $3\times3$ submatrix, with all the elements in the diagonal equal to each other, and that, while the 00 component is related to the elements in the spatial submatrix, the $0i$ terms are not related to either the $00$ elements or $ii$ ones or to each other. We will see later that the continuity and Euler equations determine some additional relations between the components of $h_{\mu\nu}$.\\
Hereafter, whenever we write $h^{\text{TT}}_{+,\times}$ it means that the space and time dependence is implicit, while whenever we write $h_{+,\times}$ it means that we have explicated the space-time dependence, in accordance with what was written earlier in Equation (\ref{eq:gw_tt}) and reported again in Equation (\ref{eq:gwTT_ridettoCap6}). In addition, for simplicity in the following we refer to $x^\rho_{\text{TT}}$ as $x^\rho$. In the analysis it would be necessary to consider an expansion both in $\epsilon$ and in the small parameter $\ell \omega/c$, where $\ell$ is the characteristic length of the system. A physical interpretation of this new small parameter is done in section \ref{sec:comparisonGW-acoustic-sec}. \\
Here below, we analyze three different types of spatial transformations of $\zeta^\mu$ and we denote them as transformations \rom{1}, \rom{2} and \rom{3}. For transformation \rom{1}, we analyze three different cases, (a), (b) and (c), where we vary the choice of $\zeta_0$. For transformation \rom{2} we only consider two different cases of $\zeta_0$: (a) and (b). Finally, we study transformation \rom{3} only with one specific choice of $\zeta_0$. It should be noted that all these coordinate system transformations are not taken from the literature, but they are derived for the first time in this work to successfully represent a propagating analogue gravitational wave.

\subsection{Transformation \rom{1}}
The first transformation that we consider has the following spatial components of $\zeta_\mu$, referring to Equation (\ref{eq:CoordTrasf_daTT}):
    \begin{equation}
    \label{eq:spatialZeta_1}
        \zeta_i=\left(\frac{1}{2}(x h^{TT}_{+}+y h^{TT}_{\times}), \frac{1}{2}(x h^{TT}_{\times}-y h^{TT}_+), 0\right).
    \end{equation}
With this choice, we get no diagonal space components, $h'_{12}=0$, but $h'_{13},h'_{23}\not=0$:
\begin{equation}
\label{eq:h13_trasf1}
    h'_{13}=-\frac{1}{2}(x\partial_z h^{TT}_+(t,z) + y\partial_z h^{TT}_{\times}(t,z))=-\frac{\omega}{2c}(x h_+ + y h_{\times})\sin(\omega(t-z/c)),
\end{equation}
\begin{equation}
\label{eq:h23_trasf1}
    h'_{23}=-\frac{1}{2}(x\partial_z h^{TT}_{\times}(t,z)-y\partial_z h^{TT}_{+}(t,z))=-\frac{\omega}{2c}(x h_{\times}-y h_{+})\sin(\omega(t-z/c)).
\end{equation}
If we consider $x\omega/c$ and $y\omega/c$ as small parameters, (the $\ell\omega/c$ parameter previously mentioned) then $h'_{13}$ and $h'_{23}$ give rise to corrections to the background metric of the order of $\mathcal{O}(\epsilon x\omega/c)$ or $\mathcal{O}(\epsilon y\omega/c)$ and thus negligible with respect to terms of $\mathcal{O}(\epsilon)$. So, to neglect $h'_{13}$ and $h'_{23}$ we need some $h'_{0\mu}$ of the order of $\epsilon$. The remaining components of $h'_{\mu\nu}$ depend on the choice of $\zeta_0$ and thus, we explore various options for $\zeta_0$ to determine whether it is possible to obtain an $h'_{\mu\nu}$ comparable to $h^{(an)}_{\mu\nu}$.

\subsubsection{Case (a)}
We start by considering 
\begin{equation}
    \zeta_0=0.
\end{equation}
With this choice we get $h'_{00}=0$ and
\begin{equation}
    h'_{01}=-\frac{1}{2c}(x\partial_t h^{TT}_+(t,z) +y\partial_t h^{TT}_{\times}(t,z))=\frac{\omega}{2c}(x h_+ +y h_{\times})\sin(\omega(t-z/c)),
\end{equation}
\begin{equation}
    h'_{02}=-\frac{1}{2c}(x\partial_t h^{TT}_{\times}(t,z)-y\partial_t h^{TT}_{+}(t,z))=\frac{\omega}{2c}(x h_{\times}-y h_{+})\sin(\omega(t-z/c)).
\end{equation}
Therefore, apart from a sign these are the same expressions reported in Equations (\ref{eq:h13_trasf1}) and (\ref{eq:h23_trasf1}). Hence, this case is not useful for our purpose.

\subsubsection{Case (b)}
Now we choose
\begin{equation}
    \zeta_0=ct(ah^{TT}_+ + b h^{TT}_{\times}),
\end{equation}
with $a$ and $b$ real numbers. In this case we get:
\begin{equation}
\begin{aligned}
     h'_{00}=-2(ah^{TT}_+(t,z) + bh^{TT}_{\times}(t,z))&-2t(a\partial_t h^{TT}_+(t,z) + b\partial_t h^{TT}_{\times}(t,z))=\\
    =-2(ah_+ + bh_{\times})\cos(\omega(t-z/c)&)+2\omega t(ah_+ + bh_{\times})\sin(\omega(t-z/c)),\\
\end{aligned}
\end{equation}
\begin{equation}
    h'_{01}=-\frac{1}{2c}(x\partial_t h^{TT}_+(t,z) +y\partial_t h^{TT}_{\times}(t,z))=\frac{\omega}{2c}(x h_+ +y h_{\times})\sin(\omega(t-z/c)),
\end{equation}
\begin{equation}
    h'_{02}=-\frac{1}{2c}(x\partial_t h^{TT}_{\times}(t,z)-y\partial_t h^{TT}_{+}(t,z))=\frac{\omega}{2c}(x h_{\times}-y h_{+})\sin(\omega(t-z/c)),
\end{equation}
\begin{equation}
    h'_{03}=-tc(a\partial_z h^{TT}_+(t,z) + b\partial_z h^{TT}_{\times}(t,z))=-\omega t(ah_+ + bh_{\times})\sin(\omega(t-z/c)).
\end{equation}
With this choice of $\zeta_0$ we can neglect $h'_{13},\,h'_{23},\,h'_{01}$ and $h'_{02}$ with respect to $h'_{00}$ and $h'_{03}$, and so in principle we can compare this $h'_{\mu\nu}$ with the analogue metric. However, we have to pay attention to the fact that we have obtained terms that are linearly dependent on time: as time lapses we expect $h'_{00}$ and $h'_{03}$ to increase, and at some point they may no longer be considered perturbations. Because of this problem with the time dependence of $h'_{00}$ and $h'_{03}$, we decide to dismiss this case.

\subsubsection{Case (c)}
As a final choice we consider 
\begin{equation}
    \label{eq:zeta0_trasf1_lin}
    \zeta_0=(lx+by+dz)(fh^{TT}_+ + g h^{TT}_{\times}),
\end{equation}
where $l,b,d,f,g$ are real numbers. We obtain:
\begin{equation}
\begin{aligned}
    h'_{00}=-\frac{2}{c}(lx+by+dz)(f\partial_t h^{TT}_+ & (t,z)+ g \partial_t h^{TT}_{\times}(t,z))=\\=
    \frac{2\omega}{c}(lx+by+dz)(fh_+ + &g h_{\times})\sin(\omega(t-z/c)),
\end{aligned}
\end{equation}
\begin{equation}
  \begin{aligned}
      h'_{01}=-\frac{1}{2c}(x\partial_t h^{TT}_+(t,z)+y\partial_t h^{TT}_{\times}(t&,z))-l(fh^{TT}_+(t,z)+g h^{TT}_{\times}(t,z))=\\
    =\frac{\omega}{2c}(xh_+ + yh_{\times})\sin(\omega(t-z/c))&-l(fh_+ + gh_{\times})\cos(\omega(t-z/c)),\\
  \end{aligned}  
\end{equation}
\begin{equation}
    \begin{aligned}
        h'_{02}=-\frac{1}{2c}(x\partial_t h^{TT}_{\times}(t,z)-y\partial_t h^{TT}_+(t&,z))-b(fh^{TT}_+(t,z)+gh^{TT}_{\times}(t,z))=\\
            =\frac{\omega}{2c}(xh_{\times} - yh_{+})\sin(\omega(t-z/c))&-b(fh_+ + gh_{\times})\cos(\omega(t-z/c)),
    \end{aligned}
\end{equation}
\begin{equation}
    \begin{aligned}
        h'_{03}=-d(fh^{TT}_+(t,z)+gh^{TT}_{\times}(t,z))-(l&x+by+dz)(f\partial_z h^{TT}_+ + g\partial_z h^{TT}_{\times})=\\=-d(fh_+ + gh_{\times})\cos(\omega(t-z/c))-\frac{\omega}{c}(lx&+by+dz)(fh_+ + gh_{\times})\sin(\omega(t-z/c)).
    \end{aligned}
\end{equation}
Because of the fact that we want to neglect $h'_{13}$ and $h'_{23}$, we shall work in a regime where $x\omega/c$ and $y\omega/c$ are small parameters, as already said. The leading terms are then $h'_{00},\,h'_{01},\, h'_{02}$ and $h'_{03}$. Therefore, this situation seems promising for our goal. However, we have to notice that what we have written until now are the $h'_{\mu\nu}(x^\rho)$ terms, so the new gravitational wave metric written in terms of the old coordinates. At this point, we want to determine $h'_{\mu\nu}(x^{\prime \rho})$. To do that, we have to invert the relations:
\begin{equation}
    \label{eq:trasfCoord_dainvertire_1}
    \begin{aligned}
        ct'=ct-\epsilon (lx+by+dz)(&fh_+ + g h_{\times})\cos(\omega(t-z/c)),\\
        x'=x+\epsilon \frac{1}{2}(x h_+ + &y h_\times) \cos(\omega(t-z/c)),\\
        y'=y+\epsilon \frac{1}{2}(xh_\times-&y h_+)\cos(\omega(t-z/c)),\\
        &z'=z.
    \end{aligned}
\end{equation}
As a first observation, we notice that since $z'=z$, the direction of wave propagation in the new reference frame is along the $z'$ axis. Consequently, $x'$ and $y'$ are perpendicular to the direction of propagation. At the leading order in $\epsilon$, the inverse relations are:
\begin{equation}
    \begin{aligned}
        &x\simeq x'-\epsilon \left[\frac{1}{2}x' h_+ \cos(\omega(t-z'/c))+\frac{1}{2}y' h_\times \cos(\omega(t-z'/c))\right],\\
        &y\simeq y'+\epsilon\left[\frac{1}{2} y' h_+ \cos(\omega(t-z'/c))-\frac{1}{2}x'h_\times \cos(\omega(t-z'/c))\right].
    \end{aligned}
\end{equation}
Substituting these results in the equation for $t'$ we get:
\begin{equation}
    \label{eq:ct_dainvertire}
    ct'=ct-\epsilon(lx'+by'+dz')(fh_++gh_\times)\cos(\omega(t-z'/c))+\mathcal{O}(\epsilon^2).
\end{equation}
To invert this equation, we substitute $t$ inside the cosine with $t'$ so that the order of expansion of $t$ inside the cosine is consistent with that of $z'$. Furthermore, if we introduce a correction of $\mathcal{O}(\epsilon)$ to $t=t'$ within the cosine term, the expansion in $\epsilon$ yields terms of $\mathcal{O}(\epsilon^3)$, which are therefore negligible. We have verified that the $\epsilon$-order analytical solution to Equation (\ref{eq:ct_dainvertire}) obtained when considering $\omega(t-z'/c)\ll 1$, is consistent with the result when replacing $t\to t'$ in Equation (\ref{eq:ct_dainvertire}). After performing the calculations, we obtain:
\begin{equation}
    \begin{aligned}
        ct=ct'+\epsilon(lx'+by'+dz')(&f h_+ +gh_\times)\cos(\omega(t'-z'/c)),\\
        x=x'-\epsilon\frac{1}{2}(x'h_++&y'h_\times)\cos(\omega(t'-z'/c)),\\
        y=y'+\epsilon \frac{1}{2}(y' h_+ -&x' h_\times)\cos(\omega(t'-z'/c)),\\
        &z=z'.
    \end{aligned}
\end{equation}
At this point we are able to write $h'_{\mu\nu}(x^{\prime \rho})$ (for the sake of clarity we specify that $x^{\prime\rho}=(c t',x',y',z')$):
\begin{equation}
    h'_{13}(x^{\prime \rho})=-\frac{\omega}{2c}(x' h_+ + y' h_{\times})\sin(\omega(t'-z'/c)),
\end{equation}
\begin{equation}
    h'_{23}(x^{\prime \rho})=-\frac{\omega}{2c}(x' h_{\times}-y' h_{+})\sin(\omega(t'-z'/c)),
\end{equation}
\begin{equation}
    h'_{00}(x^{\prime \rho})=
    \frac{2\omega}{c}(lx'+by'+dz')(fh_+ + g h_{\times})\sin(\omega(t'-z'/c)),
\end{equation}
\begin{equation}
    h'_{01}(x^{\prime \rho})=\frac{\omega}{2c}(x'h_+ + y'h_{\times})\sin(\omega(t'-z'/c))-l(fh_+ + gh_{\times})\cos(\omega(t'-z'/c)), 
\end{equation}
\begin{equation}
    h'_{02}(x^{\prime \rho})=\frac{\omega}{2c}(x'h_{\times} - y'h_{+})\sin(\omega(t'-z'/c))-b(fh_+ + gh_{\times})\cos(\omega(t'-z'/c)),
\end{equation}
\begin{equation}
    h'_{03}(x^{\prime \rho})=-d(fh_+ + gh_{\times})\cos(\omega(t'-z'/c))-\frac{\omega}{c}(lx'+by'+dz')(fh_+ + gh_{\times})\sin(\omega(t'-z'/c)),
\end{equation}
and all the other components are equal to zero. We notice that under the assumption $x'\omega/c,\,y'\omega/c \ll 1$, which we henceforth refer to as the long wavelength limit, the dominant terms are:
\begin{equation}
    \label{eq:trasf1_h00}
    h'_{00}(x^{\prime \rho})=2dz'\frac{\omega}{c}(fh_++gh_\times)\sin(\omega(t'-z'/c)),
\end{equation}
\begin{equation}
    \label{eq:trasf1_h01}
    h'_{01}(x^{\prime \rho})=-l(fh_+ + gh_{\times})\cos(\omega(t'-z'/c)), 
\end{equation}
\begin{equation}
    \label{eq:trasf1_h02}
    h'_{02}(x^{\prime \rho})=-b(fh_+ + gh_{\times})\cos(\omega(t'-z'/c)),
\end{equation}
\begin{equation}
    \label{eq:trasf1_h03}
    h'_{03}(x^{\prime \rho})=-d(fh_+ + gh_{\times})\cos(\omega(t'-z'/c))-dz'\frac{\omega}{c}(fh_+ + gh_{\times})\sin(\omega(t'-z'/c)).
\end{equation}
We will see in Section \ref{sec:comparisonGW-acoustic-sec} that the gravitational wave metric written in this reference frame and in the long wavelength limit is comparable with the acoustic metric perturbation $h^{(an)}$. Although this is the transformation that we will use in the comparison with the analogue perturbation, for illustrative purposes we present other different transformations.

\subsection{Transformation \rom{2}}
The second type of transformations that we consider have the following spatial components of $\zeta_\mu$:
\begin{equation}
    \label{eq:spatialZeta_2}
    \zeta_i=\left(\frac{1}{2}(x h^{TT}_{+}+y h^{TT}_{\times}),\frac{1}{2}(x h^{TT}_{\times}-y h^{TT}_+),-\frac{1}{4}((x^2-y^2)\partial_z h^{TT}_+ + 2xy\partial_z h^{TT}_{\times})\right).
\end{equation}
With this choice we get $h'_{12}=h'_{13}=h'_{23}=h'_{11}=h'_{22}=0$ but $h'_{33}\not =0$:
\begin{equation}
    h'_{33}=-\frac{\omega^2}{2c^2}((x^2-y^2) h_{+}+2xy h_{\times})\cos(\omega(t-z/c_s)).
\end{equation}
Now, we check if, with a specific choice of $\zeta_0$ and some approximations, it is possible to neglect $h'_{33}$ and obtain an expression with the same form of the analogue case. In doing so, we apply the same reasoning previously shown, so we only present the relevant results here.

\subsubsection{Case (a)}
We begin with investigating the case with
\begin{equation}
    \zeta_0=0.
\end{equation}
With this choice, following the same reasoning as in the previous case, it is possible to show that the old coordinates written in terms of the new ones are: 
\begin{equation}
    \begin{aligned}
        ct=c&t',\\
        x=x'-\epsilon \frac{1}{2}(x' h_+ +y'h_\times&)\cos(\omega(t'-z'/c)),\\
        y=y'+\epsilon\frac{1}{2}(y'h_+-x'h_\times &)\cos(\omega(t'-z'/c)),\\
        z=z'+\epsilon \frac{1}{4}\left((x^{\prime 2}-y^{\prime 2})\frac{\omega}{c}h_+ + 2\right.&\left. x'y' \frac{\omega}{c} h_\times\right)\sin(\omega(t'-z'/c)).
    \end{aligned}
\end{equation}
The non-linearities in the coordinates arise from the nonlinearity of $\zeta_3$, which is required to make $h'_{13}=h'_{23}=0$ but that introduces $h'_{33}$ as a second-order polynomial in $x,\,y$. Therefore, in the new reference frame the gravitational wave does not propagate exactly along $z'$. In this new gauge the $h'_{\mu\nu}(x')$ components are:
\begin{equation}
    h'_{01}(x^{\prime \rho})=\frac{\omega}{2c}(x' h_+ +y' h_{\times})\sin(\omega(t'-z'/c)),
\end{equation}
\begin{equation}
    h'_{02}(x^{\prime \rho})=\frac{\omega}{2c}(x' h_{\times}-y' h_{+})\sin(\omega(t'-z'/c)),
\end{equation}
\begin{equation}
    h'_{03}(x^{\prime \rho})=\frac{\omega^2}{4c^2}((x^{\prime 2}-y^{\prime 2}) h_++ 2x'y'  h_{\times})\cos(\omega(t'-z'/c)),
\end{equation}
\begin{equation}
    h'_{33}(x^{\prime \rho})=-\frac{\omega^2}{2c^2}((x^{\prime 2}-y^{\prime 2}) h_+ + 2x'y'  h_{\times})\cos(\omega(t'-z'/c)),
\end{equation}
and all the other terms are equal to zero. In the long wavelength limit, it is possible to obtain the gravitational wave metric in a form comparable with the acoustic perturbation one. Indeed, the dominant terms are:
\begin{equation}
    \label{eq:trasf2_h01_0}
    h'_{01}(x^{\prime \rho})=\frac{\omega}{2c}(x' h_+ + y' h_{\times})\sin(\omega(t'-z'/c)),
\end{equation}
\begin{equation}
    \label{eq:trasf2_h02_0}
    h'_{02}(x^{\prime \rho})=\frac{\omega}{2c}(x' h_{\times}-y' h_+) \sin(\omega(t'-z'/c)).
\end{equation}

\subsubsection{Case (b)}
Even though we have already found a case with the spatial choice of $\zeta_\mu$ in Equation (\ref{eq:spatialZeta_2}) that can be compared to the acoustic perturbation metric $h_{\mu\nu}^{(an)}$, for consistency with transformation \rom{1} we consider also the case in which $\zeta_0$ is a linear polynomial in the spatial coordinates:
\begin{equation}
    \zeta_0=(lx+by+dz)(fh_+^{TT}+g h_\times^{TT}).
\end{equation}
With this choice, in the long wavelength limit we recover the case (c) of transformation \rom{1}.

\subsection{Transformation \rom{3}}
The last spatial transformation that we consider is the following:
\begin{equation}
    \begin{aligned}
        \zeta_i=\biggl( \frac{1}{2}(x h_{+}^{TT}+y h_{\times}^{TT})-\frac{1}{4}\biggl(\biggl(x^2y&-\frac{y^3}{3}\biggr) \partial_z^2 h^{TT}_{\times}+\biggl(\frac{x^3}{3}-xy^2\biggr)\partial_z^2 h^{TT}_+\biggr),\\
        \frac{1}{2}(x h^{TT}_{\times}-y h^{TT}_+)-\frac{1}{4}\biggl(\biggl( xy^2&-\frac{x^3}{3}\biggr) \partial_z^2 h^{TT}_{\times}-\biggl(\frac{y^3}{3}-x^2y\biggr)\partial_z^2 h^{TT}_+\biggr),\\
        -\frac{1}{4}((x^2-y^2&)\partial_z h^{TT}_+ + 2xy\partial_z h^{TT}_{\times})\biggr).
    \end{aligned}
\end{equation}
These non-linearities in the coordinates are needed in order to obtain all the diagonal spatial terms of $h'_{\mu\nu}$ equal to each other. Here, we only focus on the case with 
\begin{equation}
    \zeta_0=-\frac{1}{4c}((x^2-y^2)\partial_t h^{TT}_+ + 2xy\partial_t h^{TT}_{\times}).
\end{equation}
From the previous cases, we have learned that to go from $h'_{\mu\nu}(x^\rho)$ to $h'_{\mu\nu}(x^{\prime \rho})$ it is sufficient to replace $x^\rho$ with $x^{\prime\rho}$. This is because we are only considering gauge transformations and metric perturbations that are linear in $\epsilon$. Indeed, taking into account corrections of $\mathcal{O}(\epsilon)$ in the definition of the old coordinates in terms of the new ones leads to corrections of at least $\mathcal{O}(\epsilon^2)$ to the background metric, and we are neglecting these types of terms. With this choice of $\zeta_\mu$, we get:
\begin{equation}
    h'_{00}(x^{\prime\rho})=-\frac{\omega^2}{2c^2} ((x^{\prime 2}-y^{\prime 2}) h_+ +2x'y' h_{\times})\cos(\omega(t'-z'/c)),
\end{equation}
\begin{equation}
    h'_{01}(x^{\prime\rho})=\frac{\omega^3}{4c^3}\biggl(\biggl(x^{\prime 2}y' -\frac{y^{\prime 3}}{3}\biggr) h_{\times}+\biggl(\frac{x^{\prime 3}}{3}-x'y^{\prime 2}\biggr) h_+\biggr)\sin(\omega(t'-z'/c)),
\end{equation}
\begin{equation}
    h'_{02}(x^{\prime\rho})=\frac{\omega^3}{4c^3}\biggl(\biggl(x'y^{\prime 2} -\frac{x^{\prime 3}}{3}\biggr)h_{\times}-\biggl(\frac{y^{\prime 3}}{3}-x^{\prime 2} y'\biggr) h_+\biggr)\sin(\omega(t'-z'/c)),
\end{equation}
\begin{equation}
    h'_{03}(x^{\prime\rho})=\frac{\omega^2}{2c^2}((x^{\prime 2}-y^{\prime 2}) h_{+}+2x'y' h_{\times})\cos(\omega(t'-z'/c)),
\end{equation}
\begin{equation}
    h'_{ij}(x^{\prime\rho})=-\frac{\omega^2}{2c^2}((x^{\prime 2}-y^{\prime 2}) h_+ + 2x'y'h_{\times})\cos(\omega(t'-z'/c))\delta_{ij},
\end{equation}
\begin{equation}
    h'_{13}(x^{\prime\rho})=-\frac{\omega^3}{4c^3}\biggl( \biggl( x^{\prime 2}y'-\frac{y^{\prime 3}}{3}\biggr) h_{\times}+\biggl( \frac{x^{\prime 3}}{3}-x'y^{\prime 2}\biggr)h_{+}\biggr)\sin(\omega(t'-z'/c)),
\end{equation}
\begin{equation}
    h'_{23}(x^{\prime\rho})=-\frac{\omega^3}{4c^3}\biggl(\biggl(x'y^{\prime 2} -\frac{x^{\prime 3}}{3}\biggr) h_{\times}-\biggl(\frac{y^{\prime 3}}{3}- x^{\prime 2}y'\biggr)
    h_{+}\biggr)\sin(\omega(t'-z'/c)),
\end{equation}
and all the other terms are equal to zero. In order to compare these results with the acoustic metric $h_{\mu\nu}^{(an)}$, we have to neglect $h'_{23}$ and $h'_{13}$. To do that, we notice that in the long wavelength limit the dominant terms are:
\begin{equation}
    h'_{00}(x^{\prime\rho})=-h'_{03}(x^{\prime\rho})=-\frac{\omega^2}{2c^2} ((x^{\prime 2}-y^{\prime 2}) h_+ +2x'y' h_{\times})\cos(\omega(t'-z'/c)),
\end{equation}
\begin{equation}
    h'_{ij}(x^{\prime\rho})=-\frac{\omega^2}{2c^2}((x^{\prime 2}-y^{\prime 2}) h_+ + 2x'y'h_{\times})\cos(\omega(t'-z'/c))\delta_{ij}.
\end{equation}

\section{Comparison between a gravitational wave and the perturbed acoustic metric}      
\label{sec:comparisonGW-acoustic-sec}
At this point we want to see if it is really possible to compare a gravitational wave metric written in a new gauge to the acoustic perturbation metric $h_{\mu\nu}^{(an)}$ written in Equation (\ref{eq:acousticPerturbMetric}). We have already stated that in order to put the background acoustic metric in the form of a Minkowski metric we need to consider a system with $v_i=0\,\,\forall i$. Now, we want to determine the other necessary system's characteristics to place it in a regime such that the perturbation of the acoustic metric resembles that of a gravitational wave. We refer to the acoustic perturbed metric written in the same form of a gravitational wave in a given gauge as the analogue gravitational wave. Throughout this procedure, we forget about the prefactor of $h_{\mu\nu}^{(an)}$, which we consider to be uniform and constant over time. Noticeably, since in the analogue model we are reproducing the same background and perturbation metric of a gravitational wave, we have that $h_{\mu\nu}^{(an)}$ satisfies the Einstein's equations. \\
In the following analysis, in order to compare the $h'_{\mu\nu}(x^{\prime\rho})$, i.e. the gravitational wave metric, with the acoustic metric $h_{\mu\nu}^{(an)}$ we make the following identifications:
\begin{itemize}
    \item The $t',x',y',z'$ coordinates of the new reference frame in which the gravitational wave is written are identified as the time and space coordinates of our physical Bose-Einstein condensate system. For simplicity, in the following we refer to $t',x',y',z'$ without the prime and we also refer to $h'_{\mu\nu}$ only with $h_{\mu\nu}$.
    \item The $\omega$ of the gravitational wave is identified as the analogue gravitational wave frequency. 
    \item The speed of light $c$ appearing in the gravitational wave metric is identified with the speed of sound $c_s$ in the analogue system. This means that our analogue of the gravitational wave moves with $c_s$.
\end{itemize}
Let us now investigate in more details the meaning of the requirements $x\omega/c_s,\, y\omega/c_s\ll 1$ in the analogue system, that is the physical system in which we are focusing on. Labelling with $L_x$ and $L_y$ the physical dimensions of the system in the $x$ and $y$ directions respectively, and posing $\Tilde{x}=x/L_x$ and $\Tilde{y}=y/L_y$, we have:
\begin{equation}
    \begin{aligned}
    &x\frac{\omega}{c_s}=\Tilde{x}\frac{L_x}{c_s}\omega=2\pi\Tilde{x}\frac{\omega}{\omega_x},\\
    &y\frac{\omega}{c_s}=\Tilde{y}\frac{L_y}{c_s}\omega=2\pi\Tilde{y}\frac{\omega}{\omega_y},
    \end{aligned}
\end{equation}
with $\omega_x$ and $\omega_y$ two characteristic frequencies of the system defined as
\begin{equation}
    \omega_x=\frac{2\pi c_s}{L_x}, \quad \omega_y=\frac{2\pi c_s}{L_y}.
\end{equation}
Thus, the assumptions $x\omega/c_s\ll 1$ and $ y\omega/c_s\ll 1$ imply:
\begin{equation}
    \label{eq:meaning_approx}
    x\frac{\omega}{c_s}\ll 1 \Leftrightarrow \frac{\omega}{\omega_x}\ll 1, \quad y\frac{\omega}{c_s}\ll 1 \Leftrightarrow \frac{\omega}{\omega_y}\ll 1.
\end{equation}
Hence, our operative conditions require that the wavelength of the analogue gravitational wave is very long with respect to $L_x$ and $L_y$. At this point, we can compare the gravitational wave $h_{\mu\nu}$ and acoustic $h_{\mu\nu}^{(an)}$ matrices. Subsequently, we perform the study for transformation \rom{1} case (c), transformation \rom{2} case (a) and transformation \rom{3}, respectively, since they are the only ones that seems promising for this purpose. 

\begin{itemize}
    \item \textbf{Focusing on transformation \rom{1} (c)}\\
We start by considering the gravitational wave metric $h_{\mu\nu}$ written in the reference frame given by the transformation \rom{1} case (c) in the long wavelength limit (see Equations (\ref{eq:trasf1_h00})-(\ref{eq:trasf1_h03})). Following the identification procedure just presented, we find that in order to simulate a gravitational wave we need to introduce:
\begin{equation}
    \label{eq:comparison_trasf1_general}
    \begin{aligned}
        \frac{\delta a}{a}=&\frac{\delta n_c}{n_c}=-dz\frac{\omega}{c_s}(fh_++gh_\times)\sin(\omega(t-z/c_s)),\\
        &\frac{\delta v_x}{c_s}=l(fh_++gh_\times)\cos(\omega(t-z/c_s)),\\
        &\frac{\delta v_y}{c_s}=b(fh_++gh_\times)\cos(\omega(t-z/c_s)),\\
        \frac{\delta v_z}{c_s}=d(fh_++gh_\times&)\cos(\omega(t-z/c_s))+dz\frac{\omega}{c_s}(fh_++gh_\times)\sin(\omega(t-z/c_s)).
    \end{aligned}
\end{equation}
The first of these equations states that the density variation is only due to the variation of the scattering length. If we consider $d=0$, thus we set equal to zero the real number $d$ of the coordinate transformation in Equation (\ref{eq:zeta0_trasf1_lin}), the use of a Feshbach resonance giving $\delta a$ would not be necessary. Indeed, with $d=0$ we get
\begin{equation}
    \label{eq:comparison_trasf1_d=0}
    \begin{aligned}
        &\frac{\delta a}{a}=\frac{\delta n_c}{n_c}=\frac{\delta v_z}{c_s}=0,\\
        \frac{\delta v_x}{c_s}=l(&fh_++gh_\times)\cos(\omega(t-z/c_s)),\\
        \frac{\delta v_y}{c_s}=b(&fh_++gh_\times)\cos(\omega(t-z/c_s)).
    \end{aligned}
\end{equation}
Given that in this case the variation of the scattering length is zero, as we will see, these fluctuations are induced by an additional interaction potential. Such interaction potential will naturally emerge in Section \ref{sec:physicalsystems-sec}, when we will investigate whether a Bose-Einstein condensate with these characteristics is physical or not, that is if the continuity and Euler equations as well as the irrotational condition hold or not.

\item \textbf{Focusing on transformation \rom{2} (a)}\\
Now we consider the gravitational wave metric $h_{\mu\nu}$ written in the reference frame given by the transformation \rom{2} case (a). By comparing $h_{\mu\nu}^{(an)}$ and the gravitational wave metric, we can determine the characteristics the system should possess in order to generate an acoustic metric that replicates the analogue of a gravitational wave. We get:
\begin{equation}
    \begin{aligned}
        \frac{\delta a}{a}&=\frac{\delta n_c}{n_c}=\frac{\delta v_z}{c_s}=0,\\
        \frac{\delta v_x}{c_s}=-\frac{\omega}{2 c_s}(&xh_++yh_\times)\sin(\omega(t-z/c_s)),\\
        \frac{\delta v_y}{c_s}=-\frac{\omega}{2 c_s}(&xh_\times -y h_+)\sin(\omega(t-z/c_s)).
    \end{aligned}
\end{equation}
As for the previous transformation taken into account, we will investigate in Section \ref{sec:physicalsystems-sec} whether a Bose-Einstein condensate with these characteristics is physical or not.

\item \textbf{Focusing on transformation \rom{3}}\\
At this point, we focus on the last transformation that we study: transformation \rom{3}. By following the same reasoning as in the previously considered cases, to reproduce the analogue of a gravitational wave, the Bose-Einstein system should exhibit the following characteristics:
\begin{equation}
    \begin{aligned}
        &\frac{\delta n_c}{n_c}=\frac{\delta v_x}{c_s}=\frac{\delta v_y}{c_s}=0,\\
        \frac{\delta a}{a}=\frac{\omega^2}{c_s^2}((x^2&-y^2)h_++2xyh_\times)\cos(\omega(t-z/c_s)),\\
        \frac{\delta v_z}{c_s}=-\frac{\omega^2}{2 c_s^2}((&x^2-y^2)h_++2xyh_\times)\cos(\omega(t-z/c_s)).
    \end{aligned}
\end{equation}
In the next section, we verify if this system is physical or not. 
\end{itemize}

\section{Physical systems}
\label{sec:physicalsystems-sec}
So far, we have found the space-time dependence of the Bose-Einstein condensate properties so that $h_{\mu\nu}^{(an)}$ in Equation (\ref{eq:acousticPerturbMetric}) represents the analogue of a gravitational wave. We now need to check whether those quantities satisfy the continuity and Euler equations. Indeed, in the hydrodynamic limit, using the Madelung representation and $\mathbf{v}=\nabla \theta/m$, where $\theta$ is the phase of the condensate wavefunction and $m$ is the mass of the considered particle, the Gross-Pitaevskii equation could be written as
\begin{equation}
    \label{eq:continuity_Euler_background}
    \begin{cases}
        &\partial_t n_c+\nabla \cdot(n_c\mathbf{v})=0\\
        &m\partial_t \mathbf{v}+\nabla\left(\frac{m v^2}{2}+V_{\text{ext}}+\frac{4\pi a \hslash^2}{m}n_c-\frac{\hslash^2}{2m}\frac{\nabla^2 \sqrt{n_c}}{\sqrt{n_c}}\right)=0,
    \end{cases}
\end{equation}
where we remind that $n_c$ is the condensate density, $\mathbf{v}$ the condensate velocity, $a$ the scattering length and $V_{\text{ext}}$ the external potential. If we consider the following perturbations:
\begin{equation}
    n_c\to n_c+\epsilon \delta n_c,\quad
    v_i\to v_i+\epsilon \delta v_i,\quad
    a\to a+\epsilon \delta a,\quad
    V_{\text{ext}}\to V_{\text{ext}}+\epsilon\delta V_{\text{ext}},
\end{equation}
then the continuity equation becomes
\begin{equation}
    \label{eq:continuity_tot}
    \partial_t n_c+\nabla \cdot(n_c\mathbf{v})+\epsilon \left[\partial_t \delta n_c+\nabla \cdot (n_c \delta \mathbf{v}+\delta n_c \mathbf{v})\right]+\mathcal{O}(\epsilon^2)=0,
\end{equation}
while the Euler's equation is
\begin{equation}
    \label{eq:Euler_tot}
    \begin{aligned}
        &m\partial_t \mathbf{v}+\nabla\left(\frac{m v^2}{2}+V_{\text{ext}}+\frac{4\pi a \hslash^2}{m}n_c-\frac{\hslash^2}{2m}\frac{\nabla^2 \sqrt{n_c}}{\sqrt{n_c}}\right)+\epsilon\left[m\partial_t \delta \mathbf{v}+\nabla\left(m\mathbf{v}\cdot \delta \mathbf{v}\right.+\delta V_{\text{ext}}+\right.\\
        &\left. \left. +\frac{4\pi \delta a \hslash^2}{m}n_c+\frac{4\pi  a \hslash^2}{m}\delta n_c-\frac{\hslash^2}{4m\sqrt{n_c}}\nabla^2 \left(\frac{\delta n_c}{\sqrt{n_c}}\right)+\frac{\hslash^2}{4m\sqrt{n_c}}\frac{\delta n_c}{n_c}\nabla^2\sqrt{n_c} \right)\right]+\mathcal{O}(\epsilon^2)=0.
    \end{aligned}
\end{equation}
Using the continuity and the Euler's equations for the background quantities (Equation (\ref{eq:continuity_Euler_background})) into Equations (\ref{eq:continuity_tot}) and (\ref{eq:Euler_tot}) respectively, we get that to $\mathcal{O}(\epsilon)$ the following equations must hold:
\begin{equation}
    \partial_t \delta n_c+\nabla \cdot (n_c \delta \mathbf{v}+\delta n_c \mathbf{v})=0
\end{equation}
\begin{equation}
    \begin{aligned}
        m\partial_t &\delta \mathbf{v}+\nabla\left(m\mathbf{v}\cdot \delta \mathbf{v}+\delta V_{\text{ext}}+\frac{4\pi \delta a \hslash^2}{m}n_c+\frac{4\pi  a \hslash^2}{m}\delta n_c+\right.\\
        &\left.-\frac{\hslash^2}{4m\sqrt{n_c}}\nabla^2 \left(\frac{\delta n_c}{\sqrt{n_c}}\right)+\frac{\hslash^2}{4m\sqrt{n_c}}\frac{\delta n_c}{n_c}\nabla^2\sqrt{n_c} \right)=0.
    \end{aligned}
\end{equation}
Because of the fact that to obtain an acoustic background metric resembling the Minkowski metric we need to require $\mathbf{v}=\mathbf{0}$, we have that under this condition the equations we need to check become:
\begin{equation}
    \label{eq:cont_euler_background}
    \begin{cases}
        &\partial_t n_c=0\\
        &\nabla\left(V_{\text{ext}}+\frac{4\pi a \hslash^2}{m}n_c-\frac{\hslash^2}{2m}\frac{\nabla^2 \sqrt{n_c}}{\sqrt{n_c}}\right)=0.
    \end{cases}
\end{equation}
To the linear order in $\epsilon$, we have:
\begin{equation}
\label{eq:continuity_Euler_perturb}
    \begin{cases}
        &\partial_t \delta n_c+\nabla(n_c \delta \mathbf{v})=0\\
        &m\partial_t \delta \mathbf{v}+\nabla\left(\delta V_{\text{ext}}+\frac{4\pi \delta a \hslash^2}{m}n_c+\frac{4\pi  a \hslash^2}{m}\delta n_c-\frac{\hslash^2}{4m\sqrt{n_c}}\nabla^2 \left(\frac{\delta n_c}{\sqrt{n_c}}\right)+\frac{\hslash^2}{4m\sqrt{n_c}}\frac{\delta n_c}{n_c}\nabla^2\sqrt{n_c} \right)=0.
    \end{cases}
\end{equation}
In the following, we will always assume that the continuity and Euler equations for the background are valid, so we will assume that Equation (\ref{eq:cont_euler_background}) holds. Instead, we verify whether Equation (\ref{eq:continuity_Euler_perturb}) is valid for the system's quantities found in section \ref{sec:comparisonGW-acoustic-sec}. Hence, we check that the condensate quantities needed to reproduce the analogue of a gravitational wave satisfy the hydrodynamic equations, to ensure that the obtained system is physical.\\
In addition, due to the fact that the velocity of the condensate is $\mathbf{v}=\nabla\theta/m$, we have that the velocity field is irrotational. Thus, we verify also whether introducing $\delta \mathbf{v}$ still maintains the validity of the irrotational condition:
\begin{equation}
    \nabla \times (\mathbf{v}+\epsilon \delta\mathbf{v})=\nabla \times \mathbf{v}+\epsilon \nabla \times \delta \mathbf{v}=0.
\end{equation}
We assume that for the background velocity $\mathbf{v}$ the irrotationality holds, so we only need to check that:
\begin{equation}
    \nabla \times \delta \mathbf{v}=0.
\end{equation}
We perform this study for the transformations \rom{1} (c), \rom{2} (a) and \rom{3}.

\begin{itemize}
\item \textbf{Focusing on transformation \rom{1} (c)}\\
\label{sec:GWtrasfSummary_checonsidero} We now want to verify whether the continuity and Euler equations hold for the characteristic quantities that the system should exhibit in order to reproduce a gravitational wave. This analysis is conducted for a gravitational wave written in the reference frame obtained through transformation \rom{1} (c) and under the long-wavelength assumption. Since in the regime described by the general case in Equation (\ref{eq:comparison_trasf1_general}) we cannot find an explicit space-time dependence for $n_c$ to verify the validity of the continuity equation in Equation (\ref{eq:continuity_Euler_perturb}), we instead focus on the specific case with $d=0$. Therefore, we consider the case in Equation (\ref{eq:comparison_trasf1_d=0}), where the linear perturbation in $\epsilon$ to the continuity equation becomes
\begin{equation}
    (\partial_x n_c)c_s l(fh_+ + gh_\times)\cos(\omega(t-z/c_s))+(\partial_y n_c) c_s b(fh_+ + gh_\times)\cos(\omega(t-z/c_s))=0.
\end{equation}
Therefore, it holds either for a spatially homogeneous $n_c$, or else if $\partial_x n_c=\alpha \partial_y n_c$ and $l=-b/\alpha$ with $\alpha\in \mathbb{R}$. At this point, we determine the external potential perturbation ($\delta V_{\text{ext}}$) needed to satisfy the linear-in-$\epsilon$ Euler's equation (the second equation in Equation (\ref{eq:continuity_Euler_perturb})) for the case under examination. The linear-in-$\epsilon$ Euler's equation in this case is as follows
\begin{equation}
\label{eq:system1}
    \begin{cases}
        &m\partial_t \delta v_x+\partial_x \delta V_{\text{ext}}=0\\
        &m\partial_t \delta v_y+\partial_y \delta V_{\text{ext}}=0\\
        &m\partial_t \delta v_z+\partial_z \delta V_{\text{ext}}=0,
    \end{cases}
\end{equation}
where we have not a priori excluded a possible perturbation $\delta v_z$, induced by the analogue gravitational wave. From the first two equations in the system in (\ref{eq:system1}) we find that the following $\delta V_{\text{ext}}$ is a solution:
\begin{equation}
    \delta V_{\text{ext}}=m\omega c_s(lx+by) (fh_++gh_\times)\sin(\omega(t-z/c_s)).
\end{equation}
The third equation in (\ref{eq:system1}) indicates that introducing this modulation of the external potential induces a perturbation of the condensate velocity along the $z$-axis of the following form:
\begin{equation}
    \label{eq:inducedDELTAVZ1}
     \frac{\delta v_z}{c_s}=\frac{\omega}{c_s}(lx+by)(f h_++gh_\times)\sin(\omega(t-z/c_s).
\end{equation}
We notice that this solution for $\delta v_z$ is consistent with the irrotationality condition. Indeed, we can demonstrate that the introduction of $\delta v_x$ and $\delta v_y$, needed to simulate a gravitational wave, induces a $\delta v_z$ due to the requirement that $\nabla\times (\mathbf{v}+\epsilon\delta \mathbf{v})=0$ must hold for our system. Through calculations, we find that the $\delta v_z$ induced in this manner is precisely the one in Equation (\ref{eq:inducedDELTAVZ1}) and thus the Euler equation and the irrotationality condition are consistent with each other. However, we observe that the perturbation $+\epsilon\delta v_z$ is of order $\mathcal{O}(\epsilon\omega/\omega_{x,y})$. Hence, within the approximation of our analysis, this term can be considered negligible.\\
To summarize, we have found that considering a coordinate frame transformation from the $\text{TT}$ gauge with a $\zeta^\mu$ given by
\begin{equation}
    \zeta^\mu=\left(-(lx+by)(fh_+^{\text{TT}} +gh_\times^{\text{TT}}),\frac{1}{2}(xh_+^{\text{TT}}+y h_\times^{\text{TT}}),\frac{1}{2}(xh_\times^{\text{TT}}-y h_+^{\text{TT}}),0\right),
\end{equation}
we get an expression of the gravitational wave metric that can be compared to the perturbation of the acoustic metric $h_{\mu\nu}^{(an)}$. To cast $h_{\mu\nu}^{(an)}$ in the gravitational wave form we can consider a Bose-Einstein condensate in a regime such that the acoustic metric is conformal to the Minkowski metric, so in particular we need to set $\mathbf{v}=\mathbf{0}$, and then introduce the following velocity perturbations
\begin{equation}
    \begin{aligned}
        &\frac{\delta v_x}{c_s}=l(fh_+ +gh_\times)\cos(\omega(t-z/c_s)),\\
        &\frac{\delta v_y}{c_s}=b(fh_+ +gh_\times)\cos(\omega(t-z/c_s)).\\
    \end{aligned}
\end{equation}
Therefore, in this case leading terms are those of the order of $\mathcal{O}(\epsilon)$ and the space-time dependence of the velocities perturbations is all contained in the factor $\cos(\omega(t-z/c_s))$. 
We have shown that it is possible to bring the Bose-Einstein condensate in a regime such that the acoustic metric it generates for the phonons resembles that of a Minkowski metric with a propagating gravitational wave perturbation written in a particular reference frame.

\item \textbf{Focusing on transformation \rom{2} (a)}\\
Now we follow the same reasoning as in the previous case, but considering the coordinate transformation \rom{2} (a). Here we find that the linear-in-$\epsilon$ continuity equation is valid either for an $n_c$ independent of $x$ and $y$, or for example for $n_c$ such that
\begin{equation}
    n_c=\frac{1}{2}((y^2-x^2)h_\times+2xy h_+) \Tilde{f}(z,\ell),
\end{equation}
with $\Tilde{f}(z,\ell)$ an arbitrary function of a given length $\ell$, and possibly of $z$, so that the dimensions of $\Tilde{f}$ are that of a $\text{length}^{-5}$. In addition, it is possible to show that at our order of approximation, the linear-in-$\epsilon$ Euler's equation is valid if we introduce a perturbation of the external potential of the form
\begin{equation}
    \delta V_{\text{ext}}=\frac{m}{4}\omega^2((x^2-y^2)h_+ +2xyh_\times)\cos(\omega(t-z/c_s))
\end{equation}
and a new velocity perturbation along the $z$-axis of the form 
\begin{equation}
    \frac{\delta v_z}{c_s}=\frac{\omega^2}{4 c_s^2}((x^2-y^2)h_+ + 2xy h_\times)\cos(\omega(t-z/c_s)).
\end{equation}
This required $\delta v_z$ is precisely the same of the one that we find from the irrotationality request. Thus, the Euler equation is consistent with irrotationality. At our order of approximation, this $\delta v_z$ induced as a response of the fluid to the gravitational wave-like perturbation is negligible. \\
To summarize, we have found that in the reference frame obtained by a coordinate transformation from the $\text{TT}$ gauge with a $\zeta^\mu$ given by
\begin{equation}
    \zeta^\mu=\left(0,\frac{1}{2}(x h^{TT}_{+}+y h^{TT}_{\times}),\frac{1}{2}(x h^{TT}_{\times}-y h^{TT}_+),-\frac{1}{4}((x^2-y^2)\partial_z h^{TT}_+ + 2xy\partial_z h^{TT}_{\times})\right),
\end{equation}
the gravitational wave is written in a form so that it can be compared to the perturbation of the acoustic metric $h_{\mu\nu}^{(an)}$. It is thus possible to express $h_{\mu\nu}^{(an)}$ in a gravitational wave form considering a Bose-Einstein condensate in a regime such that the background acoustic metric resembles the flat spacetime metric (we need to set $\mathbf{v}=\mathbf{0}$) and then introducing the following perturbations:
\begin{equation}
    \begin{aligned}
        &\frac{\delta v_x}{c_s}=-\frac{1}{2}\frac{\omega}{c_s}(xh_+ +yh_\times)\sin(\omega(t-z/c)),\\
        &\frac{\delta v_y}{c_s}=-\frac{1}{2}\frac{\omega}{c_s}(xh_\times -yh_+)\sin(\omega(t-z/c)).\\
    \end{aligned}
\end{equation}
Thus, in this case, the dominant terms are on the scale of $\mathcal{O}(\epsilon \omega/\omega_{x,y})$, and the space-time variation of velocity perturbations is more complicated compared to the previous situation.

\item \textbf{Focusing on transformation \rom{3}}\\
As the last case to analyze, we consider transformation \rom{3}. Here, we firstly look at the irrotationality of the flow. We notice that in response to the $\delta v_z$ perturbation introduced in this case, the system develops the following $\delta v_x$ and $\delta v_y$ due to the irrotationality:
\begin{equation}
    \begin{aligned}
    &\frac{\delta v_x}{c_s}=\frac{\omega}{c_s} (xh_+ +yh_\times)\sin(\omega(t-z/c_s)),\\
    &\frac{\delta v_y}{c_s}=\frac{\omega}{c_s}(xh_\times - y h_+)\sin(\omega(t-z/c_s)).
    \end{aligned}
\end{equation}
Therefore, we observe that $\delta v_x$ and $\delta v_y$ dominate over $\delta v_z$ as they result in perturbations of the order of $\mathcal{O}(\epsilon \omega/\omega_{x,y})$ to the background velocity, while $\delta v_z$ produces perturbations of the order of $\mathcal{O}(\epsilon \omega^2/\omega_{x,y}^2)$. Consequently, considering now only perturbations of the order of $\mathcal{O}(\epsilon \omega/\omega_{x,y})$, we end up with:
\begin{equation}
    \begin{aligned}
        &\frac{\delta  n_c}{n_c}=\frac{\delta a}{a}=\frac{\delta v_z}{c_s}=0,\\
        \frac{\delta v_x}{c_s}=\frac{\omega}{c_s} &(xh_+ +yh_\times)\sin(\omega(t-z/c_s)),\\
        \frac{\delta v_y}{c_s}=\frac{\omega}{c_s}&(xh_\times - y h_+)\sin(\omega(t-z/c_s)).
    \end{aligned}
\end{equation}
Hence, we observe that we have arrived at the same situation as in the previous case. Indeed, the $\delta v_x$ and $\delta v_y$ induced here by $\delta v_z$ are identical to those in the previous case except for a multiplicative $-1/2$ factor, which is a constant. Additionally, the $\delta v_z$ introduced here as the initial perturbation is exactly the same, except for the same $-1/2$ factor, as the one which arises in the previous case as a response of the fluid to the $\delta v_x$ and $\delta v_y$. Therefore, we do not delve further into the analysis of this case as we have determined it to be identical to the previous one.
\end{itemize}
\vspace{5mm}

To summarize, in this chapter we have shown how to design a system where the analogue of a gravitational wave propagates on the top of an acoustic flat metric. Remarkably, this perturbation satisfies Einstein's equations in vacuum, being it a perturbation with the same space-time dependence as the gravitational wave one propagating on the same flat background. This is the first time that the dynamics of a gravitational scenario is reproduced in an analogue model. The ability to reproduce a gravitational wave in an analogue system is a relevant achievement for its interesting implications. To start with, we can think of simulating a gravitational wave over a metric with an acoustic horizon: this is what we study in the next chapter. In Chapter \ref{chap:GWinBH} we indeed design a system where an analogue black hole is excited by an impinging gravitational wave-like perturbation. To do that we use the analogue gravitational wave found in this chapter through transformation \rom{1} (c). We choose to work with this specific case because there everything is considered up to $\mathcal{O}(\epsilon)$, while for the other transformations dominant terms are of order $\mathcal{O}(\epsilon \omega/\omega_{x,y})$.

\chapter{Acoustic Black Hole with a Gravitational Wave–like perturbation}
\label{chap:GWinBH}

\begin{chapabstract}
    \begin{adjustwidth}{1cm}{1cm}
        We present the second original result of this thesis: a gravitational wave–like perturbation on an acoustic black hole metric in a Bose-Einstein condensate.
    \end{adjustwidth}
\end{chapabstract}

\section{Method}
\begin{figure}[ht]
    \centering
    \includegraphics[width=0.6\textwidth]{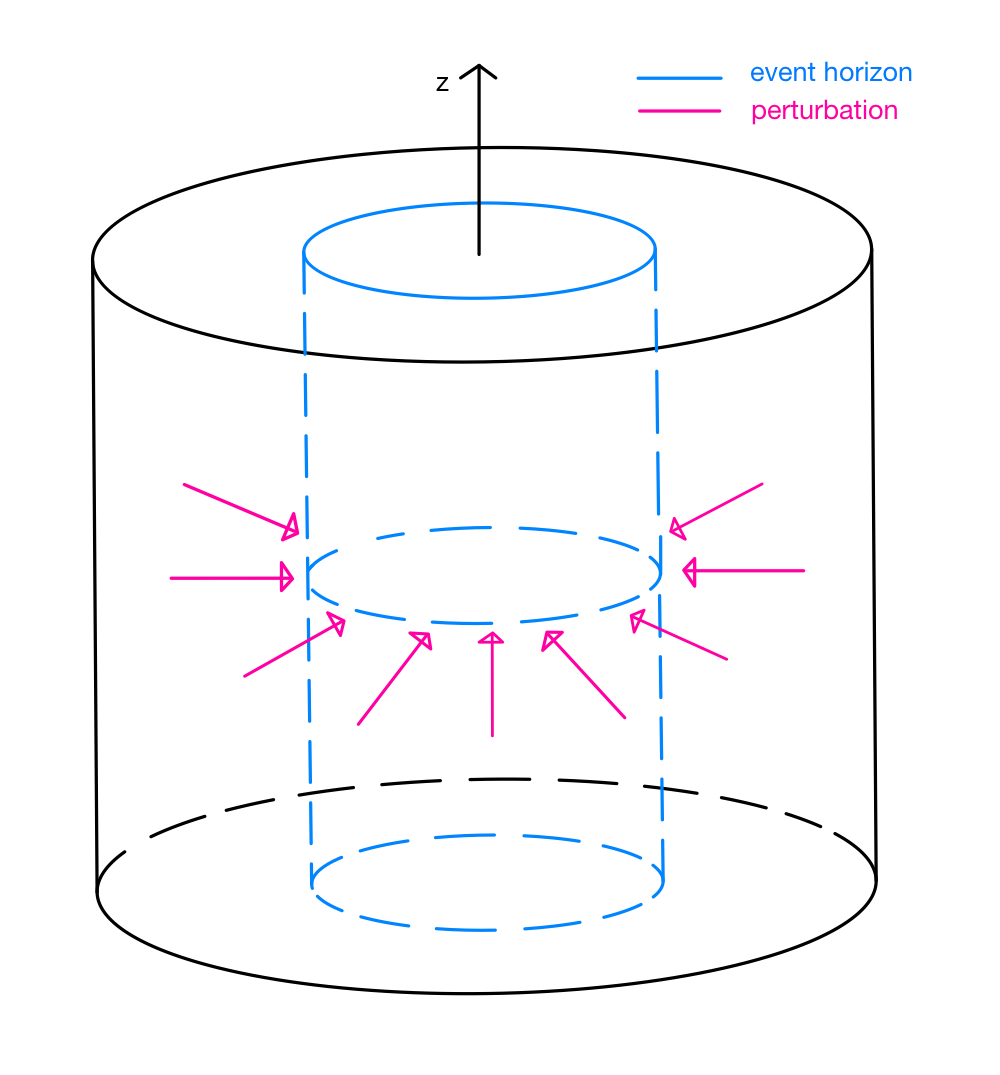}
    \caption{\textit{Gravitational wave-like perturbation in cylindrical geometry horizons.} Schematic representation of the examined system (in black). The blue cylinder represents the event horizon; while pink arrows indicate the propagation direction of the gravitational wave-like perturbation, that is radial.}
    \label{fig:ABH+GW_schematic}
\end{figure}

Our aim is to design a system with an acoustic horizon and a gravitational wave-like perturbation. To do so, we want to extend one of the solutions found in the previous chapter to the case of an acoustic black hole background metric. We choose to work with a Bose-Einstein condensate in a regime such that the acoustic metric has an event horizon and to introduce an appropriate extension in this new geometry of a gravitational wave-like perturbation, as found in Chapter \ref{chap:simulationGWMinkowski}. As previously mentioned in Chapter \ref{chap:simulationGWMinkowski}, we have opted to focus on the case \rom{1} (c). This choice is primarily motivated by the fact that we want to work within the $\mathcal{O}(\epsilon)$ approximation, and also because in this specific scenario the spatial dependencies of the condensate velocity perturbations required to replicate the gravitational wave are the most straightforward. Regarding the acoustic black hole, we choose to use the cylindrical geometry introduced in Section \ref{sec:ABH_cylindrical}. Therefore, the background metric on which we shall focus will be the one in Equation (\ref{eq:cylindricalBH_metric}), that we report also here for the reader convenience:
\begin{equation}
    \mathrm{d} s^2=-c_s^2 \mathrm{d}t^2+\left( \mathrm{d}r-\frac{A}{r}\mathrm{d} t \right)^2+\left(r \mathrm{d}\theta -\frac{B}{r}\mathrm{d}t\right)^2+\mathrm{d}z^2,
\end{equation}
with $A<0$ and the event horizon located at $r_H=|A|/c_s\;\forall z$. We choose this geometry because it serves our purpose well. We could have opted for a simpler configuration, such as a two-dimensional horizon, but the investigation of the cylindrical case is more general and might also be reproduced experimentally. Indeed, box-like Bose-Einstein condensates have already been created in laboratories \cite{BECinaBox}\cite{HadzibabicBOX}\cite{ZwierleinBOX}. It thus may be possible to reproduce a Bose-Einstein condensate confined in a trapping potential with cylindrical symmetry. Furthermore, our analysis could be extended to the limits of a cigar-shaped Bose-Einstein condensate (see Section \ref{sec:Steinhauer-sec}) or a disk-shaped Bose-Einstein condensate (see Section \ref{sec:Hu-sec}) \cite{cigarBEC_diskBEC}.\\
A schematic representation of the system under consideration is depicted in Figure \ref{fig:ABH+GW_schematic}.
The steps we need to follow in order to design such a system are the following:
\begin{enumerate}
    \item \textbf{Cylindrical perturbations}: We extend the gravitational wave-like perturbation found in Section \ref{sec:GWtrasfSummary_checonsidero} to the cylindrical geometry: we shall choose it to propagate radially. See Section \ref{sec:extendedPerturbationGW}.
    \item \textbf{Physical requirements}: We check that the irrotational condition for the condensate velocity is satisfied, as well as the continuity and Euler equations. See Section \ref{sec:physicalsystem_BH+Gw}.
    \item \textbf{Acoustic metric}: We compute the acoustic metric and its inverse. See Section \ref{sec:acousticmetricBHGW}.
    \item \textbf{Perturbed acoustic horizon}: We study the horizon's position once the analogue gravitational wave impinges on it. See Section \ref{sec:perturbedAH}.
    \item \textbf{Horizon's generators}: We compute the horizon's generators. See Section \ref{sec:horizon_generators-sec}.
\end{enumerate}
We remark that we will deal with two different types of fluctuations: the background ones, which will be related to the metric perturbations, and the phonons fluctuations, which are the modes that propagate on top of the emergent acoustic metric.

\section{Cylindrical perturbations}
\label{sec:extendedPerturbationGW}
\begin{figure}[ht]
    \centering
    \includegraphics[width=1\textwidth]{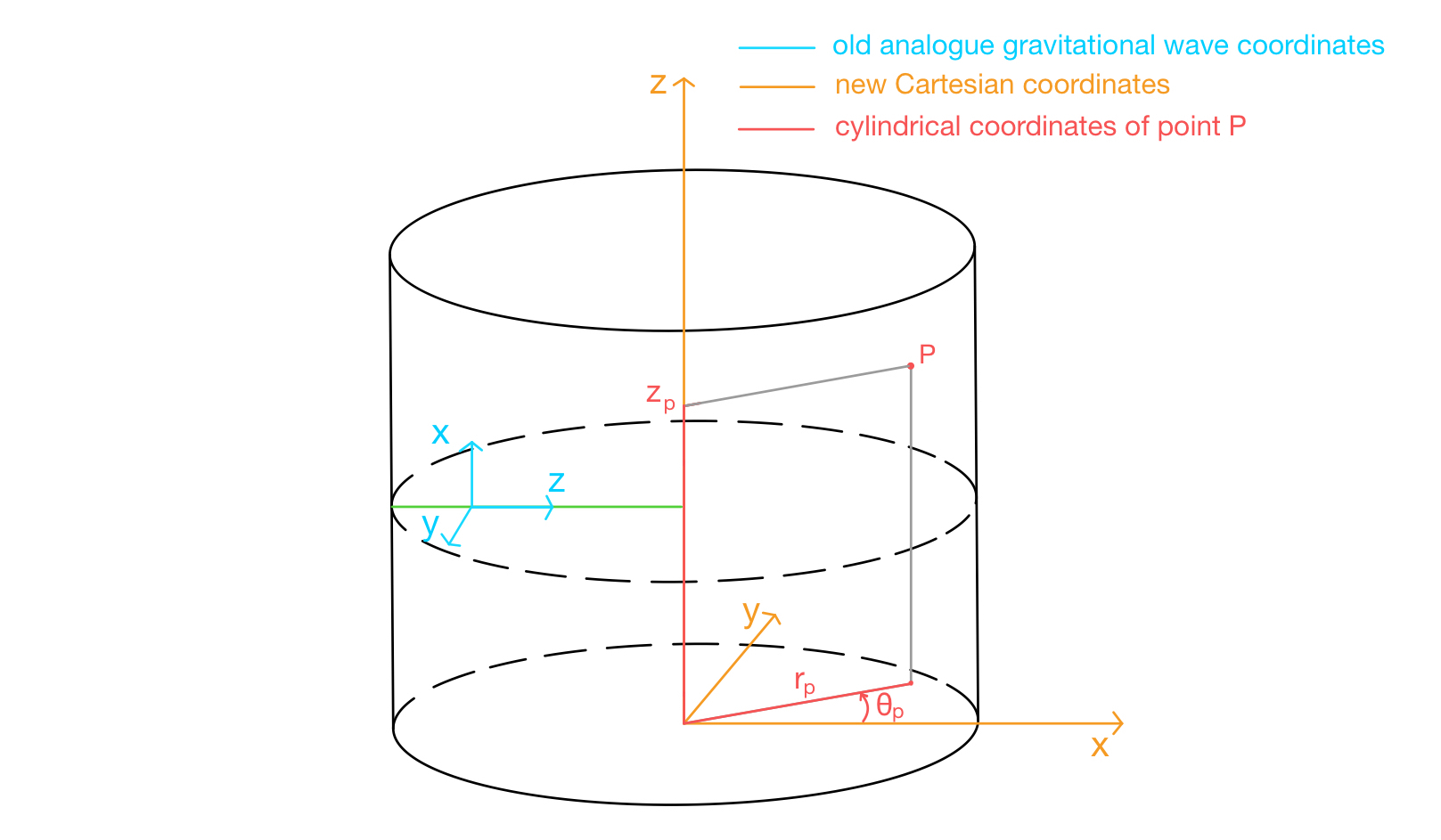}
    \caption{\textit{Old coordinates of the gravitational wave-like perturbation and new coordinates of the system.} We are concentrating on the perturbation at a fixed $\theta$ and $z$, which propagates along the green line. In blue, we present the old coordinates in which the analogue gravitational wave was described. In orange, we show the Cartesian coordinates used for our current system represented in black. Furthermore, we have marked a generic point P inside the cylinder and indicated its cylindrical coordinates in red.}
    \label{fig:oldREF-newREF}
\end{figure}

The gravitational wave-like perturbation found in Section \ref{sec:GWtrasfSummary_checonsidero} gives a contribution to the metric of the form:
\begin{equation}
    \epsilon h_{\mu\nu}\mathrm{d}x^\mu \mathrm{d}x^\nu=\epsilon \left(-2 \frac{\delta v_x}{c_s}\delta_{\mu 0}\delta_{\nu 1}-2\frac{\delta v_y}{c_s}\delta_{\mu 0}\delta_{\nu 2} \right)\mathrm{d}x^\mu \mathrm{d}x^\nu,
\end{equation}
with
\begin{equation}
    \frac{\delta v_x}{c_s}=l(fh_++gh_\times)\cos(\omega(t-z/c_s)),
\end{equation}
\begin{equation}
    \frac{\delta v_y}{c_s}=b(fh_++gh_\times)\cos(\omega(t-z/c_s)).
\end{equation}
Motivated by the plane wave solution, we then extend to a cylindrical geometry, abandoning the analogy with an astrophysical event, as the purpose of creating this system is to study the properties of an acoustic horizon. In this setup we assume that perturbations propagate radially. Therefore, we replace $\cos(\omega(t-z/c_s))$ with $\cos(\omega(t-r/c_s))$, and we identify the previous $x$ axis as the new $z$ axis (see Figure \ref{fig:oldREF-newREF}).
In our extension, we can establish the following correspondences between the old coordinates in which the analogue gravitational wave is expressed, and the new ones: $\mathrm{d}z\to \mathrm{d}r$, $\mathrm{d}x \to \mathrm{d}z$, $\mathrm{d}y\to r\mathrm{d}\theta$, resulting in $\delta v_x \to \delta v_z$ (with $\delta \mathbf{v}_z=\delta v_z \hat{z}$ the velocity perturbation on the new $z$ axis) and $\delta v_y\to (r/r_0) \delta v_\theta$ (with $\delta \mathbf{v}_\theta=\delta v_\theta \hat{\theta}$ the velocity perturbation along $\hat{\theta}$). Here, $r_0$ is a normalizing constant with the dimension of a length. As a result, the perturbations in velocities introduced in our cylindrical black hole are:
\begin{equation}
    \label{eq:deltaVZ_BH+GW}
    \frac{\delta v_z}{c_s}=l(fh_++gh_\times)\cos(\omega(t-r/c_s))
\end{equation}
\begin{equation}
    \label{eq:deltaVTHETA_BH+GW}
    \frac{\delta v_\theta}{c_s}=\frac{r_0}{r}b(fh_++gh_\times)\cos(\omega(t-r/c_s)),
\end{equation}
where for simplicity we assume $f,g>0$. In this case, this is the perturbation that serves as the analogue of a gravitational wave. This means that the analogue of the gravitational wave has no effect on the radial flow, but produces a flow along the lateral surface of the cylinder. For consistency with the regime where we found the analogue gravitational wave solution in Chapter \ref{chap:simulationGWMinkowski}, we assume that $\omega/\omega_z$ and $\omega/\omega_r$ are small parameters, where $\omega_z$ and $\omega_r$ are the characteristic frequencies of the system, defined as:
\begin{equation}
    \label{eq:characteristic_frequencies}
    \omega_z=\frac{2\pi c_s}{L_z}, \quad    \omega_r=\frac{2\pi c_s}{L_r},
\end{equation}
where $L_z$ and $L_r$ represents the physical dimensions of the system along the $z$ and $r$ direction, respectively. We emphasize that this system, i.e. what we call an acoustic black hole perturbed by a gravitational wave-like perturbation, is not intended to represent a real astrophysical system.

\section{Physical requirements}
\label{sec:physicalsystem_BH+Gw}
At this point, our aim is to verify whether the irrotational condition, as well as the continuity and Euler equations, are satisfied. As previously discussed in Section \ref{sec:ABH_cylindrical}, for the unperturbed system, these equations require the condensate density $n_c$ and the speed of sound $c_s$ to remain constant and independent of position, while $v_r$ and $v_\theta$ follow the profiles $v_r=A/r$ and $v_\theta=B/r$. Now, we investigate the effect of introducing the velocity perturbations in Equations (\ref{eq:deltaVZ_BH+GW}) and (\ref{eq:deltaVTHETA_BH+GW}). We begin by assessing the irrotationality of the flow, specifically checking if $\nabla \times \delta \mathbf{v}=\mathbf{0}$. We observe that introducing $\delta v_z$ and $\delta v_\theta$ induces a response in the form of $\delta v_r$, i.e. a perturbation to the radial velocity ($\delta \mathbf{v}_r=\delta v_r \hat{r}$), within the system. To ensure that $\delta v_r$ remains a monodromic function, we need to set $b=0$ in Equation (\ref{eq:deltaVTHETA_BH+GW}). This implies $\delta v_\theta=0$, allowed by the conditions on the background quantities. We thus have:
\begin{equation}
    \frac{\delta v_r}{c_s}=lz\frac{\omega}{c}(fh_++gh_\times)\sin(\omega(t-r/c_s)).
\end{equation}
This $\delta v_r$ induced by the analogue gravitational wave deforms the acoustic horizon. Because of that, we decide not to neglect this term, and in what follows we will consider all terms that are a perturbation of order $\epsilon$ and we will include all orders of expansion in $\omega/\omega_z$ and $\omega/\omega_r$. For clarity, we express $z$ as $\Tilde{z}L_z$ with $\Tilde{z}\in[0,1]$ and $r$ as $\Tilde{r}L_r$ with $\Tilde{r}\in[0,1]$. In this way we can write (remember that $\delta v_\theta=0$):
\begin{equation} 
    \frac{\delta v_z}{c_s}=l(fh_++gh_\times)\cos(\omega(t-2\pi \Tilde{r}/\omega_r))
\end{equation}
\begin{equation}
    \frac{\delta v_r}{c_s}=2\pi l \Tilde{z}\frac{\omega}{\omega_z}(fh_++gh_\times)\sin(\omega(t-2\pi \Tilde{r}/\omega_r)).
\end{equation}
The $\delta v_z$ and $\delta v_r$ perturbations presented here were initially determined in Cartesian coordinates for a plane wave and have now been extended to cylindrical coordinates. We demonstrate that they can still be considered solutions of the linear-in-$\epsilon$ continuity and Euler equations, with the condition that appropriate additional terms are introduced. Specifically, a source term and a perturbation to the external potential must be added, while in Section \ref{sec:GWtrasfSummary_checonsidero} we have seen that in a flat background only an external potential was needed. Now, the linear-in-$\epsilon$ continuity equation holds only when we introduce a source term $\epsilon \delta S$:
\begin{equation}
    \partial_t\delta n_c+\nabla(n_c \delta \mathbf{v}+\delta n_c \mathbf{v})=\delta S,
\end{equation}
and therefore $n_c$ is not conserved. A solution to this equation, with the previously determined $\mathbf{v}$ and $\delta \mathbf{v}$ where we set $l=\frac{1}{2\pi}$, yields the following:
\begin{equation}
    \label{eq:n_c_pert}
    \frac{\delta n_c}{n_c}=\frac{\Tilde{z}\Tilde{r}}{\Tilde{r}_H}\frac{\omega}{\omega_z}(fh_++gh_\times)\sin(\omega(t-2\pi \Tilde{r}/\omega_r))+\frac{2\pi}{3}\frac{\Tilde{z}\Tilde{r}^3}{\Tilde{r}_H^2}\frac{\omega^2}{\omega_r \omega_z}(fh_++gh_\times)\cos(\omega(t-2\pi \Tilde{r}/\omega_r))
\end{equation}
\begin{equation}
    \label{eq:S_pert}
    \delta S=\frac{2\pi}{3}n_c\frac{\Tilde{z} \Tilde{r}^3}{\Tilde{r}_H^2}\frac{\omega^3}{\omega_r \omega_z}(fh_++gh_\times)\left(-1-\frac{\Tilde{r}_H}{\Tilde{r}}\right)\sin(\omega(t-2\pi \Tilde{r}/\omega_r)).
\end{equation}
Here, $\Tilde{r}_H$ represents the position of the unperturbed acoustic horizon in radial units. Finally, we calculate the necessary external potential for the Euler equation to be valid. In doing this, we require $\delta c_s=0$ to maintain consistency with the local speed of light always being $c$. This requirement implies that we need to introduce a Feshbach resonance such that (see Equation (\ref{eq:deltaCS_funzione}))
\begin{equation}
    \label{eq:a_pert}
    \frac{\delta a}{a}=-\frac{\delta n_c}{n_c}.
\end{equation}
Introducing this $\delta a$, it is possible to show that the linear-in-$\epsilon$ Euler equation
\begin{equation}
    m\partial_t \delta \mathbf{v}+\nabla\left(\delta V_{\text{ext}}+\frac{4\pi \delta a \hslash^2}{m}n_c+\frac{4\pi  a \hslash^2}{m}\delta n_c-\frac{\hslash^2}{4m\sqrt{n_c}}\nabla^2 \left(\frac{\delta n_c}{\sqrt{n_c}}\right)+\frac{\hslash^2}{4m\sqrt{n_c}}\frac{\delta n_c}{n_c}\nabla^2\sqrt{n_c} \right)=0
\end{equation}
holds if we introduce a perturbation of the external potential, given by:
\begin{equation}
\label{eq:Vext_pert}
    \begin{aligned}
    &\delta V_{\text{ext}}=\left(\frac{3}{8\pi}\frac{\Tilde{z}(\Tilde{r}-\Tilde{r}_H)}{\Tilde{r}_H^2}\frac{\omega^2 \omega_r}{\omega_z c_s^2}    -\frac{\pi}{6}\frac{\Tilde{z}\Tilde{r}^3}{\Tilde{r}_H^2}\frac{\omega^4}{\omega_z \omega_r c_s^2}\right)\frac{\hslash^2}{m}(fh_++gh_\times)\cos(\omega(t-2\pi\Tilde{r}/\omega_r))+\\
    &-\left(\Tilde{z}\frac{c_s^2\omega}{\omega_z}m  +\frac{1}{16\pi^2}\frac{\Tilde{z}}{\Tilde{r}\Tilde{r}_H}\frac{\omega \omega_r^2}{\omega_z c_s^2}\frac{\hslash^2}{m}+\frac{\Tilde{z}\Tilde{r}}{\Tilde{r}_H^2}\left(\frac{7}{12}\Tilde{r}-\frac{1}{4}\Tilde{r}_H\right)\frac{\omega^3}{\omega_z c_s^2}\frac{\hslash^2}{m}\right)(fh_++gh_\times)\sin(\omega(t-2\pi\Tilde{r}/\omega_r)).
    \end{aligned}
\end{equation}
To summarize, we are considering a Bose-Einstein condensate with a cylindrical symmetry and with
\begin{equation}
    \mathbf{v}=\frac{A}{r}\hat{r}, \quad n_c, c_s:\,\, \text{constant and uniform},\quad \,\, A<0,
\end{equation}
resulting in an acoustic black hole with the horizon located at points where $|v_r|=c_s$. Next, we introduce the following velocity perturbation
\begin{equation} 
    \label{eq:deltavz_l1/2pi}
    \frac{\delta v_z}{c_s}=\frac{1}{2\pi}(fh_++gh_\times)\cos(\omega(t-2\pi \Tilde{r}/\omega_r)),
\end{equation}
as well as perturbations in the condensate density (in Equation (\ref{eq:n_c_pert})), the scattering length (in Equation (\ref{eq:a_pert})), the external potential (in Equation (\ref{eq:Vext_pert})), and a source term (in Equation (\ref{eq:S_pert})).
The velocity perturbation introduced in Equation (\ref{eq:deltavz_l1/2pi}) induces a radial velocity perturbation due to the irrotationality of the flow:
\begin{equation}
    \frac{\delta v_r}{c_s}=\Tilde{z}\frac{\omega}{\omega_z}(fh_++gh_\times)\sin(\omega(t-2\pi \Tilde{r}/\omega_r)).
\end{equation}
In this way we obtain a physical system (irrotationality, continuity and Euler equations are satisfied), in the geometry depicted in Figure \ref{fig:ABH+GW_schematic}, that represents an acoustic black hole perturbed by an analogue gravitational wave.

\section{Acoustic metric}
\label{sec:acousticmetricBHGW}
At this point, we compute the emergent acoustic metric tensor for phonons, determined by our system's characteristics. In the coordinates $x^\mu=(c_s t,r,\theta,z)$, we find that the metric is
\begin{equation}
    g_{\mu\nu}=\Bar{g}_{\mu\nu}+\epsilon h_{\mu\nu},
\end{equation}
with
\begin{equation}
    \Bar{g}_{\mu\nu}=\frac{n_c}{mc_s}
    \begin{pmatrix}
        -\left(1-\frac{v_r^2}{c_s^2}\right)&-\frac{v_r}{c_s} &0&0\\
        -\frac{v_r}{c_s} &1&0&0\\
        0&0&r^2&0\\
        0&0&0&1
    \end{pmatrix}
    ,
\end{equation}
and
\begin{equation}
    h_{\mu\nu}=\frac{n_c}{mc_s}
    \begin{pmatrix}
        -\left(1-\frac{v_r^2}{c_s^2}\right)\frac{\delta n_c}{n_c}+2\frac{v_r \delta v_r}{c_s^2}&-\frac{\delta v_r}{c_s}-\frac{v_r}{c_s}\frac{\delta n_c}{n_c} &0&-\frac{\delta v_z}{c_s}\\
        -\frac{\delta v_r}{c_s}-\frac{v_r}{c_s}\frac{\delta n_c}{n_c}&\frac{\delta n_c}{n_c}&0&0\\
        0&0&r^2\frac{\delta n_c}{n_c}&0\\
        -\frac{\delta v_z}{c_s}&0&0&\frac{\delta n_c}{n_c}
    \end{pmatrix}
    .
\end{equation}
Far from the horizon, this metric perturbation can be seen as a fluctuation of the Minkowski metric. Close to the horizon, due to the additional terms introduced in the continuity and Euler equations, this becomes a valid metric perturbation for the acoustic metric.\\
We also want to compute the inverse of the metric $g^{\mu\nu}$ defined by:
\begin{equation}
    \label{eq:inverse_metric_def}
    g_{\mu\nu}g^{\nu\alpha}=\delta_{\mu}{}^{\alpha}.
\end{equation}
Because of the fact that 
\begin{equation}
    \label{eq:etaeta=1}
    \Bar{g}_{\mu\nu}\Bar{g}^{\nu\alpha}=\delta_{\mu}{}^{\alpha},
\end{equation}
we assume 
\begin{equation}
    \label{eq:gnualpha1}
    g^{\nu\alpha}=\Bar{g}^{\nu\alpha}+\epsilon h^{\nu\alpha},
\end{equation}
and so we get
\begin{equation}
    \begin{aligned}
    (\Bar{g}_{\mu\nu}+\epsilon h_{\mu\nu})(\Bar{g}^{\nu\alpha}&+\epsilon h^{\nu\alpha})=\Bar{g}_{\mu\nu}\Bar{g}^{\nu\alpha}+\epsilon (h_{\mu\nu}\Bar{g}^{\nu\alpha}+\Bar{g}_{\mu\nu}h^{\nu\alpha})+\mathcal{O}(\epsilon^2)=\\
    &=\delta_\mu{}^\alpha+\epsilon(h_{\mu\nu}\Bar{g}^{\nu\alpha}+\Bar{g}_{\mu\nu}h^{\nu\alpha})+\mathcal{O}(\epsilon^2).
    \end{aligned}
\end{equation}
Comparing this last equation with Equation (\ref{eq:inverse_metric_def}), we get the condition:
\begin{equation}
    \label{eq:consition_h^munu}
    h_{\mu\nu}\Bar{g}^{\nu\alpha}+\Bar{g}_{\mu\nu}h^{\nu\alpha}=0.
\end{equation}
Multiplying Equation (\ref{eq:consition_h^munu}) for $\Bar{g}^{\beta\mu}$ we obtain:
\begin{equation}
    \label{eq:hinv_def}
    h^{\beta\alpha}=-\Bar{g}^{\beta\mu}h_{\mu\nu}\Bar{g}^{\nu\alpha}.
\end{equation}
Using the condition in Equation (\ref{eq:etaeta=1}), we get
\begin{equation}
    \Bar{g}^{\mu\nu}=\frac{m c_s}{n_c}
    \begin{pmatrix}
        -1&-\frac{v_r}{c_s} &0&0\\
        -\frac{v_r}{c_s}&1-\frac{v_r^2}{c_s^2}&0&0\\
        0&0&\frac{1}{r^2}&0\\
        0&0&0&1
    \end{pmatrix},
\end{equation}
while using the Equation (\ref{eq:hinv_def}) we obtain:
\begin{equation}
    \label{eq:h^munuvecchio}
    h^{\mu\nu}=-\frac{mc_s}{n_c}
    \begin{pmatrix}
        -\frac{\delta n_c}{n_c}& -\frac{\delta n_c}{n_c}\frac{v_r}{c_s}+\frac{\delta v_r}{c_s}&0&\frac{\delta v_z}{c_s}\\
        -\frac{\delta n_c}{n_c}\frac{v_r}{c_s}+\frac{\delta v_r}{c_s}& 2\frac{v_r \delta v_r}{c_s^2}+\frac{\delta n_c}{n_c}-\frac{v_r^2}{c_s^2}\frac{\delta n_c}{n_c}&0&\frac{v_r \delta v_z }{c_s^2}\\
        0&0&\frac{1}{r^2}\frac{\delta n_c}{n_c}&0\\
        \frac{\delta v_z}{c_s}&\frac{v_r \delta v_z }{c_s^2}&0&\frac{\delta n_c}{n_c}
    \end{pmatrix}
    .
\end{equation}
For simplicity, hereafter we use the following definition for the inverse of the metric
\begin{equation}
    \label{eq:gnualpha2}
    g^{\mu\nu}=\Bar{g}^{\mu\nu}-\epsilon h^{\mu\nu}.
\end{equation}
and thus
\begin{equation}
    h^{\mu\nu}\to -h^{\mu\nu}.
\end{equation}
In the coordinates $(c_s t, \Tilde{r}, \theta, \Tilde{z})$ we have
\begin{equation}
    \label{eq:eta_coordTILde}
    \Bar{g}_{\mu\nu}=\frac{n_c}{m c_s}\begin{pmatrix}
        -\left(1-\frac{v_r^2}{c_s^2}\right)&-\frac{v_r}{c_s}L_r&0&0\\
        -\frac{v_r}{c_s}L_r& L_r^2&0&0\\
        0&0&\Tilde{r}^2L_r^2&0\\
        0&0&0&L_z^2\\
    \end{pmatrix}
    ,
\end{equation}
\begin{equation}
    \label{eq:etaINV_coordTilde}
    \Bar{g}^{\mu\nu}=\frac{c_s m}{n_c}\begin{pmatrix}
        -1&-\frac{v_r}{c_s L_r}&0&0\\
        -\frac{v_r}{c_s L_r}&\frac{1}{L_r^2}\left(1-\frac{v_r^2}{c_s^2}\right)&0&0\\
        0&0&\frac{1}{\Tilde{r}^2L_r^2}&0\\
        0&0&0&\frac{1}{L_z^2}\\
    \end{pmatrix}
    ,
\end{equation}
\begin{equation}
    \label{eq:h_coordTILde}
    h_{\mu\nu}=\frac{n_c}{m c_s}\begin{pmatrix}
        -\left(1-\frac{v_r^2}{c_s^2}\right)\frac{\delta n_c}{n_c}+2\frac{v_r\delta v_r}{c_s^2}&L_r\left(-\frac{\delta v_r}{c_s}-\frac{v_r}{c_s}\frac{\delta n_c}{n_c}\right)&0&-\frac{\delta v_z}{c_s}L_z\\
        L_r\left(-\frac{\delta v_r}{c_s}-\frac{v_r}{c_s}\frac{\delta n_c}{n_c}\right)&\frac{\delta n_c}{n_c}L_r^2&0&0\\
        0&0&\Tilde{r}^2 L_r^2\frac{\delta n_c}{n_c}&0\\
        -\frac{\delta v_z}{c_s}L_z&0&0&\frac{\delta n_c}{n_c}L_z^2\\
    \end{pmatrix}
    ,
\end{equation}
\begin{equation}
    \label{eq:hINV_coordTilde}
    h^{\mu\nu}=\frac{mc_s}{n_c}\begin{pmatrix}
        -\frac{\delta n_c}{n_c}&\left(-\frac{\delta n_c}{n_c}\frac{v_r}{c_s}+\frac{\delta v_r}{c_s}\right)\frac{1}{L_r}&0&\frac{\delta v_z}{c_s}\frac{1}{L_z}\\
        \left(-\frac{\delta n_c}{n_c}\frac{v_r}{c_s}+\frac{\delta v_r}{c_s}\right)\frac{1}{L_r}&\left(2\frac{v_r \delta v_r}{c_s^2}+\frac{\delta n_c}{n_c}-\frac{v_r^2}{c_s^2}\frac{\delta n_c}{n_c}\right)\frac{1}{L_r^2}&0&\frac{v_r\delta v_z}{c_s^2}\frac{1}{L_r L_z}\\
        0&0&\frac{1}{\Tilde{r}^2 L_r^2}\frac{\delta n_c}{n_c}&0\\
        \frac{\delta v_z}{c_s}\frac{1}{L_z}&\frac{v_r\delta v_z}{c_s^2}\frac{1}{L_r L_z}&0&\frac{\delta n_c}{n_c}\frac{1}{L_z^2}
    \end{pmatrix}
    ,
\end{equation}
with the definition of $h^{\mu\nu}$ given in Equation (\ref{eq:gnualpha2}). To summarize, we have found the emergent acoustic metric tensor, determined by the condensate characteristics, on which the phonons move. This metric represents that of an acoustic cylindrical black hole plus an analogue gravitational wave. We have also computed the inverse of the metric as it will be required in the next sections, particularly for calculating the event horizon's generators.

\section{Perturbed acoustic horizon}
\label{sec:perturbedAH}
\begin{figure}[ht]
    \centering
    \includegraphics[width=1\textwidth]{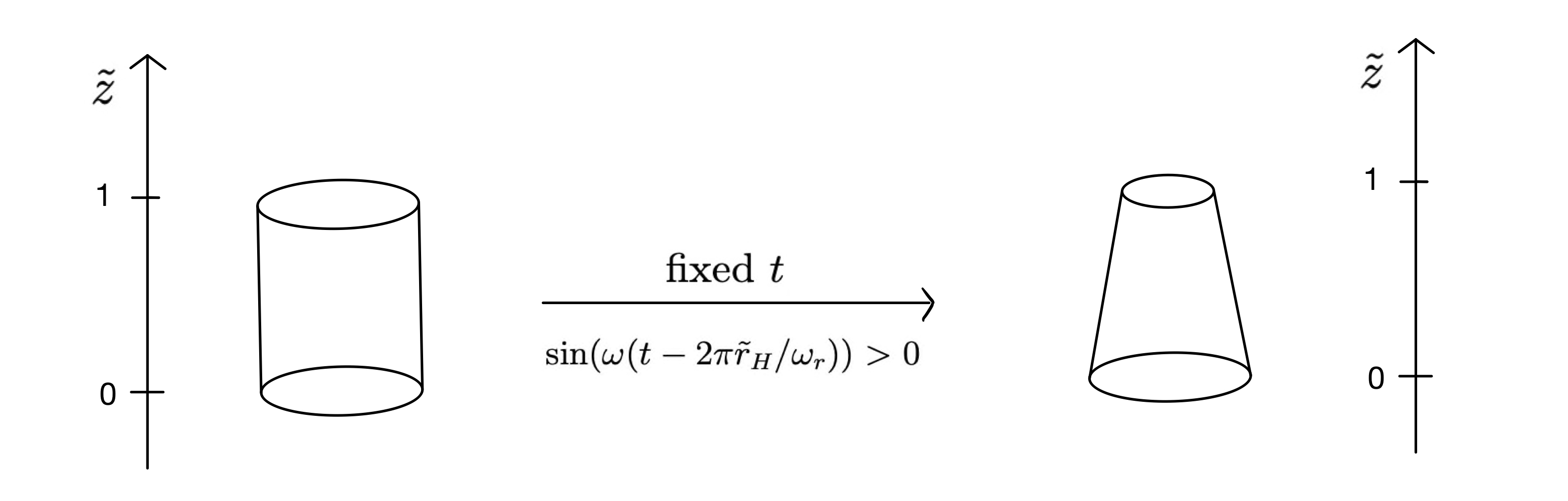}
    \caption{\textit{Schematic representation of the perturbation of the acoustic horizon}. Left: unperturbed horizon. Right: perturbed acoustic horizon at a fixed time and with $\sin(\omega(t-2\pi\Tilde{r}_H/\omega_r))>0$. The size of the perturbation is artificially enhanced for clarity.}
    \label{fig:deltarh_tfix_sin>0}
\end{figure}

Now, we want to see how the acoustic horizon is perturbed by the analogue gravitational wave. Indeed, the acoustic horizon of the unperturbed system is located where $|v_r|_{r_{H}}=c_s$, therefore at (see Equation (\ref{eq:acHOR_posit_vortex}))
\begin{equation}
    \Tilde{r}_H=\frac{|A|}{c_s L_r}.
\end{equation}
However, once we introduce the analogue gravitational wave perturbation, because of the $\delta v_r$ induced by irrotationality, the radial position of the acoustic horizon is perturbed. The perturbed acoustic horizon is located at $\Tilde{r}_H^{\text{new}}=\Tilde{r}_H+\epsilon \delta \Tilde{r}_H$, such that
\begin{equation}
    |v_r+\epsilon \delta v_r|_{\Tilde{r}_H^{\text{new}}}=c_s.
\end{equation}
It is possible to show that $\delta \Tilde{r}_H=\delta \Tilde{r}_H(t,\Tilde{r})$, with
\begin{equation}
    \label{eq:deltaRH}
    \frac{\delta \Tilde{r}_H}{\Tilde{r}_H}=-\Tilde{z}\frac{\omega}{\omega_z}(fh_++gh_\times)\sin(\omega(t-2\pi\Tilde{r}_H/\omega_r)).
\end{equation}
At fixed $t$, if $\sin(\omega(t-2\pi\Tilde{r}_H/\omega_r))>0$ then $\delta\Tilde{r}_H<0$ and it decreases as $\Tilde{z}$ increases (see Figure \ref{fig:deltarh_tfix_sin>0}).
This makes sense since in the case under examination $\delta \mathbf{v}_r=\delta v_r \hat{r}$ could be written as $\delta \mathbf{v}_r=\delta v_0 \Tilde{z} \hat{r}$ with $\delta v_0>0$: the perturbation $\delta \mathbf{v}_r$ is in the opposite direction of the background velocity $\mathbf{v}_r$, and since $\delta v_r$ is linear in $\Tilde{z}$, we have that $\mathbf{v}_r+\epsilon \delta \mathbf{v}_r$ decreases as $\Tilde{z}$ increases. As a result, the horizon's radius, which is unperturbed at $\Tilde{z}=0$ ($\Tilde{r}_H^{\text{new}}(\Tilde{z}=0)=\Tilde{r}_H$), decreases with $\Tilde{z}$ increasing, consistently with the sign of $\delta \Tilde{r}_H$ obtained in Equation (\ref{eq:deltaRH}), see Figure \ref{fig:deltarh_tfix_sin>0}. In addition, we observe that when $\sin(\omega(t-2\pi\Tilde{r}_H/\omega_r))=0$ the horizon remains unperturbed. Moreover, at fixed time $t$ and when $\sin(\omega(t-2\pi\Tilde{r}_H/\omega_r))<0$, we observe that $\delta \Tilde{r}_H$ increases linearly with $\Tilde{z}$.
\begin{figure}[ht]
    \centering
    \includegraphics[width=1\textwidth]{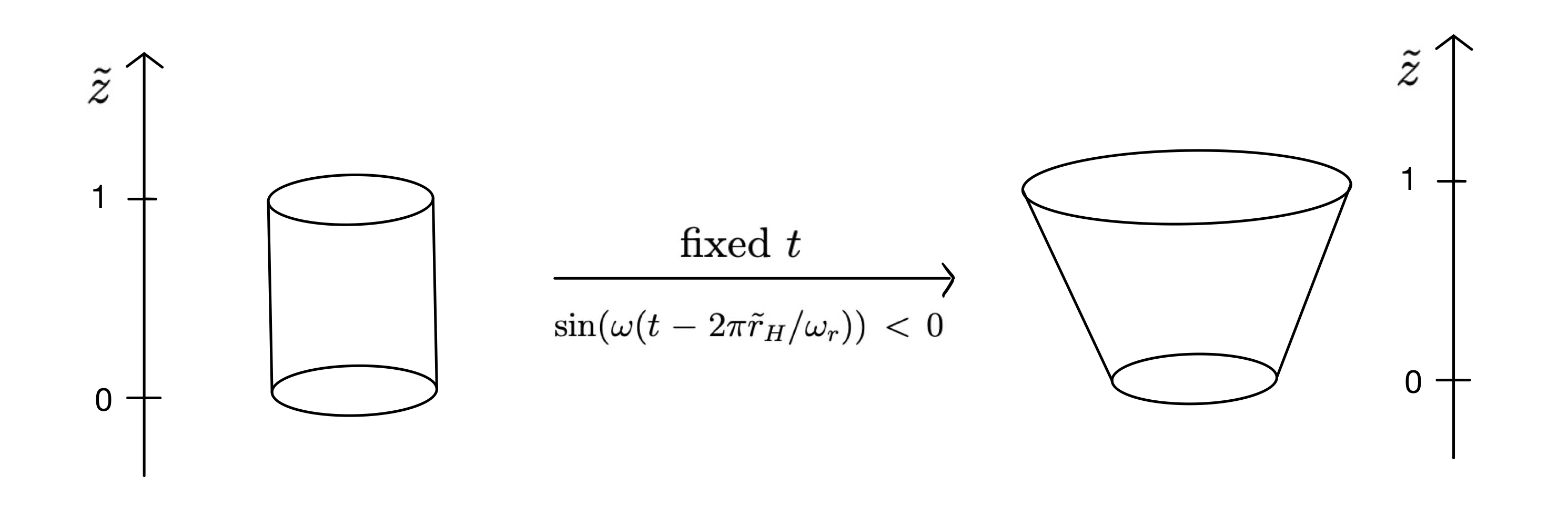}
    \caption{\textit{Schematic representation of the perturbation of the acoustic horizon}. Left: unperturbed horizon. Right: perturbed acoustic horizon at a fixed time and with $\sin(\omega(t-2\pi\Tilde{r}_H/\omega_r))<0$. The size of the perturbation is artificially enhanced for clarity.}
    \label{fig:deltarh_tfix_sin<0}
\end{figure}
In this scenario, $\delta \mathbf{v}_r=\delta v_0 \Tilde{z}\hat{r}$ with $\delta v_0<0$, indicating that the perturbation $\delta \mathbf{v}_r$ aligns with the background velocity $\mathbf{v}_r$ and thus augments it. Therefore, the radial velocity increases as $\Tilde{z}$ increases and, because of that, we expect the horizon's radius to increase with $\Tilde{z}$, see Figure \ref{fig:deltarh_tfix_sin<0}. This is consistent with the fact that in this case $\delta \Tilde{r}_H>0$ (see Equation (\ref{eq:deltaRH})). 
\begin{figure}[ht]
    \centering
    \includegraphics[width=1\textwidth]{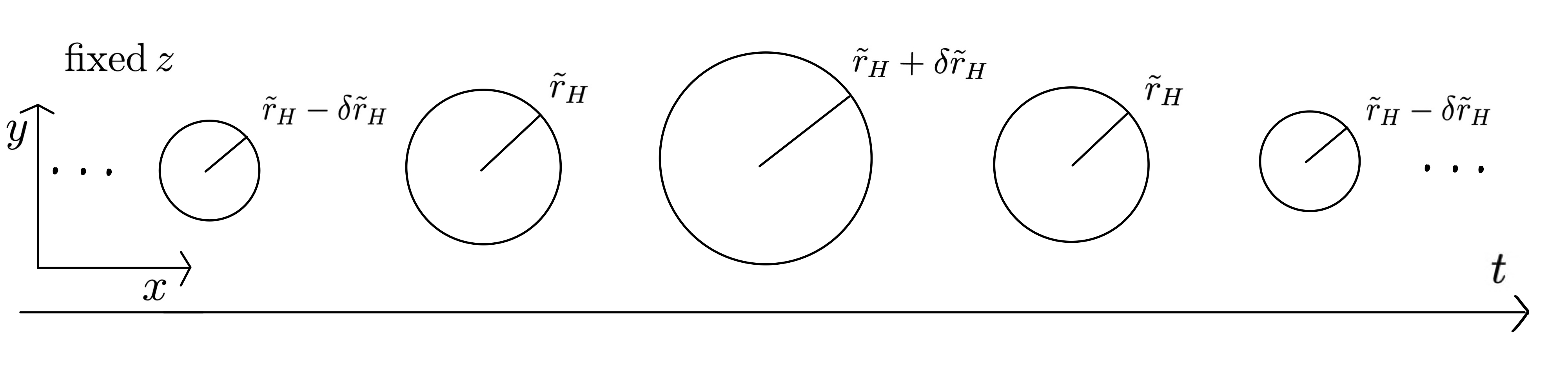}
    \caption{\textit{Schematic representation of the perturbation of the perturbed acoustic horizon}. Depiction of the acoustic horizon in the plane $xy$ at fixed $\Tilde{z}$ as time lapses.}
    \label{fig:deltarh_zfix}
\end{figure}
If we now examine the scenario where $\Tilde{z}$ is held constant and we observe the changes over time, we find the following progression: when $\delta \Tilde{r}_H<0$ at the beginning (for example), with lapsing time we have the following sequence: $\delta \Tilde{r}_H<0 \to \delta \Tilde{r}_H=0 \to \delta \Tilde{r}_H>0 \to \delta \Tilde{r}_H=0 \to \delta \Tilde{r}_H<0$ and so forth. Therefore, at fixed $\Tilde{z}$, as time flows the horizon seen in the $xy$ plane is represented in Figure \ref{fig:deltarh_zfix}. If we place ourselves in a plane at a fixed $z$, or if we consider the limiting case where our cylindrical geometry reduces to a disk-shaped system (as in Section \ref{sec:Hu-sec}), we only observe an expansion or contraction of the event horizon due to the analogue gravitational wave. However, when considering the entire cylindrical geometry, the expansion or contraction observed at a fixed $z$ leads to the tilting of the horizon. Due to this tilt, Hawking radiation phonons, which are predominantly emitted in the direction perpendicular to the horizon, experience a change in their propagation direction as compared to the unperturbed case \cite{1983.book.Shapiro}\cite{Mannarelli:emissione}. Finally, focusing for instance on the $xz$ plane at $y=0$ (refer to Figure \ref{fig:oldREF-newREF}) and considering only the half-plane $x>0$, the situation closely resembles the analysis conducted in \cite{CGMTCarXiv}, where the shear viscosity to entropy density ratio $\eta/s$ at the acoustic horizon is calculated. This suggests that further studies aiming to compute $\eta/s$ in our current analogue system may require to extend the analysis performed in \cite{CGMTCarXiv} to (3+1)-dimensions.

\section{Horizon's generators}
\label{sec:horizon_generators-sec}
As a final step, to characterize the kinematics of the horizon, we need to compute the null vector field $k^\alpha$ tangent (and also orthogonal) to the null congruence of geodesics generating the horizon, both in the unperturbed and perturbed cases. Indeed, once we know it, it becomes possible to find the expansion rate, the shear and the rotation of the null geodesics congruence, all useful to better understand the kinematics of the horizon (see Section \ref{sec:congruenceGeodesics-sec}). We know that the vector field normal to an hypersurface $S=\text{constant}$ is (see Equation (\ref{eq:vectorField_normalHyp})):
\begin{equation}
    k=\Tilde{f}(x)(g^{\mu\nu}\partial_\nu S)\frac{\partial}{\partial x^\mu}
\end{equation}
and so its components are
\begin{equation}
    k^\mu=\Tilde{f}(x)(g^{\mu\nu}\partial_\nu S),
\end{equation}
with $\Tilde{f}(x)$ an arbitrary non-zero function of the coordinates. In the unperturbed case we consider the hypersurfaces family (consider that $\Tilde{r}_H=|A|/c_s L_r$)
\begin{equation}
    S=\Tilde{r}-\Tilde{r}_H=\text{constant}
\end{equation}
and so $S=0$ is the acoustic horizon. Using Equation (\ref{eq:etaINV_coordTilde}) we find that $k$ is
\begin{equation}
    k=\Tilde{f}(x) \frac{m c_s}{n_c}\left(-\frac{1}{L_r}\frac{v_r}{c_s^2}\partial_t+\frac{1}{L_r^2}\left(1-\frac{v_r^2}{c_s^2}\right)\partial_{\Tilde{r}}\right),
\end{equation}
with components:
\begin{equation}
    \label{eq:components_unpert_withF}
    \begin{aligned}
        &k^0=\Tilde{f}(x)\frac{m c_s}{n_c}\left(-\frac{v_r}{c_s L_r}\right)\\
        &k^1=\Tilde{f}(x)\frac{m c_s}{n_c}\frac{1}{L_r^2}\left(1-\frac{v_r^2}{c_s^2}\right)\\
        &k^2=k^3=0.\\
    \end{aligned}
\end{equation}
We also obtain that (see Equation (\ref{eq:vectorField^2_normalHyp}))
\begin{equation}
    k^2=\Tilde{f}^2(x)\frac{m c_s}{n_c}\frac{1}{L_r^2}\left(1-\frac{v_r^2}{c_s^2}\right).
\end{equation}
Therefore, since $v_r^2|_{S=0}=c_s^2$, we have
\begin{equation}
    k^2|_{S=0}=0,
\end{equation}
as required by the fact that the horizon is a null hypersurface. Choosing $\Tilde{f}=n_c L_r/m$, we get ($v_r|_{S=0}=-c_s$)
\begin{equation}
    k=-\frac{v_r}{c_s}\partial_t+\frac{1}{2\pi}\omega_r\left(1-\frac{v_r^2}{c_s^2}\right)\partial_{\Tilde{r}},\quad \quad k|_{S=0}=\partial_t
\end{equation}
and
\begin{equation}
    \label{eq:component_unpert_withoutF}
    \begin{aligned}
        &k^0=-v_r\\
        &k^1=\frac{1}{2\pi}\omega_r\left(1-\frac{v_r^2}{c_s^2}\right)\\
        &k^2=k^3=0\\
    \end{aligned}
    \quad \rightarrow \quad
    \begin{aligned}
        &k^0|_{S=0}=c_s\\
        &k^1|_{S=0}=k^2|_{S=0}=k^3|_{S=0}=0.\\
    \end{aligned}
\end{equation}
Now, we want to analyze the perturbed case. We consider the following family of hypersurfaces
\begin{equation}
    S=\Tilde{r}-\left(\Tilde{r}_H+\epsilon \delta \Tilde{r}_H\right)=\text{constant},
\end{equation}
so that the perturbed horizon corresponds to $S=0$. Using Equation (\ref{eq:hINV_coordTilde}) we find:
\begin{equation}
    \begin{aligned}
    k^0=&\Tilde{f}(x)\frac{m}{2\pi n_c}\omega_r\frac{\Tilde{r}_H}{\Tilde{r}}+\epsilon \Tilde{f}(x)(fh_++gh_\times)\left[\left(-\frac{1}{3}\frac{m}{n_c}\frac{\omega^2}{\omega_z}\frac{\Tilde{r}^2\Tilde{z}}{\Tilde{r}_H}+\right.\right.\\
    &\left.\left.-\frac{m}{n_c}\Tilde{z}\Tilde{r}_H \frac{\omega^2}{\omega_z}\right)\cos(\omega(t-2\pi\Tilde{r}/\omega_r))+\frac{m}{n_c}\frac{\Tilde{z}}{\pi}\frac{\omega \omega_r}{\omega_z}\sin(\omega(t-2\pi\Tilde{r}/\omega_r))\right],
    \end{aligned}
\end{equation}
\begin{equation}
    \begin{aligned}
    k^1=\Tilde{f}(x)&\frac{m}{4\pi^2 n_c c_s}\omega_r^2\left(1-\frac{\Tilde{r}_H^2}{\Tilde{r}^2}\right)+\epsilon \Tilde{f}(x)(fh_++gh_\times)\left[\left(\frac{m}{2\pi n_c c_s}\frac{\omega^2\omega_r}{\omega_z}\frac{\Tilde{z}\Tilde{r}_H^2}{\Tilde{r}}+\right.\right.\\
    &\left.+\frac{m}{6\pi n_c c_s}
    \frac{\omega^2\omega_r}{\omega_z}\frac{\Tilde{z}\Tilde{r}}{\Tilde{r}_H^2}(-\Tilde{r}^2+\Tilde{r}_H^2)\right)\cos(\omega(t-2\pi\Tilde{r}/\omega_r))+\\  
    &\left.-\frac{m}{4\pi^2 n_c c_s}\frac{\omega\omega_r^2}{\omega_z}\frac{\Tilde{z}}{\Tilde{r}\Tilde{r}_H}(\Tilde{r}^2-3\Tilde{r}_H^2)\sin(\omega(t-2\pi\Tilde{r}/\omega_r))\right],
    \end{aligned}
\end{equation}
\begin{equation}
    k^2=0,
\end{equation}
\begin{equation}
    \begin{aligned}
    k^3=\epsilon\Tilde{f}(x)(&fh_+ +gh_\times)\left(\frac{m}{4\pi^2 n_c c_s}\omega\omega_z\Tilde{r}_H\sin(\omega(t-2\pi\Tilde{r}/\omega_r))+\right.\\
    &\left.+\frac{m}{8 \pi^3 n_c c_s}\omega_z\omega_r\frac{\Tilde{r}_H}{\Tilde{r}} \cos(\omega(t-2\pi\Tilde{r}/\omega_r))\right).
    \end{aligned}
\end{equation}
Therefore, remembering that $v_r/c_s=-\Tilde{r}_H/\Tilde{r}$, we observe that the unperturbed values of $k^\mu$ are the same as those reported in Equation (\ref{eq:components_unpert_withF}), as expected. For simplicity, we define $l^\mu=k^\mu|_{S=0}$ and choosing also in this case $\Tilde{f}=n_c L_r/m$, we get:
\begin{equation}
    \label{eq:horPertGen}
    \begin{aligned}
        &l^0=c_s+\epsilon(fh_++gh_\times)\left[3c_s\frac{\omega}{\omega_z}\Tilde{z}\sin(\omega(t-2\pi\Tilde{r}_H /\omega_r))-\frac{8\pi}{3}c_s\frac{\omega^2}{\omega_r\omega_z}\Tilde{r}_H \Tilde{z}\cos(\omega(t-2\pi\Tilde{r}_H /\omega_r))\right],\\
        &l^1=\epsilon \Tilde{r}_H \Tilde{z}\frac{\omega^2}{\omega_z}(fh_++gh_\times)\cos(\omega(t-2\pi\Tilde{r}_H /\omega_r)),\\
        &l^2=0,\\
        &l^3=\epsilon(fh_++gh_\times)\left[\frac{1}{2\pi}\frac{\omega_z\omega}{\omega_r}\Tilde{r}_H\sin(\omega(t-2\pi\Tilde{r}_H/\omega_r))+\frac{1}{4\pi^2}\omega_z\cos(\omega(t-2\pi\Tilde{r}_H /\omega_r))\right].
    \end{aligned}
\end{equation}
Thus, the $l^\mu$ terms of $\mathcal{O}(1)$ in $\epsilon$ are the same as those found in the unperturbed case (see Equation (\ref{eq:component_unpert_withoutF})). Now, we compute $g_{\mu\nu}l^\nu l^\mu$ and we find:
\begin{equation}
    l_\mu l^\mu=\epsilon 4\pi \frac{n_c c_s}{m}\frac{\omega^2}{\omega_r \omega_z}\Tilde{r}_H \Tilde{z}(fh_++gh_\times)\cos(\omega(t-2\pi \Tilde{r}_H/\omega_r)).
\end{equation}
Differently from the unperturbed case here we have a dynamical horizon and because of that, it is not required to be a null hypersurface \cite{noteLiberati}\cite{dynamicalHor}. We indeed notice that depending on the sign of the cosine we can have a null, timelike or spacelike hypersurface. However, for a Killing horizon (Section \ref{sec:killing_horizon-section}) this quantity should be zero and indeed for sufficiently low fluctuations this term vanishes: it is of order $\mathcal{O}(\epsilon \omega^2/\omega_r \omega_z)$.
\vspace{7mm}\\

To summarize, in this chapter we have developed the description of an acoustic cylindrical black hole perturbed by a gravitational wave-like perturbation. This perturbation is introduced as the cylindrical extension of the analogue gravitational wave found in Chapter \ref{chap:simulationGWMinkowski}. After having illustrated the designed physical system, we have studied how the acoustic horizon is perturbed by the impinging analogue gravitational wave, computing also its generators. This analogue model is important since it allows to study how an acoustic horizon responds to a gravitational wave-like perturbation.

\chapter*{Conclusions and future perspectives}
\markboth{Conclusions and future perspectives}{Conclusions and future perspectives}
\addcontentsline{toc}{chapter}{Conclusions and future perspectives}
In this thesis work we have obtained two original results:
\begin{enumerate}
    \item The propagation of the analogue of a gravitational wave perturbation on top of a flat background acoustic metric within a Bose-Einstein condensate.
    \item The excitation of a cylindrical acoustic black hole with a gravitational wave-like perturbation within a Bose-Einstein condensate.
\end{enumerate}
These results are shown in Chapter \ref{chap:simulationGWMinkowski} and in Chapter \ref{chap:GWinBH}, respectively. In particular, in Section \ref{sec:physicalsystems-sec} we have shown the condensate velocity perturbations and the modulation of the external trapping potential that could be included in the system to represent a gravitational wave on flat Minkowski metric. Instead, in Sections \ref{sec:perturbedAH} and \ref{sec:horizon_generators-sec} we have studied the stretching of a cylindrical acoustic horizon made by a gravitational wave-like perturbation.\\
To obtain the first result, i.e. the analogue gravitational wave, we adopted the following approach. Firstly, we expressed the general form of the acoustic metric as a Minkowski background plus a perturbation. Then, exploiting the gauge symmetry of General Relativity, i.e. the invariance under coordinate transformations, we have written a gravitational wave metric in a new gauge such that it can be compared with the analogue perturbation metric. In this way, we have found the properties that the condensate should have in order to be in a regime where the phonons satisfy the equations of motion of a massless scalar field in a Minkowski plus gravitational wave spacetime. At the same time, we identified the necessary perturbation of the external potential to ensure that hydrodynamic equations hold, making the obtained system physically valid. Therefore, we have found how to design a system that simulate a propagating analogue gravitational wave. Previous studies of analogue models were only able to represent static solutions of General Relativity, while here our analogue gravitational wave satisfies the dynamics dictated by Einstein's equations. This first result of the thesis raises the question of whether it is possible to simulate the analogue of a gravitational wave on a background metric that includes an event horizon, and here is the placement of the second result of this thesis. \\
The second outcome of this thesis, as already pointed out, comes straightforwardly from the previous result. In principle, we may have worked with a simple horizon configuration, as a two-dimensional plane. However, for illustrating the procedure of perturbing an acoustic horizon, we found more convenient to consider a cylindrical acoustic black hole, which is a more general case and has also the experimental advantage to be possibly realized by ultra-cold atoms. Clearly, this configuration has no connection with observed astrophysical black holes. We have then assumed that a gravitational wave-like perturbation impinges on the cylindrical horizon. To do that, we expressed the solution of the first outcome, i.e. the analogue gravitational wave, in this new cylindrical geometry. We checked also in this case that hydrodynamic equations hold: we have designed a perturbed black hole incorporating a perturbation that could be engineered in a laboratory setting. This perturbation is determined through an analogy with gravitational waves, which are known for deforming spacetime.\\
This second result opens various possible perspectives. A first one, is the study of the reflectivity properties of the acoustic horizon \cite{2020Oshita}. This is related to quasi-normal modes as well as to dissipative properties of the acoustic horizon \cite{maggiore2018gravitational}.\\
The shear viscosity to entropy density ratio $\eta/s$ (see Section \ref{sec:KSS-sec}) at the acoustic horizon could be computed. Indeed, having seen how the horizon stretches because of the analogue gravitational wave, it could be possible to calculate how the area of the horizon changes. The deformation of the horizon leads to the variation of the phonons worldlines, which are predominantly spontaneously emitted by the acoustic black hole in the direction orthogonal to the horizon. This is demonstrated in \cite{Mannarelli:emissione} and can also be understood by looking at \cite{2018Liberati_emission}. In \cite{2018Liberati_emission}, it is indeed demonstrated that for an ultra-compact object, a quasi-black hole, the solid angle $\Delta\Omega$ from which photons can escape the surface scales as $\Delta\Omega/2\pi=(27/8)\mu+\mathcal{O}(\mu^2)$ with $\mu=1-2M/r_\star$ ($G=c=1$), where $r_\star$ represents the quasi-black hole radius that tends to $2M$. Thus, as $\mu$ approaches 0, the escape angle from the surface narrows, allowing only photons perpendicular to the surface to escape. This behavior is expected to hold for Hawking radiation as well, although the Hawking quanta are on-shell not at the horizon but in a thin layer just outside it. Consequently, emission perpendicular to the horizon would be dominant, with a possible subleading non-perpendicular component. This conclusion is in agreement with what is found in \cite{Mannarelli:emissione} for acoustic black holes. These emitted phonons, in analogy with the membrane paradigm (see Section \ref{sec:membrane-section}), could be identified as a fluid endowed with a shear $\eta$ in a ``effectively'' (3+1)-dimensional thin layer about the horizon. Because of this identification, it could be possible to compute this shear viscosity coefficient following the reasoning made in \cite{CGMTCarXiv}. In that work, a tilted two-dimensional acoustic horizon is studied through a kinetic theory approach \cite{MannarelliTransport}\cite{CGMT_1Kinetic}. This could be a promising way of computing $\eta$, also because the geometry of our system could be interpreted as a (3+1)-dimensional extension of the (2+1) case analyzed in \cite{CGMTCarXiv}. Consequently, $\eta/s$ could be calculated, where $s$ is the entanglement entropy density of the acoustic horizon (see Section \ref{sec:openissueHawking_entropy}). Such a coefficient represents a quantity that is conjectured to exhibit universal behavior: according to the conjecture of Kovtun-Son-Starinets (KSS) $1/4\pi$ is the lower bound for $\eta/s$ of any fluid in nature. Whether this is an actual lower bound for any system is still unclear. Through the current Analogue Gravity approach, it may be possible to gain theoretical insights into the universal characteristics of hydrodynamic systems. Indeed, analogue models offer both a quantum fluid-dynamics and a microscopic description, providing valuable hints in this regard. Thus, it could be possible to check whether the KSS bound is saturated at the acoustic horizon, and, if so, to gain intuitions into the necessary conditions for the fluid to achieve the minimum $\eta/s$ value.\\
Finally, opting to work with Bose-Einstein condensates makes it feasible, given the rapid experimental progress (see Section \ref{sec:BECexperimental-sec}), to design an experiment that reproduces our theoretical setup needed to test the KSS bound.

\appendix

\chapter{Gibbons-Hawking-York term}
\label{app:Gibbons-Hawking-York}
As already stated in Chapter \ref{chap:GRandGW}, the Einstein equations can be derived by minimizing the functional variation of the action $\mathcal{S}=\mathcal{S}_{\text{grav}}+\mathcal{S}_{\text{mat}}$ where $\mathcal{S}_{\text{grav}}$ is the Einstein-Hilbert action and $\mathcal{S}_{\text{mat}}$ is the covariant integral of the lagrangian $\mathcal{L}_{\text{mat}}$, that we report also here:
\begin{equation}
    \label{eq:Einstein-Hilbert_action_appendix}
    \mathcal{S}_{\text{grav}}= \frac{c^3}{16 \pi G}\int d^4x \sqrt{|g|}R
\end{equation}
\begin{equation}
    \mathcal{S}_{\text{mat}}=\int d^4x \sqrt{|g|}\mathcal{L}_{\text{mat}}.
\end{equation}
Now, we want to analyze the boundary terms that arise from the Einstein-Hilbert action. As a first observation, we notice that in order to obtain differential equations in $g_{\mu\nu}$ of the second order, we need to have in the Einstein-Hilbert action a scalar only made by $g_{\mu\nu}$ and $\partial_{\rho} g_{\mu\nu}$. Unfortunately this is not possible, since the first non trivial scalar  is the scalar curvature $R$ that contains also terms $\partial_{\rho} \partial_{\sigma} g_{\mu\nu}$. However, we can demonstrate that those terms containing $\partial_{\rho}\partial_{\sigma} g_{\mu\nu}$ can be written as a boundary term. Thus
\begin{equation}
    \int d^4x \sqrt{|g|} R=\int d^4 x \sqrt{|g|} A(g_{\mu\nu},\partial_{\sigma} g_{\mu\nu}) + \text{boundary terms}
\end{equation}
with 
\begin{equation}
    A(g,\partial g)=\frac{1}{\sqrt{|g|}}\left( -\partial_{\mu} \left(\sqrt{|g|} g^{\nu\alpha}\right) \Gamma^{\mu}{}_{\nu\alpha}+\partial_{\alpha} \left(\sqrt{|g|} g^{\nu\alpha}\right) \Gamma^{\mu}{}_{\mu\nu}\right)+g^{\nu\alpha}\left( \Gamma^{\mu}{}_{\lambda\mu}\Gamma^{\lambda}{}_{\alpha\nu} -\Gamma^{\mu}{}_{\lambda\alpha}\Gamma^{\lambda}{}_{\mu\nu}\right).
\end{equation}
This assures that from the variation we will get second order differential equations. As a second observation, we notice that by doing $g_{\mu\nu} \to g_{\mu\nu}+\delta g_{\mu\nu}$ after some calculations it is possible to find:
\begin{equation}
    \delta \mathcal{S}_{\text{grav}}= \frac{c^3}{16 \pi G} \int d^4 x \sqrt{|g|} G_{\mu\nu} \delta g^{\mu \nu} +\frac{c^3}{16 \pi G} \int d^4 x \sqrt{|g|} \mathcal{D}_{\lambda} v^{\lambda} 
\end{equation}
with $v^{\lambda}=g^{\alpha\beta}\delta \Gamma^\lambda{}_{\alpha\beta}-g^{\alpha\lambda}\delta \Gamma^\nu{}_{\nu\alpha}$. Therefore, 
\begin{equation}
    \delta \mathcal{S}[g_{\mu\nu}]=\delta \mathcal{S}_{\text{grav}}+\mathcal{S}_{\text{mat}}=0 \implies G_{\mu\nu}=\frac{8 \pi G}{c^4} T_{\mu \nu} \iff \int d^4 x \sqrt{|g|} \mathcal{D}_{\lambda} v^{\lambda} =0.
\end{equation}
However, the condition $\int d^4 x \sqrt{|g|} \mathcal{D}_{\lambda} v^{\lambda} =0$ is not valid in general. Indeed, using the Stokes theorem (we can use it since we have the divergence of a vector) we have (with $\partial \mathcal{M}$ the boundary of the manifold)
\begin{equation}
    \int_{\mathcal{M}} d^4 x \sqrt{|g|} \mathcal{D}_{\lambda} v^{\lambda} = \int_{\partial \mathcal{M}} d\Sigma_{\lambda} v^{\lambda}.
\end{equation}
If $\partial \mathcal{M}=0$, thus the manifold has no boundary, then Einstein's equations are valid, while if $\partial \mathcal{M} \not= 0$ we have to pay attention. Indeed since 
\begin{equation}
    \delta \Gamma^{\lambda}{}_{\mu\nu}=\frac{1}{2}g^{\lambda\alpha}\left( \mathcal{D}_{\mu}\delta g_{\alpha\nu}+\mathcal{D}_{\nu}\delta g_{\mu\alpha}-\mathcal{D}_{\alpha}\delta g_{\mu\nu}\right)
\end{equation}
$\int_{\partial \mathcal{M}} d\Sigma_{\lambda} v^{\lambda}$ has inside both $\delta g_{\mu\nu}$ and $\mathcal{D}_{\alpha}\delta g_{\mu\nu}$: it is possible to show that we can rewrite \cite{noteLiberati}
\begin{equation}
    v^\lambda=-\mathcal{D}_\nu\delta g^{\lambda \nu}+g_{\alpha\beta}\mathcal{D}^\lambda\delta g^{\alpha\beta}.
\end{equation}
Typically, boundary conditions are $\delta g_{\mu\nu}|_{\partial \mathcal{M}}=0$, but this does not say anything about $\mathcal{D}_{\alpha}\delta g_{\mu\nu}|_{\partial \mathcal{M}}$. To be more precise: tangential derivatives of $\delta g_{\mu\nu}$ along $\partial \mathcal{M}$ are equal to zero (since $\delta g_{\mu\nu}|_{\partial \mathcal{M}}=0$ and thus $g_{\mu\nu}$ is constant along the boundary), while normal derivatives do not necessarily vanish. Therefore we are now facing a problem. In order to understand how to solve it, we need the geometry of hypersurfaces, that is introduced in Section \ref{sec-geometry_hyper}. Recalling Equation (\ref{eq:infinitesimal_oriented}), we can write the term
\begin{equation}
   \delta \mathcal{S}_{\text{EH}} \supset \frac{c^3}{16 \pi G} \int_{\mathcal{M}} d^4x \sqrt{|g|}\mathcal{D}_\lambda v^\lambda = \frac{c^3}{16 \pi G} \int_{\partial \mathcal{M}} d\Sigma_\lambda v^\lambda
\end{equation}
as
\begin{equation}
    \label{eq:boundary_term}
   \frac{c^3}{16 \pi G}  \int_{\partial \mathcal{M}} d\Sigma_\lambda v^\lambda=\frac{c^3}{16 \pi G} \int_{\partial \mathcal{M}} d^3 y \sqrt{|m|} \varepsilon n_\lambda v^\lambda .
\end{equation}
with $m$ the determinant of the projection operator $m_{\alpha\beta}$ given in Equation (\ref{eq:trasversemetric_geometry}). At this point we analyze the product $n_\lambda v^\lambda$:
\begin{equation}
    n_\lambda v^\lambda=n_\lambda(-\mathcal{D}_\nu\delta g^{\lambda \nu}+g_{\alpha\beta}\mathcal{D}^\lambda\delta g^{\alpha\beta})=n_\lambda g_{\alpha\beta}(\mathcal{D}^\lambda \delta g^{\alpha\beta}-\mathcal{D}^\beta \delta g^{\lambda \alpha})=n_\lambda m_{\alpha\beta}(\mathcal{D}^\lambda \delta g^{\alpha\beta}-\mathcal{D}^\beta \delta g^{\lambda \alpha})
\end{equation}
where in the last passage we have substituted the projection operator $m_{\alpha\beta}$ since its second term applied to the parenthesis vanishes. We notice that the term $m_{\alpha\beta}\mathcal{D}^\beta \delta g^{\lambda\alpha}$ vanishes: it is equivalent to calculating the covariant derivative projected onto the hypersurface, which results in a value of 0. This occurs because, in a variational principle, the field remains fixed at the boundary, causing $\delta g^{\lambda\alpha}$ to disappear on $\partial \mathcal{M}$. Therefore, the term in Equation (\ref{eq:boundary_term}) becomes
\begin{equation}
    \label{eq:boundary_term_pre_extr}
     \frac{c^3}{16 \pi G}\int_{\partial \mathcal{M}} d^3 y \sqrt{|m|}\varepsilon (n_\lambda m_{\alpha\beta}\mathcal{D}^\lambda \delta g^{\alpha\beta}).
\end{equation}
The expression within the parentheses represents the variation of the trace of the extrinsic curvature $k$ of the boundary. Indeed, using the expression in Equation (\ref{eq:extrcurv_variation}), the term in Equation (\ref{eq:boundary_term_pre_extr}) can be written as:
\begin{equation}
    \frac{c^3}{16 \pi G} 2\varepsilon \int_{\partial \mathcal{M}}d^3 y \sqrt{|m|}\delta \textit{k}.
\end{equation}
It is evident that to obtain the conventional Einstein equations, we must include a counter term to the original gravitational action, known as the Gibbons-Hawking-York term, and redefine the gravitational action as:
\begin{equation}
    \mathcal{S}_{\text{grav}}=\mathcal{S}_{\text{EH}}+\mathcal{S}_{\text{GHY}}=\frac{c^3}{16 \pi G}\int_{\mathcal{M}} d^4x \sqrt{|g|}R-\frac{2\varepsilon c^3}{16 \pi G}\int_{\partial \mathcal{M}}d^3y \sqrt{|m|}\textit{k}.
\end{equation}
Indeed, when we carry out the variation of this action, the extra term effectively cancels the term involving the variation of the Ricci tensor. It is worth noting that during the variation we should also consider the variation with respect to $\sqrt{|m|}$. However, since the full metric is constrained at the boundary and its variation is zero, the projected metric also exhibits no variation. Consequently, we can obtain Einstein's equations using this approach.

\chapter{Graviton: a spin-2 field}
\label{app:graviton_spin2}

In quantum gravity the graviton is a massless particle with helicity $\pm 2$. To see that, we look for a boson field (in quantum field theory all interactions are mediated by the exchange of bosons) which couples to the energy-momentum tensor. Indeed from General Relativity we have learnt that all forms of energy are sources of gravitation, and since the energy density is $T_{00}$, we look for couplings between the gravitational field and $T_{\mu\nu}$. We work at the linearized level, thus in $T^{\mu\nu}$ we can neglect the gravitational field contributions. Moreover, we want to recover the Newtonian gravity in the non-relativistic limit and so we want the gravitational field to be massless and coupled to a conserved tensor in order to get a long range potential. The simplest possibility is that gravity is mediated by a spin 0 massless boson, described by a scalar field $\phi$. Hence, for non derivative coupling, the interaction term in the Lagrangian would be $g\phi T^\mu{}_\mu$. Even if this possibility leads to the right Newtonian potential in the non relativistic limit, it fails because in this theory photons do not couple to gravity (since $T_{em}{}^\mu{}_\mu=0$), while experimentally the gravitational bending of light rays from massive objects is well established. The next possibility of a spin 1 field $A_\mu$ is also ruled out: for non derivative coupling it is not possible to construct gauge invariant terms, except for $A_\mu j^\mu$ (with $j^\mu$ that in the limit of point-like particle reduces to $m\frac{dx^\mu_0}{d\tau}\delta^{(3)}(\mathbf{x}-\mathbf{x}_0(t))$) but it leads to a repulsive potential between positive masses. Fields with spin $j\geq 3$ fail too: they cannot produce long-range forces. Indeed, long-range force requires a massless field coupled to a conserved tensor but, except for possibly total derivative terms, there is no conserved tensor with three or more indices \cite{1995WeinbergQFT}. It remains only the case of a spin 2 field. We notice that the simplest tensor that contains a spin 2 is the traceless symmetric tensor $S_{\mu\nu}$:
\begin{equation}
    S_{\mu\nu}\in \mathbf{0}\oplus \mathbf{1}\oplus\mathbf{2}
\end{equation}
where $\oplus$ indicates the direct sum of representations and $\mathbf{s}$ is the representation of the rotation group corresponding to a spin $s$. We can use it to describe a massless particle with spin 2: we impose local invariance to eliminate spurious degrees of freedom. Indeed a traceless symmetric tensor has 9 degrees of freedom, while massless particles -independently on their spin- have two degrees of freedom. This is because massless particles are defined as representations of the Poincaré group and of parity: the two values of the helicity are seen as two polarization states of the same particle, since helicity is a pseudoscalar under parity ($h\to -h$). We choose to start from a symmetric but not traceless tensor $h_{\mu\nu}$, which can be decomposed into its trace and traceless symmetric part:
\begin{equation}
    h_{\mu\nu}\in \mathbf{0} \oplus \left( \mathbf{0}\oplus \mathbf{1}\oplus\mathbf{2} \right)
\end{equation}
Then we impose the invariance under:
\begin{equation}
    h_{\mu\nu}(x)\to h_{\mu\nu}(x)-(\partial_\mu \zeta_\nu+\partial_\nu\zeta_\mu).
\end{equation}
which is the gauge transformation in Equation (\ref{eq:gaugetrasf_hmunu}), that is the symmetry of linearized Einstein gravity. It is possible to show \cite{Maggiore:2007ulw} that the gauge invariant action for the free theory, with a chosen overall normalization, is:
\begin{equation}
    S_2=\frac{1}{2}\int d^4x \left[-\partial_\rho h_{\mu\nu}\partial^\rho h^{\mu\nu}+2\partial_\rho h_{\mu\nu}\partial^\nu h^{\mu\rho}-2\partial_\nu h^{\mu\nu}\partial_\mu h+\partial^\mu h \partial_\mu h\right].
\end{equation}
It is called Pauli-Fierz action. We recover Einstein action of linearized theory after a rescaling $h_{\mu\nu}\to (32\pi G)^{-1/2}h_{\mu\nu}$: the linearized Einstein action is the unique action that describes a free massless particle propagating in flat spacetime with helicities $\pm 2$. In order to fine the graviton propagator, we need to add a gauge-fixing term -like what we do in electrodynamics- that is:
\begin{equation}
    S_{gf}=-\int d^4x (\partial^\nu \Bar{h}_{\mu\nu})^2.
\end{equation}
The interaction term reads:
\begin{equation}
    S_{int}=\frac{\kappa}{2}\int d^4 x h_{\mu\nu} T^{\mu\nu}.
\end{equation}
Thus:
\begin{equation}
    S=S_2+S_{gf}+S_{int}=\int d^4x \left[-\frac{1}{2}\partial_\rho h_{\mu\nu}\partial^\rho h^{\mu\nu}+\frac{1}{4}\partial^\mu h \partial_\mu h+\frac{\kappa}{2}h_{\mu\nu}T^{\mu\nu}\right]
\end{equation}
and comparing the relative equation of motion with Equation (\ref{eq:linEinstein_lorentz}) we can fix the coupling constant (in units $c=1$):
\begin{equation}
    k=(32 \pi G)^{1/2}.
\end{equation}
With this theory the Newtonian limit is recovered. However, this is not the full story, since the correct field theory of gravitation must develop a full non-linear structure. Nevertheless, because of the fact that the dimension of the coupling constant is negative in mass units, the theory is not renormalizable. It is possible to show \cite{Maggiore:2007ulw} that for the graviton there are no states with total angular momentum $j=0,1$, thus for gravitational waves there can be no monopole nor dipole radiation.

\end{document}